\documentclass[12pt]{article}
\usepackage{epsf,amsmath}

\def\bm#1{\mbox{\boldmath$#1$\unboldmath}}
\def\pslash{\rlap{\hspace{0.02cm}/}{p}}
\def\kslash{\rlap{\hspace{0.02cm}/}{k}}

\def\half{\textstyle\frac12}
\def\3half{\textstyle\frac32}

\begin{document}

\begin{titlepage}

\begin{flushright}
{\small
CLNS~03/1835\\
PITHA~03/06\\
{\tt hep-ph/0308039}\\[0.2cm]
August 4, 2003}
\end{flushright}

\vspace{0.7cm}
\begin{center}
\Large\bf\boldmath
QCD factorization for $B\to PP$ and $B\to PV$ decays
\unboldmath
\end{center}

\vspace{0.8cm}
\begin{center}
{\sc Martin Beneke$^a$ and Matthias Neubert$^b$}\\
\vspace{0.7cm}
{\sl ${}^a$Institut f\"ur Theoretische Physik E, RWTH Aachen\\
D--52056 Aachen, Germany\\
\vspace{0.3cm}
${}^b$Newman Laboratory for Elementary-Particle Physics, Cornell 
University\\
Ithaca, NY 14853, USA}
\end{center}

\vspace{1.0cm}
\begin{abstract}
\vspace{0.2cm}\noindent
A comprehensive study of exclusive hadronic $B$-meson decays into final 
states containing two pseudoscalar mesons ($PP$) or a pseudoscalar and a 
vector meson ($PV$) is presented. The decay amplitudes are calculated at 
leading power in $\Lambda_{\rm QCD}/m_b$ and at next-to-leading order in 
$\alpha_s$ using the QCD factorization approach. The calculation of the 
relevant hard-scattering kernels is completed. Important classes of power 
corrections, including ``chirally-enhanced'' terms and weak annihilation 
contributions, are estimated and included in the phenomenological analysis. 
Predictions are presented for the branching ratios of the complete set of 
the 96 decays of $B^-$, $\bar B^0$, and $\bar B_s$ mesons into $PP$ and 
$PV$ final states, and for most of the corresponding CP asymmetries. 
Several decays and observables of particular phenomenological interest are 
discussed in detail, including the magnitudes of the penguin amplitudes in 
$PP$ and $PV$ final states, an analysis of the $\pi\rho$ system, and the 
time-dependent CP asymmetry in the $K\phi$ and $K\eta'$ final states. 
\end{abstract}

\vfil
\end{titlepage}

\section{Introduction}

As the $B$ factories \cite{Aubert:2001tu,Mori:2000cg} continue to 
accumulate large data samples, an increasing number of different 
$B$-decay modes becomes accessible to investigation. Many of these modes 
carry interesting information on CP-violating interactions or hadronic 
flavor-changing neutral currents, but except for a number of decay 
channels considered ``theoretically clean'' a theoretical framework is 
required to correct for the effects of the strong interaction.

For a long time, exclusive two-body $B$-decay amplitudes have been 
estimated in the ``naive'' factorization approach (see, e.g., 
\cite{Bauer:1986bm,Neubert:1997uc,Ali:1997nh,Ali:1998eb,Chen:1999nx} and 
references therein) or modifications thereof. In many cases this approach 
provides the correct order of magnitude of branching fractions, but it 
cannot predict direct CP asymmetries due to the assumption of no strong 
rescattering. It is therefore no longer adequate for a detailed 
phenomenological analysis of $B$-factory data. Naive factorization has 
now been superseded by QCD factorization \cite{BBNS1,BBNS2}. Although not 
yet proved rigorously, this scheme provides the means to compute two-body 
decay amplitudes from first principles. Its accuracy is limited only by 
power corrections to the heavy-quark limit and the uncertainties of 
theoretical inputs such as quark masses, form factors, and light-cone 
distribution amplitudes. For the charmless decays considered in this paper 
the lowest-order approximation in QCD factorization coincides with 
``naive'' factorization.

Among the charmless hadronic $B$ decays, the modes $B\to\pi\pi$ and 
$B\to\pi K$ have been studied first and most extensively within QCD 
factorization \cite{BBNS1,BBNS3,Muta:2000ti,Du:2000ff,Du:2001hr}, because 
they are the simplest decays for which a significant interference of 
tree and penguin amplitudes is expected. Other specific final states 
that have been investigated include those with vector mesons and exotic 
mesons \cite{Yang:2000xn,Cheng:2000hv,Cheng:2001aa,diehl}, as well as 
$\eta$ or $\eta^\prime$ along with a pseudoscalar or vector kaon 
\cite{Beneke:2002jn}. In the present paper, we complete the phenomenology 
of hadronic $B$ decays into two light pseudoscalar mesons or one 
pseudoscalar and one vector meson, all from the ground-state nonet. We 
consider the complete set of 96 decay modes, including decays of $B_s$ 
mesons, and decays into mesons with flavor-singlet components. We also 
summarize in a compact notation all the required decay coefficients and 
hard-scattering kernels up to the next-to-leading order in $\alpha_s$. In 
a series of 
recent papers \cite{Du:2002up,Du:2002cf,Sun:2002rn} Du et al.\ have 
considered a subset of these decay modes, including decays of $B_s$ 
mesons. In \cite{Aleksan:2003qi,deGroot:2003hp} fits to the data on 
pseudoscalar--vector meson final states based on the generalization 
of the hard-scattering kernels of \cite{BBNS3} to these final 
states have been performed. While we agree qualitatively with 
the conclusions reached from these fits, we believe that it is 
currently more useful to investigate in more detail the dynamical 
origin of agreements and discrepancies with experimental data. In this 
respect the analysis presented in this paper is more  
complete than the fits performed previously as far as decay channels, 
error estimates, and analysis of observables are concerned.
 
The motivation for our analysis is threefold: First, the larger set of 
decay channels provides additional information on the CKM phase $\gamma$ 
and on hadronic flavor-changing currents, complementary to the $\pi\pi$ 
and $\pi K$ final states. Second, the different strong-interaction 
dynamics underlying specific decay modes provides valuable tests of the 
QCD factorization framework. Being able to compute correctly a large 
number of decays, including ``uninteresting'' ones, increases our 
confidence in the reliability of the approach where it is necessary to 
extract fundamental parameters. An example of this concerns the role of 
so-called scalar penguin operators, which are an important contribution 
to the penguin amplitude in decays to pions and kaons, but which are  
expected to be suppressed for decays into vector mesons 
\cite{Cheng:2000hv,Chen:2001pr}. Finally, our analysis provides 
up-to-date expectations for the branching fractions of decay modes yet to 
be discovered. 

The outline of the paper is as follows: In the rather technical
Section~\ref{sec:amps} we briefly review the QCD factorization approach 
and then collect the notation, definitions, and analytic formulae required 
to compute the complete set of $B\to PP$ and $B\to PV$ decay amplitudes at 
next-to-leading order. This includes a new notation for the 
parameterization of flavor amplitudes, which we find more convenient than 
the conventional notation in terms of parameters $a_i$. 
Sections~\ref{sec:inputs} and \ref{sec:strategy} contain a summary of 
the input parameters entering our analysis and an outline of the 
adopted analysis strategy for the large set of final states. Also included 
in Section~\ref{sec:strategy} is a discussion of the $B^-\to\pi^-\pi^0$ 
tree amplitude.

We then analyze separately penguin-dominated $\Delta S=1$ and $\Delta D=1$ 
decays, and tree-dominated $\Delta D=1$ decays in 
Sections~\ref{sec:penguins} and \ref{sec:trees}, respectively.  
We begin with an investigation of the magnitude of the penguin amplitude 
in $\Delta S=1$ decays, which dominates the overall decay rate. 
The factorization approach makes distinct predictions for this 
amplitude depending on whether the final state is $PP$, $PV$, or $VP$. 
The comparison with existing data provides important information 
on the usefulness of next-to-leading order calculations 
in QCD factorization and simultaneously on the magnitude of penguin 
weak annihilation. We proceed to discuss ratios of branching
fractions, and a few observations that we find difficult to
accommodate in the Standard Model: we quantify the correction to the 
measurement of 
$\sin 2\beta$ in the final states $\phi K_S$ and $\eta^\prime K_S$, and
we define a new ratio involving the $\pi^0\bar K^0$ final state
that may suggest an anomaly in the electroweak penguin sector.
We also consider penguin-dominated $\Delta D=1$ decays, which have
small branching fractions, and discuss empirical constraints on the 
size of weak annihilation contributions. 

Section~\ref{sec:trees} on tree-dominated decays focuses on 
the information that results from the measurement 
of $\pi\rho$ final states. In addition to CP-averaged branching
fractions, direct CP asymmetries, and certain ratios of branching 
fractions, we consider the five asymmetries that can 
be defined in the time-dependent study of the final states 
$\pi^\mp\rho^\pm$. We find that the asymmetry $S$ is well-suited to
constrain $\gamma$. $S$ together with the analogous quantity 
in $B\to \pi^+\pi^-$ decay imply $\gamma=70^\circ$ with an error of 
about $10^\circ$ at $1\sigma$.  In this analysis as in the rest of the
paper we adopt the quasi two-body assumption, which considers the 
vector mesons as stable, neglecting 
interference effects related to the non-negligible widths of 
the decaying vector mesons. 

In Section~\ref{sec:etas} we summarize the results for final states 
containing $\eta$ or $\eta^\prime$, which complete those of 
\cite{Beneke:2002jn}, where an additional $K$ or $K^*$ in the 
final state was assumed. We briefly discuss the decays of 
$B_s$ mesons in Section~\ref{sec:Bs}, and conclude in 
Section~\ref{sec:concl}. 
In three appendices we collect: (i)~the complete set of decay amplitudes 
expressed in terms of the amplitude parameters $\alpha_i$ (formerly 
$a_i$) and $\beta_i$, including weak annihilation; (ii)~the convolutions 
of the new hard-scattering kernels with the distribution amplitudes of 
vector mesons up to the second Gegenbauer moment; and (iii)~the results 
of the BaBar, Belle, and CLEO experiments, from which we computed the 
experimental averages used in the paper.

\subsubsection*{Guide to reading the paper}

Since, in the course of time, this paper has grown to an undue volume, 
we have structured it such that most sections can be read independently 
of each other. Most of the material in Section~2 and Appendices~A and B 
provides an 
up-to-date summary of the technical work necessary to implement QCD 
factorization at next-to-leading order. The reader familiar with the 
basic ideas of the theoretical framework may read only 
Section~\ref{sec:QCDF}. The reader interested only in specific 
phenomenological applications should consult Sections~\ref{sec:inputs} 
and \ref{sec:strategy} (omitting perhaps Section~\ref{sec:treeamplitude})
and the introduction to Section~\ref{sec:penguins}, where we specify 
the theoretical input, the general strategy of the analysis, and the 
definition of specific analysis scenarios. The analysis sections 
\ref{sec:treeamplitude} (tree amplitude in $B\to\pi\pi$ decays), 
\ref{sec:penguins} (penguin-dominated decays), \ref{sec:trees} 
(tree-dominated decays), \ref{sec:etas} (final states with $\eta$ or 
$\eta'$), and \ref{sec:Bs} ($B_s$ decays) could then be read 
independently.

\section{Two-body decay amplitudes}
\label{sec:amps}

In this section we detail the theoretical framework within which our 
results for the decay amplitudes are computed. We begin with a brief 
summary of the QCD factorization method. We then introduce a new notation 
for the basic transition operator ${\cal T}$, which allows us to describe 
pseudoscalar and vector mesons in the final two-body state in terms of a 
single expression. The decay amplitude for $\bar B\to M_1 M_2$ is 
proportional to the matrix element 
$\langle M_1 M_2|{\cal T}|\bar B\rangle$. The transition operator is 
decomposed into a complete basis of ``flavor operators'' accounting for
the different topologies of the various decay mechanisms (tree, penguin, 
annihilation, etc.). After a summary of light-cone distribution 
amplitudes we collect the results for the coefficients of the different 
terms in the transition operator. This discussion is necessarily rather 
technical, as one of our goals is to provide a reference for the decay 
amplitudes and hard-scattering kernels in QCD factorization at 
next-to-leading order. Some readers may wish to read only 
Section~\ref{sec:QCDF} and then continue with Section~\ref{sec:inputs}.

\subsection{The QCD factorization approach}
\label{sec:QCDF}

A detailed discussion of the QCD factorization approach can be found in 
\cite{BBNS1,BBNS2,BBNS3}. Here we recapitulate the basic formulae to set 
up the notation. The effective weak Hamiltonian for charmless hadronic 
$B$ decays consists of a sum of local operators $Q_i$ multiplied by 
short-distance coefficients $C_i$ and products of elements of the quark 
mixing matrix, $\lambda_p^{(D)}=V_{pb} V_{pD}^*$, where $D=d,s$ can be a 
down or strange quark depending on the decay mode under consideration, 
and $p=u,c,t$. Using the unitarity relation 
$\lambda_u^{(D)}+\lambda_c^{(D)}+\lambda_t^{(D)}=0$ we write
\begin{equation}\label{Heff}
   {\cal H}_{\rm eff} = \frac{G_F}{\sqrt2} \sum_{p=u,c} \!
   \lambda_p^{(D)} \bigg( C_1\,Q_1^p + C_2\,Q_2^p
   + \!\sum_{i=3}^{10} C_i\,Q_i + C_{7\gamma}\,Q_{7\gamma}
   + C_{8g}\,Q_{8g} \bigg) + \mbox{h.c.} \,,
\end{equation}
where $Q_{1,2}^p$ are the left-handed current--current operators arising 
from $W$-boson exchange, $Q_{3,\dots, 6}$ and $Q_{7,\dots, 10}$ are QCD 
and electroweak penguin operators, and $Q_{7\gamma}$ and $Q_{8g}$ are 
the electromagnetic and chromomagnetic dipole operators as given in 
\cite{BBNS3}. The effective Hamiltonian describes the quark transitions 
$b\to u\bar u D$, $b\to c\bar c D$, $b\to D\bar q q$ with $q=u,d,s,c,b$, 
and $b\to D g$, $b\to D\gamma$, as appropriate for decay modes with 
interference  of ``tree'' and ``penguin'' contributions. The Wilson 
coefficients are evaluated at next-to-leading order, consistent with the 
calculation of operator matrix elements described below. For the 
coefficients of electroweak penguin operators the evaluation described in 
\cite{BBNS3}, which incorporates some terms normally counted as 
next-to-next-to-leading order, is employed.

\begin{figure}
\epsfxsize=11cm
\centerline{\epsffile{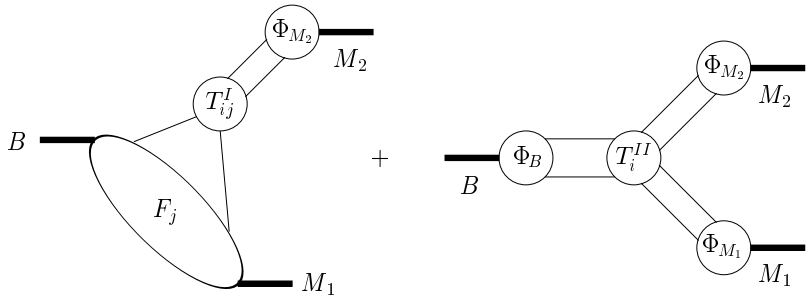}}
\centerline{\parbox{14cm}{\caption{\label{fig1}
Graphical representation of the factorization formula (\ref{fact}). Only 
one of the two form-factor terms is shown for simplicity.}}}
\end{figure}

The QCD factorization formalism allows us to compute systematically the 
matrix elements of the effective weak Hamiltonian in the heavy-quark 
limit for certain two-body final states $M_1' M_2'$. In condensed 
notation, the matrix element of every operator in the effective 
Hamiltonian is evaluated as
\begin{eqnarray}\label{fact}
   \langle M_1' M_2'|Q_i|\bar B\rangle
   &=& \sum_{\{M_1,M_2\}\in \{M_1',M_2'\}}
    F_j^{B\to M_1}\,T_{ij}^I * f_{M_2}\Phi_{M_2} \nonumber\\
   &&\mbox{}\hspace*{1cm}
    + T_i^{II} * f_B\Phi_B * f_{M_1'}\Phi_{M_1'} * f_{M_2'}\Phi_{M_2'}
    \,,
\end{eqnarray}
where $F_j^{B\to M_1}$ is an appropriate form factor, $\Phi_M$ are 
leading-twist light-cone distribution amplitudes, and the star products 
imply an integration over the light-cone momentum fractions of the 
constituent quarks inside the mesons. A graphical representation of this 
result is shown in Figure~\ref{fig1}. Whenever the spectator antiquark 
line goes from the $\bar B$ meson to one of the final-state mesons we 
call this meson $M_1$ and the other one $M_2$. The sum in the first term 
on the right-hand side of (\ref{fact}) accounts for the possibility that 
for particular final states the spectator antiquark can end up in either 
one of the two mesons. If the spectator antiquark is annihilated we use 
the convention that $M_1$ is the meson that carries away the antiquark 
from the weak decay vertex.

A justification of the factorization formula for final states in which 
$M_1$ is a heavy meson (applicable to $D$ mesons in the heavy-quark 
limit) can be found in \cite{BBNS2,Bauer:2001cu}. In this case the term
in the second line of (\ref{fact}) is absent. For final states with two 
light mesons factorization has been proved at order $\alpha_s$ 
\cite{BBNS1}, but a complete proof has not yet been given (for recent
developments in this direction, see \cite{Chay:2003ju}). The 
factorization formula reduces the complicated hadronic matrix elements of 
four-quark operators to simpler non-perturbative quantities and calculable 
hard-scattering kernels $T_{ij}^I$ and $T_i^{II}$. In this paper we 
complete the calculation of all relevant kernels at next-to-leading order 
in $\alpha_s$.
 
The light-cone expansion implies that only leading-twist distribution 
amplitudes are needed in the heavy-quark limit. There exist however a 
number of subleading quark--antiquark distribution amplitudes of 
twist~3, which have large normalization factors for pseudoscalar mesons, 
e.g.\ for the pion
\begin{equation}\label{rpidef}
   r_\chi^\pi(\mu) = \frac{2 m_\pi^2}{m_b(\mu)\,(m_u+m_d)(\mu)} 
   \sim \frac{\Lambda_{\rm QCD}}{m_b} \,.
\end{equation}
For realistic $b$-quark masses these ``chirally-enhanced'' terms are not 
much suppressed 
numerically. We therefore include in our analysis all quark--antiquark 
twist-3 amplitudes. (The quark--antiquark--gluon amplitude at twist-3 
does not have an anomalously large normalization.) In order to perform 
the same analysis for all final states we also include the 
quark--antiquark twist-3 amplitudes for vector mesons, even though there 
is no particular enhancement in this case, $r_\chi$ being replaced by 
$2m_V/m_b$ times a ratio of two decay constants (see below), with $m_V$ 
the vector-meson mass. 

The inclusion of chirally-enhanced terms is important to account for 
the large branching fractions of penguin-dominated decay modes with 
pseudoscalar final-state mesons, such as $B\to\pi K$ \cite{BBNS3}, but 
it also causes a number of conceptual problems. Factorization is not 
expected to hold at subleading order in $\Lambda_{\rm QCD}/m_b$ and, 
somewhat unfortunately, is indeed violated by some of the 
chirally-enhanced terms \cite{BBNS1}. In contrast to the leading-twist 
distribution amplitudes, the twist-3 two-particle amplitudes do not 
vanish at the endpoints but rather approach constants. The kernels 
$T_{ij}^I$ in the first term of the factorization formula also approach 
constants at the endpoints (modulo logarithms), and hence there is no 
difficulty with this term. These kernels include the 
important scalar penguin amplitude mentioned in the introduction, 
conventionally denoted by $a_6$. However, the second term in the 
factorization formula, which accounts for the interactions with the 
spectator quark, contains integrals that are dominated by the endpoint 
regions if the distribution amplitudes do not vanish at the endpoint. 
These integrals formally diverge logarithmically in a perturbative 
framework. This implies a non-factorizable soft interaction with the 
spectator quark, while $M_1$ is formed in a highly asymmetric 
configuration, in which one quark carries almost all the momentum of the 
meson. Similar factorization-breaking effects occur in weak annihilation 
contributions, which are also suppressed by at least one power of 
$\Lambda_{\rm QCD}/m_b$ in the heavy-quark expansion.

In \cite{BBNS3} we have parameterized the power corrections from 
chirally-enhanced and weak annihilation terms in an {\em ad hoc\/} way 
and included a variation of the corresponding parameters in our error 
estimate. Based on the analysis of $\pi\pi$ and $\pi K$ final states we 
concluded that these factorization-breaking terms introduce a sometimes 
substantial uncertainty into the theory but do not render the framework 
unpredictive. In particular, the data so far give no indication that 
these corrections should be outside the range specified by our error 
estimate. Given this situation, we follow our previous analysis strategy 
as regards twist-3 and annihilation effects. The inclusion of twist-3 
effects for vector mesons then allows us to estimate the impact of some 
power-suppressed effects, which turn out to be small. But we should stress 
that this is far from a complete account of such contributions. We also 
note that some other classes of power corrections have been analyzed using 
the method of light-cone QCD sum rules 
\cite{Khodjamirian:2000mi,Khodjamirian:2003eq}. 

The pseudoscalar mesons $\eta$ and $\eta^\prime$, which contain a
flavor-singlet component in their wave function, require special 
attention, because they can be formed from two gluons at leading twist. 
The new effects associated with this possibility have been worked out in 
the context of QCD factorization in \cite{Beneke:2002jn}. For these 
mesons the factorization formula (\ref{fact}) holds provided it is 
extended by an additional non-local form-factor term. At the order we are 
working the corresponding complications do not appear for vector mesons, 
since they do not have a two-gluon component. There is, however, one 
novel effect for neutral vector mesons, namely that they can be produced 
via a virtual photon. Although this effect is very small, we shall 
discuss it, because it is of the same order as other electroweak 
contributions to the decay amplitudes.

\subsection{Flavor operators}
\label{sec:decayop}

In the  literature on QCD factorization the result of computing the 
hard-scattering kernels for the various operators in the effective weak 
Hamiltonian is usually presented in terms of ``factorized operators'' 
with coefficients $a_i(M_1 M_2)$. (The matrix element of a factorized 
operator is simply proportional to a form factor times a decay constant.) 
The reasons for this are largely historical. Besides containing some 
redundancy, such a notation leads to inconveniences for vector mesons
since the factorized operator 
$\sum_q (\bar q b)_{S-P}\otimes (\bar D q)_{S+P}$ vanishes, while there 
exist non-vanishing terms at order $\alpha_s$ that one would naturally 
associate with the corresponding coefficient $a_6$. Here we introduce a 
new notation that eliminates these inconveniences, and which at the same 
time is sufficiently general and explicit to facilitate the construction 
of the decay amplitude for any particular decay with either pseudoscalar 
or vector mesons in the final state. The main point of the new notation 
is to keep track of only the flavor structure of an operator. Its Dirac 
structure, which distinguishes some of the $a_i(M_1 M_2)$ coefficients in 
the conventional notation, becomes irrelevant, since operators with 
different Dirac structure (but identical flavor structure) always 
contribute in the same combination to a particular set of final states.

We first introduce our phase convention for the flavor wave functions. We 
take the light quark triplet as $(u,d,s)$ and the antitriplet as 
$(\bar u,\bar d,\bar s)$. Consequently,
\begin{equation}\label{Bconv}
   B^-\sim \bar u b \,, \qquad 
   \bar B_d \sim \bar d b \,, \qquad
   \bar B_s\sim \bar s b \,.
\end{equation}
Assuming isospin symmetry, the non-singlet members of the light nonet of 
pseudoscalar mesons are given by
\begin{equation}\label{flavorwf}
\begin{aligned}
   \pi^0\sim \frac{1}{\sqrt2}\,(\bar u u-\bar d d), \qquad
   &\pi^-\sim \bar u d \,, \qquad \pi^+\sim \bar d u \,, \\
   \bar K^0\sim \bar d s \,, \qquad K^0\sim \bar s d \,, \qquad 
   &K^-\sim \bar u s \,, \qquad K^+\sim \bar s u \,,
\end{aligned}
\end{equation}
and analogous expressions hold for the corresponding vector mesons. The 
treatment of mesons containing a flavor-singlet component in their wave 
function has been explained in detail in the dedicated paper 
\cite{Beneke:2002jn}, to which we refer the reader for all details. The 
meson $\eta$ (and similarly $\eta^\prime$ and the vector mesons $\omega$ 
and $\phi$) can be treated as a coherent superposition of the flavor 
components 
\begin{equation}\label{etaqs}
   \eta_q\sim \frac{1}{\sqrt2}\,(\bar u u+\bar d d) \,, \qquad
   \eta_s\sim \bar s s \,.
\end{equation}
This should not be confused with the representation of the meson state as 
a superposition of flavor states, but simply means 
that these mesons have matrix elements of $\bar s s$ and 
$\bar u u$ (or $\bar d d$, these two being equal due to the assumed 
isospin symmetry) operators that are a priori unrelated. 
Consequently, each meson 
is described by two decay constants; for instance, for
the $\eta$ meson 
\begin{equation}
\label{singletdecayconstants}
\langle \eta(q)|\bar u\gamma_\mu\gamma_5 u|0\rangle = -\frac{i}
{\sqrt{2}} \,f_\eta^q q_\mu,
\qquad
\langle \eta(q)|\bar s\gamma_\mu \gamma_5 s|0\rangle = -i f_\eta^s 
q_\mu.
\end{equation}
Similarly, there are two transition form factors, two leading-twist 
quark--antiquark distribution amplitudes etc., corresponding to the two 
distinct flavor components. In addition, for the pseudoscalar case there 
exist contributions proportional to the two-gluon content of $\eta$ and 
$\eta'$. Finally, certain contributions to the $b\to D gg$ amplitude are 
conveniently interpreted in terms of $\bar c c$ flavor components  
$\eta_c$, $\eta_c^\prime$ for 
$\eta$ and $\eta'$. The QCD factorization approach presented 
here is sufficiently general to allow for a discussion of states containing 
flavor-singlet contributions without assumption of a particular 
flavor-mixing scheme. However, in 
our phenomenological analysis we will adopt the 
Feldmann--Kroll--Stech scheme for $\eta$ and $\eta'$ mixing, in which it is 
sufficient to introduce a single mixing angle in the flavor basis 
\cite{Feldmann:1998vh}. The corresponding expressions for the hadronic 
matrix elements (decay constants, form factors, etc.) needed in the 
calculations of the decay amplitudes can be found in 
\cite{Beneke:2002jn}. For the vector mesons $\omega$ and $\phi$ we will
assume ideal mixing, so that 
$\omega=\omega_q\sim(\bar u u+\bar d d)/\sqrt2$ and 
$\phi=\phi_s\sim\bar s s$. 

We match the effective weak Hamiltonian onto a transition operator such 
that its matrix element is given by 
\begin{equation}\label{Top}
   \langle M_1'M_2'|{\cal H}_{\rm eff}|\bar B\rangle
   = \sum_{p=u,c} \lambda_p^{(D)}\,
   \langle M_1' M_2'|{\cal T}_A^p + {\cal T}_B^p|\bar B\rangle \,.
\end{equation}
The two terms account for the flavor topologies of the form-factor and 
hard-scattering terms in (\ref{fact}), respectively, corresponding to the 
two diagrams of Figure~\ref{fig1}. Each term in the transition operator 
contains several operators labeled only by the flavor composition of the 
four-quark final state. Note that below we will include in the definition 
of ${\cal T}_B^p$ also some power-suppressed terms, such as weak 
annihilation contributions. 

Including electroweak penguin topologies there are six different flavor 
structures one can write down for the left diagram of Figure~\ref{fig1}. 
We define (with $D=d$ or $s$)
\begin{eqnarray}\label{alphaidef}
   {\cal T}_A^p
   &=& \delta_{pu}\,\alpha_1(M_1 M_2)\,A([\bar q_s u][\bar u D])
    + \delta_{pu}\,\alpha_2(M_1 M_2)\,A([\bar q_s D][\bar u u])
    \nonumber\\[0.2cm]
   &&\mbox{}+ \alpha_3^p(M_1 M_2)\,\sum_q A([\bar q_s D][\bar q q])
    + \alpha_4^p(M_1 M_2)\,\sum_q A([\bar q_s q][\bar q D])
    \nonumber\\[-0.3cm]
   &&\mbox{}+ \alpha_{3,\rm EW}^p(M_1 M_2)\,\sum_q\frac32\,e_q\,
    A([\bar q_s D][\bar q q])
    + \alpha_{4,\rm EW}^p(M_1 M_2)\,\sum_q\frac32\,e_q\,
    A([\bar q_s q][\bar q D]) \,, \qquad
\end{eqnarray}
where the sums extend over $q=u,d,s$, and $\bar q_s$ denotes the 
spectator antiquark. The operators 
$A([\bar q_{M_1} q_{M_1}][\bar q_{M_2} q_{M_2}])$ also contain an 
implicit sum over $q_s=u,d,s$ to cover all possible $B$-meson initial 
states. The coefficients $\alpha_i^p(M_1 M_2)$ 
contain all dynamical information, while the arguments of $A$ encode 
the flavor composition of the final state and hence determine the final
state to which a given term can contribute. We define
\begin{equation}\label{Adef}
   \langle M_1' M_2'|\alpha_i^p(M_1 M_2)\,A([\ldots][\ldots])
   |\bar B_{q_s}\rangle\equiv c\,\alpha_i^p(M_1' M_2')\,A_{M_1' M_2'} 
\end{equation}
whenever the quark flavors of the first (second) square bracket match 
those of $M_1'$ ($M_2'$). The constant $c$ is a product of three factors 
of $1$, $\pm 1/\sqrt{2}$ etc.\ from the flavor composition of the 
$\bar B$ meson and $M_{1,2}$ as specified above. The quantity 
$A_{M_1 M_2}$ is given by
\begin{equation}\label{am1m2}
   A_{M_1 M_2} = i\,\frac{G_F}{\sqrt2}\,\left\{
   \begin{array}{ll}
    m_B^2 F_0^{B\to M_1}(0) f_{M_2} \,; & \quad \mbox{if~} M_1=M_2=P, \\
    - 2m_V\,\epsilon_{M_1}^*\!\cdot p_B\,A_0^{B\to M_1}(0) f_{M_2} \,; & 
     \quad \mbox{if~} M_1=V,\, M_2=P,\\
    - 2m_V\,\epsilon_{M_2}^*\!\cdot p_B\,F_+^{B\to M_1}(0) f_{M_2} \,; & 
     \quad \mbox{if~} M_1=P,\, M_2=V.
   \end{array}\right.
\end{equation}
Here $F_{+,\,0}$ and $A_0$ denote pseudoscalar ($P$) and vector ($V$) meson 
form factors in the standard convention (see, e.g., \cite{Bauer:1986bm}).
The decay constants $f_{M_2}$ are normalized according to 
\begin{equation}\label{decayconstants}
   \langle\pi^-(q)|\bar d\gamma_\mu\gamma_5 u|0\rangle
    = -i f_\pi q_\mu \,, \qquad
   \langle\rho^-(q)|\bar d\gamma_\mu u|0\rangle = -i f_\rho m_\rho 
    \epsilon^*_\mu \,.
\end{equation}
For a vector meson, $f_V\equiv f_V^\parallel$ always refers to the decay 
constant of a longitudinally polarized meson. We neglect corrections to 
the decay amplitudes quadratic in the light meson masses, so that all form 
factors are evaluated at $q^2=0$. (At this kinematic point, the form 
factors $F_+$ and $F_0$ coincide.) Parameters referring to the $B$ meson 
depend on whether the decaying meson is $B_s$ or $B_{u,d}$. This will be 
implicitly understood in the following. The above expression can be 
simplified by replacing $2m_V\,\epsilon^*\!\cdot p_B \to m_B^2$, since 
the left-hand side squared and summed over the polarizations of the 
vector meson gives $m_B^4$ (neglecting again quadratic meson-mass 
corrections). As expected by angular momentum conservation the 
final-state vector meson is longitudinally polarized. When the 
replacement above is done the polarization sum for vector mesons must be 
omitted, and the decay rate is simply given by
\begin{equation}
   \Gamma = \frac{S}{16\pi m_B}\,
   \left|\langle M_1'M_2'|{\cal H}_{\rm eff}|\bar B\rangle\right|^2 ,
\end{equation}
where $S=1/2$ if $M_1'$ and $M_2'$ are identical, and $S=1$ otherwise. 

We do not discuss in this paper final states containing two vector 
mesons. The extension to this case is straightforward, since the leading 
amplitude in the heavy-quark limit is the one for two longitudinally 
polarized vector mesons, which in many ways behave like pseudoscalar 
mesons. 

In order to exemplify the notation consider 
the decay $\bar B_d\to \pi^0\rho^0$, for which $q_s=d$ and $D=d$. The 
spectator quark can go to either one of the two mesons, so $M_1$ can be 
$\pi^0$ or $\rho^0$. Hence, e.g.\
\begin{equation}
   \langle\pi^0\rho^0|\alpha_4^p(M_1 M_2) \sum_q A([\bar d q][\bar q d])
   |\bar B_d\rangle = - \frac12 \left[ 
   \alpha_4^p(\pi^0\rho^0)\,A_{\pi^0\rho^0}
   + \alpha_4^p(\rho^0\pi^0)\,A_{\rho^0\pi^0} \right] .
\end{equation}
The order of the arguments of $\alpha_4^p$ is relevant as will be seen 
from the explicit expressions given in Section~\ref{subsec:coefs} below. 
The factor $c=1/2$ arises from the flavor wave functions of the mesons. 
On the other hand, 
\begin{equation}
   \langle\pi^0\rho^0|\alpha_3^p(M_1 M_2) \sum_q A([\bar d d][\bar q q])
   |\bar B_d\rangle = 0 \,,
\end{equation}
since $q=u,d$ contribute equally but with opposite sign for the mesons 
$\pi^0$ and $\rho^0$.

We now discuss the flavor structure of the hard-scattering term in 
(\ref{fact}), i.e., the second diagram in Figure~\ref{fig1}. Since all 
six quarks participate in the hard scattering, the generic flavor 
operator in ${\cal T}_B^p$ is of the form 
$B([\bar q_{M_1} q_{M_1}][\bar q_{M_2} q_{M_2}][\bar q_s b])$ with 
possible sums over quarks from penguin transitions or flavor-singlet 
conversion $g\to \bar q q$. We define the matrix element of a 
$B$-operator as 
\begin{equation}\label{Bdef}
   \langle M_1 M_2|B([\ldots][\ldots]][\ldots])|\bar B_q\rangle
    \equiv c \,B_{M_1 M_2} \,, \quad \mbox{with} \quad
   B_{M_1 M_2} = \pm i\,\frac{G_F}{\sqrt{2}}\,f_{B_q} f_{M_1} f_{M_2} \,, 
\end{equation}
whenever the quark flavors of the three brackets match those of $M_1$, 
$M_2$, and $\bar B_q$. The constant $c$ is the same as in (\ref{Adef}). 
The upper sign in the definition of $B_{M_1 M_2}$ applies when both 
mesons are pseudoscalar, and the lower when one of the mesons is a vector 
meson. This matches the sign conventions for the quantities $A_{M_1 M_2}$ 
in (\ref{am1m2}). The most important case of spectator scattering is when 
the spectator-quark line goes from the $B$ meson to a final-state meson, 
which we then call $M_1$. These are the hard spectator interactions, 
which are the only terms of leading power in the heavy-quark limit. (We 
refer to all other contributions to ${\cal T}_B^p$ as annihilation.) This 
special case implies $\bar q_s=\bar q_{M_1}$, leaving six different 
amplitudes that are in one-to-one correspondence with the six 
$A$-operators defined above, because
\begin{equation}
   B([\bar q_s q_{M_1}][\bar q_{M_2} q_{M_2}][\bar q_s b])
   = \frac{B_{M_1 M_2}}{A_{M_1 M_2}}\,
   A([\bar q_s q_{M_1}][\bar q_{M_2} q_{M_2}]) \,.
\end{equation}
We therefore absorb the spectator scattering contributions into the
definition of the coefficients $\alpha_i^p(M_1 M_2)$. 

With parts of ${\cal T}_B^p$ absorbed into ${\cal T}_A^p$, the transition 
operator ${\cal T}_A^p$ now contains all scattering mechanisms except 
weak annihilation. In particular, it contains all contributions of 
leading power in the heavy-quark limit. The remaining, power-suppressed 
annihilation part of ${\cal T}_B^p$ is parameterized in its most general 
form as
\begin{eqnarray}\label{bis}
   {\cal T}_B^p 
   &=& \delta_{pu}\,b_1(M_1 M_2)\,\sum_{q'}
    B([\bar u q'][\bar q' u][\bar D b])
    + \delta_{pu}\,b_2(M_1 M_2)\,\sum_{q'}
    B([\bar u q'][\bar q' D][\bar u b]) \nonumber\\
   &+& b_3^p(M_1 M_2)\,\sum_{q,q'} B([\bar q q'][\bar q' D][\bar q b])
    + b_4^p(M_1 M_2)\,\sum_{q,q'} B([\bar q q'][\bar q' q][\bar D b])
    \nonumber\\[-0.1cm]
   &+& b_{3,\rm EW}^p(M_1 M_2)\,\sum_{q,q'} \frac{3}{2}\,e_q\,
    B([\bar q q'][\bar q' D][\bar q b])
    + b_{4,\rm EW}^p(M_1 M_2)\,\sum_{q,q'} \frac{3}{2}\,e_q\,
    B([\bar q q'][\bar q' q][\bar D b]) \nonumber\\
   &+& \delta_{pu}\,b_{S1}(M_1 M_2)\,\sum_{q'}
    B([\bar u u][\bar q'\!q'][\bar D b])
    + \delta_{pu}\,b_{S2}(M_1 M_2)\,\sum_{q'}
    B([\bar u D][\bar q'\!q'][\bar u b]) \nonumber\\
   &+& b_{S3}^p(M_1 M_2)\,\sum_{q,q'}
    B([\bar q D][\bar q'\!q'][\bar q b])
    + b_{S4}^p(M_1 M_2)\,\sum_{q,q'}
    B([\bar q q][\bar q'\!q'][\bar D b]) \\[-0.1cm]
   &+& b_{S3,\rm EW}^p(M_1 M_2)\,\sum_{q,q'} \frac{3}{2}\,e_q\,
    B([\bar q D][\bar q'\!q'][\bar q b])
    + b_{S4,\rm EW}^p(M_1 M_2)\,\sum_{q,q'} \frac{3}{2}\,e_q\,
    B([\bar q q][\bar q'\!q'][\bar D b]) \,, \nonumber
\end{eqnarray}
where the sums extend over $q,q'=u,d,s$. The sum over $q'$ arises because 
a quark--antiquark pair must be created by $g\to\bar q' q'$ after the 
spectator quark is annihilated. The definitions of the first six 
coefficients coincide with the corresponding definitions in \cite{BBNS3} 
for $\pi\pi$ and $\pi K$ final states. Note that the operator 
$\sum_{q,q'} B([\bar q q'][\bar q' D][\bar q b])$ is redundant, because 
it is equivalent to $\sum_q A([\bar q_s q][\bar q D])$ (which includes 
an implicit sum over $q_s$). We allow this redundancy to keep the 
parameterization of annihilation effects separate from the others. The 
six new coefficients with subscript `$S$' contribute only to final states 
containing flavor-singlet mesons or neutral vector mesons and were not 
needed in our previous analysis. It will be convenient to use the 
notation
\begin{equation}
   \beta_i^p(M_1 M_2)\equiv \frac{B_{M_1 M_2}}{A_{M_1 M_2}}\,
   b_i^p(M_1 M_2) 
\end{equation}
whenever $A_{M_1 M_2}$ does not vanish. However, for some pure 
annihilation decays such as $\bar B_s \to\pi^+\pi^-$, where $\beta_i^p$ 
is not defined since $F_0^{B_s\to\pi}(0)=0$, we express the decay 
amplitude in terms of the coefficients $b_i^p$. The redundancy mentioned 
above implies that $\alpha_4^p(M_1 M_2)$ and $\beta_3^p(M_1 M_2)$ always 
appear in the combination
\begin{equation}
   \hat\alpha_4^p(M_1 M_2)\equiv 
   \alpha_4^p(M_1 M_2) + \beta_3^p(M_1 M_2) \,.
\end{equation}

It is now straightforward to express the $\bar B_q\to M_1' M_2'$ decay 
amplitude in terms of linear combinations of the coefficients 
$\alpha_i^p$ and $\beta_i^p$. The results are collected in Appendix~A,  
where we also give a convenient 
master formula suitable for implementation 
in a computer program, which generates the entire set of 96 decay 
amplitudes by evaluating a single expression. 
It remains to express the flavor coefficients in 
terms of the hard-scattering kernels of the QCD factorization approach.

\subsection{Distribution amplitudes}

We now summarize the definitions of the light-cone distribution 
amplitudes for light pseudoscalar and vector mesons. The corresponding 
amplitudes for $B$ mesons have been discussed in 
\cite{Grozin:1996pq,Beneke:2000wa}. While our treatment of the 
leading-twist distributions is completely general, at the level of 
twist-3 power corrections we work in the approximation of neglecting 
the $q\bar q g$ Fock state of the meson. The motivation for this 
approximation is that it retains all effects with large 
(chirally-enhanced) normalization factors but neglects ``ordinary'' power 
corrections of order $\Lambda_{\rm QCD}/m_b$. The following discussion 
relies on the analysis of distribution amplitudes in coordinate space 
presented in \cite{Braun:1989iv, Ball:1998sk}, while the momentum-space 
projectors are taken from \cite{Beneke:2000wa}. A more detailed 
discussion of the pseudoscalar case can be found in 
\cite{BBNS3,Beneke:2002bs}.

We recall that in general the collinear approximation for the parton 
momenta can be taken only after the light-cone projection has been 
applied. We therefore assign momenta 
\begin{equation}\label{momenta2}
   k_1^\mu = x p^\mu + k_\perp^\mu
    + \frac{\vec k_\perp^2}{2x\,p\cdot\bar p}\,\bar p^\mu \,, \qquad
   k_2^\mu = \bar x p^\mu - k_\perp^\mu
    + \frac{\vec k_\perp^2}{2\bar x\,p\cdot\bar p}\,\bar p^\mu
\end{equation}
to the quark and antiquark in a light meson with momentum $p$, where 
$\bar p$ is a light-like vector whose 3-components point into the 
opposite direction of $\vec p$, and $\bar x\equiv 1-x$. The light-cone 
projection operator of a light pseudoscalar meson in momentum space, 
including twist-3 two-particle contributions, then reads
\begin{equation}\label{pimeson2}
   M_{\alpha\beta}^P = \frac{i f_P}{4} \Bigg\{
   \pslash\,\gamma_5\,\Phi_P(x) - \mu_P\gamma_5 \left(
   \Phi_p(x) - i\sigma_{\mu\nu}\,\frac{p^\mu\,\bar p^\nu}{p\cdot\bar p}\,
   \frac{\Phi'_\sigma(x)}{6}
   + i\sigma_{\mu\nu}\,p^\mu\,\frac{\Phi_\sigma(x)}{6}\,
   \frac{\partial}{\partial k_{\perp\nu}} \right)\!\Bigg\}_{\alpha\beta}
   ,
\end{equation}
where $\mu_P$ is defined as $m_b \,r_\chi^P/2$ with $r_\chi^P$ defined
as in (\ref{rpidef}). The convention for the 
projection is that one computes $\mbox{tr}\,(M^P A)$ if $\bar u A v$ is 
the scattering amplitude with an on-shell quark and antiquark. 
The derivative acts on 
the scattering amplitude $A$, and it is understood that, 
after the derivative is taken, the momenta $k_1$ and $k_2$ are set equal 
to $x p$ and $\bar x p$, respectively. The overall sign of 
(\ref{pimeson2}) corresponds to defining the projector (in coordinate space) 
through the matrix element 
$\langle P(p)|\bar q_\beta(z) q_\alpha(0)|0\rangle$ rather than the 
opposite ordering of the quark fields. (For simplicity, we suppress the 
gauge string connecting the two fermion fields.) The meson $\eta$ 
(and similarly $\eta'$, $\omega$, and $\phi$) is described by two 
quark--antiquark distribution amplitudes corresponding to the two 
independent flavor components $\eta_q$ and $\eta_s$ in (\ref{etaqs}). 

The leading-twist distribution amplitude is conventionally expanded in 
Gegenbauer polynomials, 
\begin{equation}\label{gegenbauer}
   \Phi_P(x,\mu) = 6x(1-x)\,\bigg[ 1 + \sum_{n=1}^\infty 
   \alpha_n^P(\mu)\,C_n^{(3/2)}(2x-1) \bigg] , 
\end{equation}
since the Gegenbauer moments $\alpha_n^P(\mu)$ are multiplicatively 
renormalized. When three-particle contributions are neglected, the 
twist-3 two-particle distribution amplitudes are determined completely by 
the equations of motion, which then require
\begin{equation}
   \Phi_p(x) = 1 \,, \qquad
   \frac{\Phi^\prime_\sigma(x)}{6} = \bar x-x \,, \qquad
   \frac{\Phi_\sigma(x)}{6} = x\bar x \,.
\end{equation}
It is not difficult to derive from this the simpler projector 
\cite{Geshkenbein:qn}
\begin{equation}\label{pimeson3}
   M_{\alpha\beta}^P = \frac{i f_P}{4} \left(
   \pslash\,\gamma_5\,\Phi_P(x) - \mu_P\gamma_5\, 
   \frac{\kslash_2\,\kslash_1}{k_2\cdot k_1}\,\Phi_p(x)
   \right)_{\alpha\beta} .
\end{equation}
The $\eta$ and $\eta'$ mesons also have a two-particle two-gluon 
distribution amplitude at leading twist, for which we adopt the 
convention given in \cite{Beneke:2002jn}. 

The corresponding equations for vector mesons are very similar. In 
general, the light-cone projection operator in momentum space contains 
two terms, $M^V=M^V_\parallel+M^V_\perp$. However, the transverse 
projector does not contribute to the $B\to PV$ decay amplitudes at 
leading and first subleading power in $\Lambda_{\rm QCD}/m_b$. The 
longitudinal projector is given by
\begin{eqnarray}\label{provect}
   \left( M^V_\parallel \right)_{\alpha\beta}
   &=& -\frac{i f_V}{4} \Bigg\{
    \pslash\,\Phi_V(x) + \frac{m_V f_V^\perp}{f_V} 
    \Bigg(\frac{h_\parallel'{}^{(s)}(x)}{2}
    - i\sigma_{\mu\nu}\,\frac{p^\mu\,\bar p^\nu}{p\cdot\bar p}\,
    h_\parallel^{(t)}(x) \nonumber\\
   &&\hspace{1.2cm}\mbox{}+ i\sigma_{\mu\nu}\,p^\mu\,
    \int_0^x\!dv\,\Big[ h_\parallel^{(t)}(v) - \Phi_\perp(v) \Big]
    \frac{\partial}{\partial k_{\perp\nu}} \Bigg)
    \Bigg\}_{\alpha\beta} .
\end{eqnarray}
The polarization vector has been replaced by 
$\epsilon^*_\mu\to p_\mu/m_V$, which is correct up to corrections 
quadratic in the meson mass. After this replacement no polarization sum 
must be taken after squaring the decay amplitude. The leading-twist 
distribution amplitude $\Phi_V(x)$ 
is expanded in Gegenbauer polynomials exactly as 
in (\ref{gegenbauer}), but with Gegenbauer moments $\alpha_n^V(\mu)$. When 
three-particle amplitudes are neglected, the twist-3 amplitudes that 
appear in (\ref{provect}) can all be expressed in terms of the 
leading-twist amplitude $\Phi_\perp(x)$ of a {\em transversely\/} 
polarized vector meson \cite{Ball:1998sk}. We define 
\begin{equation}\label{phivGegenb}
   \Phi_v(x) \equiv \int_0^x\!dv\,\frac{\Phi_\perp(v)}{\bar v}
   - \int_x^1\!dv\,\frac{\Phi_\perp(v)}{v}
   = 3\sum_{n=0}^\infty \alpha_{n,\perp}^V(\mu)\,P_{n+1}(2x-1) \,,
\end{equation}
where $\alpha_{0,\perp}^V=1$, and $P_n(x)$ are the Legendre polynomials.
The second equation is obtained by inserting the Gegenbauer expansion 
of $\Phi_\perp(x)$. The equations of motion then lead to the 
Wandzura--Wilczek relations \cite{Ball:1998sk}
\begin{equation}
\begin{aligned}
   \frac{h_\parallel'{}^{(s)}(x)}{2}
   &= - \Phi_v(x) \,, \qquad
    h_\parallel^{(t)}(x) = -(\bar x-x)\,\Phi_v(x) \,, \\
   &\int_0^x\!dv\,\Big[ h_\parallel^{(t)}(v) - \Phi_\perp(v) \Big]
    = -x\bar x \,\Phi_v(x) \,,
\end{aligned}
\end{equation}
which allow us to express the twist-3 two-particle projection in terms of 
the single function $\Phi_v(x)$. Comparing the vector meson projection in 
(\ref{provect}) with the pseudoscalar projection in (\ref{pimeson2}) we 
see that the function $\Phi_v(x)$ is analogous to $\Phi_p(x)$. In analogy 
with the pseudoscalar case we obtain the simpler projector 
\begin{equation}\label{provect2}
   \left( M^V_\parallel \right)_{\alpha\beta}
   = -\frac{i f_V}{4} \left( \pslash\,\Phi_V(x)
   - \frac{m_V f_V^\perp}{f_V}\, 
   \frac{\kslash_2\,\kslash_1}{k_2\cdot k_1}\,\Phi_v(x)
   \right)_{\alpha\beta} ,
\end{equation}
valid in the approximation where one neglects three-particle 
contributions. Like $\Phi_p(x)$, the function $\Phi_v(x)$ does not 
vanish at the endpoints $x=0,1$. We note, however, the vanishing of the 
integral
\begin{equation}\label{zero}
   \int_0^1\!dx\,\Phi_v(x) = 0 \,.
\end{equation}

With these remarks, we can obtain the hard-scattering kernels for vector 
mesons in the final state from those for two pseudoscalars given in 
\cite{BBNS3} by performing the replacements $\Phi_P(x)\to\Phi_V(x)$, 
$\Phi_p(x)\to\Phi_v(x)$, $f_P\to f_V$, and by interpreting $r_\chi$ as 
in (\ref{rchi}), (\ref{rchiV}) below. However, one must pay attention to 
certain 
sign changes arising from the absence of $\gamma_5$ in the vector-meson 
projector (\ref{provect2}) and in the matrix elements that define 
vector-meson form factors. This leads to the sign alternations in the 
definitions of $\alpha_3^p$, $\alpha_4^p$, $\alpha_{3,\rm EW}^p$, and 
$\alpha_{4,\rm EW}^p$ in (\ref{ais}) below, and to sign alternations in 
the weak annihilation terms to be discussed later.

\subsection{Coefficients of the decay operators in QCD factorization}
\label{subsec:coefs}

In the following we give the expressions for the coefficients of the 
decay operator up to the next-to-leading order. First, some comments 
on the treatment of electromagnetic corrections and electroweak 
penguin effects, and on the treatment of weak annihilation, are 
in order. 

\subsubsection*{\it Electroweak penguin effects and electromagnetic 
corrections}

\begin{itemize}
\item[1)] 
We neglect electromagnetic corrections to the QCD coefficients 
$\alpha_1$, $\alpha_2$, $\alpha^p_{3}$, and $\alpha^p_{4}$, since these 
are much smaller than the next-to-leading order QCD corrections. We also 
neglect corrections of order $\alpha_s C_{7-10}$ to these coefficients, 
since the Wilson coefficients $C_{7-10}$ are proportional to 
the electromagnetic coupling $\alpha$.
\item[2)] 
In leading order the electroweak penguin coefficients 
$\alpha^p_{3,\rm EW}$ and $\alpha^p_{4,\rm EW}$ involve only the Wilson 
coefficients $C_{7-10}$. We include the QCD corrections of order 
$\alpha_s C_{7-10}$ to $\alpha^p_{3,\rm EW}$ and $\alpha^p_{4,\rm EW}$. 
\item[3)] 
We include electromagnetic corrections to $\alpha^p_{3,\rm EW}$ and 
$\alpha^p_{4,\rm EW}$ only when they are proportional to the large 
Wilson coefficients $C_{1,2,7\gamma}$, i.e., we include the corrections 
of order $\alpha\,C_{1,2,7\gamma}$, but neglect those proportional to 
$\alpha\,C_{3-10}$.
\end{itemize}

\subsubsection*{\it Weak annihilation}

\begin{itemize}
\item[1)] 
We neglect weak annihilation mechanisms involving photons 
($\gamma\to\bar q' q'$), which in fact would introduce more operators 
in (\ref{bis}), i.e., we drop annihilation terms of order 
$\alpha\,C_i$, but keep terms of order $\alpha_s\,C_i$. The reason 
for this is that the former can never be CKM-enhanced when one of the 
large Wilson coefficients $C_{1,2}$ is involved.
\item[2)] 
We use an approximation where all singlet annihilation coefficients are 
set to zero except for $\beta_{S3}$, see below and \cite{Beneke:2002jn}.
\end{itemize}
 
In the following we first give the generic results for the $\alpha_i$ and 
$\beta_i$ coefficients applicable to all final states, and then discuss 
contributions specific to particular pseudoscalar or vector mesons. 

\boldmath
\subsubsection*{$A$-operators (generic results)}
\unboldmath

The coefficients of the flavor operators $\alpha_i^p$ can be expressed in
terms of the coefficients $a_i^p$ defined in \cite{BBNS1,BBNS3} as 
follows:\footnote{The numerical values of the coefficients 
$\alpha_i(M_1 M_2)$ also depend on the nature of the initial-state 
$B$ meson. This dependence is not indicated explicitly by our notation. 
The same remark applies to the annihilation coefficients $b_i^p$ defined 
below.}
\begin{eqnarray}\label{ais}
   \alpha_1(M_1 M_2) &=& a_1(M_1 M_2) \,, \nonumber\\
   \alpha_2(M_1 M_2) &=& a_2(M_1 M_2) \,, \nonumber\\
   \alpha_3^p(M_1 M_2) &=& \left\{
    \begin{array}{cl} 
     a_3^p(M_1 M_2) - a_5^p(M_1 M_2) \,;
      & \quad \mbox{if~} M_1 M_2=PP, \,VP \,, \\
     a_3^p(M_1 M_2) + a_5^p(M_1 M_2) \,;
      & \quad \mbox{if~} M_1 M_2=PV  \,, 
    \end{array}\right. \nonumber\\
   \alpha_4^p(M_1 M_2) &=& \left\{
    \begin{array}{cl} 
     a_4^p(M_1 M_2) + r_{\chi}^{M_2}\,a_6^p(M_1 M_2) \,;
      & \quad \mbox{if~} M_1 M_2=PP, \,PV \,, \\
     a_4^p(M_1 M_2) - r_{\chi}^{M_2}\,a_6^p(M_1 M_2) \,;
      & \quad \mbox{if~} M_1 M_2=VP\,,   
    \end{array}\right.\\
   \alpha_{3,\rm EW}^p(M_1 M_2) &=& \left\{
    \begin{array}{cl} 
     a_9^p(M_1 M_2) - a_7^p(M_1 M_2) \,;
      & \quad \mbox{if~} M_1 M_2=PP, \,VP \,, \\
     a_9^p(M_1 M_2) + a_7^p(M_1 M_2) \,;
      & \quad \mbox{if~} M_1 M_2=PV  \,, 
    \end{array}\right. \nonumber\\
   \alpha_{4,\rm EW}^p(M_1 M_2) &=& \left\{
    \begin{array}{cl} 
     a_{10}^p(M_1 M_2) + r_{\chi}^{M_2}\,a_8^p(M_1 M_2) \,;
      & \quad \mbox{if~} M_1 M_2=PP, \,PV \,, \\
     a_{10}^p(M_1 M_2) - r_{\chi}^{M_2}\,a_8^p(M_1 M_2) \,;
      & \quad \mbox{if~} M_1 M_2=VP\,.   
     \end{array}\right.\nonumber
\end{eqnarray}
Note that the order of the arguments in $\alpha_i^p(M_1 M_2)$ and 
$a_i^p(M_1 M_2)$ is relevant. For pions and kaons, the ratios 
$r_\chi^{M_2}$ are defined as
\begin{equation}\label{rchi}
   r_\chi^\pi(\mu) = \frac{2m_\pi^2}{m_b(\mu)\,2m_q(\mu)} \,, \qquad 
   r_\chi^K(\mu) = \frac{2m_K^2}{m_b(\mu)\,(m_q+m_s)(\mu)} \,,
\end{equation}
while their generalizations to $\eta$ and $\eta'$ can be found in 
\cite{Beneke:2002jn}. All quark masses are running masses defined in the 
$\overline{\rm MS}$ scheme, and $m_q$ denotes the average of the up and 
down quark masses. For vector mesons we have
\begin{equation}
\label{rchiV}
   r_\chi^V(\mu) = \frac{2m_V}{m_b(\mu)}\,\frac{f_V^\perp(\mu)}{f_V} \,,
\end{equation}
where the scale-dependent transverse decay constant is defined as
\begin{equation}
   \langle V(p,\varepsilon^*)|\bar{q}\sigma_{\mu\nu} q'|0\rangle 
   = f_V^\perp (p_\mu\varepsilon^*_\nu-p_\nu\varepsilon^*_\mu) \,.
\end{equation}
Note that all the terms proportional to $r_\chi^{M_2}$ are formally 
suppressed by one power of $\Lambda_{\rm QCD}/m_b$ in the heavy-quark 
limit. Numerically, however, these terms are not always small. 

The general form of the coefficients $a_i^p$ at next-to-leading order in 
$\alpha_s$ is
\begin{eqnarray}
   a_i^p(M_1 M_2) &=& \left( C_i + \frac{C_{i\pm 1}}{N_c} \right)
   N_i(M_2) \nonumber\\
  && + \,\frac{C_{i\pm 1}}{N_c}\,\frac{C_F\alpha_s}{4\pi}
   \left[ V_i(M_2) + \frac{4\pi^2}{N_c}\,H_i(M_1 M_2) \right] 
   + P_i^p(M_2) \,,
\end{eqnarray}
where the upper (lower) signs apply when $i$ is odd (even). It is
understood that the superscript `$p$' is to be omitted for $i=1,2$. The
quantities $V_i(M_2)$ account for one-loop vertex corrections, 
$H_i(M_1 M_2)$ for hard spectator interactions, and $P_i^p(M_1 M_2)$ for 
penguin contractions. We now present the explicit expressions for these 
objects, extending the results of \cite{BBNS3} to the case when $M_1$ or 
$M_2$ is a vector meson. 

\begin{figure}
\epsfxsize=4cm
\centerline{\epsffile{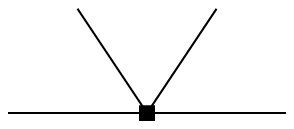}}
\centerline{\parbox{14cm}{\caption{\label{figlo} 
Leading-order contribution to the coefficients $a_i^p$. The weak decay of 
the $b$ quark through a four-fermion operator is represented by the black 
square. The $b$-quark line comes in from the left. The outgoing line to 
the right represents the quark in $M_1$. The two lines directed upward 
represent the meson $M_2$. The spectator antiquark is not drawn, because 
it does not participate in the hard scattering.}}}
\end{figure}

\paragraph{\it Leading order (Figure~\ref{figlo}).} 
The leading-order coefficient $N_i(M_2)$ is simply the normalization 
integral of the relevant light-cone distribution amplitude $\Phi_{P,V}$ 
or $\Phi_{p,v}$. It follows that
\begin{equation}\label{loterms}
   N_i(M_2) = \Bigg\{
   \begin{array}{ll}
    ~0 \,; & \quad \mbox{$i=6,8 $ and $M_2=V$,} \\
    ~1 \,; & \quad \mbox{all other cases.}
   \end{array}
\end{equation}
The special case for vector mesons arises because of the vanishing 
integral (\ref{zero}). This implies the absence of scalar penguin
contributions for vector mesons at tree level, since the coefficients 
$a_{6,8}^p$ originate from $(V-A)\otimes(V+A)$ penguin operators in the 
weak Hamiltonian, which must be Fierz-transformed into the Form 
$(-2)(S-P)\otimes(S+P)$ to match the quark flavors in the second 
$(S+P)$ current with those of the meson $M_2$.

\begin{figure}
\epsfxsize=15cm
\centerline{\epsffile{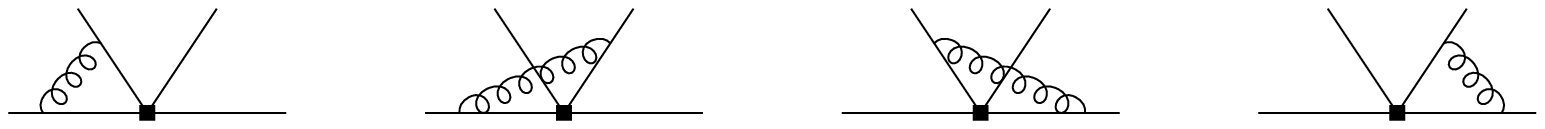}}
\centerline{\parbox{14cm}{\caption{\label{figvertex} 
Next-to-leading order vertex contribution to the coefficients $a_i^p$. 
The meaning of the external lines is the same as in 
Figure~\ref{figlo}.}}}
\end{figure}

\paragraph{\it Vertex terms (Figure~\ref{figvertex}).} 
The vertex corrections are given by 
\begin{equation}\label{vertexterms}
   V_i(M_2) = \left\{\,\,
   \begin{array}{ll}
    {\displaystyle \int_0^1\!dx\,\Phi_{M_2}(x)\,
     \Big[ 12\ln\frac{m_b}{\mu} - 18 + g(x) \Big]} \,; & \qquad
     i=\mbox{1--4},9,10, \\[0.4cm] 
   {\displaystyle \int_0^1\!dx\,\Phi_{M_2}(x)\,
     \Big[ - 12\ln\frac{m_b}{\mu} + 6 - g(1-x) \Big]} \,; & \qquad
     i=5,7, \\[0.4cm]
   {\displaystyle \int_0^1\!dx\,\Phi_{m_2}(x)\,\Big[ -6 + h(x) \Big]}
    \,; & \qquad i=6,8,
   \end{array}\right.
\end{equation}
with
\begin{equation}\label{FM}
\begin{aligned}
   g(x) &= 3\left( \frac{1-2x}{1-x}\ln x-i\pi \right) \\
   &\quad\mbox{}+ \left[ 2 \,\mbox{Li}_2(x) - \ln^2\!x
    + \frac{2\ln x}{1-x} - (3+2i\pi)\ln x - (x\leftrightarrow 1-x)
    \right] , \\
   h(x) &= 2 \,\mbox{Li}_2(x) - \ln^2\!x - (1+2\pi i)\,\ln x  
    - (x\leftrightarrow 1-x) \,.
\end{aligned}
\end{equation}
The expression for $g(x)$ has already been presented in \cite{BBNS3}, 
while the kernel $h(x)$ is new. The constants $-18$, $6$, $-6$ are 
scheme dependent and correspond to using the NDR scheme for $\gamma_5$.
The light-cone distribution amplitude $\Phi_{M_2}$ is one of the 
leading-twist amplitudes $\Phi_{P,V}$, depending on whether $M_2$ is a
pseudoscalar or vector meson, whereas $\Phi_{m_2}$ is one of the 
twist-3 amplitudes $\Phi_{p,v}$. We recall that $\Phi_p(x)=1$, so 
$\int\!dx\,\Phi_{m_2}(x)\,[-6+ h(x)]=-6$ for pseudoscalar mesons, which 
reproduces the result of \cite{BBNS3}. On the other hand, because of 
(\ref{zero}) the scheme-dependent constant $-6$ does not contribute for 
vector mesons as it should be, since there is no leading-order 
contribution that could cause a scheme dependence at next-to-leading 
order. 

\begin{figure}
\epsfxsize=9cm
\centerline{\epsffile{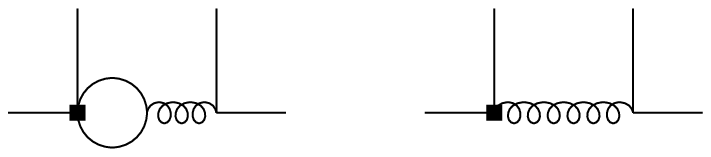}}
\centerline{\parbox{14cm}{\caption{\label{figpenguin} 
Next-to-leading order penguin contribution to the coefficients 
$a_i^p$.}}}
\end{figure}

\paragraph{\it Penguin terms (Figure~\ref{figpenguin}).} 
At order $\alpha_s$ a correction from penguin contractions is present 
only for $i=4,6$. For $i=4$ we obtain
\begin{eqnarray}\label{PK}
   P_4^p(M_2) &=& \frac{C_F\alpha_s}{4\pi N_c}\,\Bigg\{
    C_1 \!\left[ \frac43\ln\frac{m_b}{\mu}
    + \frac23 - G_{M_2}(s_p) \right]\!
    + C_3 \!\left[ \frac83\ln\frac{m_b}{\mu} + \frac43
    - G_{M_2}(0) - G_{M_2}(1) \right] \nonumber\\
   &&\hspace{1.4cm}\mbox{}+ (C_4+C_6)\!
    \left[ \frac{4n_f}{3}\ln\frac{m_b}{\mu}
    - (n_f-2)\,G_{M_2}(0) - G_{M_2}(s_c) - G_{M_2}(1) \right] \nonumber\\
   &&\hspace{1.4cm}\mbox{}- 2 C_{8g}^{\rm eff} \int_0^1 \frac{dx}{1-x}\,
    \Phi_{M_2}(x) \Bigg\} \,,
\end{eqnarray}
where $n_f=5$ is the number of light quark flavors, and $s_u=0$, 
$s_c=(m_c/m_b)^2$ are mass ratios involved in the evaluation of the 
penguin diagrams. Small electroweak corrections from $C_{7-10}$ 
are neglected within the approximations discussed above. The function 
$G_{M_2}(s)$ is given by
\begin{equation}\label{GK}
\begin{aligned}
   G_{M_2}(s) &= \int_0^1\!dx\,G(s-i\epsilon,1-x)\,\Phi_{M_2}(x) \,, \\
   G(s,x) &= -4\int_0^1\!du\,u(1-u) \ln[s-u(1-u)x] \\
   &= \frac{2(12s+5x-3x\ln s)}{9x}
    - \frac{4\sqrt{4s-x}\,(2s+x)}{3x^{3/2}}
    \arctan\sqrt{\frac{x}{4s-x}} \,. 
\end{aligned}
\end{equation}
The interpretation of $\Phi_{M_2}$ is the same as in the discussion of 
vertex corrections. For $i=6$, the result for the penguin contribution is
\begin{eqnarray}
   P_6^p(M_2) &=& \frac{C_F\alpha_s}{4\pi N_c}\,\Bigg\{
    C_1 \!\left[ \frac43\ln\frac{m_b}{\mu}
    + \frac23 - \hat G_{M_2}(s_p) \right]\!
    + C_3 \!\left[ \frac83\ln\frac{m_b}{\mu} + \frac43
    - \hat G_{M_2}(0) - \hat G_{M_2}(1) \right] \nonumber\\
   &&\hspace{-1.1cm}\mbox{}+ (C_4+C_6)\!
    \left[ \frac{4n_f}{3}\ln\frac{m_b}{\mu}
    - (n_f-2)\,\hat G_{M_2}(0) - \hat G_{M_2}(s_c) - \hat G_{M_2}(1)
    \right] - 2 C_{8g}^{\rm eff} \Bigg\} 
\end{eqnarray}
if $M_2$ is a pseudoscalar meson, and 
\begin{eqnarray}
   P_6^p(M_2) &=& - \frac{C_F\alpha_s}{4\pi N_c}\,\Bigg\{
    C_1\,\hat G_{M_2}(s_p)
    + C_3\,\Big[ \hat G_{M_2}(0) + \hat G_{M_2}(1) \Big] \nonumber\\
   &&\mbox{}+ (C_4+C_6) \left[ (n_f-2)\,\hat G_{M_2}(0)
    + \hat G_{M_2}(s_c) + \hat G_{M_2}(1) \right] \Bigg\}
\end{eqnarray}
if $M_2$ is a vector meson. In analogy with (\ref{GK}), the function 
$\hat G_{M_2}(s)$ is defined as
\begin{equation}
   \hat G_{M_2}(s) = \int_0^1\!dx\,G(s-i\epsilon,1-x)\,\Phi_{m_2}(x) \,.
\end{equation}

As mentioned above we take into account electromagnetic corrections only  
for $\alpha_{3,\rm EW}^p$  and $\alpha_{4,\rm EW}^p$, and only 
if they are proportional to the large Wilson coefficients $C_{1,2}$ and 
$C_{7\gamma}^{\rm eff}$. These corrections are present for $i=8,10$ and 
correspond to the penguin diagrams of Figure~\ref{figpenguin} with the 
gluon replaced by a photon. (An additional contribution for neutral 
vector mesons will be discussed separately below). For 
$i=10$ we obtain 
\begin{equation}\label{PKEW}
   P_{10}^p(M_2) = \frac{\alpha}{9\pi N_c}\,\left\{
   (C_1+N_c C_2) \left[ \frac{4}{3}\ln\frac{m_b}{\mu} + \frac23  
   - G_{M_2}(s_p) \right]
   - 3 C_{7\gamma}^{\rm eff} \int_0^1 \frac{dx}{1-x}\,\Phi_{M_2}(x)
   \right\} .
\end{equation}
For $i=8$ we find
\begin{equation}
   P_8^p(M_2) = \frac{\alpha}{9\pi N_c}\,\left\{
   (C_1+N_c C_2) \left[ \frac{4}{3}\ln\frac{m_b}{\mu} + \frac23  
   - \hat G_{M_2}(s_p) \right] - 3 C_{7\gamma}^{\rm eff} \right\}
\end{equation}
if $M_2$ is a pseudoscalar meson, and 
\begin{equation}
   P_8^p(M_2) = - \frac{\alpha}{9\pi N_c}\,(C_1+N_c C_2)\,
   \hat G_{M_2}(s_p)
\end{equation}
if $M_2$ is a vector meson.

\begin{figure}
\epsfxsize=8cm
\centerline{\epsffile{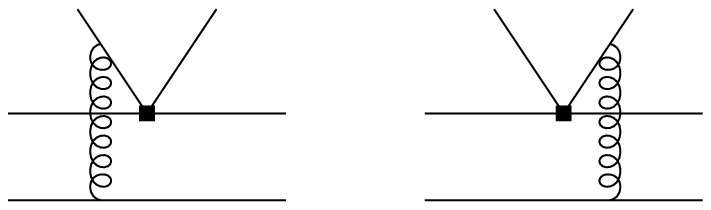}}
\centerline{\parbox{14cm}{\caption{\label{fighardspec} 
Hard spectator-scattering contribution to the coefficients $a_i^p$. The 
meaning of the external lines is the same as in Figure~\ref{figlo}, but 
the spectator-quark line is now included in the drawing.}}}
\end{figure}

\paragraph{\it Hard spectator terms (Figure~\ref{fighardspec}).} 
The correction from hard gluon exchange between $M_2$ and the spectator 
quark is given by
\begin{equation}\label{hardspecterms1}
   H_i(M_1M_2)
   = \frac{B_{M_1 M_2}}{A_{M_1 M_2}}\,\frac{m_B}{\lambda_B}\,
   \int_0^1\!dx \int_0^1\!dy \left[
   \frac{\Phi_{M_2}(x)\Phi_{M_1}(y)}{\bar x\bar y}
   + r_\chi^{M_1}\,\frac{\Phi_{M_2}(x)\Phi_{m_1}(y)}{x\bar y} \right]
\end{equation}
for $i=1$--4,9,10,  
\begin{equation}\label{hardspecterms2}
   H_i(M_1M_2)
   = - \frac{B_{M_1 M_2}}{A_{M_1 M_2}}\,\frac{m_B}{\lambda_B}\,
   \int_0^1\!dx \int_0^1\!dy \left[
   \frac{\Phi_{M_2}(x)\Phi_{M_1}(y)}{x\bar y}
   + r_\chi^{M_1}\,\frac{\Phi_{M_2}(x)\Phi_{m_1}(y)}{\bar x\bar y} 
   \right]
\end{equation}
for $i=5,7$, and $H_i(M_1M_2)=0$ for $i=6,8$. In these results 
$\lambda_B$ is defined by \cite{BBNS1}
\begin{equation}
   \int_0^1 \frac{d\xi}{\xi}\,\Phi_B(\xi)\equiv \frac{m_B}{\lambda_B}
\end{equation}
with $\Phi_B(\xi)$ one of the two light-cone distribution amplitudes of 
the $B$ meson. We recall that the term involving $r_\chi^{M_1}$ is 
suppressed by a factor of $\Lambda_{\rm QCD}/m_b$ in heavy-quark power 
counting. Since the twist-3 distribution amplitude $\Phi_{m_1}(y)$ does 
not vanish at $y=1$, the power-suppressed term is divergent. We extract 
this divergence by defining a parameter $X_H^{M_1}$ through
\begin{eqnarray}\label{XHdef}
   \int_0^1\frac{d y}{\bar y}\,\Phi_{m_1}(y) 
   &=& \Phi_{m_1}(1)\,\int_0^1\frac{d y}{\bar y}
    + \int_0^1\frac{d y}{\bar y}\,\Big[ \Phi_{m_1}(y)-\Phi_{m_1}(1)
    \Big] \nonumber\\
   &\equiv& \Phi_{m_1}(1)\,X_H^{M_1}
    + \int_0^1\frac{d y}{[\bar y]_+}\,\Phi_{m_1}(y) \, .
\end{eqnarray}
The remaining integral is finite (it vanishes for pseudoscalar mesons 
since $\Phi_p(y)=1$), but $X_H^{M_1}$ is an unknown parameter 
representing a soft-gluon interaction with the spectator quark. Since the 
divergence that appears in an attempt to compute this soft interaction 
perturbatively is regulated by a physical scale of order 
$\Lambda_{\rm QCD}$ (i.e.\ at $\bar y\sim\Lambda_{\rm QCD}/m_b$), we 
expect that $X_H^M\sim\ln(m_b/\Lambda_{\rm QCD})$, however with a 
potentially complex coefficient, since multiple soft scattering can 
introduce a strong-interaction phase. A consequence of this is that power 
corrections to the spectator interaction, including chirally-enhanced 
ones for pseudoscalar mesons, are unavoidably model dependent. As in 
\cite{BBNS3}, our model consists of varying $X_H^{M_1}$ within a certain 
range (specified later) and to treat the resulting variation of the 
coefficients $\alpha_i^p$ as an uncertainty. We also assume that 
$X_H^{M_1}$ is universal, i.e., that it does not depend on $M_1$ and on 
the index $i$ of $H_i(M_1 M_2)$.

In \cite{BBNS3} the convolution integrals relevant to $B\to PP$ decays 
were evaluated explicitly up to second non-trivial order in the 
Gegenbauer expansion. In Appendix~B we present the corresponding results
for convolution integrals involving the function $\Phi_v(x)$, which 
appear only in $B\to PV$ decays. 

\boldmath
\subsubsection*{$A$-operators (particular results)}
\unboldmath

\begin{figure}
\epsfxsize=8cm
\centerline{\epsffile{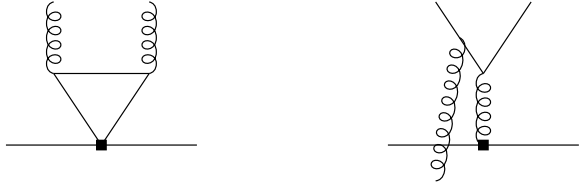}}
\centerline{\parbox{14cm}{\caption{\label{figeta} 
Additional contributions (representative diagrams shown) to the decay 
amplitude when $M_2=\eta$ or $\eta'$. Left: Charm-loop contribution to 
the $b\to D gg$ amplitude. Right: Soft spectator-scattering diagram, 
which contributes to $\Delta\alpha_3^p(M_1\eta_{q,s})$ in 
(\ref{extraal3}). The gluon to the left is soft and part of the operator 
whose matrix element defines the non-local form factor 
${\cal F}_g^{B\to M_1}$. The gluon that originates at the weak vertex is 
semi-hard.}}}
\end{figure}

In addition to the generic expressions given above, there exist 
additional contributions to the decay amplitudes for final states 
containing an $\eta$ or $\eta'$ meson (related to their two-gluon 
content), and for final states containing one of the neutral vector 
mesons $\rho^0$, $\omega$, $\phi$ (related to their coupling to the 
photon).

The contributions specific to $\eta^{(\prime)}$ mesons have been 
discussed in \cite{Beneke:2002jn}. There are two effects 
relevant to the $A$-operators. First, the leading contribution of the 
$b\to D gg$ amplitude can be interpreted in terms of a ``charm decay 
constant'' of the $\eta^{(\prime)}$ meson, which is calculable in a 
$\Lambda_{\rm QCD}/m_c$ expansion (left side of Figure~\ref{figeta}). 
To account for this effect the term 
\begin{equation}\label{ccbareffect}
   \big[ \delta_{pc}\,\alpha_2(M_1\eta_c^{(\prime)}) + 
   \alpha_3^p(M_1\eta_c^{(\prime)}) \big]\,
   A([\bar q_s D][\bar c c])
\end{equation}
is added to $ {\cal T}_A^p$ in (\ref{alphaidef}). Since this 
contribution is very small, we will only need the leading-order 
expressions for the new $\alpha_i$ coefficients, which coincide with the 
general leading-order result. We will also neglect the corresponding 
electroweak penguin effect, which is even smaller. Second, there exists 
an additional form-factor type contribution 
$\Delta\alpha_3^p(M_1\eta_{q,s}^{(\prime)})$ to the flavor-singlet 
coefficient $\alpha_3^p(M_1\eta_{q,s}^{(\prime)})$, which is given by 
\begin{equation}\label{extraal3}
   \Delta\alpha_3^p(M_1\eta_{q,s}^{(\prime)})
   = - \frac{3\alpha_s(\mu_h)}{8\pi N_c}\,C_{8g}^{\rm eff}(\mu_h)
   \left( \int_0^1\!dx\,
   \frac{\Phi_{\eta_{q,s}^{(\prime)}}(x)}{6x\bar x} + 
   \dots
   \right) \frac{{\cal F}_g^{B\to M_1}(0)}{F^{B\to M_1}(0)} \,,
\end{equation}
where $\mu_h=\sqrt{m_b\Lambda_h}$ with $\Lambda_h=0.5$\,GeV serves as a 
typical scale for the semi-hard subprocess (right side of 
Figure~\ref{figeta}), and the dots stand for a contribution from the 
leading-twist two-gluon distribution amplitude of $\eta$ or $\eta'$, 
which vanishes for asymptotic distribution amplitudes. The new feature of 
(\ref{extraal3}) is that this contribution is proportional to a non-local 
``form factor'' ${\cal F}_g^{B\to M_1}(0)$ (rather than the usual local 
form factors $F=F_0$, $A_0$) defined in terms of the $B\to M_1$ matrix
element of a bilocal quark--antiquark--gluon operator 
\cite{Beneke:2002jn}. The ratio of the two form factors in 
(\ref{extraal3}) is of order unity in the heavy-quark limit and has been 
estimated to be close to 1; however, there is a large uncertainty 
associated with this estimate.

\begin{figure}
\epsfxsize=8cm
\centerline{\epsffile{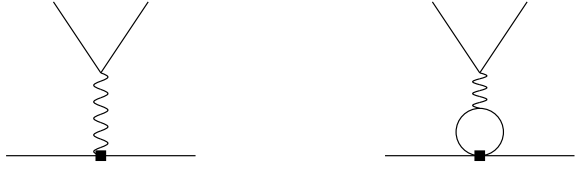}}
\centerline{\parbox{14cm}{\caption{\label{figrhopen} 
Additional contributions (representative diagrams shown) to the decay 
amplitude when $M_2=\rho^0,\omega,\phi$. Wavy lines denote a photon.}}}
\end{figure}

For the vector mesons $\rho^0$, $\omega$, $\phi$ the particular 
contributions are related to the electromagnetic penguin diagrams shown 
in Figure~\ref{figrhopen}. These contributions are very small, but it is 
interesting to discuss them from a conceptual point of view. In both 
diagrams the photon propagator is canceled. The left diagram where the 
photon originates from the electric dipole operator $Q_{7\gamma}$ is 
therefore 
effectively a local vertex. The other diagram in the figure can be 
calculated explicitly in an expansion in $\Lambda_{\rm QCD}/m_Q$ when the 
quark in the loop is heavy (with mass $m_Q$). In the case of a charm 
quark this produces logarithms of the form $\ln(m_b/m_c)$, which could be 
summed by 
introducing a $c\bar c$ decay constant and distribution amplitude for the 
vector mesons, similar to the treatment of the $b\to Dgg$ amplitude for 
$\eta$ and $\eta'$. The quark loop contains an ultraviolet divergence, 
which must be subtracted in accordance with the scheme used to define the 
Wilson coefficients. The scale and scheme dependence after subtraction is 
required to cancel the scale and scheme dependence of the electroweak 
penguin coefficients $C_{7-10}$. When the quark in the loop is massless 
(light quarks), the loop integral is infrared and ultraviolet divergent. 
The ultraviolet divergence is subtracted as before. The infrared 
divergence must be factorized into the longitudinal decay constant of the 
vector meson. As a consequence, parameters such as $f_{\rho^0}$ become 
scale-dependent.\footnote{Strictly speaking, we should then distinguish 
$f_{\rho^0}(\nu)$ from $f_{\rho^+}$, which is scale-independent. However, 
given the smallness of the effect we set the two decay constants equal 
at $\nu=\mu$.}
The reason for this is that the operator $\bar q\gamma^\mu q$ has a 
non-vanishing anomalous dimension in the presence of electromagnetic 
interactions. With these remarks it is straightforward to compute the 
additional contribution $\Delta\alpha_{3,\rm EW}^p(M_1 V)$ to the 
electromagnetic penguin amplitude $\alpha_{3,\rm EW}^p(M_1 V)$ when 
$V=\rho^0,\omega,\phi$. Within our approximation of keeping only 
the $\alpha C_{1,2,7\gamma}$ corrections we find  
\begin{equation}\label{extraal3EW}
   \Delta\alpha_{3,\rm EW}^p(M_1 V)
   = \frac{2\alpha}{27\pi}\,\left\{ 4(C_1+N_c C_2)
   \left( \frac12 - \delta_{pc}\,\ln\frac{\mu}{m_c}
   - \delta_{pu}\,\ln\frac{\mu}{\nu} \right) 
   - 9 C_{7\gamma}^{\rm eff} \right\} .
\end{equation}
Here $\mu$ refers to the renormalization scale of the Wilson coefficients 
$C_i(\mu)$, and $\nu$ to the scale of the decay constant $f_V(\nu)$. The 
$\nu$ dependence of (\ref{extraal3EW}) cancels the $\nu$ dependence of the 
leading-order decay amplitude to order $\alpha$. In our analysis we set 
$\nu=\mu$.

\boldmath
\subsubsection*{$B$-operators/weak annihilation (generic results)}
\unboldmath

\begin{figure}
\epsfxsize=15cm
\centerline{\epsffile{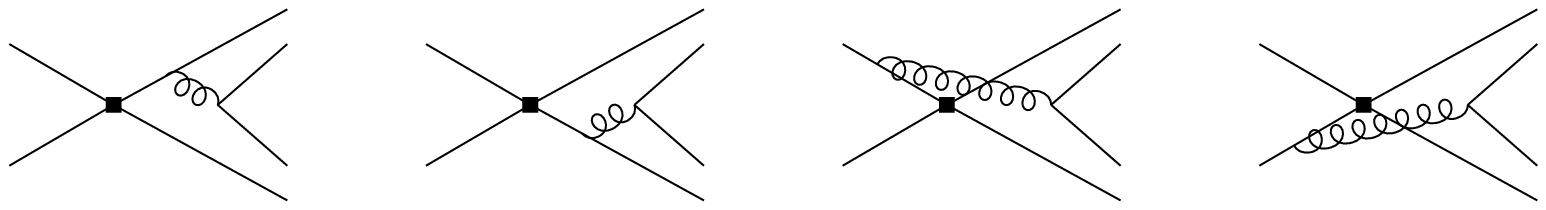}}
\centerline{\parbox{14cm}{\caption{\label{figannh} 
Weak annihilation contributions.}}}
\end{figure}

We now turn to a discussion of power-suppressed weak annihilation effects.
Except for certain sign changes for final states with vector mesons
the ``non-singlet'' annihilation 
coefficients can be taken from \cite{BBNS3}. We consider $b$-quark decay 
and use the convention that $M_1$ contains an antiquark from the weak 
vertex with longitudinal momentum fraction $\bar y$. For non-singlet 
annihilation $M_2$ then contains a quark from the weak vertex with 
momentum fraction $x$. The basic building blocks when both mesons are 
pseudoscalar are given by (omitting the argument $M_1 M_2$ for brevity)
\begin{eqnarray}\label{blocks}
   A_1^i &=& \pi\alpha_s \int_0^1\! dx dy\, 
    \left\{ \Phi_{M_2}(x)\,\Phi_{M_1}(y)
    \left[ \frac{1}{y(1-x\bar y)} + \frac{1}{\bar x^2 y} \right]
    + r_\chi^{M_1} r_\chi^{M_2}\,\Phi_{m_2}(x)\,\Phi_{m_1}(y)\,
     \frac{2}{\bar x y} \right\} ,
    \nonumber\\
   A_1^f &=& 0 \,, \nonumber\\ 
   A_2^i &=& \pi\alpha_s \int_0^1\! dx dy\, 
    \left\{ \Phi_{M_2}(x)\,\Phi_{M_1}(y)
    \left[ \frac{1}{\bar x(1-x\bar y)} + \frac{1}{\bar x y^2} \right]
    + r_\chi^{M_1} r_\chi^{M_2}\,\Phi_{m_2}(x)\,\Phi_{m_1}(y)\,
     \frac{2}{\bar x y} \right\} ,
    \nonumber\\
   A_2^f &=& 0 \,, \\ 
   A_3^i &=& \pi\alpha_s \int_0^1\! dx dy\,
    \left\{r_\chi^{M_1}\,\Phi_{M_2}(x)\,\Phi_{m_1}(y)\,
    \frac{2\bar y}{\bar x y(1-x\bar y)}
    - r_\chi^{M_2}\,\Phi_{M_1}(y)\,\Phi_{m_2}(x)\,
    \frac{2x}{\bar x y(1-x\bar y)} \right\} , \nonumber\\
   A_3^f &=& \pi\alpha_s \int_0^1\! dx dy\,
    \left\{r_\chi^{M_1}\,\Phi_{M_2}(x)\,\Phi_{m_1}(y)\,
    \frac{2(1+\bar x)}{\bar x^2 y}
    +  r_\chi^{M_2}\,\Phi_{M_1}(y)\,\Phi_{m_2}(x)\,
    \frac{2(1+y)}{\bar x y^2} \right\} . \nonumber
\end{eqnarray}
When $M_1$ is a vector meson and $M_2$ a pseudoscalar, one has to change 
the sign of the second (twist-4) term in $A_1^i$, the first (twist-2) 
term in $A_2^i$, and the second term in $A_3^i$ and $A_3^f$. When $M_2$ is 
a vector meson and $M_1$ a pseudoscalar, one only has to change the 
overall sign of $A_2^i$. 

In (\ref{blocks}) the superscripts `$i$' and `$f$' refer to gluon 
emission from the initial and final-state quarks, respectively (see
Figure~\ref{figannh}). The 
subscript `$k$' on $A_k^{i,f}$ refers to one of the three possible Dirac 
structures $\Gamma_1\otimes\Gamma_2$, which arise when the four-quark 
operators in the effective weak Hamiltonian are Fierz-transformed into 
the form $(\bar q_1 b)_{\Gamma_1}(\bar q_2 q_3)_{\Gamma_2}$, such that 
the quarks in the first bracket refer to the constituents of the $\bar B$ 
meson. Specifically, we have $k=1$ for $(V-A)\otimes(V-A)$, $k=2$ for 
$(V-A)\otimes(V+A)$, and $k=3$ for $(-2)(S-P)\otimes(S+P)$.  The power 
suppression of weak annihilation terms compared to the leading spectator 
interaction via gluon exchange is evident from the fact that annihilation 
terms are proportional to $f_B$ rather than $f_B m_B/\lambda_B$.

In terms of these building blocks the non-singlet annihilation 
coefficients are given by
\begin{equation}\label{bidef}
\begin{aligned}
   b_1 &= \frac{C_F}{N_c^2}\,C_1 A_1^i \,, \qquad
    b_3^p = \frac{C_F}{N_c^2} \Big[ C_3 A_1^i + C_5 (A_3^i+A_3^f)
    + N_c C_6 A_3^f \Big] \,, \\
   b_2 &= \frac{C_F}{N_c^2}\,C_2 A_1^i \,, \qquad
    b_4^p = \frac{C_F}{N_c^2}\,\Big[ C_4 A_1^i + C_6 A_2^i \Big] \,, \\
   b_{3,\rm EW}^p &= \frac{C_F}{N_c^2} \Big[ C_9 A_1^i
    + C_7 (A_3^i+A_3^f) + N_c C_8 A_3^f \Big] \,, \\
   b_{4,\rm EW}^p &= \frac{C_F}{N_c^2}\,\Big[ C_{10} A_1^i
    + C_8 A_2^i \Big] \,,
\end{aligned}
\end{equation}
omitting again the argument $M_1 M_2$. These coefficients correspond to 
current--current annihilation ($b_1,b_2$), penguin  annihilation 
($b_3,b_4$), and electroweak penguin annihilation 
($b_3^{\rm EW},b_4^{\rm EW}$), where within each pair the two 
coefficients correspond to different flavor structures as defined in 
(\ref{bis}).

The weak annihilation kernels exhibit endpoint divergences, which we 
treat in the same manner as the power corrections to the hard spectator 
scattering. The divergent subtractions are interpreted as 
\begin{equation}\label{XAdef}
   \int_0^1 \frac{dy}{y}\to X_A^{M_1} \,, \qquad 
   \int_0^1\!dy\,\frac{\ln y}{y}\to -\frac{1}{2}\,(X_A^{M_1})^2 \,,
\end{equation}
and similarly for $M_2$ with $y\to \bar x$. The treatment of weak 
annihilation is model-dependent in the QCD factorization approach, and 
the explicit results of this subsection are useful mainly to keep track 
of overall factors from Wilson coefficients and color. We treat $X_A^M$ 
as an unknown complex number of order $\ln(m_b/\Lambda_{\rm QCD})$ and 
make the simplifying assumption that this number is independent of the 
identity of the meson $M$ and the weak decay vertex. (The first 
assumption will be relaxed in a specific scenario, where we allow 
different $X_A$ for the three cases $PP$, $PV$, and $VP$.) Since the 
treatment of annihilation is model-dependent anyway, we further simplify 
our results by evaluating the convolution integrals with asymptotic 
distribution amplitudes $\Phi(x)=\Phi_\parallel(x)=6x\bar x$, 
$\Phi_p(x)=1$, and $\Phi_v(x)=3(x-\bar x)$. We then find the simple 
expressions 
\begin{equation}\label{XAmodel}
\begin{aligned}
   A_1^i &\approx A_2^i \approx 2\pi\alpha_s \left[\,
    9\,\bigg( X_A - 4 + \frac{\pi^2}{3} \bigg)
    + r_\chi^{M_1} r_\chi^{M_2} X_A^2 \right] , \\
   A_3^i &\approx 6\pi\alpha_s\,(r_\chi^{M_1}-r_\chi^{M_2})
    \bigg( X_A^2 - 2 X_A + \frac{\pi^2}{3} \bigg) \,, \\[0.3cm]
   A_3^f &\approx 6\pi\alpha_s\,(r_\chi^{M_1}+r_\chi^{M_2})\,
    (2 X_A^2 - X_A) \,,
\end{aligned}
\end{equation}
and $A_1^f=A_2^f=0$ when both final state mesons are pseudoscalar, 
whereas 
\begin{equation}\label{XAdefvect}
\begin{aligned}
   A_1^i &\approx -A_2^i \approx 6\pi\alpha_s \left[\,
    3\,\bigg( X_A - 4 + \frac{\pi^2}{3} \bigg)
    + r_\chi^{M_1} r_\chi^{M_2} (X_A^2-2 X_A) \right] , \\
   A_3^i &\approx 6\pi\alpha_s \left[ -3 r_\chi^{M_1}
    \bigg( X_A^2 - 2 X_A - \frac{\pi^2}{3} + 4 \bigg)
    + r_\chi^{M_2} \bigg( X_A^2 - 2 X_A + \frac{\pi^2}{3} \bigg) 
    \right] , \\[0.3cm]
   A_3^f &\approx 6\pi\alpha_s \left[\, 3 r_\chi^{M_1}\,
    (2 X_A-1) (2-X_A) - r_\chi^{M_2}\,(2 X_A^2 - X_A) \right] \,,
\end{aligned}
\end{equation}
and $A_1^f=A_2^f=0$ when $M_1$ is a vector meson and $M_2$ a 
pseudoscalar. For the opposite case of a pseudoscalar $M_1$ and a vector 
$M_2$, one exchanges $r_\chi^{M_1}\leftrightarrow r_\chi^{M_2}$ in the 
previous equations and changes the sign of $A_3^f$.

\boldmath
\subsubsection*{$B$-operators/weak annihilation (particular results)}
\unboldmath

\begin{figure}
\epsfxsize=14cm
\centerline{\epsffile{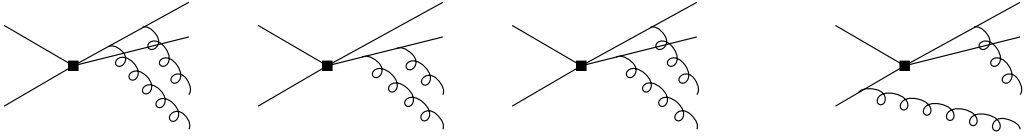}}
\centerline{\parbox{14cm}{\caption{\label{figsingannh} 
Examples of flavor-singlet annihilation graphs. Diagrams where both 
gluons attach to the constituents of the $\bar B$ meson belong to the 
$B\to \eta^{(\prime)}$ form factors. The fourth diagram and similar ones 
with one of the two gluons attached to the constituents of the $B$ meson 
do not contribute to $\beta_{S3}^p$.}}}
\end{figure}

The calculation of the singlet weak annihilation coefficients $b_{Si}$ is 
even more uncertain. Some of the diagrams that can contribute to these 
quantities are shown in Figure~\ref{figsingannh}. Since we consider 
only diagrams formally proportional to $\alpha_s$, the second meson $M_2$ 
must have a two-gluon content, hence the effect vanishes for vector 
mesons. We shall also make the approximation adopted in 
\cite{Beneke:2002jn}, in which the small electroweak singlet annihilation 
coefficients $b_{S3,\rm EW}$ and $b_{S4,\rm EW}$ are neglected, and in 
which also $b_{S1}$ and $b_{S2}$ are not computed and only $b_{S3}$ is 
kept. The reason for neglecting $b_{S1}$ and $b_{S2}$ is not that these 
terms are smaller than $b_{S3}$, but that in decay amplitudes they 
always appear together with the large tree coefficients $\alpha_1$ and 
$\alpha_2$. On the other hand, $b_{S3}$ appears in conjunction with the 
singlet penguin amplitude $\alpha_3^p$. The singlet weak annihilation 
effect is then confined to final states with an $\eta$ or $\eta'$ meson. 

With these approximations only the first three diagrams shown in the 
figure (and the corresponding crossed diagrams) have to be calculated. 
The result is \cite{Beneke:2002jn}
\begin{equation}\label{bs3}
   b_{S3}(M_1 P_{q,s}) = \frac{C_F}{N_c^2} \left(C_5+N_c C_6\right) \,
   \frac{3\pi\alpha_s}{2} \int_0^1\!dy\,
   \frac{ r_\chi^{M_1}\Phi_{m1}(y)}{y\bar y}
   \int_0^1\!dx\,\Phi_{Pg}(x)\,\frac{x-\bar x}{x^2\bar x^2} \,,
\end{equation}
where $\Phi_{Pg}(x)$ is the leading-twist two-gluon distribution 
amplitude of $P$, whose Gegenbauer expansion reads 
\begin{equation}\label{ggas}
   \Phi_{Pg}(x) = 5 B_2^{Pg}(\mu)\,x^2\bar x^2\,(x-\bar x) + \dots \,.
\end{equation}
The coefficient $B_2^{Pg}(\mu)$ as well as all higher Gegenbauer moments 
vanish for $\mu\to\infty$. Expression (\ref{bs3}) is endpoint-divergent 
as any other weak annihilation term. Introducing $X_A^{M_1}$ as before, 
and truncating the Gegenbauer expansion of $\Phi_{Pg}(x)$ after the 
leading term, we obtain the estimate
\begin{equation}\label{bs3est}
   b_{S3}(M_1 P_{q,s}) = \frac{C_F}{N_c^2} \left(C_5+N_c C_6\right) \,
   5\pi\alpha_s B_2^{Pg}(\mu_h) X_A^{M_1} \,,
\end{equation}
which we will use in our numerical analysis below. The process 
$\gamma\gamma^*\to\eta^{(\prime)}$ can in principle be used to constrain 
$B_2^{Pg}(\mu)$. In our normalization convention the analysis performed 
in \cite{KP02} gives $B_2^{Pg}(1\,\mbox{GeV})=2\pm 3$. Note, however, 
that the second Gegenbauer moment of the singlet quark--antiquark 
amplitude is much smaller, of order $-0.1$.

\section{Input parameters}
\label{sec:inputs}

The predictions obtained using the QCD factorization approach depend on 
many input parameters. First there are Standard Model parameters such as 
the elements of the CKM matrix, quark masses, and the strong coupling 
constant. Of those our results are most sensitive to the strange-quark 
mass, which sets the scale for the ratios $r_\chi$ for pseudoscalar 
mesons defined in (\ref{rchi}). (We work with a fixed ratio $m_q/m_s$, 
hence $r_\chi^\pi$ implicitly depends on $m_s$.) 
Some branching ratios, and in particular the direct CP 
asymmetries, are also very sensitive to the value of $|V_{ub}|$ and the 
weak phase $\gamma=\mbox{arg}(V_{ub}^*)$. Next there are hadronic 
parameters that 
can, in principle, be determined from experiment, such as meson decay 
constants and transition form factors. In practice, information about 
these quantities often comes from theoretical calculations such as 
light-cone QCD sum rules or lattice calculations. The corresponding 
uncertainties in the form factors are substantial and often have a large 
impact on our results. Finally, we need predictions for a variety of 
light-cone distribution amplitudes, which we parameterize by the first 
two Gegenbauer coefficients in their moment expansion. Experimental 
information can at best provide indirect constraints on these parameters. 
Fortunately, it turns out that the sensitivity of our predictions to the 
Gegenbauer coefficients is usually small. The notable exception is the 
color-suppressed tree amplitude $\alpha_2(M_1 M_2)$, which shows a 
considerable dependence on the first 
inverse moment of the $B$-meson distribution amplitude ($\lambda_B$) 
and the second Gegenbauer moment of the light mesons through the 
hard-spectator interaction. 

\begin{table}[p]
\centerline{\parbox{14cm}{\caption{\label{tab:inputs}
Summary of theoretical input parameters. All scale-dependent quantities
refer to $\mu=2$\,GeV unless indicated otherwise. Parameters related to 
$\eta$ and $\eta'$ are given in \cite{Beneke:2002jn}.}}}
\vspace{0.1cm}
\begin{center}
{\tabcolsep=0.786cm\begin{tabular}{|c|c|c|c|c|}
\hline\hline
\multicolumn{5}{|c|}{QCD scale and running quark masses [GeV]} \\
\hline
$\Lambda_{\overline{\rm MS}}^{(5)}$ & $m_b(m_b)$ & $m_c(m_b)$
 & $m_s(2\,\mbox{GeV})$ & $m_q/m_s$ \\
\hline
0.225 & 4.2 & $1.3\pm 0.2$ & $0.090\pm 0.020$ & 0.0413 \\
\hline
\end{tabular}}
{\tabcolsep=0.496cm\begin{tabular}{|c|c|c|c|c|c|}
\hline
\multicolumn{6}{|c|}{CKM parameters and $B$-meson lifetimes [ps]} \\
\hline
$|V_{cb}|$ & $|V_{ub}/V_{cb}|$ & $\gamma$ & $\tau(B^-)$ & $\tau(B_d)$
 & $\tau(B_s)$ \\
\hline
$0.041\pm 0.002$ & $0.09\pm 0.02$ & $(70\pm 20)^\circ$ & 1.67 & 1.54
 & 1.46 \\
\hline
\end{tabular}}
{\tabcolsep=1.375cm\begin{tabular}{|c|c|c|c|}
\hline
\multicolumn{4}{|c|}{Pseudoscalar-meson decay constants [MeV]} \\
\hline
$f_\pi$ & $f_K$ & $f_B$ & $f_{B_s}$ \\
\hline
131 & 160 & $200\pm 30$ & $230\pm 30$ \\
\hline
\end{tabular}}
{\tabcolsep=1.248cm\begin{tabular}{|c|c|c|c|}
\hline
\multicolumn{4}{|c|}{Vector-meson decay constants [MeV]} \\
\hline
$f_\rho$ & $f_{K^*}$ & $f_\omega$ & $f_\phi$ \\
\hline
$209\pm 1$ & $218\pm 4$ & $187\pm 3$ & $221\pm 3$ \\
\hline
\end{tabular}}
{\tabcolsep=1.145cm\begin{tabular}{|c|c|c|c|}
\hline
\multicolumn{4}{|c|}{Transverse vector-meson decay constants [MeV]} \\
\hline
$f_\rho^\perp$ & $f_{K^*}^\perp$ & $f_\omega^\perp$ & $f_\phi^\perp$ \\
\hline
$150\pm 25$ & $175\pm 25$ & $150\pm 25$ & $175\pm 25$ \\
\hline
\end{tabular}}
{\tabcolsep=1.565cm\begin{tabular}{|c|c|c|}
\hline
\multicolumn{3}{|c|}{Form factors for pseudoscalar mesons (at $q^2=0$)} \\
\hline
$F_0^{B\to\pi}$ & $F_0^{B\to K}$ & $F_0^{B_s\to\bar K}$ \\
\hline
$0.28\pm 0.05$ & $0.34\pm 0.05$ & $0.31\pm 0.05$ \\
\hline
\end{tabular}}
{\tabcolsep=0.542cm\begin{tabular}{|c|c|c|c|c|}
\hline
\multicolumn{5}{|c|}{Form factors for vector mesons (at $q^2=0$)} \\
\hline
$A_0^{B\to\rho}$ & $A_0^{B\to K^*}$ & $A_0^{B\to\omega}$
 & $A_0^{B_s\to\bar K^*}$ & $A_0^{B_s\to\phi}$ \\
\hline
$0.37\pm 0.06$ & $0.45\pm 0.07$ & $0.33\pm 0.05$ & $0.29\pm 0.05$
 & $0.34\pm 0.05$ \\
\hline
\end{tabular}}
{\tabcolsep=0.70cm\begin{tabular}{|c|c|c|c|c|}
\hline
\multicolumn{5}{|c|}{Parameters of pseudoscalar-meson distribution 
amplitudes} \\
\hline
$\alpha_2^\pi$ & $\alpha_1^{\bar K}$ & $\alpha_2^{\bar K}$
 & $\lambda_B$ [MeV]  & $\lambda_{B_s}$ [MeV] \\
\hline
$0.1\pm 0.3$ & $0.2\pm 0.2$ & $0.1\pm 0.3$ & $350\pm 150$
 & $350\pm 150$ \\
\hline
\end{tabular}}
{\tabcolsep=0.742cm\begin{tabular}{|c|c|c|c|c|}
\hline
\multicolumn{5}{|c|}{Parameters of vector-meson distribution
amplitudes} \\
\hline
$\alpha_2^\rho$, $\alpha_{2,\perp}^\rho$
 & $\alpha_1^{\bar K^*}$, $\alpha_{1,\perp}^{\bar K^*}$
 & $\alpha_2^{\bar K^*}$, $\alpha_{2,\perp}^{\bar K^*}$
 & $\alpha_2^\omega$, $\alpha_{2,\perp}^\omega$
 & $\alpha_2^\phi$, $\alpha_{2,\perp}^\phi$ \\
\hline
$0.1\pm 0.3$ & $0.2\pm 0.2$ & $0.1\pm 0.3$ & $0\pm 0.3$ & $0\pm 0.3$ \\
\hline\hline
\end{tabular}}
\end{center}
\end{table}

A summary of the input parameters entering our numerical analysis is 
given in Table~\ref{tab:inputs}. Some additional parameters related to 
$\eta$ and $\eta'$ mesons, such as their decay constants, form 
factors, and the $\eta$--$\eta'$ mixing angle in the quark-flavor basis, 
can be found in \cite{Beneke:2002jn}. Not given in this reference are 
the values of the second Gegenbauer moments of the quark--antiquark 
twist-2 distribution amplitudes for the flavor components 
$\eta_q^{(\prime)}$ and $\eta_s^{(\prime)}$, for which we take 
$\alpha_2^{\eta_{q,s}^{(\prime)}}=0\pm 0.3$.

A few additional comments are in order. The values for the decay 
constants of pseudoscalar mesons and longitudinally polarized vector 
mesons can be determined with good accuracy from experimental data on 
the leptonic decays $\pi^-\to\mu\bar\nu_\mu$, $K^-\to\mu\bar\nu_\mu$,
the semileptonic decay $\tau\to\rho^-\,\nu_\tau$, and the electromagnetic 
decays $V\to e^+ e^-$ with $V=\rho^0$, $\omega$, or $\phi$. We have 
updated the values obtained in \cite{Neubert:1997uc} by using the most
recent results for the various decay rates. We will neglect the small 
uncertainties on these parameters in our numerical analysis. The values
we take for the decay constants of the $B$ and $B_s$ mesons are in the 
ball park of many theoretical calculations using QCD sum rules and 
lattice gauge theory. The values for the heavy-to-light form factors are 
close to the results of light-cone QCD sum rules where available 
\cite{Khodjamirian:1997ub,Ball:1998kk,Ball:1998tj}. In other 
cases we base our values on a crude estimate of SU(3) flavor 
symmetry breaking effects. The $B\to\eta^{(\prime)}$ form factors 
receive an unknown two-gluon contribution. We therefore parameterize 
the form factor as \cite{Beneke:2002jn}
\begin{equation}\label{F2def}
   F_0^{B\to\eta^{(\prime)}}
   = F_0^{B\to\pi}\,\frac{f^q_{\eta^{(\prime)}}}{f_\pi}
   + F_2\,\frac{\sqrt2 f^q_{\eta^{(\prime)}} + f^s_{\eta^{(\prime)}}}
               {\sqrt3 f_\pi} \,.
\end{equation}
(For $B_s$ decay replace $F_0^{B\to\pi}\,f^q_{\eta^{(\prime)}}$ 
by  $F_0^{B\to K}\,f^s_{\eta^{(\prime)}}$.) Information on $F_2$ 
can in principle be obtained from semileptonic $B\to\eta\,l\nu$ decay 
at $q^2=0$. At present, however, the parameter $F_2$ is completely 
undetermined, and for lack of better knowledge we adopt the value 
$F_2=0$, for which $F_0^{B\to\eta}=0.23$ and $F_0^{B\to\eta'}=0.19$. The 
modes with $\eta'$ in the final state are rather sensitive to this 
choice. This introduces an additional theoretical uncertainty not taken 
into account in the error ranges given below (see \cite{Beneke:2002jn} 
for the dependence of the $B\to K^{(*)}\eta^{(\prime)}$ modes on the 
choice of $F_2$). The values for the transverse decay constants and 
Gegenbauer moments of vector 
mesons are rounded numbers taken from \cite{Ball:1998kk}, however we have 
inflated the small errors quoted there.\footnote{To facilitate 
the comparison with the results of \cite{BBNS3, Beneke:2002jn}, where they 
overlap, we note the following changes of input parameters relative 
to those papers: $m_s$ was $(110\pm 25)$\,MeV in \cite{BBNS3} and 
$(100\pm 25)$\,MeV in \cite{Beneke:2002jn}; $|V_{ub}/V_{cb}|$ was 
$0.085\pm 0.017$, $f_B$ was $(180\pm 40)$\,MeV, and $F_0^{B\to K}$ was 
$0.9 f_K/f_\pi F_0^{B\to\pi}$ in \cite{BBNS3}; $\alpha_1^{\bar K}$ was 
$0.3\pm 0.3$, $\tau(B^-)$ was 1.65\,ps, and $\tau(B_d)$ was 1.56\,ps in 
\cite{BBNS3,Beneke:2002jn}.}

The quark masses are running masses in the $\overline{\mbox{MS}}$ scheme. 
Note that the value of the charm-quark mass is given at $\mu=m_b$. The 
ratio $s_c=(m_c/m_b)^2$ needed for the calculation of the penguin 
contributions is scale independent. The values of the light quark masses 
are such that $r_\chi^K=r_\chi^\pi$. Finally, the value of the QCD scale 
parameter corresponds to $\alpha_s(M_Z)=0.118$ for the two-loop running
coupling in the $\overline{\rm MS}$ scheme. The corresponding 
results for the Wilson coefficients $C_i$ are tabulated in \cite{BBNS3}.

As discussed in detail in \cite{BBNS3}, there are large theoretical 
uncertainties related to the modeling of power corrections corresponding 
to weak annihilation effects and the chirally-enhanced power corrections 
to hard spectator scattering. As in our earlier work we parameterize 
these effects in terms of the divergent integrals $X_H$ (hard spectator 
scattering) and $X_A$ (weak annihilation) introduced in (\ref{XHdef}) and 
(\ref{XAdef}). We model these quantities by using the parameterization
\begin{equation}\label{XHparam}
   X_A = \left( 1 + \varrho_A\,e^{i\varphi_A} \right)
   \ln\frac{m_B}{\Lambda_h} \,; \qquad
   \varrho_A \le 1 \,, \qquad \Lambda_h=0.5\,\mbox{GeV} \,,
\end{equation}
and similarly for $X_H$. Here $\varphi_A$ is an arbitrary 
strong-interaction phase, which may be caused by soft rescattering. In 
other words, we assign a 100\% uncertainty to the ``default value'' 
$X_A=\ln(m_M/\Lambda_h)\approx 2.4$. Unless otherwise stated we assume 
that $X_A$ and $X_H$ are universal for all decay processes. Finally, in 
the evaluation of the hard-scattering and annihilation terms we evaluate 
the running coupling constant and the Wilson coefficients at an 
intermediate scale $\mu_h\sim(\Lambda_{\rm QCD}\,m_b)^{1/2}$ rather than 
$\mu\sim m_b$. Specifically, we use $\mu_h=\sqrt{\Lambda_h\,\mu}$.

\boldmath
\section{Analysis strategy for $B^-$ and $\bar B^0$ decays}
\unboldmath
\label{sec:strategy}

We are now in a position to discuss the phenomenological implications 
of our results, and to compare them to the experimental data compiled in 
Appendix~C. Unless otherwise stated, all results for branching fractions
and decay rates refer to an average over CP-conjugate modes, i.e., a 
result for $\mbox{Br}(\bar B\to\bar f)$ actually refers to 
\begin{equation}
   \frac12 \left[ \mbox{Br}(\bar B\to\bar f)+\mbox{Br}(B\to f) \right] .
\end{equation}
Our convention for the direct CP asymmetry follows the standard 
``$\bar B$ minus $B$'' convention
\begin{equation}\label{acpdef}
   A_{\rm CP}(\bar f\,) \equiv
   \frac{\mbox{Br}(\bar B^0\to\bar f) - \mbox{Br}(B^0\to f)}
        {\mbox{Br}(\bar B^0\to\bar f) + \mbox{Br}(B^0\to f)} \,,
\end{equation}
which is opposite to the sign convention employed in our previous paper 
\cite{BBNS3}.

\subsection{Outline}

In the following sections we will compare the data with our theoretical 
results. Besides presenting 
our default values for the branching fractions and CP asymmetries, along 
with a detailed estimation of the various sources of theoretical 
uncertainty, we will consider a series of scenarios of specific parameter 
sets which elucidate correlations between different quantities and their 
sensitivity to hadronic parameters. Sections~\ref{sec:penguins} to 
\ref{sec:etas} deal with decays of $B^-$ and $\bar B^0$ mesons (and their 
CP conjugates), while $B_s$ decays are treated in Section~\ref{sec:Bs}. 

The decay modes are categorized according to which flavor topology (tree, 
penguin, or annihilation) gives the dominant contribution to the 
amplitude. The $\Delta S=1$ decay modes are always penguin dominated 
and are discussed first, in Section~\ref{sec:penguins}. They typically 
have ``large'' branching ratios of order few times $10^{-6}$ to few times 
$10^{-5}$. The theoretical predictions for these modes often suffer from
large uncertainties due to the strange-quark mass and due to power 
corrections contributing to the weak annihilation coefficient $\beta_3^c$, 
which is part of the dominant amplitude 
$\hat\alpha_4^c=\alpha_4^c+\beta_3^c$. These two sources of uncertainty 
are almost fully correlated between the various decay channels, since 
they always contribute to the penguin coefficients $\hat\alpha_4^p$. We 
illustrate methods that allow one to extract from data the 
magnitudes and strong phases of the dominant penguin contributions in the 
three topologies for which $M_1 M_2=PP$, $PV$, and $VP$. Once the 
measurements become more precise, it will be possible to reduce 
significantly the corresponding correlated theoretical uncertainties. For 
the time being we consider different scenarios for the unknown 
annihilation contributions as representatives for modifications of the 
penguin amplitude  $\hat\alpha_4^c$ away from our default parameter 
choices. There are also some $\Delta D=1$ decays which are penguin or 
annihilation dominated. In the last part of Section~\ref{sec:penguins} we 
show how they can be used to derive constraints on annihilation parameters.

Most of the $\Delta D=1$ decay modes are dominated by tree amplitudes,
which do not 
suffer from a large sensitivity to light quark masses or chirally-enhanced 
power corrections. These decays are studied in Section~\ref{sec:trees}. In 
many cases the dominant theoretical uncertainty arises from the variation 
of CKM parameters or form factors and decay constants. The latter source 
of uncertainty can be reduced once better data on semileptonic or leptonic 
$B$ decays become available. The sensitivity to CKM parameters is not a 
theoretical limitation but rather provides access to $|V_{ub}|$ and 
$\gamma$, which is important for CKM fits.

We omit from our discussion in Sections~\ref{sec:penguins} and 
\ref{sec:trees} the decays with an $\eta$ or $\eta'$ meson in the final 
state. As discussed in \cite{Beneke:2002jn} these modes are characterized 
by a complicated interplay of many different flavor topologies and suffer 
from additional large theoretical uncertainties due to $\eta$--$\eta'$ 
mixing, the two-gluon component in the $\eta^{(\prime)}$ wave functions, 
and an annihilation contribution to the $B\to\eta^{(\prime)}$ semileptonic 
form factors. We will, however, consider modes with an $\omega$ or $\phi$ 
meson in the final states, taking the flavor wave functions 
$\omega=\omega_q\sim(\bar u u+\bar d d)/\sqrt2$ and 
$\phi=\phi_s\sim\bar s s$ corresponding to ideal mixing, and neglecting 
singlet annihilation contributions (which for vector mesons vanish within
our approximations). 
The large number of final states with $\eta$ or $\eta^\prime$ mesons is 
then analyzed briefly in Section~\ref{sec:etas}. The decays with 
additional pseudoscalar or vector kaons have already been discussed in 
\cite{Beneke:2002jn}. Here we give the results for the complete set,  
concentrating on $B^-\to\pi^-\eta^{(\prime)}$ and  
$B^-\to\rho^-\eta^{(\prime)}$ decay, which have sizable branching fractions.

\subsection{Simplified expressions for the decay amplitudes}
\label{sec:simpleampls}

Appendix~A contains the exact expressions for all 96 decay amplitudes in 
terms of the flavor parameters $\alpha_i$ and $\beta_i$. While these
results (along with the expressions for the flavor operators collected in 
Section~\ref{subsec:coefs}) are used in our numerical evaluations, for a 
phenomenological analysis is will be very useful to have simpler,
approximate expressions at hand, which capture the dominant contributions
to the amplitudes. These can be obtained by making the following 
approximations:
\begin{itemize}
\item
{\em $\Delta S=1$ Decays:~}
We neglect annihilation contributions proportional to $\lambda_u^{(s)}$, 
since they are strongly CKM suppressed with respect to the corresponding 
terms proportional to $\lambda_c^{(s)}$. This amounts to setting $\beta_1$, 
$\beta_2$, $\beta_3^u$, and $\beta_4^u$ to zero in Appendix~A. Of the 
electroweak penguin contributions we only keep $\alpha_{3,{\rm EW}}^c$, 
since all other electroweak penguin terms are strongly suppressed. We 
neglect in particular all electroweak annihilation contributions. 
\item 
{\em $\Delta D=1$ Decays:~}
We neglect all electroweak contributions, since they are never CKM 
enhanced and formally of order $\alpha$, hence most likely smaller than 
unknown QCD corrections of order $\alpha_s^2$.
\end{itemize}
Within these approximations the various contributions to the decay 
amplitudes can be classified as tree topologies ($\alpha_1$ and 
$\alpha_2$), penguin topologies ($\alpha_3^p$ and $\alpha_4^p$), an 
electroweak penguin topology ($\alpha_{3,{\rm EW}}^c$), and annihilation 
topologies ($\beta_i$). The penguin annihilation contribution $\beta_3^p$ 
is always combined with the penguin contribution $\alpha_4^p$ into the 
combination $\hat\alpha_4^p=\alpha_4^p+\beta_3^p$. The electroweak 
penguin contribution is kept only for $\Delta S=1$ decays.

\boldmath
\subsection{The $B^-\to\pi^-\pi^0$ tree amplitude}
\unboldmath
\label{sec:treeamplitude}

Before discussing the various penguin and tree-dominated $B$-meson decay 
modes it is instructive to consider the process $B^-\to\pi^-\pi^0$, which 
among the charmless modes discussed here is the single pure tree decay 
(within the approximations mentioned above). For the corresponding 
CP-averaged branching ratio we obtain 
\begin{equation}
   10^6\,\mbox{Br}(B^-\to\pi^-\pi^0) = (6.1^{\,+1.1}_{\,-0.7})
   \times\left[\frac{|V_{ub}|}{0.0037}\,\frac{F_0^{B\to\pi}(0)}{0.28}
   \right]^2 \,,
\end{equation}
where the largest sources of uncertainty come from the 
parameter $\lambda_B$ of the $B$-meson distribution amplitude, the second 
Gegenbauer moment of the pion distribution amplitude, and the quantity 
$X_H$. This value can be compared with the experimental result 
$\mbox{Br}(B^-\to\pi^-\pi^0)=(5.3\pm 0.8)\cdot 10^{-6}$. The good 
agreement of the central values indicates that the magnitude of the tree 
amplitude is obtained naturally with the default set of parameters. 
However, given the substantial uncertainty in the overall normalization 
due to $|V_{ub}|$ and $F_0^{B\to\pi}(0)$, a similarly good agreement 
could also result if the amplitude coefficients $\alpha_{1,2}(\pi\pi)$ 
were to take values rather different from our expectations
$|\alpha_1(\pi\pi)|=0.99^{\,+0.04}_{\,-0.07}$ and 
$|\alpha_2(\pi\pi)|=0.20^{\,+0.17}_{\,-0.11}$.

The comparison could be made independent of the values of $|V_{ub}|$ and 
$F_0^{B\to\pi}(0)$ if the semileptonic $B\to\pi\,l\,\nu$ rate were 
measured near $q^2=0$. One could then perform a direct measurement of the 
tree coefficients via the ratio \cite{Bjorken:kk,Neubert:2001ev} 
\begin{equation}\label{a1a2}
   \frac{\Gamma(B^-\to\pi^-\pi^0)}
        {d\Gamma(\bar B^0\to\pi^+ l^-\bar\nu)/dq^2\big|_{q^2=0}}
   = 3\pi^2 f_\pi^2\,|V_{ud}|^2\,|\alpha_1(\pi\pi)+\alpha_2(\pi\pi)|^2 \,.
\end{equation}
In QCD factorization we find 
$|\alpha_1(\pi\pi)+\alpha_2(\pi\pi)|=1.17_{\,-0.07}^{\,+0.11}$,
which yields the value $(0.66_{\,-0.08}^{\,+0.13})$\,GeV$^2$ for the 
above ratio. 

The CLEO collaboration \cite{Athar:2003yg} has measured the semileptonic 
decay spectrum in three $q^2$ bins. Using their result for the lowest bin,
\begin{equation}
   \int_0^{8\,\rm GeV^2} \hspace*{-0.3cm} dq^2\,\frac{d\mbox{Br}}{dq^2}
   (\bar B^0\to\pi^+ l^-\bar\nu) = (4.31\pm 1.06) \cdot 10^{-5} \,,
\end{equation}
and making the assumption (fulfilled in all form-factor models) 
$F_+^{B\to\pi}(q^2)>F_+^{B\to\pi}(0)$ for $q^2<8\,\mbox{GeV}^2$, we 
obtain the upper bound
\begin{equation}
   |V_{ub}|\,F_+^{B\to\pi}(0) < \sqrt{6.05\cdot 10^{-5}}\left(
   \tau(B_d) \,\frac{G_F^2}{24\pi^3}\int_0^{8\,\rm GeV^2}
   \hspace*{-0.3cm} dq^2\,|\vec{p}_\pi|^3\right)^{\!-1/2}
   = 1.22\cdot 10^{-3}
\end{equation}
at $90\%$ confidence level, which is already close to our default input 
value  $(1.03\pm 0.30)\cdot 10^{-3}$. (The central experimental value 
would give the upper bound $1.03\,\cdot 10^{-3}$.) With 
$\mbox{Br}(B^-\to\pi^-\pi^0)> 4.0\cdot 10^{-6}$ at $90\%\,$CL we find 
\begin{equation}
   |\alpha_1(\pi\pi) + \alpha_2(\pi\pi)|
   > 0.87 \quad (90\%\,\mbox{CL}) \,,
\end{equation}
or $>1.10$ for central values. Although the bound at $90\%\,$CL is not 
yet in an interesting range, the limit obtained using central values 
shows the potential of the ratio (\ref{a1a2}) for constraining 
$\alpha_{1,2}(\pi\pi)$ directly. This underlines the importance of the 
semileptonic decay spectrum for understanding the pattern of 
non-leptonic $\pi\pi$ final states. 

Significantly stronger limits can already be obtained if one assumes a 
model for the $q^2$ dependence of the form factor, which may be guided 
by QCD sum-rule calculations or lattice data at large $q^2$. A fit of 
such form factor models to the CLEO decay spectrum and lattice data has 
recently been performed \cite{Luo:2003hn}, resulting in 
\begin{equation}
   |V_{ub}|\,F_+^{B\to\pi}(0) = (0.83\pm 0.16)\cdot 10^{-3}
\end{equation}
and a small form factor $F_+^{B\to\pi}(0)=0.23\pm 0.04$. This would
require a sizable value of $|\alpha_1(\pi\pi)+\alpha_2(\pi\pi)|$ to 
account for the $B^-\to\pi^-\pi^0$ branching fraction. 

A scenario where $\alpha_2(\pi\pi)$ is large may be of interest, since 
the current experimental value of the ratio 
\begin{equation}
   \frac{\mbox{Br}(\bar B^0\to\pi^-\pi^+)}{\mbox{Br}(B^-\to\pi^-\pi^0)}
   = 0.86\pm 0.15
\end{equation}
is not in good agreement with the QCD factorization result 
$1.47^{\,+0.37}_{\,-0.43}$ (with $\gamma=70^\circ$ and 
$|V_{ub}/V_{cb}|=0.09$ fixed). This is often taken as an indication for 
destructive interference of tree and penguin contributions to 
$\bar B^0\to\pi^-\pi^+$, implying a large value of $\gamma$ or a large 
strong phase, in contradiction with factorization. However, another 
possibility could be that $\bar B^0\to\pi^-\pi^0$ is enhanced relative to 
$\bar B^0\to\pi^-\pi^+$ because the color-suppressed tree amplitude 
$\alpha_2(\pi\pi)$ is large, of order $0.4-0.5$. To keep the 
$\bar B^0\to\pi^-\pi^0$ branching ratio in agreement with its 
experimental value we must then assume that the $B\to \pi$ form factor
$F_0^{B\to \pi}(0)$ is about 0.25, on the lower edge of the range of model 
predictions, but in agreement with the trend indicated by the 
semileptonic decay spectrum. Below, when we define a set of parameter 
scenarios different from the default parameters to exhibit correlations 
among the decay modes, we shall therefore consider this 
``large-$\alpha_2$'' scenario as one of our options. 

This scenario is not as contrived as it may seem, since it can be 
realized by simply taking the the second Gegenbauer moment of the pion to 
be $\alpha_2^\pi=0.4$ and $\lambda_B=200\,$MeV, both at the boundary of our 
parameter region. Indeed, there is no rigorous information available on 
the parameter $\lambda_B$ of the $B$-meson distribution amplitude, and 
although the light-cone distribution amplitude of the pion is now believed 
to be close to the asymptotic form already at small scales (disfavoring a 
large Gegenbauer moment $\alpha_2^\pi$), the calculations of the 
low-energy hadronic processes from which this information is extracted 
may have significant theoretical uncertainties. Furthermore, the 
heavy-to-light form factors at $q^2=0$ are usually taken from QCD sum-rule 
calculations, but it now appears that these form factors exhibit a 
substantially more complex dynamics than $B\to D$ form factors or pion 
transition form factors. It remains to be investigated whether the 
dynamics of these transitions at large momentum transfer is adequately 
represented in the sum-rule framework.

\section{Penguin-dominated decays}
\label{sec:penguins}

All $B^-$ and $\bar B^0$ two-body decays with kaons in the final state 
are dominated by penguin amplitudes. The modes with a single kaon are 
based on $\Delta S=1$ transitions and have ``large'' branching ratios of 
order few times $10^{-6}$ to few times $10^{-5}$. Modes with two kaons 
proceed through $\Delta D=1$ transitions and therefore are expected to 
have smaller branching fractions. In this section we focus on decays
without $\eta$ or $\eta'$ mesons in the final state. These modes will be
studied in Section~\ref{sec:etas}.

A successful prediction for the branching ratios and CP asymmetries in 
rare $B$ decays requires theoretical control over the magnitudes and 
relative strong-interaction phases of tree topologies, penguin 
topologies, electroweak penguin topologies, and annihilation topologies. 
The fact that generically a given decay amplitudes receives several 
interfering contributions complicates the interpretation of the 
experimental data in terms of flavor topologies. We expect that our 
predictions for annihilation effects (as modeled by the complex 
parameters $\varrho_A$ and $\varphi_A$) and for the strong phases of the 
various contributions are
afflicted by the largest theoretical uncertainties. The former effects 
are incalculable within QCD factorization and so can only be estimated 
using a simple model. Strong phases are predicted  
to vanish in the heavy-quark limit. The leading contributions to these 
phases are then of order $\alpha_s$ and calculable, or 
of order $\Lambda/m_b$ and incalculable. Since in practice the two 
expansion parameters $\alpha_s$ and $\Lambda/m_b$ are 
not very different, we expect that our perturbative analysis at one-loop
order can only give a rough estimate of the values of strong phases. 
While we do trust the generic prediction that these phases are small, we 
expect significant deviations from the values obtained at order 
$\alpha_s$ from higher-order perturbative and power corrections. This 
expectation is indeed supported by an analysis of hadronic $B$ decays
using the renormalon calculus \cite{Becher:2001hu}.

Given these limitations, it is instructive to test the 
reliability of our predictions for the tree, penguin, and annihilation
topologies in decays without interference of the different topologies. 
This will probe the magnitude of these topologies but not their 
strong-interaction phases. Note that the penguin and annihilation terms
cannot be completely separated phenomenologically, 
since $\alpha_4^p$ and $\beta_3^p$ 
always enter in the combination $\hat\alpha_4^p=\alpha_4^p+\beta_3^p$. 
However, some other annihilation contributions can be probed directly.

A discussion of the tree contribution to the decay $B^-\to\pi^-\pi^0$
has already been given in Section~\ref{sec:treeamplitude}.
In the following we present a detailed study of penguin and 
annihilation terms, using penguin-dominated decay processes based on
$\Delta S=1$ and $\Delta D=1$ transitions. For each class of decays
we give explicit expressions for the decay amplitudes in terms
of flavor parameters, adopting the approximations described earlier.
We then present numerical predictions
for the CP-averaged branching fractions and CP asymmetries and compare
them with the world-average experimental data compiled in 
Appendix~C.
These results are always obtained using the exact representations of
the decay amplitudes in terms of flavor parameters as given in Appendix~A. 

In the tables we first present our ``default results'' (column labeled
``Theory'') along with detailed error estimates corresponding to the 
different types of theoretical
uncertainties detailed in Section~\ref{sec:inputs}. The first error shown 
corresponds to the variation of the CKM parameters $|V_{cb}|$, $|V_{ub}|$, 
and $\gamma$ (``CKM''), the second error refers to the variation of the 
renormalization scale, quark masses, decay constants (except for 
transverse ones), form factors, and (in later sections) the 
$\eta$--$\eta'$ mixing angle (``hadronic 1''). The third error 
corresponds to the uncertainty due to the Gegenbauer moments in the 
expansion of the light-cone distribution amplitudes, and also 
includes the scale-dependent transverse decay constants for vector mesons
(``hadronic 2''). Finally, the last error reflects our estimate of power 
corrections parameterized by the quantities $X_A$ and $X_H$ (``power''), 
for which we adopt the form (\ref{XHparam}) with $\varrho_{A,H}\le 1$ and 
arbitrary strong phases $\varphi_{A,H}$. 

In order to illustrate correlations between errors we explore four
parameter scenarios in which certain parameters are changed within their 
error ranges while all others take their 
default values. Specifically, these scenarios are defined as follows:
\begin{itemize}
\item
Scenario S1 (``large $\gamma$''):\\
To study the dependence of the various observables on the CP-violating
phase $\gamma$ we use the default parameter values
but set $\gamma=110^\circ$ instead of $70^\circ$.
\item
Scenario S2 (``large $\alpha_2$''):\\
As discussed in Section~\ref{sec:treeamplitude},
the experimental branching ratios for $\bar B\to\pi\pi$ modes can be
reproduced in QCD factorization either by using a large value of 
$\gamma>90^\circ$, or by enhancing the ratio 
$\alpha_2(\pi\pi)/\alpha_1(\pi\pi)$ with respect to its default value of
about 0.2. This can be done using  the form factors 
$F_0^{B\to\pi}(0)=0.25$ and $F_0^{B\to K}(0)=0.31$, the 
Standard Model parameters $|V_{ub}/V_{cb}|=0.08$ and $m_s=70$\,MeV, 
as well as the Gegenbauer moments $\lambda_B=200$\,MeV and 
$\alpha_2^\pi=0.4$, all of which are within the error ranges specified
earlier.
\item
Scenario S3 (``universal annihilation''):\\
The dominant penguin coefficient $\hat\alpha_4^p$, which includes the
annihilation contribution $\beta_3^p$, is rather sensitive to the 
parameter $X_A$. While $B$ decays into two 
pseudoscalar mesons are well described by a small annihilation 
contribution, certain $B\to PV$ decay amplitudes favor a penguin 
coefficient that is larger than its default value. Here we 
adopt the simplest model in which a moderate enhancement of 
$\hat\alpha_4^p$ is attributed to weak annihilation. Specifically, we 
treat $X_A$ as a universal parameter obtained by using $\varrho_A=1$ and 
a strong phase $\varphi_A=-45^\circ$. The sign of this phase is not 
predicted, but is chosen such that the sign of the direct CP asymmetry 
$A_{\rm CP}(\pi^+ K^-)$ agrees with the data.
\item
Scenario S4 (``combined''):\\
Here we consider a combination of S2 and S3, but with the more moderate 
parameter values $m_s=80$\,MeV and $\alpha_2^\pi=0.3$, as well as 
non-universal annihilation phases $\varphi_A=-55^\circ$ ($PP$), 
$\varphi_A=-20^\circ$ ($PV$), and $\varphi_A=-70^\circ$ ($VP$). The signs 
of these phases are not predicted. As a result, our predictions for the 
signs of CP asymmetries in this scenario must be taken with caution. 
While each of the previous scenarios fails to describe well some classes 
of decay modes, scenario S4 is an attempt to combine certain parameter 
choices (all within our theoretical error ranges) in such as way as to
obtain a good description of all currently available experimental 
data. In view of this, this is our currently favored scenario.
\end{itemize}
In Section~\ref{sec:largeXA} we will also study a modification of scenario
S3 in which we use the large value $\varrho_A=2$ with a universal phase 
$\varphi_A=-60^\circ$.

\boldmath
\subsection{Penguin-dominated $\Delta S=1$ decays}
\unboldmath

We start by presenting in Tables~\ref{tab:pred1BR} and \ref{tab:pred1CP} 
our results for the CP-averaged branching fractions and direct CP 
asymmetries for the decays $\bar B\to\pi\bar K$, 
$\bar B\to\pi\bar K^*$, $\bar B\to\bar K\rho$, as well as for the modes
$\bar B\to\bar K\omega$ and $\bar B\to\bar K\phi$. The 
$\bar B\to\pi\bar K$ modes have the largest branching fractions, of order
(1--2)$\cdot 10^{-5}$. The data show that the corresponding rates for the 
$PV$ modes $\bar B\to\pi\bar K^*$ and $\bar B\to\bar K\rho$ are smaller by 
about a factor of two, indicating a sizable suppression of the penguin
amplitudes in the cases with a final-state vector meson. QCD factorization
predicts small direct CP asymmetries for most decay modes, however with a 
few exceptions. All predictions for CP asymmetries agree with 
the data within errors.

\begin{table}[t]
\centerline{\parbox{14cm}{\caption{\label{tab:pred1BR}
CP-averaged branching ratios (in units of $10^{-6}$) of penguin-dominated 
$\bar B\to PP$ decays (top) and $\bar B\to PV$ decays (bottom) with 
$\Delta S=1$. The theoretical errors shown in the first column correspond 
(in this order) to the uncertainties referred to as ``CKM'', 
``hadronic~1'', ``hadronic~2'', and ``power'' in the text. The numbers
shown in the remaining columns correspond to different scenarios of 
hadronic parameters explained in the text.}}}
\vspace{0.1cm}
\begin{center}
\begin{tabular}{|l|c|cccc|c|}
\hline\hline
\multicolumn{1}{|c|}{Mode} & Theory & S1 & S2 & S3 & S4 & Experiment \\
\hline\hline
$B^-\to\pi^-\bar K^0$
 & $19.3_{\,-1.9\,-\phantom{1}7.8\,-2.1\,-\phantom{1}5.6}^{\,+1.9\,+11.3\,+1.9\,+13.2}$
 & 18.8 & 20.7 & 24.8 & 20.3
 & $20.6\pm 1.3$ \\
$B^-\to\pi^0 K^-$
 & $11.1_{\,-1.7\,-4.0\,-1.0\,-3.0}^{\,+1.8\,+5.8\,+0.9\,+6.9}$
 & 14.0 & 11.9 & 14.0 & 11.7
 & $12.8\pm 1.1$ \\
$\bar B^0\to\pi^+ K^-$
 & $16.3_{\,-2.3\,-6.5\,-1.4\,-\phantom{1}4.8}^{\,+2.6\,+9.6\,+1.4\,+11.4}$
 & 20.3 & 18.8 & 21.0 & 18.4
 & $18.2\pm 0.8$ \\
$\bar B^0\to\pi^0\bar K^0$
 & $7.0_{\,-0.7\,-3.2\,-0.7\,-2.3}^{\,+0.7\,+4.7\,+0.7\,+5.4}$
 & 6.5 & 8.3 & 9.3 & 8.0
 & $11.2\pm 1.4$ \\
\hline\hline
$B^-\to\pi^-\bar K^{*0}$
 & $3.6_{\,-0.3\,-1.4\,-1.2\,-2.3}^{\,+0.4\,+1.5\,+1.2\,+7.7}$
 & 3.4 & 2.2 & 7.3 & 8.4
 & $13.0\pm 3.0$ \\
$B^-\to\pi^0 K^{*-}$
 & $3.3_{\,-1.0\,-0.9\,-0.6\,-1.4}^{\,+1.1\,+1.0\,+0.6\,+4.4}$
 & 5.5 & 2.6 & 5.4 & 6.5
 & $<31$ \\
$\bar B^0\to\pi^+ K^{*-}$
 & $3.3_{\,-1.2\,-1.2\,-0.8\,-1.6}^{\,+1.4\,+1.3\,+0.8\,+6.2}$
 & 5.9 & 2.4 & 6.6 & 8.1
 & $15.3\pm 3.8$ \\
$\bar B^0\to\pi^0\bar K^{*0}$
 & $0.7_{\,-0.1\,-0.4\,-0.3\,-0.5}^{\,+0.1\,+0.5\,+0.3\,+2.6}$
 & 0.6 & 0.4 & 2.1 & 2.5
 & $<3.6$ \\
\hline
$B^-\to\bar K^0\rho^-$
 & $5.8_{\,-0.6\,-3.3\,-1.3\,-\phantom{1}3.2}^{\,+0.6\,+7.0\,+1.5\,+10.3}$
 & 5.6 & 13.6 & 10.8 & 9.7
 & $<48$ \\
$B^-\to K^-\rho^0$
 & $2.6_{\,-0.9\,-1.4\,-0.6\,-1.2}^{\,+0.9\,+3.1\,+0.8\,+4.3}$
 & 1.3 & 6.0 & 4.7 & 4.3
 & $<6.2$ \\
$\bar B^0\to K^-\rho^+$
 & $7.4_{\,-1.9\,-3.6\,-1.1\,-\phantom{1}3.5}^{\,+1.8\,+7.1\,+1.2\,+10.7}$
 & 4.3 & 13.9 & 12.5 & 10.1
 & $8.9\pm 2.2$ \\
$\bar B^0\to\bar K^0\rho^0$
 & $4.6_{\,-0.5\,-2.1\,-0.7\,-2.1}^{\,+0.5\,+4.0\,+0.7\,+6.1}$
 & 5.0 & 8.4 & 7.5 & 6.2
 & $<12$ \\
\hline
$B^-\to K^-\omega$
 & $3.5_{\,-1.0\,-1.6\,-0.9\,-1.6}^{\,+1.0\,+3.3\,+1.4\,+4.7}$
 & 1.9 & 7.9 & 5.8 & 5.9
 & $5.3\pm 0.8$ \\
$\bar B^0\to\bar K^0\omega$
 & $2.3_{\,-0.3\,-1.3\,-0.8\,-1.3}^{\,+0.3\,+2.8\,+1.3\,+4.3}$
 & 1.9 & 6.6 & 4.5 & 4.9
 & $5.1\pm 1.1$ \\
$B^-\to K^-\phi$
 & $4.5_{\,-0.4\,-1.7\,-2.1\,-\phantom{1}3.3}^{\,+0.5\,+1.8\,+1.9\,+11.8}$
 & 4.4 & 2.5 & 10.1 & 11.6
 & $9.2\pm 1.0$ \\
$\bar B^0\to\bar K^0\phi$
 & $4.1_{\,-0.4\,-1.6\,-1.9\,-\phantom{1}3.0}^{\,+0.4\,+1.7\,+1.8\,+10.6}$
 & 4.0 & 2.3 & 9.1 & 10.5
 & $7.7\pm 1.1$ \\
\hline\hline
\end{tabular}
\end{center}
\end{table}

\begin{table}[t]
\centerline{\parbox{14cm}{\caption{\label{tab:pred1CP}
Direct CP asymmetries (in units of $10^{-2}$) of penguin-dominated 
$\bar B\to PP$ decays (top) and $\bar B\to PV$ decays (bottom) with 
$\Delta S=1$. See text for explanations.}}}
\vspace{0.1cm}
\begin{center}
\begin{tabular}{|l|c|cccc|c|}
\hline\hline
\multicolumn{1}{|c|}{Mode} & Theory & S1 & S2 & S3 & S4 & Experiment \\
\hline\hline
$B^-\to\pi^-\bar K^0$
 & $0.9_{\,-0.3\,-0.3\,-0.1\,-0.5}^{\,+0.2\,+0.3\,+0.1\,+0.6}$
 & 0.9 & 0.8 & 0.4 & 0.3
 & $-2\pm 9$ \\
$B^-\to\pi^0 K^-$
 & $7.1_{\,-1.8\,-2.0\,-0.6\,-9.7}^{\,+1.7\,+2.0\,+0.8\,+9.0}$
 & 5.7 & 6.3 & $-1.3$ & $-3.6$
 & $1\pm 12$ \\
$\bar B^0\to\pi^+ K^-$
 & $4.5_{\,-1.1\,-2.5\,-0.6\,-9.5}^{\,+1.1\,+2.2\,+0.5\,+8.7}$
 & 3.6 & 3.0 & $-3.6$ & $-4.1$
 & $-9\pm 4$ \\
$\bar B^0\to\pi^0\bar K^0$
 & $-3.3_{\,-0.8\,-1.6\,-1.0\,-3.3}^{\,+1.0\,+1.3\,+0.5\,+3.4}$
 & $-3.5$ & $-3.6$ & $-1.2$ & 0.8
 & $3\pm 37$ \\
\hline\hline
$B^-\to\pi^-\bar K^{*0}$
 & $1.6_{\,-0.5\,-0.5\,-0.4\,-1.0}^{\,+0.4\,+0.6\,+0.5\,+2.5}$
 & 1.7 & 1.6 & 0.8 & 0.8
 & --- \\
$B^-\to\pi^0 K^{*-}$
 & $8.7_{\,-2.6\,-4.3\,-3.4\,-44.2}^{\,+2.1\,+5.0\,+2.9\,+41.7}$
 & 5.2 & 8.7 & $-20.4$ & $-6.5$
 & --- \\
$\bar B^0\to\pi^+ K^{*-}$
 & $2.1_{\,-0.7\,-7.9\,-5.8\,-64.2}^{\,+0.6\,+8.2\,+5.1\,+62.5}$
 & 1.2 & 1.7 & $-33.8$ & $-12.1$
 & $26\pm 35$ \\
$\bar B^0\to\pi^0\bar K^{*0}$
 & $-12.8_{\,-3.2\,-7.0\,-4.0\,-35.3}^{\,+4.0\,+4.7\,+2.7\,+31.7}$
 & $-15.5$ & $-17.7$ & 1.5 & 1.0
 & --- \\
\hline
$B^-\to\bar K^0\rho^-$
 & $0.3_{\,-0.1\,-0.4\,-0.1\,-1.3}^{\,+0.1\,+0.3\,+0.2\,+1.6}$
 & 0.3 & 0.4 & 0.4 & 0.8
 & --- \\
$B^-\to K^-\rho^0$
 & $-13.6_{\,-5.7\,-4.4\,-3.1\,-55.4}^{\,+4.5\,+6.9\,+3.7\,+62.7}$
 & $-27.3$ & $-9.3$ & 26.6 & 31.7
 & --- \\
$\bar B^0\to K^-\rho^+$
 & $-3.8_{\,-1.4\,-2.7\,-1.6\,-32.7}^{\,+1.3\,+4.4\,+1.9\,+34.5}$
 & $-6.6$ & $-2.6$ & 18.3 & 20.0
 & $26\pm 15$ \\
$\bar B^0\to\bar K^0\rho^0$
 & $7.5_{\,-2.1\,-2.0\,-0.4\,-8.7}^{\,+1.7\,+2.3\,+0.7\,+8.8}$
 & 6.9 & 5.5 & 1.4 & $-2.8$
 & --- \\
\hline
$B^-\to K^-\omega$
 & $-7.8_{\,-3.0\,-3.6\,-1.9\,-38.0}^{\,+2.6\,+5.9\,+2.4\,+39.8}$
 & $-14.5$ & $-5.5$ & 18.4 & 19.3
 & $0\pm 12$ \\
$\bar B^0\to\bar K^0\omega$
 & $-8.1_{\,-2.0\,-3.3\,-1.4\,-12.9}^{\,+2.5\,+3.0\,+1.7\,+11.8}$
 & $-9.6$ & $-4.3$ & 0.0 & 3.7
 & --- \\
$B^-\to K^-\phi$
 & $1.6_{\,-0.5\,-0.5\,-0.3\,-1.2}^{\,+0.4\,+0.6\,+0.5\,+3.0}$
 & 1.6 & 1.7 & 0.6 & 0.7
 & $3\pm 7$ \\
$\bar B^0\to\bar K^0\phi$
 & $1.7_{\,-0.5\,-0.5\,-0.3\,-0.8}^{\,+0.4\,+0.6\,+0.5\,+1.4}$
 & 1.7 & 1.9 & 0.9 & 0.8
 & $19\pm 68$ \\
\hline\hline
\end{tabular}
\end{center}
\end{table}

Adopting the approximations described in the previous section, the 
$\bar B\to\pi\bar K$ decay amplitudes are given by
\begin{equation}\label{ApiK}
\begin{aligned}
   {\cal A}_{B^-\to\pi^-\bar K^0}
   &= A_{\pi\bar K}\,\hat\alpha_4^p \,, \\
   \sqrt2\,{\cal A}_{B^-\to\pi^0 K^-}
   &= A_{\pi\bar K} \Big[ \delta_{pu}\,\alpha_1 + \hat\alpha_4^p \Big] 
    + A_{\bar K\pi} \Big[ \delta_{pu}\,\alpha_2 
    + \delta_{pc}\,\3half\alpha_{3,{\rm EW}}^c \Big] \,, \\
   {\cal A}_{\bar B^0\to\pi^+ K^-}
   &= A_{\pi\bar K} \Big[ \delta_{pu}\,\alpha_1 + \hat\alpha_4^p \Big]
    , \\
   \sqrt2\,{\cal A}_{\bar B^0\to\pi^0\bar K^0}
   &= A_{\pi\bar K}\,\Big[ - \hat\alpha_4^p \Big]
    + A_{\bar K\pi} \Big[ \delta_{pu}\,\alpha_2 
    + \delta_{pc}\,\3half\alpha_{3,{\rm EW}}^c \Big] \,,
\end{aligned}
\end{equation}
where it is understood that each term must be multiplied with 
$\lambda_p^{(s)}$ and summed over $p=u,c$. The order of the arguments of
the coefficients $\alpha_i^p(M_1 M_2)$ and $\beta_i^p(M_1 M_2)$ is 
determined by the order of the arguments of the $A_{M_1 M_2}$ prefactors.
The expressions for the 
$\bar B\to\pi\bar K^*$ and $\bar B\to\bar K\rho$ amplitudes are obtained 
by replacing $(\pi,\bar K)\to (\pi,\bar K^*)$ and $(\rho,\bar K)$, 
respectively. The remaining amplitudes take the form
\begin{equation}\label{AKw}
\begin{aligned}
   \sqrt2\,{\cal A}_{B^-\to K^-\omega}
   &= A_{\bar K\omega} \Big[ \delta_{pu}\,\alpha_2 + 2\alpha_3^p
    + \delta_{pc}\,\half\alpha_{3,{\rm EW}}^c \Big] 
    + A_{\omega\bar K} \Big[ \delta_{pu}\,\alpha_1 + \hat\alpha_4^p
    \Big] \,, \\
   \sqrt2\,{\cal A}_{\bar B^0\to\bar K^0\omega}
   &= A_{\bar K\omega} \Big[ \delta_{pu}\,\alpha_2 + 2\alpha_3^p
    + \delta_{pc}\,\half\alpha_{3,{\rm EW}}^c \Big] 
    + A_{\omega\bar K}\,\hat\alpha_4^p \,, \\
   {\cal A}_{B^-\to K^-\phi}
   &= A_{\bar K\phi} \Big[ \alpha_3^p + \hat\alpha_4^p
    - \delta_{pc}\,\half\alpha_{3,{\rm EW}}^c \Big] \,, \\
   {\cal A}_{\bar B^0\to\bar K^0\phi}
   &= A_{\bar K\phi} \Big[ \alpha_3^p + \hat\alpha_4^p
    - \delta_{pc}\,\half\alpha_{3,{\rm EW}}^c \Big] \,. 
\end{aligned}
\end{equation}
An important feature of all these modes is that, within our
approximations, annihilation effects enter the decay amplitudes only 
through the effective penguin coefficients 
$\hat\alpha_4^c=\alpha_4^c+\beta_3^c$. As a result, these effects 
can be readily constrained using data. This explains why the 
$\bar B\to\pi\bar K$ decay modes have been employed frequently (by many 
authors and using various analysis strategies) as a way of constraining
the phase $\gamma$. 

Consider first our default results labeled ``Theory''. While the 
experimental results for the $\bar B\to\pi\bar K$ modes are well 
reproduced within errors, it appears that in the case of the
modes $\bar B\to\pi\bar K^*$ and $\bar B\to\bar K\phi$
the central values for the branching fractions obtained from QCD
factorization are consistently lower than the data by a factor 2 to 3.
No significant discrepancy (at the present level of precision) is seen in
the modes $\bar B\to\bar K\rho$ and $\bar B\to\bar K\omega$,
in which the dominant penguin coefficient is $\hat\alpha_4^c(VP)$ instead
of $\hat\alpha_4^c(PV)$ (we may call these modes $B\to VP$). We will come
back to this observation below.

In all cases the theoretical 
predictions suffer from large uncertainties due to the strange-quark mass 
(largest contribution to the second error), and due to power corrections 
contributing to the weak annihilation coefficient $\beta_3^c$, which is 
part of the dominant penguin amplitude $\hat\alpha_4^c$ (largest 
contribution to the last error). We stress, however, that these two 
sources of uncertainty are almost fully correlated between the various 
decay channels, since they always contribute to the penguin coefficients 
$\hat\alpha_4^p$. Below we will discuss how these coefficients can be 
directly determined from the data. Once the measurements become more 
precise it will be possible to reduce significantly the corresponding 
correlated theoretical uncertainties.

As a final comment, let us add that for some observables in the decays 
under consideration here (especially $\bar B\to\bar K\phi$ and 
$\bar B\to\pi\bar K$) the present experimental values are inconsistent 
with Standard Model predictions at the level of 2 to 3 standard 
deviations.
We will discuss these discrepancies at length later. While we will not 
pursue New Physics explanations of these discrepancies in the present
work, the reader should bear in mind that, unless these discrepancies
disappear with more precise data, one might expect large deviations also
in other observables in these decays, such as their CP-averaged branching 
fractions. Therefore, deviations of our predictions from experiment do
not necessarily imply a failure of the QCD factorization approach.

\subsubsection{Magnitudes and phases of penguin coefficients}
\label{subsec:PTratios}

The decay amplitudes for the three sets of decays $\bar B\to\pi\bar K$, 
$\bar B\to\pi\bar K^*$, and $\bar B\to\rho\bar K$ shown in (\ref{ApiK})
are particularly simple. In the $B\to PP$ modes various ratios of 
CP-averaged decay rates have been used to derive information about the 
weak phase $\gamma$ as well as the strong phase of the ratio 
$\hat\alpha_4^c/\alpha_1$ 
\cite{BBNS3,Gronau_Rosner,Fleischer_Mannel,Neubert_Rosner,Buras_Fleischer,Neubert}.

The amplitudes for the decays $B^-\to\pi^-\bar K^0$, 
$B^-\to\rho^-\bar K^0$, and $B^-\to\pi^-\bar K^{*0}$ are determined in
terms of the coefficients $\hat\alpha_4^p$. QCD factorization predicts 
that $\hat\alpha_4^c\approx\hat\alpha_4^u$ to a good approximation. Given 
that $|\lambda_u^{(s)}/\lambda_c^{(s)}|\approx 0.02$, it then follows 
that the contribution with $p=u$ can be neglected at the present level of 
accuracy of the experimental data. (This assertion would be invalidated 
if a significant direct CP asymmetry was found in any of these modes.)
It follows that 
$\Gamma(\bar B^-\to\pi^-\bar K^0)\propto|\hat\alpha_4^c(\pi K)|^2$ and
$A_{\rm CP}(\bar B^-\to\pi^-\bar K^0)\approx 0$ to a very good 
approximation, and similar relations hold for the other two decays.
These decays therefore provide a direct test of the coefficient
$\hat\alpha_4^c$, which contains the penguin annihilation parameter
$\beta_3^c$. These penguin annihilation contributions, as well as potential 
other non-perturbative contributions to the penguin coefficient 
$\hat\alpha_4^c$, are sometimes referred to as ``charming penguins'' 
\cite{Ciuchini:1997hb}.

Uncertainties related to heavy-to-light form factors can be eliminated by
normalizing the pure-penguin rates to the pure-tree rate for the decay
$B^-\to\pi^-\pi^0$ discussed in Section~\ref{sec:treeamplitude}. 
Specifically, the magnitude of the penguin coefficient 
$\hat\alpha_4^c(\pi\bar K)$ in the decays $B\to\pi K$ can be probed by 
considering the ratio
\begin{equation}\label{a4piK}
   \left| \frac{\hat\alpha_4^c(\pi\bar K)}
               {\alpha_1(\pi\pi)+\alpha_2(\pi\pi)} \right|
   = \left| \frac{V_{ub}}{V_{cb}} \right| \frac{f_\pi}{f_K}
   \left[ \frac{\Gamma(B^-\to\pi^-\bar K^0)}
               {2\Gamma(B^-\to\pi^-\pi^0)} \right]^{1/2}\! 
   = 0.103\pm 0.008\pm 0.023 \,,
\end{equation}
where the numerical value uses the experimental measurements of the decay 
rates. The first error is experimental while the 
second reflects the uncertainty in $|V_{ub}|$. This ratio, which is 
inversely proportional to the quantity called 
$\varepsilon_{\rm exp}$ in \cite{BBNS3}, provides a crucial  
test of the QCD factorization approach, since it directly probes 
the magnitude of the dominant penguin-to-tree ratio in $B\to PP$ decays. 
With default parameters, QCD factorization predicts the values
$|\alpha_1(\pi\pi)+\alpha_2(\pi\pi)|=1.17$ and 
$|\alpha_4^c(\pi\bar K)+\beta_3^c(\pi\bar K)|=|(-0.099-0.013i)-0.009|
=0.108$, yielding 0.093 for the ratio of coefficients on the left-hand 
side, which is in excellent agreement with the data. Note that without 
the weak annihilation contribution from $\beta_3^c$ we would obtain 0.085
for this ratio, which is still in good agreement with the data. For 
comparison, this ratio equals 0.061 in naive factorization. It 
follows that the dominant penguin amplitude in $B\to PP$ decays is 
correctly predicted in QCD factorization without any adjustment of
parameters, and that the (incalculable) penguin annihilation contribution
to this amplitude is a small effect, which in fact is consistent with
the magnitude expected for a power correction. Specifically, we find that
$|\beta_3^c/\alpha_4^c|=0.09_{\,-0.09}^{\,+0.32}$. At least for the case 
of $\bar B\to\pi\bar K$ modes there is thus very little room for large 
non-perturbative contributions to the dominant penguin amplitude besides 
those already included in QCD factorization. 

\begin{figure}[t]
\epsfxsize=15cm
\centerline{\epsffile{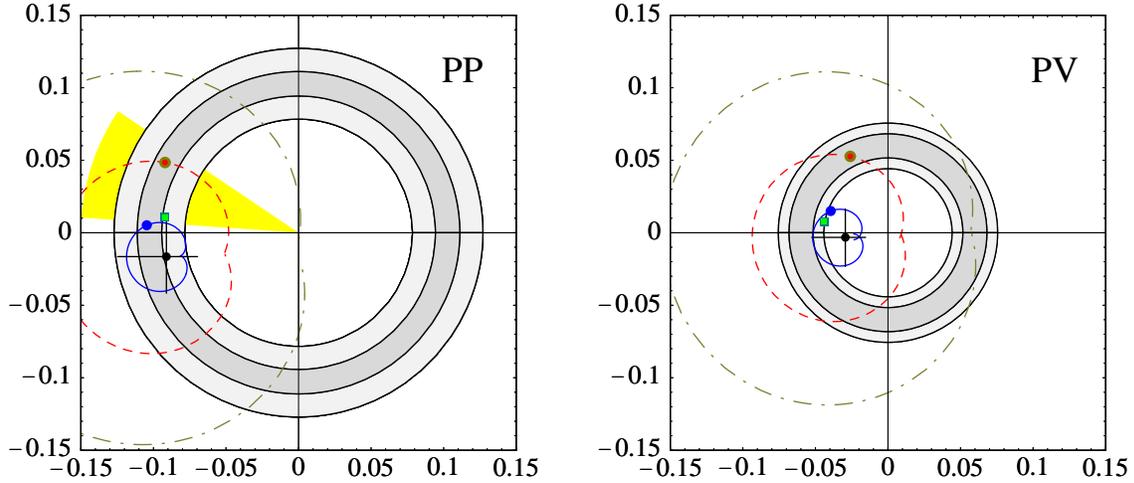}}
\centerline{\parbox{14cm}{\caption{\label{fig:eps32}
Complex ratios $\hat\alpha_4^c(M_1 M_2)/(\alpha_1+\alpha_2)(\pi\pi)$ for 
the cases $M_1 M_2=\pi\bar K$ ($PP$) and $\pi\bar K^*$ ($PV$) (horizontal 
axis = real part, vertical axis = imaginary part). The rings show the 
experimental values of the magnitudes of the two ratios with (light) and 
without (dark) including the uncertainty from $|V_{ub}/V_{cb}|$. The 
three contour lines in each plot correspond to $\varrho_A=1$ (solid), 2 
(dashed), and 3 (dashed-dotted). The triangular-shaped area in the first 
plot shows the constraint on the strong phase derived from the direct CP 
asymmetry in $B^-\to\pi^-\bar K^0$ decays. The various symbols indicate 
theoretical points corresponding to different parameter scenarios: the 
dots on the 
solid contours correspond to S3, the squares correspond to S4, and the 
circles on the dashed contours correspond to the ``large annihilation 
scenario'' with $\varrho_A=2$ and $\varphi_A=-60^\circ$.}}}
\end{figure}

This observation can be used to place a constraint on the quantity
$X_A$ in (\ref{XHparam}), which we use to parameterize our model estimate 
for the annihilation terms. In the first plot in Figure~\ref{fig:eps32} 
we compare our theoretical results for the ratio 
$\hat\alpha_4^c(\pi\bar K)/(\alpha_1+\alpha_2)(\pi\pi)$ with 
the experimental value of its magnitude 
obtained from (\ref{a4piK}). The point with error 
bars gives our central value with errors from parameter variations 
(including $X_A$ and $X_H$). The solid ``onion-shaped'' curve indicates 
the variation allowed with $\varrho_A=1$ and arbitrary annihilation phase 
$\varphi_A$. This is the region allowed by the error analysis of 
\cite{BBNS3}. The larger onion-shaped curves correspond to values of the 
annihilation parameters outside our usual error analysis, namely 
$\varrho_A=2$ (dashed curve) and $\varrho_A=3$ (dashed-dotted curve). They 
are included here to indicate that
for such large values of $\varrho_A$ agreement with the data requires 
fine-tuning of the annihilation phase $\varphi_A$. We will see 
below that such large $\varrho_A$ values are all but excluded by current 
data. In fact, the $\bar B\to\pi\bar K$ decays alone ($PP$ case) do not 
allow for values of $\varrho_A$ above about 2.5 once data from direct CP 
asymmetries are included (triangular shape, see below).

The penguin coefficients in the $B\to PV$ modes $B\to\pi K^*$ and
$B\to K\rho$ can be probed via relations analogous to (\ref{a4piK}). For
the case of $B^-\to\pi^-\bar K^{*0}$ data exist, and the corresponding 
result is shown in the second plot in Figure~\ref{fig:eps32}. In that
case the ratio $f_\pi/f_K$ in (\ref{a4piK}) must be replaced by 
$f_\pi/f_{K^*}$, but there is still no dependence on hadronic form 
factors. We observe that the central value for the magnitude of the 
penguin coefficient $\hat\alpha_4^c(\pi\bar K^*)$ obtained in QCD 
factorization is significantly
smaller than the experimental value, and the two barely overlap when 
theoretical uncertainties are taken into account. Unfortunately,
the branching fraction for the decay $B^-\to\rho^-\bar K^0$ has not yet
been measured. We therefore cannot test whether the penguin coefficient 
in the $B\to VP$ modes $\bar B\to\rho\bar K$ is correctly predicted in QCD
factorization.

It is also instructive to determine ratios of penguin coefficients that 
are independent of $|V_{ub}/V_{cb}|$, using the relations
\begin{equation}
\begin{aligned}
   \left| \frac{\hat\alpha_4^c(\pi\bar K^*)}
               {\hat\alpha_4^c(\pi\bar K)} \right|
   &= \frac{f_K}{f_{K^*}} \left[
    \frac{\Gamma(B^-\to\pi^-\bar K^{*0})}{\Gamma(B^-\to\pi^-\bar K^0)}
    \right]^{1/2}
    = 0.58\pm 0.07 \quad (\mbox{exp.}) \,, \\
   \left| \frac{\hat\alpha_4^c(\rho\bar K)}
               {\hat\alpha_4^c(\pi\bar K)} \right|
   &= \frac{F_0^{B\to\pi}(0)}{A_0^{B\to\rho}(0)} \left[
    \frac{\Gamma(B^-\to\rho^-\bar K^0)}{\Gamma(B^-\to\pi^-\bar K^0)}
    \right]^{1/2} \quad \mbox{(not yet observed)} \,,
\end{aligned}
\end{equation}
which can be used to relate the dominant penguin coefficients 
$\hat\alpha_4^c$ in the $PV$ modes $\bar B\to\pi\bar K^*$ and 
$\bar B\to\rho\bar K$ to the corresponding coefficient in the $PP$ mode 
$\bar B\to\pi\bar K$. While
unfortunately no data are available yet that would allow us to deduce
the relative magnitude of the penguin coefficients in the $B\to PV$ and 
$B\to VP$ modes, the first ratio above provides a first clue about the
magnitude of the penguin coefficient in $B\to PV$ modes, which is seen
to be significantly smaller than the corresponding coefficient in 
$B\to PP$ decays. Qualitatively, this reduction can be understood in 
terms of the fact that the quantity $a_6^p$ in the relation 
$\alpha_4^p=a_4^p+r_\chi\,a_6^p$ in (\ref{ais}) vanishes at tree level 
for the case where $M_2$ is a vector meson. In fact, using default 
parameters we predict an even more drastic reduction of the penguin 
coefficient, $|\alpha_4^c(\pi\bar K^*)+\beta_3^c(\pi\bar K^*)|
=|(-0.030-0.002i)-0.005|=0.035$, where the power-suppressed twist-3
term in the projector (\ref{provect2}) contributes about 10\% to the
result for $\alpha_4^c$. This leads to a ratio 
$|\hat\alpha_4^c(\pi\bar K^*)/\hat\alpha_4^c(\pi\bar K)|=0.32$. Note 
that the annihilation contribution is now potentially much bigger, 
$|\beta_3^c/\alpha_4^c|=0.16_{\,-0.14}^{\,+0.89}$, and also much more
uncertain. The main reason for the enhancement of this ratio is the
reduction of $\alpha_4^c$ compared with the case of $B\to PP$ decays. We
expect a similar situation to hold in the case of $VP$ modes, for which 
we expect $|\alpha_4^c(\rho\bar K)+\beta_3^c(\rho\bar K)|
=|(0.037+0.003i)+0.007|=0.044$ and 
$|\beta_3^c/\alpha_4^c|=0.18_{\,-0.18}^{\,+0.83}$.

Additional information can be obtained by combining the observables for 
the decays $\bar B^0\to\pi^+ K^-$ and $B^-\to\pi^-\bar K^0$, as well as for
the corresponding $B\to PV$ modes \cite{Fleischer_Mannel,Buras_Fleischer}.
Neglecting the small parameter $\hat\alpha_4^u$ (which is justified at 
the level of a few percent) and introducing the tree-to-penguin ratio 
$r_{\rm FM}=|\lambda_u^{(s)}/\lambda_c^{(s)}|
\cdot(-\alpha_1(\pi\bar K)/\hat\alpha_4^c(\pi\bar K))\approx 0.2$ , we 
obtain
\begin{equation}\label{FMratios}
\begin{aligned}
   R_{\rm FM} 
   = \frac{\Gamma(\bar B^0\to\pi^+ K^-)}{\Gamma(B^-\to\pi^-\bar K^0)}
   &= 1 - 2\cos\gamma\,\mbox{Re}\,r_{\rm FM} + |r_{\rm FM}|^2 \,, \\
   R_{\rm FM}\cdot A_{\rm CP}(\bar B^0\to\pi^+ K^-)
   &= -2\sin\gamma\,\mbox{Im}\,r_{\rm FM} \,.
\end{aligned}
\end{equation}
In the $\bar B\to\pi\bar K$ sector the experimental values of these 
ratios are $R_{\rm FM}=0.96\pm 0.07$ and
$ R_{\rm FM}\cdot A_{\rm CP}(\bar B^0\to\pi^+ K^-)=-(8.2\pm 3.8)\%$.
The second relation in (\ref{FMratios}) provides useful information on the
relative strong-interaction phase of the ratio 
$\alpha_1(\pi\bar K)/\hat\alpha_4^c(\pi\bar K)$.
Assuming that $\alpha_1(\pi\bar K)$ has about the same phase as 
$\alpha_1(\pi\pi)$ (both are expected to be almost real and close to 1), 
we can then extract an estimate for the strong phase of the penguin 
coefficient $\hat\alpha_4^c(\pi\bar K)$. The result is illustrated by the 
triangular-shaped area in the first plot in Figure~\ref{fig:eps32}.

We stress at this point that, in the future, the analogous ratios
in the $\bar B\to\pi\bar K^*$ and  $\bar B\to\rho\bar K$ systems will
provide interesting information about tree--penguin interference in $PV$ 
and $VP$ modes. The corresponding ratios $r_{\rm FM}$ in these two systems 
are expected to be larger than $r_{\rm FM}(\pi\bar K)$ by roughly a factor 
of 2, thereby enhancing significantly the interference terms in 
(\ref{FMratios}) and the sensitivity to $\gamma$. Note that the sign of 
the interference term is expected to be different in the $\rho\bar K$ 
modes ($\mbox{Re}(r_{\rm FM})<0$) as compared with the $\pi\bar K^{(*)}$ 
modes ($\mbox{Re}(r_{\rm FM})>0$).  

The results for the penguin coefficients illustrated in 
Figure~\ref{fig:eps32} motivated our scenarios S3 to S4 explained earlier.
The values of the penguin-to-tree ratios obtained in these scenarios are
indicated by the dots on the solid contours (S3) and by the squares (S4).
From Tables~\ref{tab:pred1BR} and \ref{tab:pred1CP} we observe that 
these scenarios are rather consistent with the data within errors.
The best match is obtained for scenario S4.
A notable exception is the decay $\bar B^0\to\pi^0\bar K^0$, whose 
measured branching fraction exceeds the values obtained using QCD
factorization. We will discuss below that this fact cannot be attributed
to uncertainties of the factorization approach.

\subsubsection{Other decays}

Within the approximations described earlier the amplitudes for the
decays $B^-\to K^-\phi$ and $\bar B^0\to\bar K^0\phi$ coincide and are 
determined in terms of the combination
$\alpha_3^c+\hat\alpha_4^c-\half\alpha_{3,{\rm EW}}^c$, 
where once again the terms with $p=u$ can be neglected to a good 
approximation. The experimental value of the ratio
\begin{equation}
   \frac{\Gamma(B^-\to K^-\phi)}{\Gamma(\bar B^0\to\bar K^0\phi)}
   = 1.10\pm 0.18 \,,
\end{equation}
is indeed consistent with unity. In QCD factorization we find the central 
values $\alpha_3^c\approx 0.002$, 
$\hat\alpha_4^c\approx -0.038$, and $-\half\alpha_{3,{\rm EW}}^c\approx 
0.004$, suggesting that the amplitude is dominated by the penguin 
coefficient $\hat\alpha_4^c$. Under this assumption, and using the 
average of these two decay rates, we then obtain
\begin{equation}
   \frac{F_0^{B\to K}(0)}{F_0^{B\to\pi}(0)}
   \left| \frac{\hat\alpha_4^c(\bar K\phi)}{\hat\alpha_4^c(\pi\bar K^*)}
   \right|
   \simeq \frac{f_{K^*}}{f_\phi} \left[
   \frac{\Gamma(\bar B\to\bar K\phi)}{\Gamma(B^-\to\pi^-\bar K^{*0})}
   \right]^{1/2} = 0.82\pm 0.10 \,.
\end{equation}
Our theoretical value $1.33_{\,-0.36}^{\,+0.43}$ for the left-hand side 
of this equation is in marginal agreement with the 
data. Leaving aside the possibility that the $B\to K$ form factor is
much smaller than assumed in our analysis, this might indicate a more
modest penguin enhancement in the PV modes $\bar B\to\bar K\phi$ than in 
$\bar B\to\pi\bar K^*$.

\subsection{Hints of departures from the Standard Model}

Some measurements related to the penguin-dominated
$\Delta S=1$ transitions are difficult to explain theoretically. We
will now study the corresponding observables. The
quantities we will focus on are the mixing-induced CP asymmetries in the 
decays  $B\to\phi K_S$ and $B\to\eta' K_S$, and the CP-averaged 
$\bar B^0\to\pi^0\bar K^0$ branching ratio.

\boldmath 
\subsubsection{Mixing-induced CP asymmetries in $B\to\phi K_S$ and 
$B\to\eta' K_S$ decays}
\unboldmath

In neutral $B$-meson decays into a CP eigenstate $f$, one can define a
time-dependent CP asymmetry
\begin{equation}\label{Sfdef}
   \frac{\mbox{Br}(\bar B^0(t)\to f) - \mbox{Br}(B^0(t)\to f)}
        {\mbox{Br}(\bar B^0(t)\to f) + \mbox{Br}(B^0(t)\to f)}
   \equiv  S_f\sin(\Delta m_B\,t) - C_f\cos(\Delta m_B\,t) \,,
\end{equation}
where $\Delta m_B>0$ denotes the mass difference of the two neutral $B$
eigenstates. The notation $B^0(t)$ refers to a state that was $B^0$ at 
time $t=0$, and $\bar B^0(t)$ refers to a state that was $\bar B^0$ at 
time $t=0$. The identity of these states at $t=0$ is determined by 
tagging the ``other $B$'', using the fact that $B^0$ and $\bar B^0$ 
are produced in a coherent state at the $\Upsilon(4S)$. The quantity $S_f$ 
is referred to as ``mixing-induced CP asymmetry'', while $-C_f$ is the 
direct CP asymmetry.

If all subdominant amplitudes are neglected, the mixing-induced CP 
asymmetries $S_f$ for the three final states $J/\psi K_S$, $\phi K_S$,
and $\eta' K_S$ all equal $\sin\phi_d$, where $\phi_d=2\beta$ is 
the $B_d$--$\bar B_d$ 
mixing phase. As is well-known the correction to this result is smallest 
for the final state $J/\psi K_S$, since the subleading amplitude is 
CKM-suppressed and suppressed by a penguin-to-tree ratio. The correction 
is somewhat larger for the other two final states, where one can rely 
only on CKM suppression. The measurements of $S_f$ for 
$f=\phi K_S$ and $\eta'K_S$ will become more accurate during the coming 
years, so that it is interesting to quantify the correction.

We can parameterize the $\bar B_d$ decay amplitude such that it is 
proportional to $1+e^{-i\gamma}\,d_f\,e^{i\theta_f}$,
where $d_f\geq 0$ includes the small ratio 
$\left|V_{ud} V_{ub}\right|/\left|V_{cd} V_{cb}\right|$ of CKM elements, 
and $\theta_f$ is the strong-interaction phase difference between the two
amplitudes with different weak phases. The CP-asymmetry $S_f$ is then 
given by
\begin{eqnarray}
   S_f
   &=& \frac{\sin\phi_d+2 d_f\cos\theta_f\sin(\phi_d+\gamma)
             + d_f^2\sin(\phi_d+2\gamma)}
            {1+2 d_f\cos\theta_f\cos\gamma+d_f^2} \nonumber\\
   &=& \sin\phi_d + 2 d_f\cos\theta_f\cos\phi_d\sin\gamma + O(d_f^2) \,.
\end{eqnarray}
For the two modes of interest we obtain 
\begin{equation}
   2 d_f\cos\theta_f = \left\{
   \begin{array}{ll}
    3.9\pm 0.9_{\,-0.1\,-0.2\,-1.6}^{\,+0.2\,+0.3\,+1.4}\,\% \,;
    & \quad f = \phi K_S \,, \\[0.2cm]
    1.8\pm 0.4\pm 0.4_{\,-1.2\,-0.8}^{\,+0.8\,+0.9}\,\% \,;
    & \quad f = \eta' K_S \,.
   \end{array} \right.
\end{equation}
where the errors have the same meaning as in previous sections. The 
various scenarios give very similar values, ranging from 3.2--3.5\% for
$\phi K_S$ and 0.5--1.7\% for $\eta' K_S$. 

As expected the correction is very small, since the QCD factorization
approach does not contain any large enhancement mechanism that could
compensate the CKM suppression. The theoretical uncertainties are under
control in both cases, including those due to the power corrections 
parameterized by the quantities $X_A$ and $X_H$. Using that
$\cos\phi_d\sin\gamma=0.64_{\,-0.12}^{\,+0.04}$ for 
$\gamma=(70\pm 20)^\circ$, we conclude that in the Standard Model
\begin{equation}\label{Spreds}
\begin{aligned}
   S_{\phi K_S} - S_{J/\psi K_S}
   &= 0.025\pm 0.012\pm 0.010 \,, \\
   S_{\eta' K_S} - S_{J/\psi K_S}
   &= 0.011\pm 0.009\pm 0.010 \,,
\end{aligned}
\end{equation}
where the second error estimates the uncertainty in 
$S_{J/\psi K_S}$. Our theoretical results obtained using QCD factorization 
agree with simple model estimates obtained in \cite{Grossman:1996ke}. The
correction for $\eta' K_S$ is smaller than for $\phi K_S$ since the 
color-suppressed tree and penguin amplitudes partially cancel each other.
The precision of the predictions (\ref{Spreds}) is higher than in a scheme 
where only 
SU(3) flavor symmetry is employed \cite{Grossman:2003qp}, in which case the 
two differences above can only be bound to be less than about 30\%. The 
present experimental values of the two differences are
\begin{equation}
\begin{aligned}
   S_{\phi K_S} - S_{J/\psi K_S} &= - 1.11\pm 0.41 \,, \\
   S_{\eta' K_S} - S_{J/\psi K_S} &= - 0.40\pm 0.34 \,.
\end{aligned}
\end{equation}
In the first case there is a 2.8$\sigma$ discrepancy between the data
and the theoretical prediction obtained in the Standard Model. If 
confirmed with more data, this would be a clean signal of physics 
beyond the Standard Model.

\boldmath
\subsubsection{The large $\bar B^0\to\pi^0\bar K^0$ decay rate}
\unboldmath

The experimental data for the $\bar B^0\to\pi^0\bar K^0$ decay rate are 
significantly larger than predictions obtained using QCD factorization. 
Sometimes this is interpreted as evidence for large rescattering effects. 
The purpose of this section is to point out that rescattering
(or, more generally, any other source of hadronic uncertainty) cannot be
invoked to explain the data.

\begin{figure}
\epsfxsize=10cm
\centerline{\epsffile{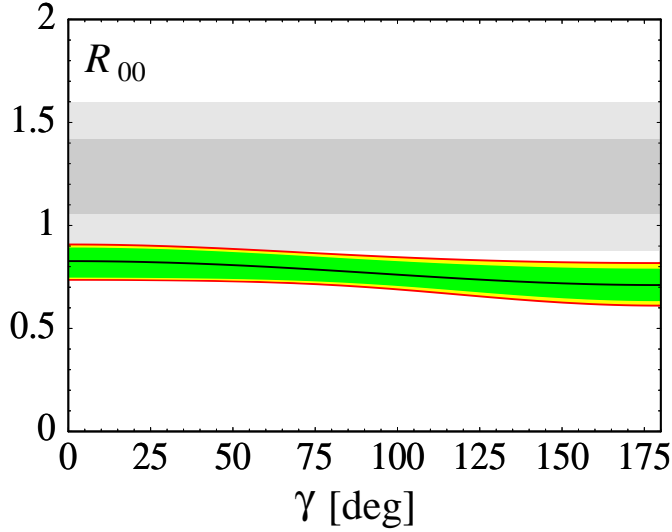}}
\centerline{\parbox{14cm}{\caption{\label{fig:R00}
Theoretical prediction for the ratio $R_{00}$ as a function of $\gamma$. 
The central value is shown by the solid line. The inner (dark) band 
corresponds to the variation of the theory input parameters, while the 
outer (light) band includes in addition the uncertainty from weak 
annihilation and twist-3 hard-spectator contributions parameterized by 
$X_A$ and $X_H$. The gray bands show the experimental result for $R_{00}$
with its 1$\sigma$ (dark) and $2\sigma$ (light) errors.}}}
\end{figure}

In many analyses the $\bar B^0\to\pi^0\bar K^0$ decay rate is normalized 
to the rate for $\bar B^0\to\pi^+ K^-$ in order to eliminate the 
dependence on the form factor \cite{BBNS3,Buras_Fleischer}. The resulting 
ratio is a strong function of $\gamma$, and in fact its high value can be 
explained by choosing a low value of $\gamma$. Here we propose to consider 
instead the ratio
\begin{equation}
   R_{00} = \frac{2\Gamma(\bar B^0\to\pi^0\bar K^0)}
                 {\Gamma(B^-\to\pi^-\bar K^0)}
\end{equation}
of CP-averaged rates. The experimental value of this ratio is 
$R_{00}^{\rm exp}=1.18\pm 0.17$. The theoretical prediction obtained using
QCD factorization is 
$R_{00}^{\rm th}=0.79\pm 0.02\pm 0.06_{\,-0.01}^{\,+0.03}\pm 0.04$,
essentially independent of $\gamma$. The largest uncertainties 
are due to the strange-quark mass, the $B\to\pi$ and $B\to K$ form 
factors, and $X_A$. The very weak dependence of this result on $\gamma$ is 
illustrated in Figure~\ref{fig:R00}. The results obtained in the various 
scenarios are: $R_{00}^{\rm th}=0.75$ (S1), 0.87 (S2), 0.81 (S3), and 0.86 
(S4). Even the largest theoretical values are about 2 standard deviations 
away from the central experimental result. 

The theoretical interpretation of the ratio $R_{00}$ in the Standard 
Model is very clean. Based on the amplitude 
parameterizations in (\ref{ApiK}) we define the ratios
\begin{eqnarray}\label{rratios}
   r_{\rm EW}
   &=& \frac32\,R_{\pi K}\,
    \frac{\alpha_{3,{\rm EW}}^c(\pi\bar K)}{\hat\alpha_4^c(\pi\bar K)}
    \approx 0.12 - 0.01i \,, \nonumber\\
   r_C
   &=& - R_{\pi K}\,\left|\frac{\lambda_u^{(s)}}{\lambda_c^{(s)}}\right|\,
    \frac{\alpha_2(\pi\bar K)}{\hat\alpha_4^c(\pi\bar K)}
    \approx 0.03 - 0.02i \,, \nonumber\\
   r_P
   &=& \left|\frac{\lambda_u^{(s)}}{\lambda_c^{(s)}}\right|\,
    \frac{\hat\alpha_4^u(\pi\bar K)}{\hat\alpha_4^c(\pi\bar K)}
    \approx 0.02 - 0.01i \,,
\end{eqnarray}
where $R_{\pi K}=(f_\pi/f_K)\cdot(F_0^{B\to K}/F_0^{B\to\pi})\approx 1$.
The numerical values correspond to the central results obtained in
QCD factorization and are meant only to illustrate the smallness of the 
various ratios. The ratio $r_{\rm EW}$ determines the relative magnitude
of electroweak to QCD penguin contributions. In the Standard Model, the
theoretical prediction for this quantity is to a large extent free of
hadronic uncertainties. To see this, note that $r_{\rm EW}$ can be written
as the product of the two ratios 
$\frac32\,R_{\pi K}\,\alpha_{3,{\rm EW}}^c/(\alpha_1+\alpha_2)$ and 
$(\alpha_1+\alpha_2)/\hat\alpha_4^c$. Using Fierz identities and top-quark
dominance in the electroweak penguin diagrams the first of these ratios 
can be calculated in a model-independent way up to corrections that break 
V-spin ($s\leftrightarrow u$) symmetry \cite{Neubert_Rosner}. The second
ratio can be determined experimentally using the strategies described in
Section~\ref{subsec:PTratios}. Next, the quantity $r_C$ determines the 
ratio of the color-suppressed tree amplitude to the dominant penguin 
amplitude. This ratio is doubly CKM suppressed. The magnitude of the 
penguin coefficient $\hat\alpha_4^c(\pi\bar K)$ is determined by data, 
and the magnitude of the tree coefficient $\alpha_2$ is expected to be 
of order 0.2 (though it is larger in the scenarios S2 and S4). It then 
follows that $r_C$ cannot be larger than a few percent in magnitude. 
Finally, the ratio $r_P$ determines the relative magnitude of up and 
charm-penguin contributions weighted by their CKM factors. Being again 
doubly CKM suppressed, this ratio is bound to be of order a few percent. 
Moreover, to first order this ratio cancels between numerator and 
denominator in the ratio $R_{00}$.

It is then justified 
to work to first order in the small ratios $r_C$ and $r_P$ and to neglect
cross terms of order $r_{\rm EW} r_C$ and  $r_{\rm EW} r_P$. 
Parametrically, this amounts to neglecting terms of 
$O(\lambda^4,\alpha\lambda^2,\alpha^2)$,
where $\lambda\approx 0.22$ is the Cabibbo angle and $\alpha$ the 
electromagnetic coupling. This gives 
\begin{equation}\label{R00formula}
   R_{00} = |1-r_{\rm EW}|^2 + 2\cos\gamma\,\mbox{Re}\,r_C + \dots \,,
\end{equation}
where the neglected terms are safely below 1\% in magnitude. For 
$\gamma\approx 70^\circ$ the $\cos\gamma$-dependent term is about $+0.02$,
which is negligible in view of the present experimental error in $R_{00}$.
The fact that $R_{00}^{\rm th}<1$ can be understood in 
terms of a negative interference between QCD and electroweak penguin
contributions. The value $r_{\rm EW}\approx 0.12$ predicted by QCD
factorization indeed leads to $R_{00}^{\rm th}\approx 0.79$ in 
(\ref{R00formula}), which explains our numerical result. In the Standard 
Model the result $R_{00}<1$ is a practically model-independent 
prediction. In this regard the
theoretical analysis of the observable $R_{00}$ is almost as clean as
that of the CP asymmetry $S_{\phi K_S}$ discussed in the previous section.
If the experimental
finding of $R_{00}>1$ is confirmed, this would be a clean signal of
physics beyond the Standard Model.\footnote{The conclusion that the 
current $\pi K$ data may point to an electroweak penguin contribution 
at variance with the Standard Model value of $\alpha_{3,{\rm EW}}^c$ 
has been reached independently in \cite{Yoshikawa:2003hb} 
through an analysis of a larger set of ratios.}

The concerned reader may ask whether the conclusion just obtained could
be affected by contributions to the decay amplitudes neglected in our 
approximation scheme. Using the exact expressions for the amplitudes
collected in Appendix~A
but dropping again terms of $O(\lambda^4,\alpha\lambda^2,\alpha^2)$,
we find that the above analysis remains valid provided the ratios 
$r_{\rm EW}$ and $r_C$ are redefined as
\begin{eqnarray}
   r_{\rm EW}
   &=& \frac32\,
    \frac{R_{\pi K}\,\alpha_{3,{\rm EW}}^c(\pi\bar K)
          + \beta_{3,{\rm EW}}^c(\pi\bar K)}
         {\hat\alpha_4^c(\pi\bar K)} \,, \nonumber\\
   r_C
   &=& - \left|\frac{\lambda_u^{(s)}}{\lambda_c^{(s)}}\right|\,
    \frac{R_{\pi K}\,\alpha_2(\pi\bar K) + \beta_2(\pi\bar K)}
         {\hat\alpha_4^c(\pi\bar K)} \,.
\end{eqnarray}
The additional annihilation and electroweak annihilation contributions 
are so small that they cannot affect the analysis presented above.

Since (\ref{R00formula}) contains a term linear in the parameter 
$r_{\rm EW}$, one might try to increase the value of $R_{00}$ (without 
changing much the decay rates for other $B\to\pi K$ modes) by adding an 
O(1) New Physics contribution to the electroweak penguin coefficient 
$\alpha_{3,{\rm EW}}^c(\pi\bar K)$. Such large non-standard effects with 
the flavor structure of electroweak penguins can indeed arise in a large 
class of extensions of the Standard Model \cite{Grossman:1999av}. 
However, there is an alternative way of displaying the ``$\pi^0 K^0$ 
anomaly'' which points in a different direction. Using the isospin
properties of the effective weak Hamiltonian and of the $(\pi K)$ final 
states it can be shown that the ratio
\begin{equation}
   R_L = \frac{2\Gamma(\bar B^0\to\pi^0\bar K^0)
               + 2\Gamma(B^-\to\pi^0 K^-)}
              {\Gamma(B^-\to\pi^-\bar K^0)
               + \Gamma(\bar B^0\to\pi^+ K^-)}
\end{equation}
equals 1 up to corrections that are at least quadratic in small amplitude 
ratios \cite{Lipkin:1998ab,Gronau:1998ep}. In terms of the amplitude 
ratio $r_{\rm EW}$ defined in (\ref{rratios}) and a new ratio 
\begin{equation}
   r_T = - \left|\frac{\lambda_u^{(s)}}{\lambda_c^{(s)}}\right|\,
   \frac{\alpha_1(\pi\bar K)}{\hat\alpha_4^c(\pi\bar K)} 
   \approx 0.18-0.02i \,,
\end{equation}
we find
\begin{equation}\label{RLformula}
   R_L = 1 + |r_{\rm EW}|^2 - \cos\gamma\,\mbox{Re}(r_T\,r_{\rm EW}^*)
   + \dots \,,
\end{equation}
where we have neglected smaller second-order terms involving the ratios 
$r_C$ and $r_P$. Because the corrections are of second order in small 
ratios the theoretical expectation for $R_L$ lies very close to 1. We find
$R_L^{\rm th}=1.01\pm 0.01\pm 0.01_{\,-0.00}^{\,+0.01}\pm 0.01$ using our
default error analysis, while the different scenarios predict: 
$R_L^{\rm th}=1.03$ (S1), 1.02 (S2), 1.01 (S3), and 1.02 (S4). The current
experimental result, $R_L^{\rm exp}=1.24\pm 0.10$, deviates from the
theoretical expectation by about 2$\sigma$, as recently emphasized in 
\cite{Gronau:2003kj}. To account for the experimental value 
requires that the magnitude of $r_{\rm EW}$ be many times larger than in 
the Standard Model (since $r_T$ is restricted by data to lie close to its
theoretical value). Specifically, from (\ref{R00formula}) and 
(\ref{RLformula}) it follows that in order to reproduce the central 
experimental data one needs $r_{\rm EW}\approx\pm 0.5i$ with a large
(weak or strong) phase.

Before proceeding, we stress that ratios analogous to $R_{00}$ and $R_L$ 
can also be considered for vector--pseudoscalar final states, for which 
the final states $\pi\bar K$ are replaced by $\rho\bar K$ or $\pi\bar K^*$. 
Let us briefly consider the case of $R_{00}$ in more detail. To leading 
order in small quantities relation (\ref{R00formula}) holds also in these 
cases once the ratio $R_{\pi K}\approx 1.0$ is replaced by corresponding 
ratios 
$R_{\rho K}=(f_\rho/f_K)\cdot(F_+^{B\to K}/A_0^{B\to\rho})\approx 1.2$ 
and $R_{\pi K^*}=(f_\pi/f_{K^*})\cdot(A_0^{B\to K^*}/F_+^{B\to\pi})
\approx 1.0$. Due to the suppression of the penguin coefficient 
$\hat\alpha_4^c$ the values of the ratios $r_{\rm EW}$ and $r_C$ are 
larger for the $PV$ modes than for the $PP$ modes. In addition, the sign 
of $r_{\rm EW}$ in the $\rho\bar K$ modes is negative and opposite to the 
cases of $\pi\bar K$ and $\pi\bar K^*$. This qualitatively explains our 
results $R_{00}^{\rm th}(\rho\bar K)=1.73_{\,-0.48}^{\,+0.62}$ and
$R_{00}^{\rm th}(\pi\bar K^*)=0.41_{\,-0.26}^{\,+0.25}$.
Obviously the uncertainties are significantly larger than in the $\pi\bar K$ 
case, but nevertheless it would be interesting to probe the qualitative 
features of these predictions.
Unfortunately, the neutral $B$ decays into $\rho^0\bar K^0$ and 
$\pi^0\bar K^{*0}$ have not been seen yet experimentally. However, in the
case of $\pi\bar K^*$ we may derive the experimental upper bound 
$R_L^{\rm exp}(\pi\bar K^*)<0.9$ at 90\% confidence level, which agrees with
the prediction that this ratio should be smaller than 1.

\boldmath
\subsection{Penguin and annihilation-dominated $\Delta D=1$ decays}
\unboldmath

The last set of decays from which interesting information on penguin
and annihilation terms can be deduced is $\bar B\to\bar K K$ and the 
corresponding $PV$ modes. The decay amplitudes for $\bar B^0\to K^-K^+$,
$\bar B^0\to K^-K^{*+}$, and $\bar B^0\to K^{*-}K^+$ are particularly
interesting, since they are the only decay modes that receive only weak
annihilation contributions. The corresponding simplified amplitudes 
are given by
\begin{equation}
\begin{aligned}
   {\cal A}_{B^-\to K^- K^0}
   &= A_{\bar K K} \Big[ \delta_{pu}\,\beta_2 + \hat\alpha_4^p \Big]
    \,, \\
   {\cal A}_{\bar B^0\to K^- K^+}
   &= A_{\bar K K} \Big[ \delta_{pu}\,\beta_1 + \beta_4^p \Big]
    + B_{K\bar K}\,b_4^p \,, \\
   {\cal A}_{\bar B^0\to\bar K^0 K^0}
   &= A_{\bar K K} \Big[ \hat\alpha_4^p + \beta_4^p \Big]
    + B_{K\bar K}\,b_4^p \,,
\end{aligned}
\end{equation}
where it is understood that each term must be multiplied with 
$\lambda_p^{(d)}$ and summed over $p=u,c$. The expressions for the 
$\bar B\to\bar K K^*$ and $\bar B\to\bar K^* K$ amplitudes are obtained 
by replacing $(\bar K K)\to (\bar K K^*)$ and $(\bar K^* K)$, 
respectively. We will also consider the modes $\bar B\to\pi\phi$, whose 
amplitudes are extremely simple and given by
\begin{equation}
   {\cal A}_{B^-\to\pi^-\phi}
   = - \sqrt2\,{\cal A}_{\bar B^0\to\pi^0\phi}
   = A_{\pi\phi}\,\alpha_3^p \,. 
\end{equation}
Our default theoretical predictions for these modes are shown in 
Tables~\ref{tab:pred2BR}, \ref{tab:pred2CP} and \ref{tab:pred3BR}.

\begin{table}[p]
\centerline{\parbox{14cm}{\caption{\label{tab:pred2BR}
CP-averaged branching ratios (in units of $10^{-6}$) of penguin-dominated
$\bar B\to PP$ decays (top) and $\bar B\to PV$ decays (bottom) with
$\Delta D=1$.}}}
\vspace{0.1cm}
\begin{center}
\begin{tabular}{|l|c|cccc|c|}
\hline\hline
\multicolumn{1}{|c|}{Mode} & Theory & S1 & S2 & S3 & S4 & Experiment \\
\hline\hline
$B^-\to K^- K^0$
 & $1.36_{\,-0.39\,-0.49\,-0.15\,-0.40}^{\,+0.45\,+0.72\,+0.14\,+0.91}$
 & 2.20 & 1.52 & 1.71 & 1.46
 & $<2.2$ \\
$\bar B^0\to\bar K^0 K^0$
 & $1.35_{\,-0.36\,-0.48\,-0.15\,-0.45}^{\,+0.41\,+0.71\,+0.13\,+1.09}$
 & 2.12 & 1.53 & 1.83 & 1.58
 & $<1.6$ \\
\hline\hline
$B^-\to K^- K^{*0}$
 & $0.30_{\,-0.09\,-0.10\,-0.09\,-0.19}^{\,+0.11\,+0.12\,+0.09\,+0.57}$
 & 0.50 & 0.21 & 0.54 & 0.66
 & $<5.3$ \\
$\bar B^0\to\bar K^0 K^{*0}$
 & $0.26_{\,-0.07\,-0.09\,-0.08\,-0.15}^{\,+0.08\,+0.10\,+0.08\,+0.46}$
 & 0.41 & 0.18 & 0.45 & 0.56
 & --- \\
$B^-\to K^0 K^{*-}$
 & $0.30_{\,-0.07\,-0.18\,-0.07\,-0.17}^{\,+0.08\,+0.41\,+0.08\,+0.58}$
 & 0.45 & 0.75 & 0.61 & 0.55
 & --- \\
$\bar B^0\to K^0\bar K^{*0}$
 & $0.29_{\,-0.09\,-0.17\,-0.07\,-0.17}^{\,+0.10\,+0.39\,+0.08\,+0.60}$
 & 0.47 & 0.71 & 0.61 & 0.54
 & --- \\
\hline
$B^-\to\pi^-\phi$
 & $\approx 0.005$
 & 0.008 & 0.010 & 0.005 & 0.009
 & $<0.4$ \\
$\bar B^0\to\pi^0\phi$
 & $\approx 0.002$
 & 0.004 & 0.005 & 0.002 & 0.004
 & $<5.0$ \\
\hline\hline
\end{tabular}
\end{center}
%
\centerline{\parbox{14cm}{\caption{\label{tab:pred2CP}
Direct CP asymmetries (in units of $10^{-2}$) of penguin-dominated 
$\bar B\to PP$ decays (top) and $\bar B\to PV$ decays (bottom) with 
$\Delta D=1$. We only consider modes with branching fractions larger than
$10^{-7}$.}}}
\vspace{0.1cm}
\begin{center}
\begin{tabular}{|l|c|cccc|c|}
\hline\hline
\multicolumn{1}{|c|}{Mode} & Theory & S1 & S2 & S3 & S4 & Experiment \\
\hline\hline
$B^-\to K^- K^0$
 & $-16.3_{\,-3.7\,-5.7\,-1.7\,-13.3}^{\,+4.7\,+5.0\,+1.6\,+11.3}$
 & $-10.0$ & $-14.7$ & $-6.5$ & $-4.3$
 & --- \\
$\bar B^0\to\bar K^0 K^0$
 & $-16.7_{\,-3.7\,-5.1\,-1.7\,-3.6}^{\,+4.7\,+4.5\,+1.5\,+4.6}$
 & $-10.6$ & $-15.0$ & $-12.7$ & $-11.5$
 & --- \\
\hline\hline
$B^-\to K^- K^{*0}$
 & $-23.5_{\,-5.7\,-9.0\,-6.5\,-36.8}^{\,+6.9\,+7.8\,+5.5\,+25.2}$
 & $-13.9$ & $-22.5$ & $-7.7$ & $-9.6$
 & --- \\
$\bar B^0\to\bar K^0 K^{*0}$
 & $-26.7_{\,-5.7\,-9.0\,-6.9\,-13.4}^{\,+7.4\,+7.2\,+5.7\,+10.9}$
 & $-16.9$ & $-26.0$ & $-16.8$ & $-13.6$
 & --- \\
$B^-\to K^0 K^{*-}$
 & $-13.4_{\,-3.0\,-3.5\,-4.7\,-36.7}^{\,+3.7\,+7.8\,+4.2\,+27.4}$
 & $-8.9$ & $-12.7$ & $-13.9$ & $-21.1$
 & --- \\
$\bar B^0\to K^0\bar K^{*0}$
 & $-13.1_{\,-3.0\,-2.9\,-5.2\,-7.4}^{\,+3.8\,+5.4\,+4.5\,+5.8}$
 & $-8.0$ & $-12.2$ & $-8.8$ & $-10.0$
 & --- \\
\hline\hline
\end{tabular}
\end{center}
%
\centerline{\parbox{14cm}{\caption{\label{tab:pred3BR}
CP-averaged branching ratios (in units of $10^{-6}$) of 
annihilation-dominated $\bar B\to PP$ decays (top) and $\bar B\to PV$ 
decays (bottom) with $\Delta D=1$.}}}
\vspace{0.1cm}
\begin{center}
{\tabcolsep=0.18cm\begin{tabular}{|l|c|cccc|c|}
\hline\hline
\multicolumn{1}{|c|}{Mode} & Theory & S1 & S2 & S3 & S4 & Experiment \\
\hline\hline
$\bar B^0\to K^- K^+$
 & $0.013_{\,-0.005\,-0.005\,-0.000\,-0.011}^{\,+0.005\,+0.008\,+0.000\,+0.087}$
 & 0.007 & 0.014 & 0.079 & 0.070
 & $<0.6$ \\
\hline\hline
$\bar B^0\to K^- K^{*+}$
 & $0.014_{\,-0.006\,-0.006\,-0.000\,-0.012}^{\,+0.007\,+0.010\,+0.000\,+0.106}$
 & 0.016 & 0.011 & 0.095 & 0.094
 & --- \\
$\bar B^0\to K^+ K^{*-}$
 & $0.014_{\,-0.006\,-0.006\,-0.000\,-0.012}^{\,+0.007\,+0.010\,+0.000\,+0.106}$
 & 0.016 & 0.011 & 0.095 & 0.056
 & --- \\
\hline\hline
\end{tabular}}
\end{center}
\end{table}

In principle, the modes $\bar B^0\to K^- K^+$ (and the corresponding $PV$
modes) can provide interesting information about the weak annihilation 
amplitudes. Unfortunately, however, their branching fractions are too 
small to be observed in the near future. The other decay modes are penguin 
dominated. Using this information we predict
\begin{equation}
   \frac{\Gamma(B^-\to K^- K^0)}{\Gamma(\bar B^0\to\bar K^0 K^0)}
    \approx 1 \,, \quad
   \frac{\Gamma(B^-\to K^- K^{*0})}{\Gamma(\bar B^0\to\bar K^0 K^{*0})}
    \approx 1 \,, \quad
   \frac{\Gamma(B^-\to K^0 K^{*-})}{\Gamma(\bar B^0\to K^0\bar K^{*0})}
    \approx 1 \,.
\end{equation}
The relevant branching fractions are found to be of order few times 
$10^{-7}$, which should be within the long-term reach of the $B$ 
factories. If these equalities could be established, this would be 
another indication that annihilation contributions are suppressed with 
respect to the dominant penguin amplitudes governed by $\alpha_4^p$.

In the approximations described above, the decays $B^-\to\pi^-\phi$ and 
$\bar B^0\to\pi^0\phi$ are determined by the singlet penguin coefficients 
$\alpha_3^p$, which are predicted to be highly suppressed in the QCD 
factorization approach. In addition these modes are CKM suppressed. 
An experimental upper bounds exists for $B^-\to\pi^-\phi$, which 
however lies two orders of magnitude above the theoretical predictions. 
It will therefore not be possible to extract useful information from 
these modes is the foreseeable future.

\subsection{Bounds on large annihilation contributions}
\label{sec:largeXA}

Our discussion in this section was confined to the treatment of 
annihilation effects suggested in \cite{BBNS3}, in which the value
of the model parameter $\varrho_A$ is limited to 
$\varrho_A\le 1$, corresponding to a 100\% uncertainty on the default
value for the quantity $X_A$. While we still believe that this is a 
reasonable assumption, which so far has not been invalidated by the data,
one might ask what would happen if larger values of $\varrho_A$ were
considered. Note that the annihilation kernels in (\ref{XAmodel}) and
(\ref{XAdefvect}) are quadratically dependent on the quantity $X_A$,
so that increasing the value of $\varrho_A$ can have a dramatic effect
on the relative strength of annihilation terms as compared with the 
leading penguin amplitude.

\begin{table}
\centerline{\parbox{14cm}{\caption{\label{tab:large_rhoA}
CP-averaged branching ratios (in units of $10^{-6}$) of some decays that 
imply bounds on the weak annihilation parameter $X_A$. The theoretical 
results refer to the choice $\varrho_A=2$ with strong phase
$\varphi_A=-60^\circ$.}}}
\vspace{0.1cm}
\begin{center}
\begin{tabular}{|l|cc|c|}
\hline\hline
\multicolumn{1}{|c|}{Mode} & Default & Large Annihilation & Experiment \\
\hline\hline
$\bar B^0\to\pi^0\bar K^{*0}$
 & 0.7 & 6.0 & $<3.6$ \\
$B^-\to K^-\rho^0$
 & 2.6 & 9.0 & $<6.2$ \\
$\bar B^0\to K^-\rho^+$
 & 7.4 & 19.3 & $8.5\pm 2.1$ \\
$B^-\to K^-\phi$
 & 4.5 & 22.4 & $9.2\pm 0.7$ \\
$\bar B^0\to\bar K^0\phi$
 & 4.1 & 20.2 & $7.7\pm 1.1$ \\
\hline\hline
$B^-\to K^- K^0$ & 1.36 & 1.65 & $<2.2$ \\
$B^-\to K^{*-} K^0$ & 0.30 & 1.14 & --- \\
$B^-\to K^- K^{*0}$ & 0.30 & 0.95 & $<5.3$ \\
$\bar B^0\to K^- K^+$ & 0.01 & 0.21 & $<0.6$ \\
$\bar B^0\to K^{*-} K^+$ & 0.01 & 0.26 & --- \\
$\bar B^0\to K^- K^{*+}$ & 0.01 & 0.26 & --- \\
\hline\hline
\end{tabular}
\end{center}
\end{table}

It is apparent from Figure~\ref{fig:eps32} that for values of $\varrho_A$ 
significantly larger than 1 a fine-tuning of the phase $\varphi_A$ is 
required so as not to be in conflict with the experimental constraints on
the magnitudes of the $\hat\alpha_4^c$ parameters in the $\pi\bar K$ and 
$\pi\bar K^*$ systems. Let us assume for a moment that this fine-tuning is 
indeed realized in nature. To be specific, we set $\varrho_A=2$ and adjust 
the phase $\varphi_A=-60^\circ$ such that the resulting $\hat\alpha_4^c$ 
values fall in the center of the dark rings in Figure~\ref{fig:eps32} (see 
the dots on the dashed curves). All other parameters 
are set to their default values. We might then ask whether such a 
scenario is consistent with the experimental data. 

Interestingly, we find that for several decay channels the results 
obtained with such large annihilation contributions are already in 
conflict with the data. In the upper part of Table~\ref{tab:large_rhoA} 
we collect results for branching ratios where significant discrepancies
appear. In the lower portion of the table we give results for  
$K^{(*)}\bar K^{(*)}$ modes, some of which are also very sensitive to 
annihilation effects. We conclude that the data support the assumption 
that a reasonable estimate of weak annihilation effects can be obtained 
by varying the parameter $\varrho_A$ between 0 and 1, allowing for an 
arbitrary strong phase $\varphi_A$. Values of $\varrho_A$ larger than 1 
are strongly disfavored by the data, and values $\varrho_A\ge 2$ can 
already be excluded, at least if universal annihilation is a reasonable
approximation. It has occasionally been argued that one should allow for 
much larger values of $\varrho_A<8$ to account for the theoretical 
uncertainties related to power corrections \cite{Ciuchini:2002uv}. 
Table~\ref{tab:large_rhoA} shows that such large weak annihilation 
effects cannot be tolerated by the data.

\section{Tree-dominated decays}
\label{sec:trees}

Many of the decays with $\Delta D=1$ are dominated by tree amplitudes 
and do not suffer from a large sensitivity to light quark masses or 
chirally-enhanced power corrections. We now analyze this class 
of decays, including the time evolution of the $\pi\rho$ final states. 

We first summarize the decay amplitudes simplified according to the 
discussion of Section~\ref{sec:simpleampls}. There are three independent 
amplitudes for the decays $\bar B\to\pi\rho$ given by
\begin{equation}
\label{pirhoampsimp}
\begin{aligned}
   \sqrt2\,{\cal A}_{B^-\to\pi^-\rho^0}
   &= A_{\pi\rho} \Big[ \delta_{pu}\,(\alpha_2 - \beta_2)
    - \hat\alpha_4^p \Big] 
    + A_{\rho\pi} \Big[ \delta_{pu}\,(\alpha_1 + \beta_2)
    + \hat\alpha_4^p \Big] \,, \\
   {\cal A}_{\bar B^0\to\pi^+\rho^-}
   &= A_{\pi\rho} \Big[ \delta_{pu}\,\alpha_1 + \hat\alpha_4^p
    + \beta_4^p \Big] 
    + A_{\rho\pi} \Big[ \delta_{pu}\,\beta_1 + \beta_4^p \Big] \,, \\
   -2\,{\cal A}_{\bar B^0\to\pi^0\rho^0}
   &= A_{\pi\rho} \Big[ \delta_{pu}\,(\alpha_2 - \beta_1) 
    - \hat\alpha_4^p - 2\beta_4^p \Big]
    + A_{\rho\pi} \Big[ \delta_{pu}\,(\alpha_2 - \beta_1) 
    - \hat\alpha_4^p - 2\beta_4^p \Big] \,. 
\end{aligned}
\end{equation}
These expressions must be multiplied with 
$\lambda_p^{(d)}$ and summed over $p=u,c$. The order of the arguments of 
the coefficients $\alpha_i^p(M_1 M_2)$ and $\beta_i^p(M_1 M_2)$ is 
determined by the order of the arguments of the $A_{M_1 M_2}$ prefactors.
The amplitudes for $B^-\to\pi^0\rho^-$ and $\bar B^0\to\pi^-\rho^+$ are
obtained from the first two expressions by interchanging 
$\pi\leftrightarrow\rho$ everywhere, including the arguments of the 
amplitude parameters. The expressions for the $\bar B\to\pi\pi$ 
amplitudes are obtained by setting $\rho\to\pi$, in which case they 
simplify to
\begin{equation}
\begin{aligned}
   \sqrt2\,{\cal A}_{B^-\to\pi^-\pi^0}
   &= A_{\pi\pi}\,\delta_{pu}\,(\alpha_1 + \alpha_2) \,, \\
   {\cal A}_{\bar B^0\to\pi^+\pi^-}
   &= A_{\pi\pi} \Big[ \delta_{pu}\,(\alpha_1 + \beta_1)
    + \hat\alpha_4^p + 2\beta_4^p \Big] \,, \\
   -{\cal A}_{\bar B^0\to\pi^0\pi^0}
   &= A_{\pi\pi} \Big[ \delta_{pu}\,(\alpha_2 - \beta_1) 
    - \hat\alpha_4^p - 2\beta_4^p \Big] \,. 
\end{aligned}
\end{equation}
Bose symmetry implies that the two-pion final state must have isospin 
$I=0,2$ but not $I=1$, which is the reason why the $\bar B\to\pi\pi$ 
decay amplitudes are simpler than those for $\bar B\to\pi\rho$. Finally, 
\begin{equation}
\begin{aligned}
   \sqrt2\,{\cal A}_{B^-\to\pi^-\omega}
   &= A_{\pi\omega} \Big[ \delta_{pu}\,(\alpha_2 + \beta_2)
    + 2\alpha_3^p + \hat\alpha_4^p \Big] 
    + A_{\omega\pi} \Big[ \delta_{pu}\,(\alpha_1 + \beta_2)
    + \hat\alpha_4^p \Big] \,, \\
   -2\,{\cal A}_{\bar B^0\to\pi^0\omega}
   &= A_{\pi\omega} \Big[ \delta_{pu}\,(\alpha_2 - \beta_1)
    + 2\alpha_3^p + \hat\alpha_4^p \Big]
    + A_{\omega\pi} \Big[ \delta_{pu}\,(-\alpha_2 - \beta_1)
    + \hat\alpha_4^p \Big]\,. 
\end{aligned}
\end{equation}

\begin{table}
\centerline{\parbox{14cm}{\caption{\label{tab:amppars}
Magnitudes of the amplitude parameters for $\pi\pi$ and $\pi\rho$ 
final states in the different scenarios. ``Default'' refers to the 
default input parameters. Scenarios not shown agree with the default. 
A minus sign indicates that a parameter is negative in naive 
factorization or (in case of the annihilation parameters) if 
$\varrho_A=0$. Theoretical errors are added in quadrature.}}}
\vspace{-0.4cm}
\begin{center}
$$
{\arraycolsep=0.2cm\begin{array}{|r|c|cc|r|c|ccc|}
\hline\hline
\multicolumn{1}{|c|}{} & \mbox{Default} &  \mbox{S2} & 
\mbox{S4} &
\multicolumn{1}{|c|}{} & \mbox{Default} &  \mbox{S2} & \mbox{S3} &
\mbox{S4}\\
\hline\hline
\alpha_1(\pi\pi) & 0.99^{+0.04}_{-0.07} & 0.84 & 0.88 & 
\beta_1(\pi\pi) & 0.025^{+0.046}_{-0.018} & 0.032 & 0.062 & 0.071 \\
\alpha_1(\pi\rho) & 0.99^{+0.04}_{-0.06} & 0.89 & 0.90 &
\beta_1(\pi\rho) & 0.020^{+0.037}_{-0.013} & 0.023 & 0.051 & 0.063 \\
\alpha_1(\rho\pi) & 1.01^{+0.04}_{-0.05} & 0.94 & 0.95 &
\beta_1(\rho\pi) & 0.024^{+0.045}_{-0.015} & 0.024 & 0.061 & 0.053 \\
\hline
\alpha_2(\pi\pi) & 0.20^{+0.17}_{-0.11} & 0.57 & 0.48 &
-\beta_2(\pi\pi) & 0.010^{+0.018}_{-0.007} & 0.012 & 0.024 & 0.027 \\
\alpha_2(\pi\rho) & 0.20^{+0.16}_{-0.10} & 0.45 & 0.41 &
-\beta_2(\pi\rho) & 0.008^{+0.015}_{-0.005} & 0.009 & 0.019 & 0.024 \\
\alpha_2(\rho\pi) & 0.16^{+0.13}_{-0.09} & 0.31 &  0.30 &
-\beta_2(\rho\pi) & 0.009^{+0.018}_{-0.006} & 0.009 & 0.023 & 0.020 \\
\hline
-\alpha_4^c(\pi\pi) & 0.102^{+0.020}_{-0.014} & 0.110 & 0.102 &
-\beta_3(\pi\pi) & 0.009^{+0.032}_{-0.009} & 0.013 & 0.035 & 0.041 \\
-\alpha_4^c(\pi\rho) & 0.032^{+0.006}_{-0.006} & 0.027 & 0.028 &
-\beta_3(\pi\rho) & 0.005^{+0.024}_{-0.004} & 0.007 & 0.025 & 0.034 \\
\alpha_4^c(\rho\pi) & 0.035^{+0.020}_{-0.013} & 0.062 & 0.049 &
\beta_3(\rho\pi) & 0.007^{+0.031}_{-0.007} & 0.008 & 0.032 & 0.028 \\
\hline\hline
\end{array}}
$$
\end{center}
\end{table}

To understand the pattern of branching fractions and asymmetries 
it is useful to bear in mind the magnitudes of the amplitude 
parameters as given in Table~\ref{tab:amppars}. The table also 
shows the values of the parameters in the four scenarios 
defined in Section~\ref{sec:penguins}. Generically $\alpha_2$ is
large in scenarios S2 and S4 (and $\alpha_1$ is 
somewhat reduced), while the annihilation coefficients 
are increased in S3 and S4. Because of the cancellation 
of vector and scalar penguin contributions, the combination
$\alpha_4^p(\rho\pi)=a_4^p(\rho\pi)-r_\chi^\pi a_6^p(\rho\pi) $ is
very sensitive to the strange-quark mass. Except for the purely 
neutral final states the most 
important interference of amplitudes with different weak phases occurs
between the tree coefficient $\alpha_1$ and the effective penguin 
amplitude $\hat\alpha_4^c$. Comparing the $\pi\rho$ final states 
to $\pi\pi$ we note two important differences: first, the penguin 
amplitudes are smaller for $\pi\rho$. This implies reduced sensitivity 
to the angle $\gamma$ in CP-averaged branching fractions and smaller 
direct CP violation, but it also implies a reduced ``penguin
pollution'' in time-dependent studies of CP violation. Second, while 
$\hat \alpha_4^c(\pi\pi)$ and $\alpha_1(\pi\pi)$ have opposite signs 
(always assuming small relative phases), which implies constructive 
tree-penguin interference for $\gamma<90^\circ$, both relative signs 
occur for $\pi\rho$.

\subsection{Branching fractions and direct CP asymmetries}

Our results for the $\pi\pi$, $\pi\rho$, and $\pi\omega$ branching 
fractions and direct CP asymmetries are shown in 
Tables~\ref{tab:pred4BR} and \ref{tab:pred4CP}.
In many cases the dominant theoretical uncertainty arises
from the variation of CKM parameters or form factors.
The latter source of uncertainty can be reduced once better data on
semileptonic or leptonic $B$ decays become available. The sensitivity to 
CKM parameters is not a theoretical limitation but rather provides access
to $|V_{ub}|$ and $\gamma$.

\begin{table}
\centerline{\parbox{14cm}{\caption{\label{tab:pred4BR}
CP-averaged branching ratios (in units of $10^{-6}$) of tree-dominated 
$\bar B\to PP$ decays (top) and $\bar B\to PV$ decays (bottom) with 
$\Delta D=1$. The errors and scenarios have the same meaning as 
explained in Section~\ref{sec:penguins}.}}}
\vspace{0.1cm}
\begin{center}
\begin{tabular}{|l|c|cccc|c|}
\hline\hline
\multicolumn{1}{|c|}{Mode} & Theory & S1 & S2 & S3 & S4 & Experiment \\
\hline\hline
$B^-\to\pi^-\pi^0$
 & $6.0_{\,-2.4\,-1.8\,-0.5\,-0.4}^{\,+3.0\,+2.1\,+1.0\,+0.4}$
 & 5.8 & 5.5 & 6.0 & 5.1
 & $5.3\pm 0.8$ \\
$\bar B^0\to\pi^+\pi^-$
 & $8.9_{\,-3.4\,-3.0\,-1.0\,-0.8}^{\,+4.0\,+3.6\,+0.6\,+1.2}$
 & 6.0 & 4.6 & 9.5 & 5.2
 & $4.6\pm 0.4$ \\
$\bar B^0\to\pi^0\pi^0$
 & $0.3_{\,-0.2\,-0.1\,-0.1\,-0.1}^{\,+0.2\,+0.2\,+0.3\,+0.2}$
 & 0.7 & 0.9 & 0.4 & 0.7
 & $1.6\pm 0.7$ \\[-0.15cm]
 & & & & & & ($<3.6$) \\
\hline\hline
$B^-\to\pi^-\rho^0$
 & $11.9_{\,-5.0\,-3.1\,-1.2\,-1.1}^{\,+6.3\,+3.6\,+2.5\,+1.3}$
 & 14.2 & 12.6 & 12.2 & 12.3
 & $9.1\pm 1.1$ \\
$B^-\to\pi^0\rho^-$
 & $14.0_{\,-5.5\,-4.3\,-0.6\,-0.7}^{\,+6.5\,+5.1\,+1.0\,+0.8}$
 & 10.7 & 10.4 & 14.2 & 10.3
 & $11.0\pm 2.7$ \\
$\bar B^0\to\pi^+\rho^-$
 & $21.2_{\,-\phantom{1}8.4\,-7.2\,-2.3\,-1.6}^{\,+10.3\,+8.7\,+1.3\,+2.0}$
 & 18.6 & 11.0 & 22.2 & 11.8
 & $13.9\pm 2.7$ \\
$\bar B^0\to\pi^-\rho^+$
 & $15.4_{\,-6.4\,-4.7\,-1.3\,-1.3}^{\,+8.0\,+5.5\,+0.7\,+1.9}$
 & 17.5 & 10.8 & 16.4 & 11.8
 & $8.9\pm 2.5$ \\
$\bar B^0\to\pi^\pm\rho^\mp$
 & $36.5_{\,-14.7\,-\phantom{1}8.6\,-3.5\,-2.9}^{\,+18.2\,+10.3\,+2.0\,+3.9}$
 & 36.1 & 21.8 & 38.6 & 23.6
 & $24.0\pm 2.5$ \\
$\bar B^0\to\pi^0\rho^0$
 & $0.4_{\,-0.2\,-0.1\,-0.3\,-0.3}^{\,+0.2\,+0.2\,+0.9\,+0.5}$
 & 0.3 & 1.7 & 0.3 & 1.1
 & $<2.5$ \\
\hline
$B^-\to\pi^-\omega$
 & $8.8_{\,-3.5\,-2.2\,-0.9\,-0.9}^{\,+4.4\,+2.6\,+1.8\,+0.8}$
 & 8.6 & 9.1 & 8.4 & 8.4
 & $5.9\pm 1.0$ \\
$\bar B^0\to\pi^0\omega$
 & $0.01_{\,-0.00\,-0.00\,-0.00\,-0.00}^{\,+0.00\,+0.02\,+0.02\,+0.03}$
 & 0.01 & 0.07 & 0.01 & 0.01
 & $<1.9$ \\
\hline\hline
\end{tabular}
\end{center}
\end{table}

\begin{table}
\centerline{\parbox{14cm}{\caption{\label{tab:pred4CP}
Direct CP asymmetries (in units of $10^{-2}$) of tree-dominated 
$\bar B\to PP$ decays (top) and $\bar B\to PV$ decays (bottom) with 
$\Delta D=1$. We only consider modes with branching fractions larger than
$10^{-7}$.}}}
\vspace{0.1cm}
\begin{center}
\begin{tabular}{|l|c|cccc|c|}
\hline\hline
\multicolumn{1}{|c|}{Mode} & Theory & S1 & S2 & S3 & S4 & Experiment \\
\hline\hline
$B^-\to\pi^-\pi^0$
 & $-0.02_{\,-0.01\,-0.05\,-0.00\,-0.01}^{\,+0.01\,+0.05\,+0.00\,+0.01}$
 & $-0.02$ & $-0.02$ & $-0.02$ & $-0.02$
 & $-7\pm 14$ \\
$\bar B^0\to\pi^+\pi^-$
 & $-6.5_{\,-2.1\,-2.8\,-0.3\,-12.8}^{\,+2.1\,+3.0\,+0.1\,+13.2}$
 & $-9.6$ & $-9.1$ & 5.6 & 10.3
 & $51\pm 23$ \\
$\bar B^0\to\pi^0\pi^0$
 & $45.1_{\,-12.8\,-13.8\,-14.1\,-61.6}^{\,+18.4\,+15.1\,+\phantom{1}4.3\,+46.5}$
 & 23.0 & 21.7 & 5.6 & $-19.0$
 & --- \\
\hline\hline
$B^-\to\pi^-\rho^0$
 & $4.1_{\,-0.9\,-2.0\,-0.7\,-18.8}^{\,+1.3\,+2.2\,+0.6\,+19.0}$
 & 3.4 & 4.6 & $-13.3$ & $-11.0$
 & $-17\pm 11$ \\
$B^-\to\pi^0\rho^-$
 & $-4.0_{\,-1.2\,-2.2\,-0.4\,-17.7}^{\,+1.2\,+1.8\,+0.4\,+17.5}$
 & $-5.3$ & $-6.3$ & 12.2 & 9.9
 & $23\pm 17$ \\
$\bar B^0\to\pi^+\rho^-$
 & $-1.5_{\,-0.4\,-1.3\,-0.3\,-8.4}^{\,+0.4\,+1.2\,+0.2\,+8.5}$
 & $-1.7$ & $-1.8$ & 6.6 & 3.9
 & $-11\pm 17$ \\
$\bar B^0\to\pi^-\rho^+$
 & $0.6_{\,-0.1\,-1.6\,-0.1\,-11.7}^{\,+0.2\,+1.3\,+0.1\,+11.5}$
 & 0.5 & 1.7 & $-10.3$ & $-12.9$
 & $-62\pm 27$  \\
$\bar B^0\to\pi^0\rho^0$
 & $-15.7_{\,-4.7\,-14.0\,-12.9\,-25.8}^{\,+4.8\,+12.3\,+11.0\,+19.8}$
 & $-20.9$ & $-9.5$ & $-10.6$ & 10.7
 & --- \\
\hline
$B^-\to\pi^-\omega$
 & $-1.8_{\,-0.5\,-3.3\,-0.7\,-2.2}^{\,+0.5\,+2.7\,+0.8\,+2.1}$
 & $-1.8$ & 0.6 & $-2.1$ & $-6.0$
 & $9\pm 21$ \\
\hline\hline
\end{tabular}
\end{center}
\end{table}

From the numbers shown in the tables one observes reasonable global 
agreement between theory and data within their respective error 
ranges. Since many of the theoretical errors are correlated among the 
various final states we shall consider below certain ratios of
observables, which provide more sensitive probes of the underlying
hadronic amplitudes. With default parameters the decay rates to 
the charged final states $\pi^+\pi^-$ and $\pi^\pm\rho^\mp$ are predicted
to be significantly larger than the data. The discrepancy disappears for
the two-pion final state if $\gamma$ is large (scenario S1) or 
$\alpha_2$ is large (scenario S2). However, the table shows 
that the $\pi\rho$ modes do not favor the large-$\gamma$ scenario. 
Both S2 and S4, which combines elements of the
large-$\alpha_2$ scenarios with a moderate increase of weak
annihilation, describe the branching fractions very
well, but S4 is singled out if we add the information from
penguin-dominated decays discussed in the previous section. 

The data on direct CP asymmetries are still too uncertain to draw 
conclusions from the comparison with theory. We note a 2--3 standard 
deviation discrepancy for the $\pi^+\pi^-$ final state. The current central
value 0.51 of the CP asymmetry is certainly too large to be understood 
in the QCD factorization framework, and appears even more puzzling in 
view of the small asymmetries for the $\pi K$ final states. Unless we 
allow for very large SU(3)-breaking effects the only difference between 
$\pi^+\pi^-$ and $\pi^+ K^-$ occurs in the annihilation term 
$\delta_{pu}\beta_1+2\beta_4^p$, but it is hard to see how this could 
cause such a dramatic effect on the CP asymmetry, since $\beta_4^p$ is 
always very small in our framework and never exceeds a quarter (and 
often less) of the other penguin annihilation coefficient $\beta_3^p$. 
In the future this combination of annihilation amplitudes can be 
constrained independently using $\bar B^0\to K^- K^+$ decays (up to 
SU(3)-breaking effects). 

With the exception of the neutral final states the direct CP asymmetries 
are all predicted to be small, although with large uncertainties due to 
weak annihilation that affect even the signs of the asymmetries. 
However, the analysis of Section~\ref{sec:penguins} has shown that for 
the largest values of weak annihilation allowed by our error definition 
not all values of the annihilation phase are compatible with data on 
branching fractions and CP asymmetries of penguin-dominated decays 
(see Figure~\ref{fig:eps32}). This might favor the signs displayed in
scenario S3 and S4, although this conclusion must be regarded with
great caution since it is mainly based on the direct CP asymmetry in 
the $\pi^\mp K^\pm$ mode. 

It is worth emphasizing that the correlations between various 
asymmetries are predicted more reliably in the factorization framework 
than the asymmetries themselves, at least if the annihilation parameter 
$\varrho_A$ is not too different for the $PP$, $PV$, and $VP$ amplitudes, 
or if annihilation is a small effect altogether. Table~\ref{tab:pred4CP}
shows that this is the case for all our scenarios. Since in the case of
a small penguin-to-tree ratio the direct CP asymmetry is approximately
given by $2\sin\gamma\,|\lambda_c^{(d)}/\lambda_u^{(d)}|\,
\mbox{Im}(\hat\alpha_4^c/\alpha_1)$, the relative magnitudes of the CP 
asymmetries in $\pi^+\pi^-$, $\pi^-\rho^0$, $\pi^0\rho^-$, $\pi^-\rho^+$, 
and $\pi^+\rho^-$ provide direct access to the magnitudes and signs 
(phases) of the penguin 
amplitudes $\hat \alpha_4^c$ for $\pi\pi$, $\pi\rho$, and $\rho\pi$,
which are predicted to be rather different (see Table~\ref{tab:amppars}). 
We note specifically the case of the $B^-\to\pi^-\rho^0$ and 
$B^-\to\pi^0\rho^-$ amplitudes, where according to (\ref{pirhoampsimp})
the $PV$ and $VP$ penguin amplitudes appear as 
$A_{\rho\pi}\,\hat\alpha_4^c-A_{\pi\rho}\,\hat\alpha_4^c$ but with 
opposite sign relative to $\alpha_1$, and the charged final states, 
which probe the magnitude and sign of the $PV$ and $VP$ penguin 
amplitudes independently. A verification of these sign 
patterns alone would provide an impressive confirmation of the relevance 
of factorization to the calculation of direct CP asymmetries.

\subsection{Ratios of decay rates}
\label{sec:brfrac}

We now discuss a number of ratios of CP-averaged $\pi\pi$ and $\pi\rho$ 
decay rates, which are either sensitive to the CKM angle $\gamma$ or to 
particular aspects of the hadronic amplitudes. In addition to 
$R_{\pi\pi}=\Gamma(\bar B^0\to\pi^+\pi^-)/(2\Gamma(B^-\to\pi^-\pi^0))$ 
we consider the ratios:
\begin{equation}\label{rhoratios}
\begin{aligned}
   R_1 &\equiv \frac{\Gamma(\bar B^0\to \pi^+\rho^-)}
                    {\Gamma(\bar B^0\to \pi^+\pi^-)} \,, 
    \qquad\quad 
   &R_2 &\equiv \frac{\Gamma(\bar B^0\to \pi^+\rho^-)+
                     \Gamma(\bar B^0\to \pi^-\rho^+)}
                    {2\Gamma(\bar B^0\to \pi^+\pi^-)} \,, \\ 
   R_3 &\equiv \frac{\Gamma(\bar B^0\to \pi^+\rho^-)}
               {\Gamma(\bar B^0\to \pi^-\rho^+)} \,, && \\
   R_4 &\equiv \frac{2\,\Gamma(B^-\to \pi^-\rho^0)}
                    {\Gamma(\bar B^0\to \pi^-\rho^+)} - 1 \,,
   &R_5 &\equiv \frac{2\,\Gamma(B^-\to \pi^0\rho^-)}
                    {\Gamma(\bar B^0\to \pi^+\rho^-)} - 1 \,.
\end{aligned}
\end{equation}
The theoretical and experimental results for these ratios are summarized 
in Table~\ref{tab:rhoratios}. The ratio $R_{\pi\pi}$ has already been 
discussed in part in 
Section~\ref{sec:treeamplitude} as one of the motivations for 
exploring a scenario with large $\alpha_2$. Since QCD factorization
does not contain a mechanism to generate a large strong phase between 
the two amplitudes with different weak phases contributing 
to $\bar B^0\to\pi^+\pi^-$, $R_{\pi\pi}$ is described well only 
if $\gamma$ is significantly larger than $100^\circ$, or if 
$\alpha_2$ is large and $A_{\pi\pi}|\lambda^{(d)}_u|$ is small.

\begin{table}
\centerline{\parbox{14cm}{\caption{\label{tab:rhoratios}
Results for the ratios $R_{\pi\pi}$ and $R_{1-5}$.}}}
\vspace{0.1cm}
\begin{center}
\begin{tabular}{|c|c|rrrr|c|}
\hline\hline
\multicolumn{1}{|c|}{} & Theory & S1~ & S2~ & S3~ & S4~ & Experiment \\
\hline\hline
$R_{\pi\pi}$ & 
$0.80^{+0.12\,+0.05\,+0.13\,+0.14}_{-0.12\,-0.06\,-0.20\,-0.11}$ & 
$0.56$ & $0.45$ & $0.85$ & $0.55$ & $0.47\pm 0.08$ \\ 
$R_1$ & 
$2.39^{+0.31\,+0.04\,+0.15\,+0.05}_{-0.25\,-0.08\,-0.12\,-0.11}$ & 
$3.11$ & $2.39$ & $2.33$ & $2.28$ & $3.02\pm 0.64$ \\ 
$R_2$ & 
$2.06^{+0.40\,+0.53\,+0.12\,+0.03}_{-0.30\,-0.36\,-0.09\,-0.06}$ & 
$3.02$ & $2.37$ & $2.03$ & $2.28$ & $2.61\pm 0.35$\\ 
$R_3$ & 
$1.38^{+0.18\,+0.82\,+0.03\,+0.02}_{-0.17\,-0.59\,-0.04\,-0.05}$ & 
$1.06$ & $1.01$ & $1.35$ & $0.99$ &  $1.56\pm 0.53$ \\ 
$R_4$ & 
$0.42_{-0.04\,-0.11\,-0.21\,-0.20}^{+0.04\,+0.15\,+0.45\,+0.23}$ & 
$0.49$ & $1.14$ & $0.37$ & $0.92$ & $0.88\pm 0.57$ \\ 
$R_5$ & 
$0.22_{-0.08\,-0.06\,-0.12\,-0.12}^{+0.07\,+0.08\,+0.23\,+0.14}$ & 
$0.06$ & $0.75$ & $0.18$ & $0.61$ & $0.46\pm 0.46$ \\ 
\hline\hline
\end{tabular}
\end{center}
\end{table}

We find that the ratio $R_1$ is theoretically rather clean and yet 
sensitive to $\gamma$,
because it is almost independent of the $B$-meson form factors, 
and it is independent of the uncertain color-suppressed tree
coefficient $\alpha_2$ as seen from (\ref{pirhoampsimp}). In fact, 
if the tree coefficient $\alpha_1$ dominated the amplitude 
and were universal, $R_1$ would equal $(f_\rho/f_\pi)^2=2.55$, which is 
not far from the complete result. In our
framework the largest theoretical uncertainty comes from the 
pion and $\rho$-meson light-cone distribution amplitudes, since 
these cause the largest non-universality of the amplitude parameters. 
From Table~\ref{tab:rhoratios} we see that $R_1$ is nearly constant,
except in the scenario where $\gamma=110^\circ$. The sensitivity to 
$\gamma$ arises from the fact that while in both cases the
penguin--tree interference is constructive for $\gamma<90^\circ$, the
effect is more pronounced for $\pi\pi$ than for $\pi\rho$ due to the
larger penguin amplitude in the former case. We should note, however, 
that the 
conclusion that $R_1$ is theoretically clean hinges on the assumption 
that the phase of the penguin annihilation amplitude is not very
different for the $PP$ and $PV$ amplitudes, or, if it is different,
that the magnitude of penguin annihilation is not too large. The 
current experimental value of $R_1$ is somewhat higher than the 
theoretical prediction for $\gamma=70^\circ$, but the difference is 
only one standard deviation.

Since the sum of the $\pi^+\rho^-$ and $\pi^-\rho^+$ decay rates is
measured more accurately than the individual rates, we also consider the 
second ratio $R_2$. In this case there is a substantial uncertainty from 
the ratio of form factors, $A_0^{B\to\rho}(0)/F_0^{B\to\pi}(0)$, 
evidenced by the second error 
of the ``Theory'' column in Table~\ref{tab:rhoratios}. Perhaps the
only thing that can be said at present is that theory and experiment
agree within their respective errors, which can be
taken as a qualitative  argument in favor of factorization. 

The ratio $R_3$ of $\bar B^0\to \pi^+\rho^-$ to 
$\bar B^0\to \pi^-\rho^+$ is mainly sensitive to the form-factor 
ratio $A_0^{B\to\rho}(0)/F_0^{B\to\pi}(0)$, and to a lesser extent to 
$\gamma$. Since the information provided by $R_3$ is largely 
equivalent to the
one from the parameter $\Delta C$ in the time-dependent analysis 
of the $\pi^\mp\rho^\pm$ final states (see Section~\ref{sec:time} below), 
we do not discuss $R_3$ further here. The experimental result is 
again in good agreement with factorization. 

Finally, the quantities $R_4$ and $R_5$ are constructed such that the
dominant tree amplitude cancels out in the numerator. This leaves 
as the leading term
\begin{equation}
   R_4 = 2\,\frac{A_{\pi\rho}}{A_{\rho\pi}}\,
   \mbox{Re}\left( \frac{\alpha_2(\pi\rho)}{\alpha_1(\rho\pi)} \right)
   + \dots \,,
\end{equation}
where the dots denote the interference terms with the penguin amplitude 
and other smaller contributions. For $R_5$ a similar approximation (with 
somewhat larger corrections) holds with $\pi\leftrightarrow\rho$ exchanged 
everywhere. Hence $R_{4,5}$ are expected to be 
nearly independent of $\gamma$, and in first approximation they access 
the real part of the color-suppressed tree amplitude. 
These observables may therefore be interesting for assessing the 
viability of the large-$\alpha_2$ scenarios S2 and S4. However, the 
current experimental values of $R_4$ and $R_5$ are not sufficiently 
precise to support or disfavor these scenarios.

\subsection{\boldmath 
Time-dependent rates in the $\pi^\mp\rho^\pm$ system}
\label{sec:time}

We now analyze the asymmetries in time-dependent measurements of $B^0$ 
and $\bar B^0$ decays into the $\pi^\mp\rho^\pm$ final states. We
begin with setting up conventions and notation for a general final 
state $f$ and its CP conjugate $\bar f$ \cite{Aleksan:1990ts}.

We define ${\cal A}_f$, ${\cal\bar A}_f$, ${\cal A}_{\bar f}$, and 
${\cal \bar A}_{\bar f}$ to be the amplitudes for the four decay modes, 
where the bar on ${\cal A}$ refers to the decay of the $\bar B^0$ meson. 
In the standard approximation, which neglects CP violation in the $
B^0$--$\bar B^0$ mixing matrix and the width difference of the two mass 
eigenstates, the decay amplitude squared at time $t$ of the state that 
was a pure $B^0$ at time $t=0$ can be parameterized by 
\begin{equation}
\label{timedep}
   |{\cal A}_f(t)|^2\equiv |\langle f|B(t)\rangle|^2 = 
   \frac{e^{-\Gamma t}}{2}\big(|{\cal A}_f|^2+|{\cal\bar A}_f|^2\big)\,
   \Big\{ 1+C_f\cos(\Delta m_B\,t)-S_f\sin(\Delta m_B\,t) \Big\} \,,
\end{equation} 
where $\Delta m_B>0$ denotes the mass difference, and $\Gamma$ the
common total width of the $B$-meson eigenstates. For an initial 
$\bar B^0$ the signs of the cos and sin terms are  reversed. For decays 
to the CP-conjugate final state one replaces $f$ by $\bar f$. 

For the following discussion we adopt the phase convention 
$\mbox{CP}|B^0\rangle=-|\bar B^0\rangle$ and define the amplitude ratios
\begin{equation}
   \rho_f=\frac{{\cal\bar A}_f}{{\cal A}_f} \,, \qquad 
   \rho_{\bar f}=\frac{{\cal\bar A}_{\bar f}}{{\cal A}_{\bar f}} \,.
\end{equation}
In terms of these 
\begin{equation}
   C_f=\frac{1-|\rho_f|^2}{1+|\rho_f|^2} \,, \qquad 
   S_f=-2\,\frac{\mbox{Im}\,(e^{-2 i\beta}\rho_f)}{1+|\rho_f|^2} \,,
\end{equation}
which are phase-convention independent. (The CKM angle $\beta$ is
defined according to the convention of the Particle Data Group.) 
The system of four decay modes
defines five asymmetries, $C_f$, $S_f$, $C_{\bar f}$, $S_{\bar f}$ -- 
alternatively parametrized as $C\equiv\frac12\,(C_f+C_{\bar f})$,  
$S\equiv\frac12\,(S_f+S_{\bar f})$, $\Delta C\equiv\frac12\,(C_f-C_{\bar f})$  
and $\Delta S\equiv\frac12\,(S_f-S_{\bar f})$ -- together with the global
charge asymmetry related to the overall normalization of 
(\ref{timedep}):
\begin{equation}
   \frac{1+A_{\rm CP}}{1-A_{\rm CP}} \equiv
   \frac{|{\cal A}_f|^2+|{\cal\bar A}_f|^2}
        {|{\cal A}_{\bar f}|^2+|{\cal\bar A}_{\bar f}|^2} \,,
\end{equation}
or
\begin{equation}
   A_{\rm CP} = \frac{|{\cal A}_f|^2+|{\cal\bar A}_f|^2
                      -|{\cal A}_{\bar f}|^2-|{\cal\bar A}_{\bar f}|^2}
                     {|{\cal A}_f|^2+|{\cal\bar A}_f|^2
                      +|{\cal A}_{\bar f}|^2+|{\cal\bar A}_{\bar f}|^2} \,.
\end{equation}
Under a CP transformation $\rho_f$ goes into $1/\rho_{\bar f}$; hence 
$\Delta C$ and $\Delta S$ are CP-even, but $A_{\rm CP}$, $C$, and $S$ 
are CP-odd. For the special case that $f$ is a CP eigenstate 
there are only two different amplitudes since $f=\bar f$, and 
$A_{\rm CP}$, $\Delta C$ and $\Delta S$ vanish. 

For the $\pi\rho\,$ system we identify $f=\pi^-\rho^+$. We write the 
amplitudes as 
\begin{equation}\label{rhopiamps}
\begin{aligned}
   \bar {\cal A}_f &= \bar {\cal A}_{\pi^-\rho^+}
    = A_{\rho\pi}\,|\lambda_u^{(d)}|\,T_{\rho\pi}\,\bigg( e^{-i\gamma}
    + \frac{P_{\rho\pi}}{T_{\rho\pi}} \bigg) \,, \\
   {\cal A}_f &= {\cal A}_{\pi^-\rho^+}
    = -\,A_{\pi\rho}\,|\lambda_u^{(d)}|\,T_{\pi\rho}\,\bigg( e^{i\gamma}
    + \frac{P_{\pi\rho}}{T_{\pi\rho}} \bigg) \,.
\end{aligned}
\end{equation}
For $\bar f=\pi^+\rho^-$ one must exchange $\pi\leftrightarrow\rho$ on 
$A$, $T$, 
and $P$. These equations define the ``tree'' amplitudes $T$ and the
``penguin-to-tree'' ratios, while the factors $A_{\pi\rho}$ and 
$A_{\rho\pi}$ are given by (\ref{am1m2}). From (\ref{pirhoampsimp}) we 
see that $\bar {\cal A}_{\pi^-\rho^+}$ is proportional to $A_{\rho\pi}$ 
up to the small weak annihilation contribution 
$\delta_{pu}\beta_1+\beta_4^p$, which suggests to extract this factor
as we have done. The sign in the last amplitude is related to the CP 
convention for the $B$-meson state and the convention that all decay
constants and form factors are taken to be 
positive.\footnote{
The phases of the $B$-meson decay amplitudes are obtained as
follows. Let $\mbox{C}\,|\pi^-\rangle = \xi_\pi|\pi^+\rangle$ 
and $\mbox{C}\,|\rho^-\rangle = \xi_\rho|\rho^+\rangle$. Our CP 
convention for the $B$ meson implies $\mbox{C}|\bar B^0\rangle = 
|B^0\rangle$, i.e., $\xi_B=1$. Isospin invariance suggests 
to choose $\xi_\pi=1$ and $\xi_\rho=-1$, since $\pi^0$ 
is C-even and $\rho^0$ is C-odd, but in the following we shall 
keep the charge-conjugation phases for the charged pions and $\rho$
mesons arbitrary. It follows from this that 
$|\pi^-\pi^+\rangle$ is CP-even, but 
$\mbox{CP}\,|\pi^-(\vec{p}\,)\rho^+(-\vec{p}\,)\rangle
= \xi_\pi\xi_\rho^*|\pi^+(-\vec{p}\,)\rho^-(\vec{p}\,)\rangle
= - \xi_\pi\xi_\rho^*|\pi^+(\vec{p}\,)\rho^-(-\vec{p}\,)\rangle$. 
The second equality assumes that we form a wave-packet state by
integrating over the relative momentum $\vec{p}$ and uses that 
the $\pi^-\rho^+$ is in a P-wave state. Using $\mbox{CP}\,{\cal O} 
\,(\mbox{CP})^\dagger ={\cal O}^\dagger$ for the operators in the 
effective weak Hamiltonian and dropping the momentum labels that are 
understood to be equal on the left and right-hand sides, we obtain
\begin{displaymath}
\langle \pi^-\pi^+|{\cal O}^\dagger|B^0\rangle = 
- \langle \pi^-\pi^+|{\cal O}|\bar B^0\rangle \,, 
\qquad 
\langle \pi^-\rho^+|{\cal O}^\dagger|B^0\rangle = 
\xi_\pi\xi_\rho^*\,\langle \pi^+\rho^-|{\cal O}|\bar B^0\rangle \,. 
\end{displaymath}
This can be used to show that $\rho_{\pi^+\pi^-}=-e^{-2 i\gamma}$, 
and hence $S_{\pi\pi}=\sin 2\alpha$  
in the absence of the penguin contribution. Similarly, one finds 
$S_{\pi^0\rho^0}=\sin 2\alpha$ for 
the neutral $\pi\rho$ final state. 

The conclusion is more subtle in the case of a non-CP eigenstate such 
as $\pi^\mp\rho^\pm$, since after applying the CP transformation 
$\rho_{\pi^-\rho^+}$ involves the $\bar B$ 
matrix elements of two different final states. To clarify the issue 
we evaluate the matrix elements in naive factorization before and
after applying the CP transformation, neglecting in 
addition the penguin contribution. This gives 
$$
\rho_{\pi^-\rho^+} = 
e^{-2i\gamma}\,\frac{f_{\pi^-} A_0^{\bar
    B\to\rho^+}}{f_{\rho^+} F_0^{B\to\pi^-}} = 
e^{-2i\gamma}\,\xi_\pi\xi_\rho^*\,\frac{f_{\pi^-} A_0^{\bar
    B\to\rho^+}}{f_{\rho^-} F_0^{\bar B\to\pi^+}} \,.
$$
The standard C and P transformations of the vector and
axial-vector currents together with the CP convention for 
the $B$ meson  imply $f_{\pi^+}=\xi_\pi f_{\pi^-}$, 
$f_{\rho^+}=-\xi_\rho f_{\rho^-}$, 
$F_0^{B\to\pi^-}=-\xi_\pi^* F_0^{\bar B\to\pi^+}$, and 
$A_0^{B\to\rho^-}=\xi_\rho^* A_0^{\bar B\to\rho^+}$. 
The previous equation is consistent with
this. However, the relative phase (actually, only a sign) 
of $f_B$ times the form factor to the light-meson decay
constant cannot be inferred from symmetry considerations
alone. Neither can it be determined from the semileptonic 
decay rates, which provide access only to the magnitudes of these 
quantities. The expression after the second equality makes it clear
that we cannot simply assume all decay constants and form factors to
be positive numbers.

To fix the sign we have to resort to theoretical input such as
lattice calculations or QCD sum rules. We then find that the 
(re-phasing invariant) ratios $f_B F_0^{\bar B\to\pi^+}\!/f_{\pi^+}$, 
$f_B A_0^{\bar B\to\rho^+}\!/f_{\rho^+}$ are positive (see, for 
instance, \cite{Khodjamirian:1998ji}), hence 
$f_B F_0^{B\to\pi^-}\!/f_{\pi^-}$ and 
the ratio of form factors and decay
constants in $\rho_{\pi^-\rho^+}$ is negative. However, in this 
paper we have ignored the possible phase and sign conventions for the
decay constants and form factors, always assuming them
positive. Hence, we must write 
$$
\rho_{\pi^-\rho^+} = 
- e^{-2i\gamma}\,\frac{A_{\rho\pi}}{A_{\pi\rho}}
$$
in the naive factorization approximation. In QCD factorization these 
considerations carry over to the full amplitude. 
This explains the minus sign in the $B$-meson decay amplitude of 
(\ref{rhopiamps}).}

\begin{table}[t]
\centerline{\parbox{14cm}{\caption{\label{tab:TandPs}
Magnitudes of the amplitude parameters in $B\to\pi\pi$ and $B\to\pi\rho$ 
decays.}}}
\vspace{0.1cm}
\begin{center}
\begin{tabular}{|c|c|rrr|}
\hline\hline
\multicolumn{1}{|c|}{} & Theory & S2~ & S3~ & S4~ \\
\hline\hline
$|T_{\pi\pi}|$ & $0.91^{+0.05}_{-0.07}$ & 0.75 & 0.92 & 0.81 \\
$|P_{\pi\pi}/T_{\pi\pi}|$ & $0.32^{+0.16}_{-0.09}$ & 0.49 & 0.37 & 0.48 \\
\hline
$|T_{\pi\rho}|$ & $0.98^{+0.04}_{-0.07}$ & 0.88 & 0.99 & 0.88 \\
$|P_{\pi\rho}/T_{\pi\rho}|$ & $0.10^{+0.06}_{-0.04}$ & 0.12 & 0.12 & 0.20 \\
\hline
$|T_{\rho\pi}|$ & $1.07^{+0.09}_{-0.07}$ & 1.03 & 1.11 & 1.07 \\
$|P_{\rho\pi}/T_{\rho\pi}|$ & $0.10^{+0.09}_{-0.05}$ & 0.19 & 0.14 & 0.15 \\
\hline\hline
\end{tabular}
\end{center}
\end{table}

The magnitudes of the ``tree'' and ``penguin-to-tree'' parameters in 
QCD factorization are given in Table~\ref{tab:TandPs}, where we also 
show the corresponding quantities for the $\pi\pi$ 
case. Note that the sign of $P_{\rho\pi}/T_{\rho\pi}$ is negative in naive 
factorization, where all amplitudes are real, while the other two 
penguin-to-tree ratios are positive. The largest uncertainty on 
the penguin-to-tree ratios is caused by $|V_{ub}|$, $m_s$, and weak 
annihilation. The important point is that the penguin amplitudes 
are smaller for the $\pi\rho$ final states compared to 
$\pi\pi$. Furthermore, $|P_{\pi\rho}/T_{\pi\rho}|$ and 
$|P_{\rho\pi}/T_{\rho\pi}|$ are similar in magnitude and 
interfere with an opposite sign with the tree amplitude. We should note
that the near equality of the two penguin-to-tree ratios is the result
of intricate dynamics specific to factorization. 
$|P_{\pi\rho}/T_{\pi\rho}|$ is explained by the fact that 
$a_4^c(\pi\rho)+r_\chi^\rho a^c_6(\pi\rho)\approx a_4^c(\pi\rho)$, 
since the scalar penguins have a small effect on $PV$ amplitudes. On the
other hand, for $|P_{\rho\pi}/T_{\rho\pi}|$ the scalar penguin 
amplitude $a_6^c(\rho\pi)$ is large and determines the sign of 
$a_4^c(\rho\pi)-r_\chi^\pi a^c_6(\rho\pi)$, such that the 
$PV$ and $VP$ amplitudes have similar magnitude but opposite sign. 

\begin{table}[t]
\centerline{\parbox{14cm}{\caption{\label{tab:timeAsym}
Parameters of the time-dependent $B\to \pi^\mp\rho^\pm$ decay rate 
asymmetries as defined in the text. $S$ and $\Delta S$ are computed 
for $\beta=23.6^\circ$, corresponding to $\sin(2\beta)=0.734$.}}}
\vspace{0.1cm}
\begin{center}
\begin{tabular}{|c|c|rrrr|r|}
\hline\hline
\multicolumn{1}{|c|}{} & Theory & S1~ & S2~ & S3~ & S4~ & Experiment \\
\hline\hline
$A_{\rm CP}$ & $\phantom{-}0.01^{\,+0.00\,+0.01\,+0.00\,+0.10}_{\,-0.00\,
   -0.01\,-0.00\,-0.10}$ & $0.01$ & $0.02$ & $-0.08$ & $-0.08$ & 
   $-0.21\pm 0.08$\\ 
$C$ & $\phantom{-}0.00^{\,+0.00\,+0.01\,+0.00\,+0.02}_{\,-0.00\,
   -0.01\,-0.00\,-0.02}$ & $0.01$ & $0.00$ & $0.02$ & $0.05$ &
   $0.36\pm 0.18$ \\ 
$S$ & $\phantom{-}0.13^{\,+0.60\,+0.04\,+0.02\,+0.02}_{\,-0.65\,
   -0.03\,-0.01\,-0.01}$ & $1.00$ & $0.20$ & $0.14$ & $0.09$ &
   $0.19\pm 0.24$ \\ 
$\Delta C$ & $\phantom{-}0.16^{\,+0.06\,+0.23\,+0.01\,+0.01}_{\,-0.07\,
   -0.26\,-0.02\,-0.02}$ & $0.03$ & $0.01$ & $0.15$ & $0.00$ &
   $0.28\pm 0.19$ \\ 
$\Delta S$ & $-0.02^{\,+0.01\,+0.00\,+0.00\,+0.01}_{\,-0.00\,
   -0.01\,-0.00\,-0.01}$ & $0.00$ & $-0.02$ & $0.00$ & $-0.03$ &
   $0.15\pm 0.25$ \\ 
\hline\hline
\end{tabular}
\end{center}
\end{table}

The results for the time-dependent asymmetry parameters are 
given in Table~\ref{tab:timeAsym} including again our default
prediction with errors and the four standard scenarios to exhibit
possible correlations. Since the penguin-to-tree ratios are small, it
is instructive to compare the complete results with an 
expansion of the asymmetries in these ratios. Defining 
$P_{\pi\rho}/T_{\pi\rho}=a\,e^{i\delta_a}$,
$P_{\rho\pi}/T_{\rho\pi}=-b\,e^{i\delta_b}$ (such that $\delta_a$, 
$\delta_b$ are small and $a$, $b$ are positive), and 
$A_{\rho\pi} T_{\rho\pi}/(A_{\pi\rho} T_{\pi\rho})=R\,e^{i\delta_T}$, and 
treating $a$, $b$, and $\delta_T$ as small, the leading terms read
\begin{equation}\label{asymexpansions}
\begin{aligned}
   A_{\rm CP} &= \frac{2}{1+R^2}\,\left(a\sin\delta_a+R^2\,
    b\sin\delta_b\right)\sin\gamma + \dots \,, \\
   C &= \frac{4 R^2}{(1+R^2)^2}\left(a\sin\delta_a-b\sin\delta_b\right)
    \sin\gamma + \dots \,, \\
   \Delta C &= \frac{1-R^2}{1+R^2} + 
    \frac{4 R^2}{(1+R^2)^2}\left(a\cos\delta_a+b\cos\delta_b\right)
    \cos\gamma + \dots \,, \\
   S &= \frac{2 R}{1+R^2}\,\sin2\alpha -
   \frac{2 R}{1+R^2}\,\bigg\{ a\cos\delta_a
   \left(\frac{2\sin 2\alpha}{1+R^2}\cos\gamma+
   \sin(2\beta+\gamma)\right) \\
  &\phantom{=} \,-\,b\cos\delta_b
   \left(\frac{2 R^2\sin 2\alpha}{1+R^2}\cos\gamma+
   \sin(2\beta+\gamma)\right) \bigg\} + \dots \,, \\
  \Delta S &= \frac{2 R}{1+R^2}\,\cos 2\alpha\sin\delta_T
   -\frac{2 R}{1+R^2}\,\bigg\{
   a\sin\delta_a \left(\frac{2\sin 2\alpha}{1+R^2}\sin\gamma+
   \cos(2\beta+\gamma)\right) \\
  &\phantom{=} \,+\,
   b\sin\delta_b \left(\frac{2 R^2\sin 2\alpha}{1+R^2}\sin\gamma+
   \cos(2\beta+\gamma)\right) \bigg\} + \dots \,,
\end{aligned}
\end{equation}
with $\alpha=\pi-\beta-\gamma$. The numerical values of $a$ and $b$ are 
given in Table~\ref{tab:TandPs}. With our input parameters the 
tree-to-tree ratio is given by $R=0.91^{\,+0.26}_{\,-0.21}$ with phase 
$\delta_T=(1\pm 3)^\circ$, where the sizable error on $R$ is 
entirely due to the form factors.

The asymmetries $A_{\rm CP}$, $C$, and $\Delta S$ are suppressed by the 
penguin-to-tree ratios and the sin of a strong phase, hence they are 
always small in QCD factorization. This can be seen explicitly in 
Table~\ref{tab:timeAsym}. Note that $\Delta S$ has a
potentially large coefficient in front of $\sin\delta_T$, however a 
large relative phase of the $PV$ and $VP$ tree amplitudes would
constitute a rather unexpected failure of QCD factorization. The 
data show a large value of $C$, which is related to the large 
direct CP asymmetry in $\bar B\to \pi^-\rho^+$, since 
\begin{equation}
   A_{\rm CP}(\pi^-\rho^+)
   = \frac{A_{\rm CP}\,(1-\Delta C)-C}{1-\Delta C-A_{\rm CP}\,C} \,,
   \qquad
   A_{\rm CP}(\pi^+\rho^-)
   = -\frac{A_{\rm CP}\,(1+\Delta C)+C}{1+\Delta C+A_{\rm CP}\,C} \,.
\end{equation}
It seems impossible to accommodate a penguin--tree interference as 
large as needed to reproduce the central experimental values of 
$A_{\rm CP}$ and $C$ within QCD factorization.
 
\begin{figure}
\epsfxsize=9cm
\centerline{\epsffile{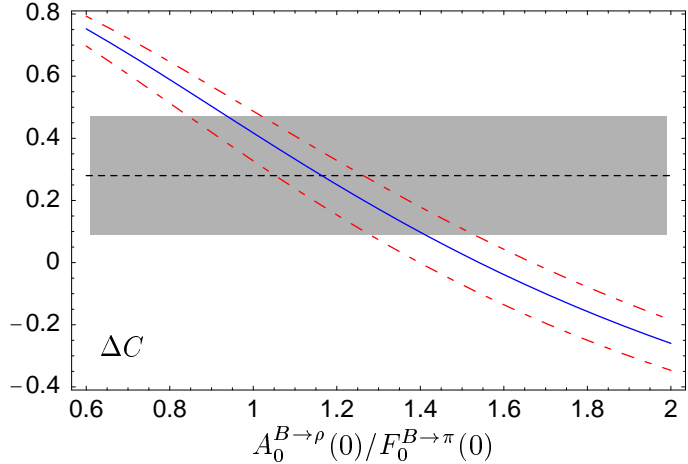}}
\centerline{\parbox{14cm}{\caption{\label{fig:delC}
Dependence of $\Delta C$ on the ratio of $B$-meson form factors (for fixed
$F^{B\to\pi}(0)=0.28$) for $\gamma=70^\circ$ (center solid), 
$\gamma=40^\circ$ (upper dashed) and $\gamma=100^\circ$ (lower dashed). 
The experimental value with its $1\sigma$ error is also shown. Our central 
input value for the form-factor ratio is 1.32.}}}
\end{figure}

In first approximation $\Delta C$ is determined by the ratio 
$R$ alone, with the largest uncertainty due to the form 
factor ratio $A_0^{B\to\rho}(0)/F_0^{B\to\pi}(0)$. Since 
$R$ is close to 1, the second term in the expansion in 
the penguin-to-tree ratios is also important and introduces some 
dependence on $\gamma$. This ratio is graphically displayed in 
Figure~\ref{fig:delC}.

\begin{figure}[t]
\epsfxsize=9cm
\centerline{\epsffile{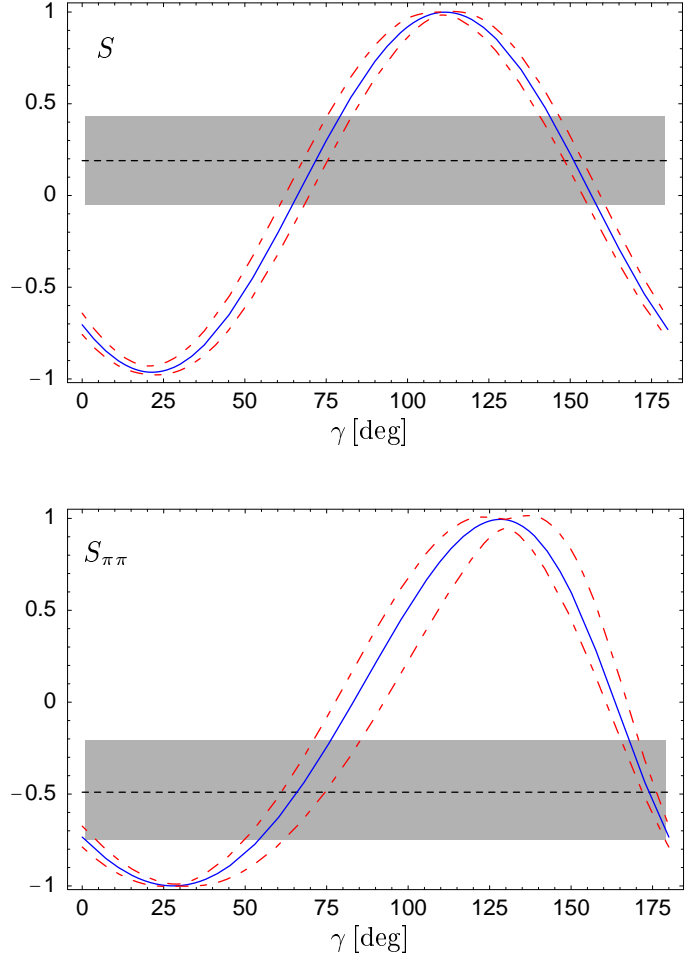}}
\centerline{\parbox{14cm}{\caption{\label{fig:S}
Dependence of $S$ on the angle $\gamma$ for $\beta=(23.6\pm 2.4)^\circ$. 
Upper panel: $S$ from the $\pi^\mp\rho^\pm$ final states. Lower panel: 
$S_{\pi\pi}$ from the $\pi^-\pi^+$ final state. The
experimental values of $S$ and $S_{\pi\pi}$ with their $1\sigma$ errors 
are overlaid. The dashed curves specify the theoretical error.}}}
\end{figure}

The most interesting asymmetry is $S$. The numerical values of the 
parameters $R$, $a$, $b$, $\delta_a$, and $\delta_b$ are such that 
the first correction to $S$ in (\ref{asymexpansions}) nearly cancels, 
hence $S$ is nearly proportional to $\sin(2\alpha)$. Furthermore, 
the expression $2 R/(1+R^2)$ is not very sensitive to $R$ near
$R=1$, leaving little uncertainty from the $B$-meson form factors. 
This suggests that $\alpha$ or, more precisely, $\gamma$ (given the 
$B$--$\bar B$ mixing phase) can be accurately determined 
from this asymmetry. In the upper panel of 
Figure~\ref{fig:S} we overlay the experimental $1\sigma$ band for 
$S$ to the theoretical prediction given as a function 
of $\gamma$ for $\beta=23.6^\circ$, which (up to a discrete ambiguity) 
corresponds to the current measurement of the $B$--$\bar B$ 
mixing phase. The figure also contains two further curves, which 
define the theory error band, including the error $\pm 2.4^\circ$ on 
$\beta$. Since this observable might be used 
to put a constraint on $\gamma$, we adopt an estimate of the 
theoretical error where the parameters $\varrho_A$ that model 
the size of weak annihilation are allowed to vary independently 
for the $PV$ and $VP$ annihilation coefficients. This is more conservative 
than the standard error estimate, since it allows for a significant 
difference in the magnitudes of $P_{\pi\rho}/T_{\pi\rho}$ and 
$P_{\rho\pi}/T_{\rho\pi}$, which breaks the near cancellation 
in the correction term to $S$ in (\ref{asymexpansions}). The figure 
shows that the theoretical error remains small even with this 
conservative estimate. Using $S_{\rm exp}=0.19\pm 0.24$ we 
obtain the allowed ranges
\begin{equation}
   \gamma = (72\pm 11)^\circ \quad\mbox{or}\quad
   \gamma=(151 \pm 10)^\circ \qquad \mbox{(from $S$)} \,,
\end{equation} 
where the limiting values are defined by the intersection of the 
theory error band with the $1\sigma$ experimental error band. The 
error given is dominated by the experimental error. The first range is 
in reasonable agreement with the standard unitarity-triangle 
fit. Note that if we allow for the second solution for $\beta$, given by 
$\beta\to \pi/2-\beta$, we obtain the above ranges with $\gamma\to 
180^\circ-\gamma$ to a very good approximation. 

It is interesting to compare the constraint on $\gamma$ from $S$ with the 
constraint obtained from the analogous asymmetry in the decay to the 
$\pi^-\pi^+$ final state. Here the correction due to the penguin 
amplitude is larger. The current average experimental value 
$S_{\pi\pi}=-0.49\pm 0.27$ yields
\begin{equation}
   \gamma = (66^{\,+19}_{\,-16})^\circ\quad\mbox{or}\quad
   \gamma=(174^{\,+9}_{\,-8} )^\circ \qquad \mbox{(from $S_{\pi\pi}$)} \,,
\end{equation} 
as shown in the lower panel of Figure~\ref{fig:S}. Despite the 
rather different values of the asymmetry, the allowed values of
$\gamma$ are consistent with each other, in particular for the first 
range that is also compatible with the standard unitarity-triangle fit. 
This provides a strong argument for the validity of the theoretical 
framework underlying the computation of the hadronic amplitudes. 
While this is reassuring, one should keep in mind that there are 
several aspects of the current $\pi\pi$ and $\pi\rho$ data that cannot
be accommodated in QCD factorization, in particular the large direct 
CP asymmetries for $\pi^-\pi^+$ and $\pi^-\rho^+$. Other pieces 
of data, such as the measurements of $S_{\pi\pi}$, are not compatible 
between the BaBar and Belle experiments. It remains to be 
seen whether these anomalies disappear without affecting the apparent
consistency of the remaining data.

The analysis of the $\pi\rho$ final states has been performed under 
the assumption that the $\rho$-meson width is negligible. The 
experimental error from the quasi two-body assumption on the 
extraction of the parameters of 
the time-evolution of $B,\bar B\to\pi^\mp\rho^\pm$ decays has been 
estimated as 0.08 for the CP-violating parameters $A_{\rm CP}$, 
$C$, and $S$, and as 0.03 for $\Delta C$ and $\Delta S$ \cite{hoecker}. 
This leaves enough room for an improvement of the experimental error, 
so that we can look forward to more comprehensive unitarity-triangle 
constraints from the $\pi\rho$ system once higher statistics data 
samples become available.

\boldmath
\section{Final states containing $\eta$ or $\eta'$}
\unboldmath
\label{sec:etas}

In this section we give results for the 23 $B^-$ and $\bar B^0$ decays 
into final states containing the mesons $\eta$ or $\eta^\prime$. A 
dedicated investigation of the $\Delta S=1$ decays in QCD
factorization has already been performed in \cite{Beneke:2002jn}. 
For comparison of the different scenarios that we defined 
in the present paper we repeat the results for these decays in 
Tables~\ref{tab:pred5BR} and \ref{tab:pred5CP}. Small differences 
in the ``Theory'' column relative to the ``Default'' in 
\cite{Beneke:2002jn} result from slight changes in the input 
parameters as mentioned in Section~\ref{sec:inputs}.

\begin{table}[t]
\centerline{\parbox{14cm}{\caption{\label{tab:pred5BR}
CP-averaged branching ratios (in units of $10^{-6}$) of penguin-dominated 
$\bar B\to PP$ decays (top) and $\bar B\to PV$ decays (bottom) with 
$\Delta S=1$. The errors and scenarios have the same meaning as 
explained in Section~\ref{sec:penguins}.}}}
\vspace{0.1cm}
\begin{center}
\begin{tabular}{|l|c|cccc|c|}
\hline\hline
\multicolumn{1}{|c|}{Mode} & Theory &  S1 & S2 & S3 & S4 & Experiment \\
\hline\hline
$B^-\to\eta K^-$
 & $1.9_{\,-0.5\,-1.6\,-0.6\,-0.7}^{\,+0.5\,+2.4\,+0.5\,+1.6}$
 & 1.1 & 1.6 & 2.4 & 1.6
 & $3.1\pm 0.7$ \\
$\bar B^0\to\eta\bar K^0$
 & $1.1_{\,-0.1\,-1.3\,-0.5\,-0.5}^{\,+0.1\,+2.0\,+0.4\,+1.3}$
 & 1.0 & 1.1 & 1.6 & 1.1
 & $<4.6$ \\
$B^-\to\eta' K^-$
 & $49.1_{\,-4.9\,-16.3\,-7.4\,-14.6}^{\,+5.1\,+26.5\,+13.6\,+33.6}$
 & 52.1 & 78.3 & 64.6 & 76.1
 & $77.6\pm 4.6$ \\
$\bar B^0\to\eta'\bar K^0$
 & $46.5_{\,-4.4\,-15.4\,-6.8\,-13.5}^{\,+4.7\,+24.9\,+12.3\,+31.0}$
 & 46.0 & 72.8 & 60.7 & 70.3
 & $60.6\pm 7.0$ \\
\hline\hline
$B^-\to\eta K^{*-}$
 & $10.8_{\,-1.7\,-4.4\,-1.3\,-5.5}^{\,+1.9\,+8.1\,+1.8\,+16.5}$
 & 14.0 & 19.4 & 19.1 & 19.9
 & $25.4\pm 5.3$ \\
$\bar B^0\to\eta\bar K^{*0}$
 & $10.7_{\,-1.0\,-4.3\,-1.2\,-5.5}^{\,+1.1\,+7.8\,+1.4\,+16.2}$
 & 10.7 & 18.2 & 18.6 & 18.6
 & $16.4\pm 3.0$ \\
$B^-\to\eta' K^{*-}$
 & $5.1_{\,-1.0\,-3.8\,-3.0\,-3.3}^{\,+0.9\,+7.5\,+2.1\,+6.7}$
 & 3.5 & 7.6 & 7.1 & 2.2
 & $<35$ \\
$\bar B^0\to\eta'\bar K^{*0}$
 & $3.9_{\,-0.4\,-3.3\,-2.5\,-2.9}^{\,+0.4\,+6.6\,+1.8\,+6.2}$
 & 3.7 & 6.7 & 6.0 & 1.9
 & $<13$ \\
\hline\hline
\end{tabular}
\end{center}
\end{table}

\begin{table}
\centerline{\parbox{14cm}{\caption{\label{tab:pred5CP}
Direct CP asymmetries (in units of $10^{-2}$) of penguin-dominated 
$\bar B\to PP$ decays (top) and $\bar B\to PV$ decays (bottom) with 
$\Delta S=1$.}}}
\vspace{0.1cm}
\begin{center}
\begin{tabular}{|l|c|cccc|c|}
\hline\hline
\multicolumn{1}{|c|}{Mode} & Theory & S1 & S2 & S3 & S4 & Experiment \\
\hline\hline
$B^-\to\eta K^-$
 & $-18.9_{\,-6.9\,-17.5\,-8.5\,-21.8}^{\,+6.4\,+11.7\,+4.8\,+25.3}$
 & $-33.3$ & $-19.6$ & 2.7 & 9.6
 & $-32\pm 20$ \\
$\bar B^0\to\eta\bar K^0$
 & $-9.0_{\,-2.1\,-12.6\,-6.2\,-7.8}^{\,+2.8\,+\phantom{1}5.4\,+2.8\,+8.2}$
 & $-10.2$ & $-10.3$ & $-3.5$ & 0.5
 & --- \\
$B^-\to\eta' K^-$
 & $2.4_{\,-0.7\,-0.8\,-0.4\,-3.5}^{\,+0.6\,+0.6\,+0.3\,+3.4}$
 & 2.3 & 1.6 & $-0.6$ & $-0.8$
 & $2\pm 4$ \\
$\bar B^0\to\eta'\bar K^0$
 & $1.8_{\,-0.5\,-0.3\,-0.2\,-0.8}^{\,+0.4\,+0.3\,+0.1\,+0.8}$
 & 1.9 & 1.4 & 1.2 & 0.7
 & $8\pm 18$ \\
\hline\hline
$B^-\to\eta K^{*-}$
 & $3.5_{\,-0.9\,-2.7\,-0.8\,-20.5}^{\,+0.9\,+1.9\,+0.8\,+20.7}$
 & 2.7 & 3.0 & $-9.2$ & $-5.7$
 & $-5\pm 28$ \\
$\bar B^0\to\eta\bar K^{*0}$
 & $3.8_{\,-1.1\,-0.8\,-0.2\,-3.5}^{\,+0.9\,+1.1\,+0.2\,+3.8}$
 & 3.8 & 2.9 & 1.5 & 0.8
 & $17\pm 27$ \\
$B^-\to\eta' K^{*-}$
 & $-14.2_{\,-4.2\,-13.8\,-14.6\,-26.1}^{\,+4.7\,+8.5\,+4.9\,+27.5}$
 & $-20.8$ & $-9.8$ & 4.5 & 22.1
 & --- \\
$\bar B^0\to\eta'\bar K^{*0}$
 & $-5.5_{\,-1.3\,-5.1\,-5.9\,-7.0}^{\,+1.6\,+3.1\,+1.8\,+6.2}$
 & $-5.8$ & $-4.1$ & $-2.4$ & 1.7
 & --- \\
\hline\hline
\end{tabular}
\end{center}
\end{table}

We shall not repeat here the discussion of the complex dynamics of 
decays to mesons with flavor-singlet components that is particularly
important for the penguin-dominated $\Delta S=1$ decays. Let us briefly
summarize the conclusions reached in \cite{Beneke:2002jn}.
The singlet penguin amplitude is presumably small and plays no important 
role in the enhancement of the $\bar B\to\eta'\bar K$ branching 
fractions. Rather, the particular pattern of the 
$\bar B\to\eta^{(\prime)}\bar K^{(*)}$ 
decay rates is caused by the interference of different non-singlet penguin 
amplitudes.
In particular, the large $\bar B\to\eta'\bar K$ branching fractions are 
obtained naturally in QCD factorization. 
There are, however, large uncertainties in applying QCD factorization to 
$\eta^{(\prime)}$ final states related to $\eta$--$\eta'$ mixing, a possible 
singlet contribution to the $B\to\eta^{(\prime)}$ form factors, and a 
novel, soft spectator-scattering term.
The numerical results given in the tables and their errors corroborate
these findings. 

\begin{table}
\centerline{\parbox{14cm}{\caption{\label{tab:pred6BR}
CP-averaged branching ratios (in units of $10^{-6}$) of tree-dominated 
$\bar B\to PP$ decays (top) and $\bar B\to PV$ decays (bottom) with 
$\Delta D=1$.}}}
\vspace{0.1cm}
\begin{center}
\begin{tabular}{|l|c|cccc|c|}
\hline\hline
\multicolumn{1}{|c|}{Mode} & Theory &  S1 & S2 & S3 & S4 & Experiment \\
\hline\hline
$B^-\to\pi^-\eta$
 & $4.7_{\,-1.7\,-1.5\,-0.3\,-0.3}^{\,+1.9\,+1.8\,+0.6\,+0.4}$
 & 2.7 & 4.1 & 4.7 & 3.8
 & $3.9\pm 0.9$ \\
$\bar B^0\to\pi^0\eta$
 & $0.28_{\,-0.08\,-0.26\,-0.02\,-0.08}^{\,+0.09\,+0.43\,+0.02\,+0.19}$
 & 0.45 & 0.31 & 0.35 & 0.30
 & $<2.9$ \\
$B^-\to\pi^-\eta'$
 & $3.1_{\,-1.2\,-1.0\,-0.3\,-0.3}^{\,+1.3\,+1.2\,+0.6\,+0.3}$
 & 1.8 & 3.1 & 3.1 & 2.9
 & $<7$ \\
$\bar B^0\to\pi^0\eta'$
 & $0.17_{\,-0.05\,-0.16\,-0.05\,-0.06}^{\,+0.05\,+0.28\,+0.10\,+0.14}$
 & 0.27 & 0.36 & 0.24 & 0.35
 & $<5.7$ \\
\hline
$\bar B^0\to\eta\eta$
 & $0.16_{\,-0.03\,-0.18\,-0.03\,-0.05}^{\,+0.03\,+0.43\,+0.09\,+0.10}$
 & 0.13 & 0.24 & 0.23 & 0.27
 & --- \\
$\bar B^0\to\eta\eta'$
 & $0.16_{\,-0.04\,-0.16\,-0.06\,-0.05}^{\,+0.04\,+0.59\,+0.14\,+0.07}$
 & 0.10 & 0.30 & 0.20 & 0.31
 & --- \\
$\bar B^0\to\eta'\eta'$
 & $0.06_{\,-0.01\,-0.05\,-0.03\,-0.03}^{\,+0.01\,+0.23\,+0.07\,+0.06}$
 & 0.03 & 0.13 & 0.10 & 0.16
 & --- \\
\hline\hline
$B^-\to\eta\rho^-$
 & $9.4_{\,-3.7\,-3.0\,-0.4\,-0.7}^{\,+4.6\,+3.6\,+0.7\,+0.7}$
 & 8.3 & 6.3 & 9.1 & 6.3
 & $<6.2$ \\
$\bar B^0\to\eta\rho^0$
 & $0.03_{\,-0.01\,-0.10\,-0.01\,-0.02}^{\,+0.02\,+0.16\,+0.02\,+0.05}$
 & 0.05 & 0.04 & 0.06 & 0.09
 & $<5.5$ \\
$B^-\to\eta'\rho^-$
 & $6.3_{\,-2.5\,-2.0\,-0.3\,-0.5}^{\,+3.1\,+2.4\,+0.5\,+0.5}$
 & 6.0 & 4.3 & 6.1 & 4.2
 & $<33$ \\
$\bar B^0\to\eta'\rho^0$
 & $0.01_{\,-0.00\,-0.06\,-0.00\,-0.01}^{\,+0.01\,+0.11\,+0.02\,+0.03}$
 & 0.01 &  0.03 & 0.03 & 0.06
 & $<12$ \\
\hline
$\bar B^0\to\eta\omega$
 & $0.31_{\,-0.12\,-0.11\,-0.14\,-0.16}^{\,+0.14\,+0.16\,+0.35\,+0.22}$
 & 0.23 & 0.53 & 0.44 & 0.65
 & --- \\
$\bar B^0\to\eta'\omega$
 & $0.20_{\,-0.08\,-0.05\,-0.10\,-0.11}^{\,+0.10\,+0.15\,+0.25\,+0.15}$
 & 0.17 & 0.36 & 0.29 & 0.44
 & --- \\
$\bar B^0\to\eta\phi$
 & $\approx 0.001$
 & 0.001 & 0.001 & 0.000 & 0.001
 & --- \\
$\bar B^0\to\eta'\phi$
 & $\approx 0.001$
 & 0.002 & 0.002 & 0.002 & 0.003 
 & --- \\
\hline\hline
\end{tabular}
\end{center}
\end{table}

\begin{table}[t]
\centerline{\parbox{14cm}{\caption{\label{tab:pred6CP}
Direct CP asymmetries (in units of $10^{-2}$) of tree-dominated 
$\bar B\to PP$ decays (top) and $\bar B\to PV$ decays (bottom) with 
$\Delta D=1$. We only consider modes with branching fractions larger than
$10^{-7}$.}}}
\vspace{0.1cm}
\begin{center}
{\tabcolsep=0.19cm\begin{tabular}{|l|c|cccc|c|}
\hline\hline
\multicolumn{1}{|c|}{Mode} & Theory &  S1 & S2 & S3 & S4 & Experiment \\
\hline\hline
$B^-\to\pi^-\eta$
 & $-14.9_{\,-5.4\,-7.4\,-0.8\,-17.3}^{\,+4.9\,+8.3\,+1.3\,+17.4}$
 & $-25.9$ & $-15.4$ & 0.7 & 5.6 
 & $-51\pm 19$ \\
$\bar B^0\to\pi^0\eta$
 & $-17.9_{\,-4.1\,-14.1\,-1.4\,-32.9}^{\,+5.2\,+\phantom{1}7.9\,+1.2\,+33.4}$
 & $-11.0$ & $-15.8$ & 3.8 & 8.5
 & --- \\
$B^-\to\pi^-\eta'$
 & $-8.6_{\,-3.1\,-\phantom{1}9.0\,-0.7\,-20.4}^{\,+2.8\,+10.5\,+0.7\,+20.4}$
 &  $-14.9$ & $-10.5$ & 8.6 & 11.1
 & --- \\
$\bar B^0\to\pi^0\eta'$
 & $-19.2_{\,-4.3\,-7.8\,-3.3\,-35.8}^{\,+5.5\,+7.7\,+4.1\,+35.7}$
 & $-11.9$ & $-12.1$ & 1.9 & 6.1
 & --- \\
\hline
$\bar B^0\to\eta\eta$
 & $-62.8_{\,-13.1\,-22.2\,-\phantom{1}9.5\,-16.2}^{\,+18.1\,+65.6\,+19.0\,+23.4}$
 &  $-79.1$ & $-41.3$ & $-45.1$ & $-20.1$
 & --- \\
$\bar B^0\to\eta\eta'$
 & $-56.3_{\,-16.6\,-\phantom{1}16.1\,-15.8\,-16.3}^{\,+17.1\,+141.0\,+20.7\,+13.7}$
 & $-87.2$ & $-34.3$ & $-46.8$ & $-28.0$ 
 & --- \\
$\bar B^0\to\eta'\eta'$
 & $-46.0_{\,-14.9\,-\phantom{1}11.0\,-17.5\,-34.0}^{\,+14.6\,+138.4\,+22.0\,+40.2}$
 & $-75.3$ & $-27.8$ & $-18.4$ & $-3.5$
 & --- \\
\hline\hline
$B^-\to\eta\rho^-$
 & $-2.4_{\,-0.7\,-6.3\,-0.4\,-0.2}^{\,+0.7\,+6.3\,+0.4\,+0.2}$
 & $-2.7$ & $-1.9$ & $ -2.4$ & $-6.0$
 & --- \\
$B^-\to\eta'\rho^-$
 & $4.1_{\,-1.1\,-6.9\,-0.8\,-7.0}^{\,+1.2\,+7.9\,+0.5\,+7.0}$
 & 4.3 & 5.2 & 4.1 & 1.1
 & --- \\
\hline
$\bar B^0\to\eta\omega$
 & $-33.4_{\,-\phantom{1}9.5\,-55.8\,-21.4\,-20.8}^{\,+10.0\,+65.3\,+20.9\,+19.2}$
 & $-44.9$ & $-10.5$ & $-31.4$ & $-27.5$
 & --- \\
$\bar B^0\to\eta'\omega$
 & $0.2_{\,-0.1\,-76.5\,-11.5\,-20.1}^{\,+0.1\,+53.0\,+11.6\,+20.4}$
 & 0.2 & 3.7 & $-1.9$ & $-15.2$
 & --- \\
\hline\hline
\end{tabular}}
\end{center}
\end{table}

The $\Delta D=1$ decay modes all have small branching fractions (and,
perhaps, large CP asymmetries) unless they involve the color-favored
tree coefficient $\alpha_1$. The corresponding results are summarized
in Tables~\ref{tab:pred6BR} and \ref{tab:pred6CP}. 
We discuss specifically here only the $\Delta D=1$ decays to 
final states with one pion or $\rho$ meson. 
The two independent decay amplitudes, simplified according to 
Section~\ref{sec:simpleampls}, are given by
\begin{eqnarray}
\label{pietasimp}
   \sqrt2\,{\cal A}_{B^-\to\pi^-\eta}
   &=& A_{\pi\eta_q} \Big[ 
    \delta_{pu}\,(\alpha_2 + \beta_2)
    + 2\hat\alpha_3^p + \hat \alpha_4^p\Big] 
    + \sqrt{2} A_{\pi\eta_s} \hat \alpha_3^p 
    + \sqrt{2} A_{\pi\eta_c} \delta_{pc}\,\alpha_2 
    \nonumber\\
   &+& A_{\eta_q\pi} \Big[ \delta_{pu}\,(\alpha_1 + \beta_2)
    + \hat \alpha_4^p \Big] \,, 
    \nonumber\\
   -2\,{\cal A}_{\bar B^0\to\pi^0\eta}
   &=& A_{\pi\eta_q} \Big[ 
    \delta_{pu}\,(\alpha_2 - \beta_1)
    + 2\hat\alpha_3^p + \hat \alpha_4^p\Big] 
    + \sqrt{2} A_{\pi\eta_s} \hat \alpha_3^p 
    + \sqrt{2} A_{\pi\eta_c} \delta_{pc}\,\alpha_2 
    \nonumber\\
   &+& A_{\eta_q\pi} \Big[ \delta_{pu}\,(-\alpha_2 - \beta_1)
    + \hat \alpha_4^p \Big] \,,
\end{eqnarray}
where $\hat \alpha_3^p\equiv\alpha_3^p+\beta_{S3}^p$. The amplitudes for 
$\bar B\to\pi \eta'$ and $\bar B\to\rho\eta^{(\prime)}$ are obtained 
from these results by replacing $(\pi,\eta)\to(\pi,\eta')$ and 
$(\pi,\eta)\to(\rho,\eta^{(\prime)})$, respectively. We find that the 
singlet coefficient $\hat\alpha_3^p$ and the ``charm content'' in the
$\eta^{(\prime)}$ are small. The 
amplitudes can then be further approximated by setting $\hat\alpha_3^p$ 
and $A_{M_1\eta_c^{(\prime)}}$ to zero.

The neutral $B$-meson decays to $\pi^0\eta^{(\prime)}$ and 
$\rho^0\eta^{(\prime)}$ have 
small branching fractions, because the color-suppressed tree amplitudes 
proportional to $\alpha_2$ tend to cancel each other. A consequence of 
this is that the $\bar B^0\to\rho^0\eta^{(\prime)}$ decay rates are 
predicted to be 
much smaller than the $\bar B^0\to\pi^0\eta^{(\prime)}$ rates, because 
the residual $PV$ and $VP$ penguin amplitudes are smaller. The charged 
decays have branching fractions of a few times $10^{-6}$. We note from 
(\ref{pietasimp}) that the penguin coefficient $\hat\alpha_4^p$ enters 
both parts of the amplitude with equal sign. We should therefore expect
larger penguin--tree interference in $B^-\to\pi^-\eta$ than in 
$\bar B^0\to\pi^-\pi^+$. Data are available for $B^-\to\pi^-\eta$, 
whose branching fraction is in good agreement with our result. 
The upper limit on the $\eta\rho^-$ final state is already below 
or near our prediction, and hence we expect this decay to be discovered 
soon. 

It should be noted that some of the decay rates for final states 
containing an $\eta'$ meson, in particular the modes 
$B^-\to\pi^-\eta'$, $\rho^-\eta'$, are very sensitive to the singlet 
contribution to the $B\to\eta'$ form factor in (\ref{F2def}), which we 
simply put to zero. For instance,
setting $F_2=0.1$ increases $\mbox{Br}(B^-\to\pi^-\eta')$ to about
$7\times 10^{-6}$, and $\mbox{Br}(B^-\to\rho^-\eta')$ to
$16\times 10^{-6}$.

The direct CP asymmetries in Table~\ref{tab:pred6CP} confirm the 
possibility of significant penguin--tree interference in the decays
$B^-\to\pi^-\eta^{(\prime)}$. However, the measurement of 
$A_{\rm CP}(\pi^-\eta)=(-51\pm 19)\%$, together with a large CP asymmetry 
$A_{\rm CP}(\pi^-\pi^+)=(51\pm 23)\%$ of opposite sign and a large 
asymmetry $A_{\rm CP}(\eta K^-)=(-32\pm 20)\%$ of the same sign, is 
difficult to understand in QCD factorization and presumably in any 
framework (see, for instance, the comments in \cite{Chiang:2003rb}), 
unless one can accommodate very large SU(3) flavor-symmetry breaking in 
the amplitudes pertaining to the $\pi\pi$, $\pi\eta^{(\prime)}$, and 
$\bar K\eta^{(\prime)}$ final states. Indeed, if the 
dominant interference is between $\hat\alpha_4^c$ and $\alpha_1$, then 
the two CP asymmetries must have the same sign, barring very large 
final-state dependence of these coefficients. Possible corrections
could come from the singlet penguin amplitude $\hat\alpha_3^c$, but for 
this amplitude to have an effect as significant as indicated by 
the data, the estimates in \cite{Beneke:2002jn} or 
\cite{Chiang:2003rb} would have to be 
grossly in error. This leaves $\alpha_2$, which if large could 
affect the phase of the penguin-to-tree ratio by a 
noticeable amount. However, for this to reverse the sign of a large 
CP asymmetry the color-suppressed tree coefficient would have to 
exceed the color-allowed one ($\alpha_1$), which would affect all 
color-suppressed $B$ decays.

\boldmath
\section{Results for $\bar B_s$ decays}
\unboldmath
\label{sec:Bs}

\subsection{Simplified expressions for the decay amplitudes}

So far we have focused our attention on the decays of $B^-$ and 
$\bar B^0$ mesons, which are currently under investigation at the 
$B$ factories. Our formalism applies equally
to the decays of $B_s$ mesons, which cannot be studied at these 
facilities. These decay modes will soon become accessible at hadronic
$B$ factories operated at the Tevatron Run-II, and in the longer term
at LHC-b and BTeV. In this section we present results
for the branching ratios and CP asymmetries of $B_s$ decays into $PP$
and $PV$ final states. Our predictions are collected in 
Tables~\ref{tab:pred7BR}--\ref{tab:pred10CP}.

\begin{table}[p]
\centerline{\parbox{14cm}{\caption{\label{tab:pred7BR}
CP-averaged branching ratios (in units of $10^{-6}$) of penguin-dominated 
$\bar B_s\to PP$ decays (top) and $\bar B_s\to PV$ decays (bottom) with 
$\Delta S=1$. The errors and scenarios have the same meaning as in 
Section~\ref{sec:penguins}.}}}
\begin{center}
\begin{tabular}{|l|c|cccc|}
\hline\hline
\multicolumn{1}{|c|}{Mode} & Theory & S1 & S2 & S3 & S4 \\
\hline\hline
$\bar B_s\to K^+ K^-$
 & $22.7_{\,-3.2\,-\phantom{1}8.4\,-2.0\,-\phantom{1}9.1}^{\,+3.5\,+12.7\,+2.0\,+24.1}$
 & 28.0 & 33.4 & 34.3 & 36.1 \\
$\bar B_s\to K^0\bar K^0$
 & $24.7_{\,-2.4\,-\phantom{1}9.2\,-2.9\,-\phantom{1}9.8}^{\,+2.5\,+13.7\,+2.6\,+25.6}$
 & 24.0 & 35.6 & 36.7 & 38.3 \\
\hline
$\bar B_s\to\eta\eta$
 & $15.6_{\,-1.5\,-6.8\,-2.5\,-\phantom{1}5.5}^{\,+1.6\,+9.9\,+2.2\,+13.5}$
 & 14.9 & 22.7 & 21.4 & 22.5 \\
$\bar B_s\to\eta\eta'$
 & $54.0_{\,-5.2\,-22.4\,-6.4\,-16.7}^{\,+5.5\,+32.4\,+8.3\,+40.5}$
 & 52.4 & 79.5 & 72.3 & 77.7 \\
$\bar B_s\to\eta'\eta'$
 & $41.7_{\,-4.0\,-17.2\,-\phantom{1}8.5\,-15.4}^{\,+4.2\,+26.3\,+15.2\,+36.6}$
 & 41.4 & 63.4 & 60.2 & 65.5 \\
\hline\hline
$\bar B_s\to K^+ K^{*-}$
 & $4.1_{\,-1.5\,-1.3\,-0.9\,-2.3}^{\,+1.7\,+1.5\,+1.0\,+9.2}$
 & 7.4 & 4.3 & 9.0 & 13.7 \\
$\bar B_s\to K^0\bar K^{*0}$
 & $3.9_{\,-0.4\,-1.4\,-1.4\,-\phantom{1}2.8}^{\,+0.4\,+1.5\,+1.3\,+10.4}$
 & 3.8 & 4.2 & 9.1 & 14.3 \\
$\bar B_s\to K^- K^{*+}$
 & $5.5_{\,-1.4\,-2.6\,-0.7\,-\phantom{1}3.6}^{\,+1.3\,+5.0\,+0.8\,+14.2}$
 & 3.3 & 9.9 & 13.0 & 9.0 \\
$\bar B_s\to\bar K^0 K^{*0}$
 & $4.2_{\,-0.4\,-2.2\,-0.9\,-\phantom{1}3.2}^{\,+0.4\,+4.6\,+1.1\,+13.2}$
 & 4.1 & 8.7 & 11.4 & 7.9 \\
\hline
$\bar B_s\to\eta\omega$
 & $0.012_{\,-0.004\,-0.003\,-0.006\,-0.006}^{\,+0.005\,+0.010\,+0.028\,+0.025}$
 & 0.017 & 0.011 & 0.010 & 0.009 \\
$\bar B_s\to\eta'\omega$
 & $0.024_{\,-0.009\,-0.006\,-0.010\,-0.015}^{\,+0.011\,+0.028\,+0.077\,+0.042}$
 & 0.024 & 0.024 & 0.039 & 0.033 \\
$\bar B_s\to\eta\phi$
 & $0.12_{\,-0.02\,-0.14\,-0.12\,-0.13}^{\,+0.02\,+0.95\,+0.54\,+0.32}$
 & 0.15 & 1.02 & 0.24 & 1.47 \\
$\bar B_s\to\eta'\phi$
 & $0.05_{\,-0.01\,-0.17\,-0.08\,-0.04}^{\,+0.01\,+1.10\,+0.18\,+0.40}$
 & 0.05 & 1.08 & 0.07 & 2.10 \\
\hline\hline
\end{tabular}
\end{center}
%
\vspace{-0.15cm}
\centerline{\parbox{14cm}{\caption{\label{tab:pred7CP}
Direct CP asymmetries (in units of $10^{-2}$) of penguin-dominated 
$\bar B_s\to PP$ decays (top) and $\bar B_s\to PV$ decays (bottom) with 
$\Delta S=1$. We only consider modes with branching fractions larger than
$10^{-7}$.}}}
\begin{center}
\begin{tabular}{|l|c|cccc|}
\hline\hline
\multicolumn{1}{|c|}{Mode} & Theory & S1 & S2 & S3 & S4 \\
\hline\hline
$\bar B_s\to K^+ K^-$
 & $4.0_{\,-1.0\,-2.3\,-0.5\,-11.3}^{\,+1.0\,+2.0\,+0.5\,+10.4}$
 & 3.2 & 3.0 & $-4.5$ & $-4.7$ \\
$\bar B_s\to K^0\bar K^0$
 & $0.9_{\,-0.2\,-0.2\,-0.1\,-0.3}^{\,+0.2\,+0.2\,+0.1\,+0.2}$
 & 0.9 & 0.7 & 0.6 & 0.6 \\
\hline
$\bar B_s\to\eta\eta$
 & $-1.6_{\,-0.4\,-0.6\,-0.7\,-2.2}^{\,+0.5\,+0.6\,+0.4\,+2.2}$
 & $-1.7$ & $-1.1$ & $-0.4$ & $-0.1$ \\
$\bar B_s\to\eta\eta'$
 & $0.4_{\,-0.1\,-0.3\,-0.1\,-0.3}^{\,+0.1\,+0.3\,+0.1\,+0.4}$
 & 0.4 & 0.4 & 0.3 & 0.2 \\
$\bar B_s\to\eta'\eta'$
 & $2.1_{\,-0.6\,-0.4\,-0.3\,-1.2}^{\,+0.5\,+0.4\,+0.2\,+1.1}$
 & 2.1 & 1.7 & 1.3 & 1.0 \\
\hline\hline
$\bar B_s\to K^+ K^{*-}$
 & $2.2_{\,-0.7\,-8.0\,-5.9\,-71.0}^{\,+0.6\,+8.4\,+5.1\,+68.6}$
 & 1.2 & 1.8 & $-34.8$ & $-10.0$ \\
$\bar B_s\to K^0\bar K^{*0}$
 & $1.7_{\,-0.5\,-0.5\,-0.4\,-0.8}^{\,+0.4\,+0.6\,+0.5\,+1.4}$
 & 1.7 & 1.5 & 1.0 & 0.8 \\
$\bar B_s\to K^- K^{*+}$
 & $-3.1_{\,-1.1\,-2.6\,-1.3\,-45.0}^{\,+1.0\,+3.8\,+1.6\,+47.5}$
 & $-5.2$ & $-2.4$ & 18.8 & 26.6 \\
$\bar B_s\to\bar K^0 K^{*0}$
 & $0.2_{\,-0.1\,-0.3\,-0.1\,-0.1}^{\,+0.0\,+0.2\,+0.1\,+0.2}$
 & 0.2 & 0.3 & 0.1 & 0.1 \\
\hline
$\bar B_s\to\eta\phi$
 & $-8.4_{\,-2.1\,-71.2\,-44.7\,-59.7}^{\,+2.0\,+30.1\,+14.6\,+36.3}$
 & $-6.5$ & 3.8 & $-8.7$ & $-9.5$ \\
$\bar B_s\to\eta'\phi$
 & $-62.2_{\,-10.2\,-\phantom{1}84.2\,-46.8\,-\phantom{1}49.9}^{\,+15.9\,+132.3\,+80.8\,+122.4}$
 & $-61.1$ & $-8.9$ & $-34.0$ & 7.5 \\
\hline\hline
\end{tabular}
\end{center}
\end{table}

\begin{table}[t]
\centerline{\parbox{14cm}{\caption{\label{tab:pred8BR}
CP-averaged branching ratios (in units of $10^{-6}$) of tree-dominated 
$\bar B_s\to PP$ decays (top) and $\bar B_s\to PV$ decays (bottom) with 
$\Delta S=1$.}}}
\vspace{0.1cm}
\begin{center}
\begin{tabular}{|l|c|cccc|}
\hline\hline
\multicolumn{1}{|c|}{Mode} & Theory & S1 & S2 & S3 & S4 \\
\hline\hline
$\bar B_s\to\pi^0\eta$
 & $0.075_{\,-0.012\,-0.025\,-0.010\,-0.007}^{\,+0.013\,+0.030\,+0.008\,+0.010}$
 & 0.097 & 0.069 & 0.077 & 0.073 \\
$\bar B_s\to\pi^0\eta'$
 & $0.11_{\,-0.02\,-0.04\,-0.01\,-0.01}^{\,+0.02\,+0.04\,+0.01\,+0.01}$
 & 0.15 & 0.10 & 0.10 & 0.10 \\
\hline\hline
$\bar B_s\to\pi^0\omega$
 & $\approx 0.0005$
 & 0.0004 & 0.0004 & 0.0033 & 0.0024 \\
$\bar B_s\to\pi^0\phi$
 & $0.12_{\,-0.02\,-0.04\,-0.01\,-0.01}^{\,+0.03\,+0.04\,+0.01\,+0.02}$
 & 0.16 & 0.12 & 0.12 & 0.12 \\
\hline
$\bar B_s\to\rho^0\eta$
 & $0.17_{\,-0.03\,-0.06\,-0.02\,-0.01}^{\,+0.03\,+0.07\,+0.02\,+0.02}$
 & 0.23 & 0.17 & 0.18 & 0.17 \\
$\bar B_s\to\rho^0\eta'$
 & $0.25_{\,-0.05\,-0.08\,-0.02\,-0.02}^{\,+0.06\,+0.10\,+0.02\,+0.02}$
 & 0.35 & 0.24 & 0.24 & 0.24 \\
\hline\hline
\end{tabular}
\end{center}
%
\centerline{\parbox{14cm}{\caption{\label{tab:pred8CP}
Direct CP asymmetries (in units of $10^{-2}$) of tree-dominated 
$\bar B_s\to PP$ decays (top) and $\bar B_s\to PV$ decays (bottom) with 
$\Delta S=1$. We only consider modes with branching fractions larger than
$10^{-7}$.}}}
\vspace{0.1cm}
\begin{center}
\begin{tabular}{|l|c|cccc|}
\hline\hline
\multicolumn{1}{|c|}{Mode} & Theory & S1 & S2 & S3 & S4 \\
\hline\hline
$\bar B_s\to\pi^0\eta'$
 & $27.8_{\,-7.1\,-5.7\,-2.0\,-27.2}^{\,+6.0\,+9.6\,+2.0\,+24.7}$
 & 19.5 & 26.1 & 36.6 & 35.3 \\
\hline\hline
$\bar B_s\to\pi^0\phi$
 & $27.2_{\,-6.8\,-5.6\,-2.4\,-37.1}^{\,+6.1\,+9.8\,+2.7\,+32.0}$
 & 20.0 & 25.2 & 27.2 & 24.8 \\
\hline
$\bar B_s\to\rho^0\eta$
 & $27.8_{\,-6.7\,-5.7\,-2.2\,-28.4}^{\,+6.4\,+9.1\,+2.6\,+25.9}$
 & 21.0 & 25.1 & 16.4 & 15.6 \\
$\bar B_s\to\rho^0\eta'$
 & $28.9_{\,-7.5\,-\phantom{1}6.3\,-1.8\,-27.5}^{\,+6.1\,+10.3\,+1.5\,+24.8}$
 & 20.0 & 26.1 & 36.7 & 32.2 \\
\hline\hline
\end{tabular}
\end{center}
\end{table}

\begin{table}[t]
\centerline{\parbox{14cm}{\caption{\label{tab:pred9BR}
CP-averaged branching ratios (in units of $10^{-6}$) of 
annihilation-do\-minated $\bar B_s\to PP$ decays (top) and 
$\bar B_s\to PV$ decays (bottom) with $\Delta S=1$.}}}
\vspace{0.1cm}
\begin{center}
\begin{tabular}{|l|c|cccc|}
\hline\hline
\multicolumn{1}{|c|}{Mode} & Theory & S1 & S2 & S3 & S4 \\
\hline\hline
$\bar B_s\to\pi^+\pi^-$
 & $0.024_{\,-0.003\,-0.012\,-0.000\,-0.021}^{\,+0.003\,+0.025\,+0.000\,+0.163}$
 & 0.027 & 0.032 & 0.149 & 0.155 \\
$\bar B_s\to\pi^0\pi^0$
 & $0.012_{\,-0.001\,-0.006\,-0.000\,-0.011}^{\,+0.001\,+0.013\,+0.000\,+0.082}$
 & 0.014 & 0.016 & 0.075 & 0.078 \\
\hline\hline
$\bar B_s\to\pi^+\rho^-$
 & $\approx 0.003$
 & 0.002 & 0.003 & 0.019 & 0.014 \\
$\bar B_s\to\pi^0\rho^0$
 & $\approx 0.003$
 & 0.002 & 0.003 & 0.019 & 0.017 \\
\hline
$\bar B_s\to\pi^-\rho^+$
 & $\approx 0.003$
 & 0.002 & 0.003 & 0.019 & 0.015 \\
\hline\hline
\end{tabular}
\end{center}
\end{table}

\boldmath
\subsubsection{Decays with $\Delta S=1$}
\unboldmath

Tables~\ref{tab:pred7BR} and \ref{tab:pred7CP} contain results for 
penguin-dominated $B_s$ decays with $\Delta S=1$. The modes with two 
pseudoscalars in the final state have large branching fractions, which in 
fact are among the largest of all rare $B$ decays studied in this work. 
The expressions for the $\bar B_s\to\bar K K$ amplitudes, simplified 
according to the approximations described in 
Section~\ref{sec:simpleampls}, are given by
\begin{equation}
\begin{aligned}
   {\cal A}_{\bar B_s\to\bar K^0 K^0}
   &= B_{\bar K K}\,\delta_{pc}\,b_4^c
    + A_{K\bar K} \Big[ \hat\alpha_4^p + \delta_{pc}\,\beta_4^c \Big]
    \,, \\
   {\cal A}_{\bar B_s\to K^- K^+}
   &= B_{\bar K K}\,\delta_{pc}\,b_4^c
    + A_{K\bar K} \Big[ \delta_{pu}\,\alpha_1 + \hat\alpha_4^p
    + \delta_{pc}\,\beta_4^c \Big] \,.
\end{aligned}
\end{equation}
The expressions on the right-hand side must be multiplied with 
$\lambda_p^{(s)}$ and summed over $p=u,c$. The amplitudes for 
$\bar B_s\to\bar K K^*$ and $\bar B_s\to\bar K^* K$ are obtained from these 
expressions by interchanging $K\leftrightarrow K^*$ or 
$\bar K\leftrightarrow\bar K^*$ everywhere. Apart from the small 
annihilation terms parameterized by $\beta_4^c$ these modes are 
characterized by a simple pattern of tree--penguin interference, which 
resembles that in the decays $B^-\to\pi^-\bar K^0$ and 
$\bar B^0\to\pi^+ K^-$ in (\ref{ApiK}). This means that many of the analysis 
strategies discussed in Section~\ref{subsec:PTratios} can be applied also
here, once experimental data on the decays $\bar B_s\to\bar K K$ (and the
corresponding $PV$ modes) will become available. The expressions for modes 
involving $\eta$ or $\eta'$ are more complicated and will not be given 
here. The exact decay amplitudes for these modes can be found in 
Appendix~A. Note from Table~\ref{tab:pred7BR} that the branching fractions
for the decays $\bar B_s\to\eta^{(\prime)}\omega$ and 
 $\bar B_s\to\eta^{(\prime)}\phi$ are several orders of magnitude smaller
than the corresponding branching fractions for the decays
$\bar B_s\to\eta^{(\prime)}\eta^{(\prime)}$. For the 
$\eta^{(\prime)}\omega$ final states this occurs because the $\omega$ 
meson is assumed to have no strange component. For the case of the 
$\eta^{(\prime)}\phi$ final states there is a strong cancellation between 
the $PV$ and $VP$ penguin amplitudes $\hat\alpha_4^c(\eta^{(\prime)}\phi)$
and $\hat\alpha_4^c(\phi\eta^{(\prime)})$. The corresponding branching
fractions can be enhanced by an order of magnitude by choosing a different
value for $m_s$ or giving up the assumption of universal annihilation.
The direct CP asymmetries of
the penguin-dominated $B_s$ decays with $\Delta S=1$ are predicted to be
small except for the modes with very small branching fractions.

Tables~\ref{tab:pred8BR} and \ref{tab:pred8CP} contain results for 
tree-dominated $B_s$ decays with $\Delta S=1$. The corresponding branching
fractions are very small, typically of order few times $10^{-7}$. 
Accordingly, the direct CP asymmetries can be large in all cases but will 
hardly be observable in the near future. As an example we quote the 
simplified amplitude expressions ${\cal A}_{\bar B_s\to\pi^0\omega}=0$ and
\begin{equation}
   \sqrt2\,{\cal A}_{\bar B_s\to\pi^0\phi}
   = A_{\phi\pi} \Big[ \delta_{pu}\,\alpha_2 
   + \delta_{pc}\,\3half\alpha_{3,{\rm EW}}^c \Big] \,. 
\end{equation}
The amplitudes for modes with $\eta$ or $\eta'$ are given in Appendix~A.

Some of the $B_s$ decays with $\Delta S=1$ receive only annihilation
contributions. As shown in Table~\ref{tab:pred9BR} these modes have
tiny branching fractions of order few times $10^{-9}$ to few times 
$10^{-8}$. The simplified expressions for the corresponding decay 
amplitudes are
\begin{equation}
   {\cal A}_{\bar B_s\to\pi^+\rho^-}
   = {\cal A}_{\bar B_s\to\pi^0\rho^0}
   = \delta_{pc} \Big[ B_{\pi\rho}\,b_4^c
    + B_{\rho\pi}\,b_4^c \Big] \,.
\end{equation}
The amplitude for $\bar B_s\to\pi^-\rho^+$ is obtained from the first 
expression by interchanging $\pi\leftrightarrow\rho$ everywhere. The 
expressions for the $\bar B_s\to\pi\pi$ amplitudes are obtained by 
setting $\rho\to\pi$.

\begin{table}
\centerline{\parbox{14cm}{\caption{\label{tab:pred10BR}
CP-averaged branching ratios (in units of $10^{-6}$) of
$\bar B_s\to PP$ decays (top) and $\bar B_s\to PV$ decays (bottom) with 
$\Delta D=1$.}}}
\vspace{0.1cm}
\begin{center}
\begin{tabular}{|l|c|cccc|}
\hline\hline
\multicolumn{1}{|c|}{Mode} & Theory & S1 & S2 & S3 & S4 \\
\hline\hline
$\bar B_s\to\pi^- K^+$
 & $10.2_{\,-3.9\,-3.2\,-1.2\,-0.7}^{\,+4.5\,+3.8\,+0.7\,+0.8}$
 & 6.8 & 8.1 & 10.4 & 8.3 \\
$\bar B_s\to\pi^0 K^0$
 & $0.49_{\,-0.24\,-0.14\,-0.14\,-0.17}^{\,+0.28\,+0.22\,+0.40\,+0.33}$
 & 0.95 & 0.68 & 0.60 & 0.61 \\
\hline
$\bar B_s\to K^0\eta$
 & $0.34_{\,-0.16\,-0.27\,-0.07\,-0.08}^{\,+0.19\,+0.64\,+0.21\,+0.16}$
 & 0.65 & 0.42 & 0.38 & 0.37 \\
$\bar B_s\to K^0\eta'$
 & $2.0_{\,-0.3\,-1.1\,-0.3\,-0.6}^{\,+0.3\,+1.5\,+0.6\,+1.5}$
 & 2.6 & 3.0 & 2.7 & 2.9 \\
\hline\hline
$\bar B_s\to\pi^- K^{*+}$
 & $8.7_{\,-3.7\,-2.9\,-1.0\,-0.7}^{\,+4.6\,+3.5\,+0.7\,+0.8}$
 & 10.1 & 6.7 & 9.0 & 6.8 \\
$\bar B_s\to\pi^0 K^{*0}$
 & $0.25_{\,-0.08\,-0.06\,-0.14\,-0.14}^{\,+0.08\,+0.10\,+0.32\,+0.30}$
 & 0.15 & 0.39 & 0.36 & 0.33 \\
\hline
$\bar B_s\to K^+\rho^-$
 & $24.5_{\,-\phantom{1}9.7\,-7.8\,-3.0\,-1.6}^{\,+11.9\,+9.2\,+1.8\,+1.6}$
 & 21.3 & 19.2 & 24.6 & 19.8 \\
$\bar B_s\to K^0\rho^0$
 & $0.61_{\,-0.26\,-0.15\,-0.38\,-0.36}^{\,+0.33\,+0.21\,+1.06\,+0.56}$
 & 0.82 & 0.58 & 0.70 & 0.68 \\
\hline
$\bar B_s\to K^0\omega$
 & $0.51_{\,-0.18\,-0.11\,-0.23\,-0.25}^{\,+0.20\,+0.15\,+0.68\,+0.40}$
 & 0.33 & 0.50 & 0.58 & 0.63 \\
$\bar B_s\to K^0\phi$
 & $0.27_{\,-0.08\,-0.14\,-0.06\,-0.18}^{\,+0.09\,+0.28\,+0.09\,+0.67}$
 & 0.43 & 0.54 & 0.64 & 0.46 \\
\hline
$\bar B_s\to\eta K^{*0}$
 & $0.26_{\,-0.13\,-0.22\,-0.05\,-0.15}^{\,+0.15\,+0.49\,+0.15\,+0.57}$
 & 0.52 & 0.29 & 0.54 & 0.57 \\
$\bar B_s\to\eta' K^{*0}$
 & $0.28_{\,-0.04\,-0.24\,-0.10\,-0.15}^{\,+0.04\,+0.46\,+0.23\,+0.29}$
 & 0.23 & 0.24 & 0.39 & 0.67 \\
\hline\hline
\end{tabular}
\end{center}
\end{table}

\begin{table}
\centerline{\parbox{14cm}{\caption{\label{tab:pred10CP}
Direct CP asymmetries (in units of $10^{-2}$) of 
$\bar B_s\to PP$ decays (top) and $\bar B_s\to PV$ decays (bottom) with 
$\Delta D=1$.}}}
\vspace{0.1cm}
\begin{center}
\begin{tabular}{|l|c|cccc|}
\hline\hline
\multicolumn{1}{|c|}{Mode} & Theory & S1 & S2 & S3 & S4 \\
\hline\hline
$\bar B_s\to\pi^- K^+$
 & $-6.7_{\,-2.2\,-2.9\,-0.4\,-15.2}^{\,+2.1\,+3.1\,+0.2\,+15.5}$
 & $-10.0$ & $-8.5$ & 7.7 & 10.9 \\
$\bar B_s\to\pi^0 K^0$
 & $41.6_{\,-12.0\,-13.3\,-14.5\,-51.0}^{\,+16.6\,+14.3\,+\phantom{1}7.8\,+40.9}$
 & 21.4 & 33.6 & 14.3 & 4.6 \\
\hline
$\bar B_s\to K^0\eta$
 & $46.8_{\,-13.2\,-32.2\,-12.5\,-45.6}^{\,+18.5\,+28.6\,+\phantom{1}5.2\,+34.6}$
 & 24.1 & 39.3 & 28.6 & 24.2 \\
$\bar B_s\to K^0\eta'$
 & $-36.6_{\,-8.2\,-7.4\,-2.5\,-17.3}^{\,+8.6\,+6.0\,+3.8\,+19.3}$
 & $-28.8$ & $-29.5$ & $-22.2$ & $-18.2$ \\
\hline\hline
$\bar B_s\to\pi^- K^{*+}$
 & $0.6_{\,-0.1\,-1.7\,-0.1\,-20.1}^{\,+0.2\,+1.4\,+0.1\,+19.9}$
 & 0.5 & 1.6 & $-18.5$ & $-22.0$ \\
$\bar B_s\to\pi^0 K^{*0}$
 & $-45.7_{\,-16.0\,-11.6\,-28.0\,-59.7}^{\,+14.3\,+13.0\,+28.4\,+80.0}$
 & $-79.3$ & $-41.9$ & 0.3 & 15.4 \\
\hline
$\bar B_s\to K^+\rho^-$
 & $-1.5_{\,-0.4\,-1.4\,-0.3\,-12.1}^{\,+0.4\,+1.2\,+0.2\,+12.1}$
 & $-1.7$ & $-1.7$ & 10.1 & 6.2 \\
$\bar B_s\to K^0\rho^0$
 & $24.7_{\,-5.2\,-12.4\,-17.7\,-52.3}^{\,+7.1\,+14.0\,+22.8\,+51.3}$
 & 18.3 & 24.5 & $-11.8$ & 11.6 \\
\hline
$\bar B_s\to K^0\omega$
 & $-43.9_{\,-13.4\,-18.2\,-30.2\,-49.3}^{\,+13.6\,+18.0\,+30.6\,+57.7}$
 & $-67.5$ & $-40.9$ & $-9.6$ & $-30.1$ \\
$\bar B_s\to K^0\phi$
 & $-10.3_{\,-2.4\,-3.0\,-4.1\,-7.5}^{\,+3.0\,+4.7\,+3.7\,+5.0}$
 & $-6.4$ & $-10.5$ & $-6.3$ & $-7.4$ \\
\hline
$\bar B_s\to\eta K^{*0}$
 & $40.2_{\,-11.5\,-30.8\,-14.0\,-96.3}^{\,+17.0\,+24.6\,+7.8\,+65.9}$
 & 20.4 & 36.3 & $-11.7$ & 0.6 \\
$\bar B_s\to\eta' K^{*0}$
 & $-58.6_{\,-11.9\,-11.7\,-13.9\,-35.7}^{\,+16.9\,+41.4\,+19.9\,+44.9}$
 & $-70.8$ & $-54.5$ & $-24.9$ & $-32.7$ \\
\hline\hline
\end{tabular}
\end{center}
\end{table}

\boldmath
\subsubsection{Decays with $\Delta D=1$}
\unboldmath

Most $B_s$ decays with $\Delta D=1$ are dominated by tree topologies.
The branching fractions and direct CP asymmetries of
these decays are given in Tables~\ref{tab:pred10BR} and 
\ref{tab:pred10CP}. The decays $\bar B_s\to\pi^- K^+$, $\pi^- K^{*+}$,
and $\rho^- K^+$ have large branching fractions of order 
(1--2)$\cdot 10^{-5}$. The corresponding neutral modes have much smaller 
rates. The direct CP asymmetries are predicted to be of
moderate, sometimes even large magnitude, ranging from order 10\% for 
the charged modes to significantly larger values for the neutral modes.
The simplified expressions for the corresponding decay amplitudes are
\begin{equation}
\begin{aligned}
   {\cal A}_{\bar B_s\to\pi^- K^+}
   &= A_{K\pi} \Big[ \delta_{pu}\,\alpha_1 + \hat\alpha_4^p \Big] \,, \\
   \sqrt2\,{\cal A}_{\bar B_s\to\pi^0 K^0}
   &= A_{K\pi} \Big[ \delta_{pu}\,\alpha_2 - \hat\alpha_4^p \Big] \,.
\end{aligned}
\end{equation}
The right-hand sides of the expressions must be multiplied 
with $\lambda_p^{(d)}$ and summed over $p=u,c$. The amplitudes for 
$\bar B_s\to\pi K^*$ and $\bar B_s\to\rho K$ are obtained 
by interchanging $(\pi,K)\leftrightarrow(\pi,K^*)$ or 
$(\pi,K)\leftrightarrow(\rho,K)$ everywhere. Note that these decays are
governed by a relatively simple pattern of tree--penguin interference, 
which allows extractions of the penguin coefficients 
$\hat\alpha_4^c$ and gives sensitivity to $\gamma$.

The branching fractions for the remaining $B_s$ decays with $\Delta D=1$ 
are smaller, typically of order few times $10^{-7}$ (except for 
$\bar B_s\to K^0\eta'$). The simplified expressions for the 
$\bar B_s\to K^0\omega$, $K^0\phi$ decay amplitudes are
\begin{equation}
\begin{aligned}
   \sqrt2\,{\cal A}_{\bar B_s\to K^0\omega}
   &= A_{K\omega} \Big[ \delta_{pu}\,\alpha_2 + 2\alpha_3^p 
    + \hat\alpha_4^p \Big] \,, \\
   {\cal A}_{\bar B_s\to K^0\phi}
   &= A_{K\phi}\,\alpha_3^p + A_{\phi K}\,\hat\alpha_4^p \,. 
\end{aligned}
\end{equation}
The amplitudes for modes with $\eta$ or $\eta'$ are given in Appendix~A.
In the first case the tree contribution is color suppressed, while the
second process is a pure penguin decay. This explains the small branching 
fractions. Correspondingly, we predict generically large direct CP 
asymmetries for these modes.

\section{Conclusions}
\label{sec:concl}

In this paper we extended our previous analysis \cite{BBNS3} of 
$B$-meson decays to $\pi\pi$ and $\pi K$ final states and to final 
states with $\eta^{(\prime)}$ mesons \cite{Beneke:2002jn} within 
the QCD factorization approach to all two-body final states with 
pseudoscalar mesons or one pseudoscalar and one vector meson, including 
also decays of $B_s$ mesons. The main motivation for performing a 
comprehensive analysis of all 96 final states is to obtain a global 
assessment of the phenomenology of QCD factorization with a consistent 
common input to all decay modes, and to display possible correlations 
between the various modes.

The theoretical analysis follows \cite{BBNS3}. The computation is 
performed at next-to-leading order in $\alpha_s$ for the 
hard-scattering kernels at leading order in $1/m_b$, and similarly for 
the kernels that multiply the subleading twist-3 quark--antiquark 
distribution amplitudes. Spectator scattering effects at subleading 
power in $1/m_b$ (including weak annihilation), which do not generally 
factorize, are estimated by a phenomenological model and assigned a 
100\% uncertainty (including an arbitrary strong-interaction phase). On 
the technical side, the generalization of the decay amplitudes computed 
in \cite{BBNS3} to pseudoscalar--vector final states is for most parts 
straightforward, involving only a few sign changes in the decay 
amplitudes and a few new hard-scattering kernels. For the analysis of 
this paper only the kernels for the twist-3 quark--antiquark amplitudes 
of vector mesons needed to be computed anew. We also discussed a new 
electroweak penguin effect that contributes only to neutral vector 
mesons. 

We have taken this comprehensive analysis as the occasion to summarize 
in a unified notation all results available at next-to-leading order 
in QCD factorization. The matrix elements of the effective weak 
Hamiltonian are decomposed according to their flavor structure. A complete 
list of all decay amplitudes expressed in terms of the flavor coefficients 
(which generalizes and simplifies the more familiar $a_i$ and $b_i$ 
notation) is given in Appendix~A. The next-to-leading order results for 
the coefficients are given in Section~\ref{sec:amps}, including the 
results from \cite{BBNS3} for completeness.

The comparison of theory with data shows many interesting effects, which 
we summarize here. We should note, however, that no experimental 
information exists to date for the majority of the 96 decay modes 
considered in this paper. This will allow further tests of the theory in 
the future. When passed successfully, this implies a rich source of 
information on flavor-changing transitions in the quark sector from 
purely hadronic decays.

\boldmath 
\subsubsection*{Results related to the $\pi\pi$, $\pi K$ final states}
\unboldmath 

A detailed discussion of these modes can be found in \cite{BBNS3}. Since 
2001 the experimental errors have been reduced by almost a factor of two. 
QCD factorization continues to provide a natural explanation for the 
magnitudes of the tree and penguin amplitudes relevant to these decays. 
The theory with default input parameters does not fare 
well on the $\pi^-\pi^+$ decay mode, which it predicts too large, and on 
the $\pi^0\bar K^0$ mode, which it predicts too small. The former 
discrepancy is often interpreted as evidence of a large value of 
$\gamma$ or large rescattering. We find that an alternative ``hadronic 
physics'' explanation is possible if the $B\to\pi$ form factor is about 
15\% smaller than usually assumed, the strange-quark mass is at the 
lower end of the currently favored range, and if the color-suppressed 
tree amplitude is enhanced, for example by a sizable spectator-scattering 
effect. While the evidence for this scenario is not conclusive now, 
crucial input will be provided by a measurement of the semileptonic 
decay rate near $q^2=0$. Interestingly, this scenario is also favored by 
measurements of the decay rates for $\bar B^0\to\pi^\mp\rho^\pm$. 
With regard to the ``$\pi^0\bar K^0$ anomaly'', we find that the large
experimental $\bar B^0\to\pi^0\bar K^0$ rate 
cannot be explained by a different 
value of $\gamma$, and that ``hadronic physics'' explanations appear 
extremely unlikely. While a significant modification of the electroweak 
penguin amplitude due to ``New Physics'' may explain the effect, we 
consider it to be more likely a statistical fluctuation. 

It will require more data to arrive at a conclusive picture for the 
direct CP asymmetries. While it appears now certain that the asymmetries 
are small for the $\pi K$ modes, in agreement with the predictions of QCD 
factorization, a quantitative comparison needs better statistics and a 
more accurate theoretical computation. The experimental situation for the 
$\pi^-\pi^+$ final state is still unsettled. If a direct CP asymmetry of 
order 50\% is confirmed in the future, the factorization 
framework would be in trouble. Similar comments apply to other 
observations of large direct CP asymmetries, which at present all have 
large experimental errors. 

\boldmath
\subsubsection*{Results related to the $\pi\rho$ final states}
\unboldmath 

The $\pi\rho$ system discussed in Section~\ref{sec:trees} exhibits some 
advantages for studies of CP violation that render it highly 
interesting even within the limitations of the quasi two-body 
assumption. We defined and discussed several ratios of 
$\pi\rho$ branching fractions that should shed light on the magnitudes 
of the hadronic amplitudes underlying this class of decays. The 
theoretical predictions for these ratios are in good agreement with the 
available data. 
In particular, QCD factorization predicts that the two distinct 
penguin-to-tree ratios in the $\pi\rho$ system are about a factor of 
three smaller than the corresponding ratio for $\pi\pi$, and have 
smaller errors. If this can be confirmed, it implies that $\gamma$ can 
be determined relatively accurately from time-dependent CP violation
measurements. We considered the five quantities that parameterize the 
time-dependent asymmetries in the decays $B^0,\bar B^0\to\pi^\mp\rho^\pm$ 
in detail. If data and theory are taken at face value, we determine 
$\gamma\approx 70^\circ$ with an error of about $10^\circ$ from the 
asymmetry $S$. This value is consistent with a less accurate 
result obtained from the corresponding quantity in the $\pi\pi$ system 
(if one takes the average of the two experiments, which are mutually 
incompatible). This appears intriguing but should be taken with some 
reservation, since the central values of the direct CP asymmetries 
$A_{\rm CP}$ and $C$ are once more larger than theoretically expected. 
We anticipate that, with more data available soon, the $\pi\rho$ system 
will play an important role in the understanding of hadronic decays and 
CP violation. 

\boldmath
\subsubsection*{Results related to the penguin-dominated $PV$ final 
states}
\unboldmath

The comparison of the penguin-dominated final states $\pi K$, $\pi K^*$ 
and $\rho K$ allows for crucial tests of the factorization framework, 
since the $PP$, $PV$, and $VP$ penguin amplitudes take very different 
values for reasons specific to factorization. The $\pi K^*$ data are 
indeed consistent with a smaller penguin amplitude; however, as has been 
noted before \cite{Du:2002cf,Aleksan:2003qi,deGroot:2003hp}, the 
reduction is not as large as predicted by theory. With default 
parameters one underestimates the amplitude by about 40\%, but the 
theoretical error on the penguin amplitude for vector-meson final states 
is large, in particular the one from weak annihilation. This taken into 
account, the prediction may be in agreement with the measured branching 
fractions within their errors. However, the situation is not 
satisfactory, since one would wish to have an explanation of the data 
that does not invoke weak annihilation. 

The estimate of weak annihilation is necessarily model-dependent in 
the QCD factorization approach. In particular, the error range 
(technically implemented by requiring $\varrho_A<1$) has to be specified 
as an ``educated guess''. It is therefore desirable to derive 
experimental constraints on weak annihilation. It was already found in 
\cite{BBNS3} that the $\pi K$ branching fractions do not favor a 
sizable annihilation amplitude, since this would require a fine-tuning 
of its strong phase to keep the branching fraction small enough. On the 
other hand, such a coincidence cannot be excluded. Assuming this to 
happen and further assuming universal annihilation amplitudes, we find 
that several branching fractions of pseudoscalar--vector final states, 
in particular $K\phi$, are much above the data for $\varrho_A\ge 2$. 
This, together with an estimate of the pure annihilation mode 
$\bar B_d\to D_s^+ K^-$ \cite{Battaglia:2003in}, seem to imply that 
weak annihilation cannot be much larger than the upper limit defined by 
the phenomenological treatment adopted in the present analysis and in 
\cite{BBNS3}. 

As a by-product of the analysis of penguin-dominated $PV$ modes, we
also obtain an estimate of the difference of the time-dependent CP 
asymmetries in $J/\psi K_S$, $\phi K_S$, and $\eta' K_S$ decays. As 
may have been expected, the factorization approach does not contain any 
mechanism that could enhance the CKM-suppressed amplitudes with a 
different weak phase, limiting the CP-asymmetry differences to a few 
percent.

\subsubsection*{Acknowledgments}

We would like to thank A.~H\"ocker for correspondence and for making 
\cite{hoecker} accessible to us. We are grateful to M.~Gronau, A.~Kagan,
U.~Nierste, and J.~Rosner for useful comments, and to J.~Alexander, 
H.~Jawahery, and H.~Yamamoto for help with collecting the references to 
the experimental results compiled in Appendix~C. We are grateful to the 
Aspen Center of Physics, where this work was completed. The work of M.B.\ 
is supported in part by the Bundesministerium f\"ur Bildung und Forschung, 
Project 05~HT1PAB/2, and by the DFG Sonder\-forschungsbereich/Transregio~9 
``Computer-gest\"utzte Theoretische Teilchenphysik''. The research of 
M.N.\ is supported by the U.S.\ National Science Foundation under Grant 
PHY-0098631.

\subsubsection*{Note added in proof}

After submission of this paper the following experimental results 
relevant to our analysis have been published: (1) The branching fraction 
for the decay $\bar B^0\to\pi^0\pi^0$ now reads 
$(1.9\pm 0.5)\cdot 10^{-6}$ \cite{Aubert:2003hf,Abe:2003yy}, which is
significantly larger than expectations. (2) The branching fraction for 
the decay $B^-\to K^-\rho^0$ has been measured by Belle to be 
$(3.9\pm 0.8)\cdot 10^{-6}$ \cite{belle338}, in good agreement with the 
predicted suppression of the $VP$ penguin amplitude discussed in 
Section~5.1. (3) The branching fraction for the decay 
$B^-\to\pi^-\bar K^{*0}$ from Belle now reads 
$(8.5\pm 1.3)\cdot 10^{-6}$ \cite{belle338}, less than half as large as 
before but in good agreement with the previous CLEO measurement. If this 
central value were confirmed by BaBar, this would remove one of the major 
discrepancies between data and the results of QCD factorization, see 
Section~5.1.

\newpage
\section*{Appendix~A: Explicit results for the decay amplitudes}

The results for the large set of $\bar B_q\to PP,PV$ decay amplitudes can 
be expressed most concisely in terms of traces over flavor matrices. We 
collect the three $B$-meson states into a row vector 
$\bm{B}=(B^-,\bar B^0,\bar B_s)$ and represent the final-state 
pseudoscalar and vector mesons by matrices
\begin{equation}
\begin{aligned}
   \bm{P} &= \left( \begin{array}{ccc} 
    \frac{\pi^0}{\sqrt2} + \frac{\eta_q}{\sqrt 2}
     + \frac{\eta'_q}{\sqrt 2} & \pi^- & K^- \\
    \pi^+ & - \frac{\pi^0}{\sqrt2} + \frac{\eta_q}{\sqrt 2}
     + \frac{\eta'_q}{\sqrt 2} & \bar K^0 \\
    K^+ & K^0 & \eta_s + \eta'_s \end{array} \right) , \\
   \bm{V} &= \left( \begin{array}{ccc}
    \frac{\rho^0}{\sqrt2} + \frac{\omega_q}{\sqrt 2}
     + \frac{\phi_q}{\sqrt 2} & \rho^- & K^{*-} \\
    \rho^+ & - \frac{\rho^0}{\sqrt2} + \frac{\omega_q}{\sqrt 2}
     + \frac{\phi_q}{\sqrt 2} & \bar K^{*0} \\
    K^{*+} & K^{*0} & \omega_s + \phi_s \end{array} \right) ,
\end{aligned}
\end{equation}
respectively. (One should also add a $\bar c c$ component for $\eta$ and 
$\eta'$, whose contribution is given by (\ref{ccbareffect}).) In 
addition, we define a column vector
\begin{equation}
   \bm{\Lambda}_p = \left( \begin{array}{c}
    0 \\ \lambda_p^{(d)} \\ \lambda_p^{(s)}
   \end{array} \right)
\end{equation}
containing CKM matrix elements, as well as matrices
\begin{equation}
   \bm{U}_p = \left( \begin{array}{ccc}
    \delta_{pu} & 0 & ~0 \\
    0 & 0~ & ~0 \\
    0 & 0~ & ~0
   \end{array} \right) ,
   \qquad 
   \bm{\hat Q} = \frac32\,\bm{Q} = \left( \begin{array}{ccc}
    1 & 0 & 0 \\
    0 & -\frac12 & 0 \\
    0 & 0 & -\frac12
   \end{array} \right) .
\end{equation}
Finally, we use the definitions of the quantities $A_{M_1 M_2}$, 
$\alpha_i(M_1 M_2)$ and $\beta_i(M_1 M_2)$ given in (\ref{am1m2}), 
(\ref{alphaidef}) and (\ref{bis}). Then the entity of all decay 
amplitudes ${\cal A}_{\bar B\to M_1 M_2}$ is reproduced by evaluating the 
master expression
\begin{eqnarray}
   &&\hspace{-1.0truecm}
    \sum_{p=u,c}\,A_{M_1 M_2}\,\bigg\{
    \bm{B}\bm{M_1} \left( \alpha_1\,\bm{U}_p + \alpha_4^p
     + \alpha_{4,{\rm EW}}^p\,\bm{\hat Q} \right)\bm{M_2}\,
     \bm{\Lambda}_p \nonumber\\
   &&\qquad\mbox{}+ \bm{B}\bm{M_1}\bm{\Lambda}_p\cdot
    \mbox{Tr}\left[\left( \alpha_2\,\bm{U}_p + \alpha_3^p
    + \alpha_{3,{\rm EW}}^p\,\bm{\hat Q} \right) \bm{M_2}\right] \nonumber\\
   &&\qquad\mbox{}+ \bm{B} \left( \beta_2\,\bm{U}_p + \beta_3^p
    + \beta_{3,{\rm EW}}^p\,\bm{\hat Q} \right)
    \bm{M_1}\bm{M_2}\bm{\Lambda}_p \nonumber\\
   &&\qquad\mbox{}+ \bm{B}\bm{\Lambda}_p\cdot
    \mbox{Tr}\left[\left( \beta_1\,\bm{U}_p + \beta_4^p
    + b_{4,{\rm EW}}^p\,\bm{\hat Q} \right) \bm{M_1}\bm{M_2}\right]
    \nonumber\\
   &&\qquad\mbox{}+ \bm{B} \left( \beta_{S2}\,\bm{U}_p + \beta_{S3}^p
    + \beta_{S3,{\rm EW}}^p\,\bm{\hat Q} \right)
    \bm{M_1}\bm{\Lambda}_p\cdot\mbox{Tr}\bm{M_2} \nonumber\\
   &&\qquad\mbox{}+ \bm{B}\bm{\Lambda}_p\cdot
    \mbox{Tr}\left[\left( \beta_{S1}\,\bm{U}_p + \beta_{S4}^p
    + b_{S4,{\rm EW}}^p\,\bm{\hat Q} \right) \bm{M_1}\right]\cdot
    \mbox{Tr}\bm{M_2} \bigg\} \,,
\end{eqnarray}
where $\alpha_i\equiv\alpha_i(M_1 M_2)$ and 
$\beta_i\equiv\beta_i(M_1 M_2)$. We recall that the $\alpha_i$ terms 
include vertex, penguin and hard spectator contributions, whereas the 
$\beta_i$ terms result from weak annihilation. Contributions proportional 
to $\bm{U}_p$ are from the current--current operators, those 
proportional to the unit matrix are from QCD penguins, and those 
proportional to the charge matrix are from electroweak penguins.

To determine the amplitude for a particular decay, for instance 
$\bar B^0\to\pi^+ K^-$, one inserts $\bm{P}$ for $\bm{M_1}$ and 
$\bm{M_2}$ and extracts the terms corresponding to 
$\pi^+ K^-$. When $\pi^+$ comes from $\bm{M_1}$ the prefactor 
is $A_{\pi K}$, otherwise it is $A_{K\pi}$. For pseudoscalar--vector 
final states both possibilities, $\bm{M_1}=\bm{P}$, $\bm{M_2}=
\bm{V}$ and $\bm{M_1}=\bm{V}$, $\bm{M_2}=\bm{P}$, must be summed and 
the $A_{M_1 M_2}$ prefactor is determined as above. For some pure 
annihilation amplitudes the form factor $\bar B\to M_1$ does 
not exist, and $A_{M_1 M_2}$ is not defined. In this case 
the expressions $A_{M_1 M_2} \,\beta_i$ must be replaced by 
$B_{M_1 M_2} \,b_i$. See Section~\ref{sec:decayop} for the definition 
of the relevant quantities.
 
We now list our results for the various decay amplitudes expressed in
terms of the $\alpha_i$ and $\beta_i$ parameters. Several sets of 
amplitudes have the same representation in terms of flavor parameters, 
apart from obvious substitutions of labels.

\boldmath
\subsection*{A.1\quad Decays with $\Delta S=1$}
\unboldmath

In this section the expressions for decay amplitudes must be 
multiplied with $\lambda_p^{(s)}$ and summed over $p=u,c$.  Throughout, 
the order of the arguments of the $\alpha_i^p(M_1 M_2)$ and 
$\beta_i^p(M_1 M_2)$ coefficients is determined by the order of the 
arguments of the $A_{M_1 M_2}$ prefactors. There is a 
total of 17 $\bar B\to PP$ and 31 $\bar B\to PV$ amplitudes, which split 
up as (4,4,9) and (8,8,15) into the flavor states 
$(B^-,\bar B^0,\bar B_s)$.

\subsubsection*{\it $\bar B\to\pi\bar K^{(*)}$ and 
$\bar B\to\rho\bar K$ decay amplitudes}

There are four independent amplitudes, given by
\begin{eqnarray}
   {\cal A}_{B^-\to\pi^-\bar K^0}
   &=& A_{\pi\bar K} \left[ \delta_{pu}\,\beta_2 
    + \alpha_4^p - \half\alpha_{4,{\rm EW}}^p + \beta_3^p
    + \beta_{3,{\rm EW}}^p \right] , \nonumber\\
   \sqrt2\,{\cal A}_{B^-\to\pi^0 K^-}
   &=& A_{\pi\bar K} \left[ \delta_{pu}\,(\alpha_1+\beta_2) 
    + \alpha_4^p + \alpha_{4,{\rm EW}}^p + \beta_3^p
    + \beta_{3,{\rm EW}}^p \right] \nonumber\\
   &+& A_{\bar K\pi} \left[ \delta_{pu}\,\alpha_2 
    + \3half\alpha_{3,{\rm EW}}^p \right] , \nonumber\\
   {\cal A}_{\bar B^0\to\pi^+ K^-}
   &=& A_{\pi\bar K} \left[ \delta_{pu}\,\alpha_1 
    + \alpha_4^p + \alpha_{4,{\rm EW}}^p + \beta_3^p
    - \half\beta_{3,{\rm EW}}^p \right] , \nonumber\\
   \sqrt2\,{\cal A}_{\bar B^0\to\pi^0\bar K^0}
   &=& A_{\pi\bar K} \left[ -\alpha_4^p + \half\alpha_{4,{\rm EW}}^p
    - \beta_3^p + \half\beta_{3,{\rm EW}}^p \right] \nonumber\\
   &+& A_{\bar K\pi} \left[ \delta_{pu}\,\alpha_2 
    + \3half\alpha_{3,{\rm EW}}^p \right] .
\end{eqnarray}
Isospin symmetry implies that
\begin{equation}
   \sqrt2\,{\cal A}_{\bar B^0\to\pi^0\bar K^0}
   = - {\cal A}_{B^-\to\pi^-\bar K^0}
   + \sqrt2\,{\cal A}_{B^-\to\pi^0 K^-}
   - {\cal A}_{\bar B^0\to\pi^+ K^-} \,.
\end{equation}
The expressions for the $\bar B\to\pi\bar K^*$ and $\bar B\to\rho\bar K$
amplitudes are obtained by setting $(\pi\bar K)\to(\pi\bar K^*)$ and
$(\pi\bar K)\to(\rho\bar K)$, respectively.

\subsubsection*{\it $\bar B\to\bar K^{(*)}\eta^{(\prime)}$ and 
$\bar B\to\bar K\omega/\phi$ decay amplitudes}

There are two independent amplitudes, given by
\begin{eqnarray}
   \sqrt2\,{\cal A}_{B^-\to K^-\eta}
   &=& A_{\bar K\eta_q} \left[ 
    \delta_{pu}\,(\alpha_2+2\beta_{S2})
    + 2\alpha_3^p + \half\alpha_{3,{\rm EW}}^p
    + 2\beta_{S3}^p + 2 \beta_{S3,{\rm EW}}^p\right] 
  \nonumber\\
  &+& \sqrt{2} A_{\bar K\eta_s} \Big[\delta_{pu}\,(\beta_2+\beta_{S2}) + 
    \alpha_3^p + \alpha_4^p - \half\alpha_{3,{\rm EW}}^p  
    -\half\alpha_{4,{\rm EW}}^p +\beta_3^p +\beta_{3,{\rm EW}}^p 
   \nonumber\\[-0.1cm]  
   &&\hspace*{1.5cm}
    +\,\beta_{S3}^p +\beta_{S3,{\rm EW}}^p \Big] 
  \nonumber\\
  &+& \sqrt{2} A_{\bar K\eta_c} \left[ 
    \delta_{pc}\,\alpha_2 + \alpha_3^p \right] 
  \nonumber\\
  &+&  A_{\eta_q\bar K} \left[ \delta_{pu}\,(\alpha_1+\beta_2)+ 
    \alpha_4^p+\alpha_{4,{\rm EW}}^p  
    + \beta_3^p +\beta_{3,{\rm EW}}^p\right],
\nonumber\\
   \sqrt2\,{\cal A}_{\bar B^0\to\bar K^0\eta}
   &=& A_{\bar K\eta_q} \left[ 
    \delta_{pu}\,\alpha_2
    + 2\alpha_3^p + \half\alpha_{3,{\rm EW}}^p
    + 2\beta_{S3}^p - \beta_{S3,{\rm EW}}^p\right] 
  \nonumber\\
  &+& \sqrt{2} A_{\bar K\eta_s} \Big[ 
    \alpha_3^p + \alpha_4^p- \half\alpha_{3,{\rm EW}}^p  
    -\half\alpha_{4,{\rm EW}}^p + \beta_3^p - \half\beta_{3,{\rm EW}}^p
   + \beta_{S3}^p - \half\beta_{S3,{\rm EW}}^p\Big] 
  \nonumber\\
  &+& \sqrt{2} A_{\bar K\eta_c} \left[ 
    \delta_{pc}\,\alpha_2 + \alpha_3^p \right] 
  \nonumber\\
  &+&  A_{\eta_q\bar K} \left[\alpha_4^p -\half\alpha_{4,{\rm EW}}^p  
    + \beta_3^p -\half\beta_{3,{\rm EW}}^p\right].
\end{eqnarray}
The amplitudes for $\bar B^0\to\bar K\eta'$, 
$\bar B^0\to\bar K^*\eta^{(\prime)}$, and $\bar B^0\to\bar K\omega/\phi$ 
are obtained from this result by replacing $(\bar K\eta)\to(\bar K\eta')$, 
$(\bar K\eta)\to(\bar K^*\eta^{(\prime)})$, and 
$(\bar K\eta)\to(\bar K\omega/\phi)$, 
respectively. When ideal mixing for $\omega$ and $\phi$ is assumed, set 
$A_{\bar K\omega_s}$ and $A_{\bar K\phi_q}$ to zero. Furthermore, with our 
approximations $A_{\bar K\omega_c}=A_{\bar K\phi_c}=0$.

\subsubsection*{\it $\bar B_s\to\pi\pi$ and $\bar B_s\to\pi\rho$ decay 
amplitudes}

There are two independent amplitudes, given by
\begin{eqnarray}
   {\cal A}_{\bar B_s\to\pi^+\rho^-}
   &=& B_{\pi\rho} \left[ b_4^p - \half b_{4,{\rm EW}}^p \right]
    \nonumber\\
   &+& B_{\rho\pi} \left[ \delta_{pu}\,b_1 + b_4^p
    + b_{4,{\rm EW}}^p \right] , \nonumber\\
   2\,{\cal A}_{\bar B_s\to\pi^0\rho^0}
   &=& B_{\pi\rho} \left[ \delta_{pu}\,b_1 + 2 b_4^p
    + \half b_{4,{\rm EW}}^p \right] \nonumber\\
   &+& B_{\rho\pi} \left[ \delta_{pu}\,b_1 + 2 b_4^p
    + \half b_{4,{\rm EW}}^p \right] .
\end{eqnarray}
The amplitudes for $\bar B_s\to\pi^-\rho^+$ is obtained from the first 
expression by interchanging $\pi\leftrightarrow\rho$ everywhere. In the 
limit of isospin symmetry the following relation holds:
\begin{equation}
   2\,{\cal A}_{\bar B_s\to\pi^0\rho^0}
   = {\cal A}_{\bar B_s\to\pi^+\rho^-} 
   + {\cal A}_{\bar B_s\to\pi^-\rho^+} \,.
\end{equation}
The expressions for the $\bar B_s\to\pi\pi$ amplitudes are obtained by
setting $\rho\to\pi$.

\subsubsection*{\it $\bar B_s\to\bar K^{(*)} K^{(*)}$ decay amplitudes}

There are two independent amplitudes, given by
\begin{eqnarray}
   {\cal A}_{\bar B_s\to\bar K^0 K^0}
   &=& B_{\bar K K} \left[b_4^p - \half b_{4,{\rm EW}}^p
    \right] \nonumber\\
   &+& A_{K\bar K} \left[ \alpha_4^p - \half\alpha_{4,{\rm EW}}^p
    + \beta_3^p + \beta_4^p - \half\beta_{3,{\rm EW}}^p
    - \half\beta_{4,{\rm EW}}^p \right] , \nonumber\\
   {\cal A}_{\bar B_s\to K^- K^+}
   &=& B_{\bar K K} \left[ \delta_{pu}\, b_1 + b_4^p
    + b_{4,{\rm EW}}^p \right] \nonumber\\
   &+& A_{K\bar K} \left[ \delta_{pu}\,\alpha_1 + \alpha_4^p
    + \alpha_{4,{\rm EW}}^p + \beta_3^p + \beta_4^p
    - \half\beta_{3,{\rm EW}}^p - \half\beta_{4,{\rm EW}}^p \right] .
\end{eqnarray}
The amplitudes for $\bar B_s\to\bar K K^*$ and $\bar B_s\to\bar K^* K$ are 
obtained from these expressions by replacing $(\bar K K)\to(\bar K K^*)$ 
and $(\bar K K)\to(\bar K^* K)$, respectively.

\subsubsection*{\it $\bar B_s\to\pi\eta^{(\prime)}$, 
$\bar B_s\to\pi\omega/\phi$, and $\bar B_s\to\rho\eta^{(\prime)}$ decay 
amplitudes}

There is only one independent amplitude, given by
\begin{eqnarray}
   2\,{\cal A}_{\bar B_s\to \pi^0\eta}
   &=& B_{\pi\eta_q} \left[ 
    \delta_{pu}\,(b_1+2 b_{S1})
    + \3half b_{4,{\rm EW}}^p + 3 b_{4S,{\rm EW}}^p\right] 
  \nonumber\\
  &+& \sqrt{2} B_{\pi\eta_s} \left[
    \delta_{pu}\,b_{S1} + \3half b_{4S,{\rm EW}}^p\right] 
  \nonumber\\
  &+&  B_{\eta_q \pi} \left[
    \delta_{pu}\,b_1 + \3half b_{4,{\rm EW}}^p\right] 
  \nonumber\\
  &+& \sqrt2 A_{\eta_s \pi} \left[\delta_{pu}\,\alpha_2 + 
    \3half\alpha_{3,{\rm EW}}^p \right]. 
\end{eqnarray}
The amplitudes for $\bar B_s\to\pi\eta'$, $\bar B_s\to\pi\omega/\phi$, and
$\bar B_s\to\rho\eta^{(\prime)}$ are obtained from these results by 
replacing $(\pi\eta)\to(\pi\eta')$, $(\pi\eta)\to(\pi\omega/\phi)$, and 
$(\pi\eta)\to(\rho\eta^{(\prime)})$, respectively. When ideal mixing 
for $\omega$ and $\phi$ is assumed, set $B_{\pi\omega_s}$ and 
$B_{\pi\phi_q}$ to zero. 

\subsubsection*{\it $\bar B_s\to\eta^{(\prime)}\eta^{(\prime)}$ and 
$\bar B_s\to\eta^{(\prime)}\omega/\phi$ 
decay amplitudes}

There is only one independent amplitude, given by
\begin{eqnarray}
2\,{\cal A}_{\bar B_s\to\eta\eta^\prime}
  &=&  B_{\eta_q\eta_q^\prime} \left[
      \delta_{pu}\,(b_1 + 2 b_{S1}) + 2 b_4^p 
      +\half b_{4,{\rm EW}}^p 
      + 4 b_{S4}^p + b_{S4,{\rm EW}}^p \right] 
  \nonumber\\
  &+& \sqrt2 B_{\eta_q\eta_s^\prime} \left[
      \delta_{pu}\,b_{S1} + 2 b_{S4}^p 
      + \half b_{S4,{\rm EW}}^p \right] 
  \nonumber\\
  &+& \sqrt2 A_{\eta_s\eta_q^\prime} \left[ 
    \delta_{pu}\,\alpha_2 
    + 2\alpha_3^p + \half\alpha_{3,{\rm EW}}^p + 2\beta_{S3}^p 
    + 2\beta_{S4}^p -\beta_{S3,{\rm EW}}^p 
    -\beta_{S4,{\rm EW}}^p \right] 
  \nonumber\\
  &+& 2 A_{\eta_s\eta_s^\prime} \Big[ 
    \alpha_3^p + \alpha_4^p - \half\alpha_{3,{\rm EW}}^p - 
    \half\alpha_{4,{\rm EW}}^p +\beta_3^p + \beta_4^p 
    -\half\beta_{3,{\rm EW}}^p 
    -\half\beta_{4,{\rm EW}}^p 
    \nonumber\\[-0.1cm] 
    &&\hspace*{1cm}+\,\beta_{S3}^p + \beta_{S4}^p
    -\half\beta_{S3,{\rm EW}}^p  
    -\half\beta_{S4,{\rm EW}}^p \Big] 
  \nonumber\\
  &+& 2 A_{\eta_s\eta_c^\prime} \left[ 
    \delta_{pc}\,\alpha_2 + \alpha_3^p \right]
    \nonumber
\end{eqnarray}
\begin{eqnarray}
  \phantom{2\,{\cal A}_{\bar B_s\to\eta\eta^\prime}}
  &+&  B_{\eta_q^\prime\eta_q} \left[
      \delta_{pu}\,(b_1 + 2 b_{S1}) + 2 b_4^p 
      +\half b_{4,{\rm EW}}^p 
      + 4 b_{S4}^p + b_{S4,{\rm EW}}^p \right] 
  \nonumber\\
  &+& \sqrt2 B_{\eta_q^\prime\eta_s} \left[
      \delta_{pu}\,b_{S1} + 2 b_{S4}^p 
      + \half b_{S4,{\rm EW}}^p \right] 
  \nonumber\\
  &+& \sqrt2 A_{\eta_s^\prime\eta_q} \left[ 
    \delta_{pu}\,\alpha_2 
    + 2\alpha_3^p + \half\alpha_{3,{\rm EW}}^p + 2\beta_{S3}^p 
    + 2\beta_{S4}^p -\beta_{S3,{\rm EW}}^p  
    -\beta_{S4,{\rm EW}}^p \right] 
  \nonumber\\
  &+& 2 A_{\eta_s^\prime\eta_s} \Big[ 
    \alpha_3^p + \alpha_4^p - \half\alpha_{3,{\rm EW}}^p - 
    \half\alpha_{4,{\rm EW}}^p +\beta_3^p + \beta_4^p
    -\half\beta_{3,{\rm EW}}^p  
    -\half\beta_{4,{\rm EW}}^p 
    \nonumber\\[-0.1cm] 
    &&\hspace*{1cm}+\,\beta_{S3}^p + \beta_{S4}^p 
    -\half\beta_{S3,{\rm EW}}^p 
    -\half\beta_{S4,{\rm EW}}^p \Big] 
  \nonumber\\
  &+& 2 A_{\eta_s^\prime\eta_c} \left[ 
    \delta_{pc}\,\alpha_2 + \alpha_3^p \right].
\end{eqnarray}
The amplitudes for $\bar B_s\to\eta\eta$, $\bar B_s\to\eta'\eta'$, and  
$\bar B_s\to\eta^{(\prime)}\omega/\phi$ are obtained from this result by 
replacing $(\eta\eta')\to(\eta\eta)$, $(\eta\eta')\to(\eta'\eta')$, and
$(\eta\eta')\to(\eta^{(\prime)}\omega/\phi)$, respectively. When ideal 
mixing for $\omega$ and $\phi$ is assumed, set 
$A_{\eta_s^{(\prime)}\omega_s}$, $A_{\eta_s^{(\prime)}\phi_q}$, 
$B_{\eta_q^{(\prime)}\omega_s}$, $B_{\eta_q^{(\prime)}\phi_q}$ to zero. 
Furthermore, with our approximations 
$A_{\eta_s^{(\prime)}\omega_c}=A_{\eta_s^{(\prime)}\phi_c}=0$.

\boldmath
\subsection*{A.2\quad Decays with $\Delta D=1$}
\unboldmath

In this subsection the expressions for decay amplitudes must be 
multiplied with $\lambda_p^{(d)}$ and summed over $p=u,c$. Throughout, 
the order of the arguments of the $\alpha_i^p(M_1 M_2)$ and 
$\beta_i^p(M_1 M_2)$ coefficients is determined by the order of the 
arguments of the $A_{M_1 M_2}$ prefactors. There is a total of 17 
$\bar B\to PP$ and 31 $\bar B\to PV$ amplitudes, which split up as 
$(4,9,4)$ and $(8,15,8)$ into the flavor states 
$(B^-,\bar B^0,\bar B_s)$.

\subsubsection*{\it $\bar B\to\pi\pi$ and $\bar B\to\pi\rho$ decay 
amplitudes}

There are three independent amplitudes, given by
\begin{eqnarray}
   \sqrt2\,{\cal A}_{B^-\to\pi^-\rho^0}
   &=& A_{\pi\rho} \left[ \delta_{pu}\,(\alpha_2 - \beta_2)
    - \alpha_4^p + \3half\alpha_{3,{\rm EW}}^p
    + \half\alpha_{4,{\rm EW}}^p  - \beta_3^p - \beta_{3,{\rm EW}}^p
    \right] \nonumber\\
   &+& A_{\rho\pi} \left[ \delta_{pu}\,(\alpha_1 + \beta_2)
    + \alpha_4^p + \alpha_{4,{\rm EW}}^p + \beta_3^p 
    + \beta_{3,{\rm EW}}^p \right] , \nonumber\\
   {\cal A}_{\bar B^0\to\pi^+\rho^-}
   &=& A_{\pi\rho} \left[ \delta_{pu}\,\alpha_1 + \alpha_4^p
    + \alpha_{4,{\rm EW}}^p + \beta_3^p + \beta_4^p
    - \half\beta_{3,{\rm EW}}^p - \half\beta_{4,{\rm EW}}^p \right]
    \nonumber\\
   &+& A_{\rho\pi} \left[ \delta_{pu}\,\beta_1 + \beta_4^p 
    + \beta_{4,{\rm EW}}^p \right] , \nonumber\\
   -2\,{\cal A}_{\bar B^0\to\pi^0\rho^0}
   &=& A_{\pi\rho} \Big[ \delta_{pu}\,(\alpha_2 - \beta_1) 
    - \alpha_4^p + \3half\alpha_{3,{\rm EW}}^p
    + \half\alpha_{4,{\rm EW}}^p - \beta_3^p - 2\beta_4^p    
    \nonumber\\[-0.1cm] 
    &&\hspace*{1cm}
    +\, \half\beta_{3,{\rm EW}}^p - \half\beta_{4,{\rm EW}}^p \Big]
    \nonumber\\
   &+& A_{\rho\pi} \Big[ \delta_{pu}\,(\alpha_2 - \beta_1) 
    - \alpha_4^p + \3half\alpha_{3,{\rm EW}}^p
    + \half\alpha_{4,{\rm EW}}^p - \beta_3^p - 2\beta_4^p
    \nonumber\\[-0.1cm] 
    &&\hspace*{1cm}
    + \,\half\beta_{3,{\rm EW}}^p - \half\beta_{4,{\rm EW}}^p \Big] . 
\end{eqnarray}
The amplitudes for $B^-\to\pi^0\rho^-$ and $\bar B^0\to\pi^-\rho^+$ are
obtained from the first two expressions by interchanging 
$\pi\leftrightarrow\rho$ everywhere. In the limit of isospin symmetry the 
following relation holds:
\begin{equation}
   2\,{\cal A}_{\bar B^0\to\pi^0\rho^0}
   = {\cal A}_{\bar B^0\to\pi^+\rho^-}
    + {\cal A}_{\bar B^0\to\pi^-\rho^+}
    - \sqrt2\,\Big( {\cal A}_{B^-\to\pi^-\rho^0}
    + {\cal A}_{B^-\to\pi^0\rho^-} \Big) \,.
\end{equation}
The expressions for the $\bar B\to\pi\pi$ amplitudes are obtained by
setting $\rho\to\pi$.

\subsubsection*{\it $\bar B\to\bar K^{(*)} K^{(*)}$ decay amplitudes}

There are three independent amplitudes, given by
\begin{eqnarray}
   {\cal A}_{B^-\to K^- K^0}
   &=& A_{\bar K K} \left[ \delta_{pu}\,\beta_2 + \alpha_4^p
    - \half\alpha_{4,{\rm EW}}^p + \beta_3^p + \beta_{3,{\rm EW}}^p
    \right] , \nonumber\\
   {\cal A}_{\bar B^0\to K^- K^+}
   &=& A_{\bar K K} \left[ \delta_{pu}\,\beta_1 + \beta_4^p
    + \beta_{4,{\rm EW}}^p \right] \nonumber\\
   &+& B_{K\bar K} \left[b_4^p - \half b_{4,{\rm EW}}^p
    \right] , \nonumber\\
   {\cal A}_{\bar B^0\to\bar K^0 K^0}
   &=& A_{\bar K K} \left[ \alpha_4^p - \half\alpha_{4,{\rm EW}}^p
    + \beta_3^p + \beta_4^p - \half\beta_{3,{\rm EW}}^p
    - \half\beta_{4,{\rm EW}}^p \right] \nonumber\\
   &+& B_{K\bar K} \left[b_4^p - \half b_{4,{\rm EW}}^p
    \right] .
\end{eqnarray}
The amplitudes for $\bar B\to\bar K K^*$ and $\bar B\to\bar K^* K$ are 
obtained from these by replacing $(\bar K K)\to(\bar K K^*)$ and 
$(\bar K K)\to(\bar K^* K)$, respectively.

\subsubsection*{\it $\bar B\to\pi\eta^{(\prime)}$, 
$\bar B\to\rho\eta^{(\prime)}$ and $\bar B\to\pi\omega/\phi$ decay amplitudes}

There are two independent amplitudes, given by
\begin{eqnarray}
\sqrt2\,{\cal A}_{B^-\to\pi^-\eta}
  &=& A_{\pi\eta_q} \Big[ 
    \delta_{pu}\,(\alpha_2 + \beta_2 + 2\beta_{S2})
    + 2\alpha_3^p + \alpha_4^p + \half\alpha_{3,{\rm EW}}^p
    - \half\alpha_{4,{\rm EW}}^p 
    \nonumber\\[-0.1cm] 
    &&\hspace*{1cm}+\, \beta_3^p + \beta_{3,{\rm EW}}^p
    + 2\beta_{S3}^p + 2\beta_{S3,{\rm EW}}^p \Big] 
  \nonumber\\
  &+& \sqrt{2} A_{\pi\eta_s} \left[ 
    \delta_{pu}\,\beta_{S2} + \alpha_3^p - \half\alpha_{3,{\rm EW}}^p
    + \beta_{S3}^p + \beta_{S3,{\rm EW}}^p
    \right] 
  \nonumber\\
  &+& \sqrt{2} A_{\pi\eta_c} \left[ 
    \delta_{pc}\,\alpha_2 + \alpha_3^p \right] 
  \nonumber\\
  &+& A_{\eta_q\pi} \left[ \delta_{pu}\,(\alpha_1 + \beta_2)
    + \alpha_4^p + \alpha_{4,{\rm EW}}^p + \beta_3^p
    + \beta_{3,{\rm EW}}^p \right] , 
\nonumber\\
   -2\,{\cal A}_{\bar B^0\to\pi^0\eta}
   &=& A_{\pi\eta_q} \Big[ 
    \delta_{pu}\,(\alpha_2 - \beta_1 - 2\beta_{S1})
    + 2\alpha_3^p + \alpha_4^p + \half\alpha_{3,{\rm EW}}^p
    - \half\alpha_{4,{\rm EW}}^p 
    \nonumber\\[-0.1cm] 
    &&\hspace*{1cm}+\, \beta_3^p -\half\beta_{3,{\rm EW}}^p
    -\3half\beta_{4,{\rm EW}}^p + 2\beta_{S3}^p - \beta_{S3,{\rm EW}}^p 
    -3  \beta_{S4,{\rm EW}}^p \Big] 
  \nonumber\\
  &+& \sqrt{2} A_{\pi\eta_s} \left[ 
     - \delta_{pu}\,\beta_{S1} + \alpha_3^p - \half\alpha_{3,{\rm EW}}^p
   + \beta_{S3}^p - \half\beta_{S3,{\rm EW}}^p
    - \3half\beta_{S4,{\rm EW}}^p\right] 
   \nonumber\\
  &+& \sqrt{2} A_{\pi\eta_c} \left[ 
    \delta_{pc}\,\alpha_2 + \alpha_3^p \right] 
   \\
  &+& A_{\eta_q\pi} \left[ \delta_{pu}\,(-\alpha_2 - \beta_1)
    + \alpha_4^p -\3half\alpha_{3,{\rm EW}}^p -\half\alpha_{4,{\rm EW}}^p  
    + \beta_3^p -\half\beta_{3,{\rm EW}}^p -\3half\beta_{4,{\rm EW}}^p
    \right]\hspace*{-0.025cm}. \nonumber
\end{eqnarray}
The amplitudes for $\bar B\to\pi\eta'$, $\bar B\to\rho\eta^{(\prime)}$, 
and $\bar B\to\pi\omega/\phi$ are obtained from these results by 
replacing $(\pi\eta)\to(\pi\eta')$, $(\pi\eta)\to(\rho\eta^{(\prime)})$, 
and $(\pi\eta)\to(\pi\omega/\phi)$, respectively. When ideal mixing 
for $\omega$ and $\phi$ is assumed, set $A_{\pi\omega_s}$ and 
$A_{\pi\phi_q}$ to zero. Furthermore, with our approximations 
$A_{\pi\omega_c}=A_{\pi \phi_c}=0$.

\subsubsection*{\it $\bar B\to\eta^{(\prime)}\eta^{(\prime)}$ and 
$\bar B\to\eta^{(\prime)}\omega/\phi$ decay 
amplitudes}

There is only one independent amplitude, given by
\begin{eqnarray}
2\,{\cal A}_{\bar B^0\to\eta\eta^\prime}
  &=& A_{\eta_q\eta_q^\prime} \Big[ 
    \delta_{pu}\,(\alpha_2 + \beta_1 + 2\beta_{S1})
    + 2\alpha_3^p + \alpha_4^p + \half\alpha_{3,{\rm EW}}^p
    - \half\alpha_{4,{\rm EW}}^p 
    \nonumber\\[-0.1cm] 
    &&\hspace*{1cm}+\, \beta_3^p + 2\beta_4^p - \half\beta_{3,{\rm EW}}^p
    + \half\beta_{4,{\rm EW}}^p+ 2\beta_{S3}^p 
    + 4\beta_{S4}^p -\beta_{S3,{\rm EW}}^p 
    +\beta_{S4,{\rm EW}}^p \Big] 
  \nonumber\\
  &+& \sqrt2 A_{\eta_q\eta_s^\prime} \left[ 
    \delta_{pu}\,\beta_{S1}
    + \alpha_3^p - \half\alpha_{3,{\rm EW}}^p +\beta_{S3}^p 
   + 2\beta_{S4}^p  -\half\beta_{S3,{\rm EW}}^p 
    +\half\beta_{S4,{\rm EW}}^p \right] 
  \nonumber\\
  &+& \sqrt{2} A_{\eta_q\eta_c^\prime} \left[ 
    \delta_{pc}\,\alpha_2 + \alpha_3^p \right] 
  \nonumber\\
  &+& \sqrt2 B_{\eta_s\eta_q^\prime} \left[
     2 b_{S4}^p -b_{S4,{\rm EW}}^p \right] 
  \nonumber\\
  &+& 2 B_{\eta_s\eta_s^\prime} \left[
    b_4^p -\half b_{4,{\rm EW}}^p + b_{S4}^p 
    -\half b_{S4,{\rm EW}}^p \right] 
  \nonumber \\
  &+& A_{\eta_q^\prime\eta_q} \Big[ 
    \delta_{pu}\,(\alpha_2 + \beta_1 + 2\beta_{S1})
    + 2\alpha_3^p + \alpha_4^p + \half\alpha_{3,{\rm EW}}^p
    - \half\alpha_{4,{\rm EW}}^p 
    \nonumber\\[-0.1cm] 
    &&\hspace*{1cm}+\, \beta_3^p + 2\beta_4^p - \half\beta_{3,{\rm EW}}^p
    + \half\beta_{4,{\rm EW}}^p+ 2\beta_{S3}^p 
    + 4\beta_{S4}^p -\beta_{S3,{\rm EW}}^p 
    +\beta_{S4,{\rm EW}}^p \Big] 
  \nonumber\\
  &+& \sqrt2 A_{\eta_q^\prime\eta_s} \left[ 
    \delta_{pu}\,\beta_{S1}
    + \alpha_3^p - \half\alpha_{3,{\rm EW}}^p +\beta_{S3}^p 
    + 2\beta_{S4}^p -\half\beta_{S3,{\rm EW}}^p  
    +\half\beta_{S4,{\rm EW}}^p \right] 
  \nonumber\\
  &+& \sqrt{2} A_{\eta_q^\prime\eta_c} \left[ 
    \delta_{pc}\,\alpha_2 + \alpha_3^p \right] 
  \nonumber\\
  &+& \sqrt2 B_{\eta_s^\prime\eta_q} \left[
     2 b_{S4}^p - b_{S4,{\rm EW}}^p \right] 
  \nonumber\\
  &+& 2 B_{\eta_s^\prime\eta_s} \left[
    b_4^p -\half b_{4,{\rm EW}}^p + b_{S4}^p 
    -\half b_{S4,{\rm EW}}^p \right].
\end{eqnarray}
The amplitudes for $\bar B^0\to\eta\eta$, $\bar B^0\to\eta'\eta'$, and  
$\bar B^0\to\eta^{(\prime)}\omega/\phi$ are obtained from this result by 
replacing $(\eta\eta')\to(\eta\eta)$, $(\eta\eta')\to(\eta'\eta')$, 
and $(\eta\eta')\to(\eta^{(\prime)}\omega/\phi)$, respectively. When 
ideal mixing for $\omega$ and $\phi$ is assumed, set 
$A_{\eta_q^{(\prime)}\omega_s}$, $A_{\eta_q^{(\prime)}\phi_q}$, 
$B_{\eta_s^{(\prime)}\omega_s}$, $B_{\eta_s^{(\prime)}\phi_q}$ to zero. 
Furthermore, with our approximations
$A_{\eta_q^{(\prime)}\omega_c}=A_{\eta_q^{(\prime)}\phi_c}=0$.

\subsubsection*{\it $\bar B_s\to\pi K^{(*)}$ and $\bar B_s\to\rho K$ 
decay amplitudes}

There are two independent amplitudes, given by
\begin{eqnarray}
   {\cal A}_{\bar B_s\to\pi^- K^+}
   &=& A_{K\pi} \left[ \delta_{pu}\,\alpha_1 + \alpha_4^p
    + \alpha_{4,{\rm EW}}^p + \beta_3^p - \half\beta_{3,{\rm EW}}^p
    \right] , \nonumber\\
   \sqrt2\,{\cal A}_{\bar B_s\to\pi^0 K^0}
   &=& A_{K\pi} \left[ \delta_{pu}\,\alpha_2 - \alpha_4^p
    + \3half\alpha_{3,{\rm EW}}^p + \half\alpha_{4,{\rm EW}}^p
    - \beta_3^p + \half\beta_{3,{\rm EW}}^p \right] .
\end{eqnarray}
The expressions for the $\bar B_s\to\pi K^*$ and $\bar B_s\to\rho K$ 
amplitudes are obtained by setting $(\pi K)\to(\pi K^*)$ and
$(\pi K)\to(\rho K)$, respectively.

\subsubsection*{\it $\bar B_s\to K^{(*)}\eta^{(\prime)}$ and 
$\bar B_s\to K\omega/\phi$ decay amplitudes}

There is only one independent amplitude, given by
\begin{eqnarray}
   \sqrt2\,{\cal A}_{\bar B_s\to K^0\eta}
   &=& A_{K\eta_q} \Big[ 
    \delta_{pu}\,\alpha_2
    + 2\alpha_3^p + \alpha_4^p + \half\alpha_{3,{\rm EW}}^p
    - \half\alpha_{4,{\rm EW}}^p 
    +\beta_3^p -\half\beta_{3,{\rm EW}}^p
    \nonumber\\[-0.1cm] 
    &&\hspace*{1cm}
    +\, 2\beta_{S3}^p - \beta_{S3,{\rm EW}}^p\Big] 
  \nonumber\\
  &+& \sqrt{2} A_{K\eta_s} \left[ 
    \alpha_3^p - \half\alpha_{3,{\rm EW}}^p
    + \beta_{S3}^p - \half\beta_{S3,{\rm EW}}^p\right] 
  \nonumber\\
  &+& \sqrt{2} A_{K\eta_c} \left[ 
    \delta_{pc}\,\alpha_2 + \alpha_3^p \right] 
  \nonumber\\
  &+&  \sqrt{2} A_{\eta_s K} \left[\alpha_4^p -\half\alpha_{4,{\rm EW}}^p  
    + \beta_3^p -\half\beta_{3,{\rm EW}}^p\right] .
\end{eqnarray}
The amplitudes for $\bar B_s\to K\eta'$, 
$\bar B_s\to K^*\eta^{(\prime)}$, and $\bar B_s\to K\omega/\phi$ are 
obtained from this result by replacing $(K\eta)\to(K\eta')$, 
$(K\eta)\to(K^*\eta^{(\prime)})$, and $(K\eta)\to(K\omega/\phi)$, 
respectively. When ideal mixing for $\omega$ and $\phi$ is assumed, set 
$A_{K\omega_s}$ and $A_{K\phi_q}$ to zero. Furthermore, with our 
approximations $A_{K\omega_c}=A_{K\phi_c}=0$.

\newpage
\section*{Appendix~B: Convolution integrals}

The convolution integrals of the hard-scattering kernels with meson 
light-cone distribution amplitudes can be evaluated using expansions of 
the distribution amplitudes in terms of Gegenbauer polynomials. In 
\cite{BBNS3} the corresponding expressions for final states containing 
two  pseudoscalar mesons were given including the first three terms in 
this expansion. Here we list the corresponding results for the 
convolution integrals involving the twist-3 distribution amplitude 
$\Phi_v(x)$, which are needed for pseudoscalar--vector final states. We 
include the first three terms in the Gegenbauer expansion 
(\ref{phivGegenb}). 

The convolution integral entering the vertex corrections for the 
coefficients $a_{6,8}^p$ is
\begin{equation}
   \int_0^1\!dx\,\Phi_v(x)\,h(x)
   = 9 - 6i\pi + \left( \frac{19}{6} - i\pi \right) \alpha_{2,\perp}^V
   + \dots \,,
\end{equation}
with $h(x)$ as given in (\ref{FM}).
The hard spectator contributions involve the divergent integral
\begin{equation}
   \int_0^1\!dx\,\frac{\Phi_v(x)}{1-x}
   = \Phi_v(1)\,X_H
   - (6 + 9 \alpha_{1,\perp}^V + 11\alpha_{2,\perp}^V + \dots) \,,
\end{equation}
where $\Phi_v(1)=3\sum_n\,\alpha_{n,\perp}^V$. The penguin 
contributions involve the convolution
\begin{equation}
   \widehat G_V(s) = \int_0^1\!dx\,\Phi_v(x)\,G(s-i\epsilon,1-x) \,,
\end{equation}
where $G(s,x)$ is the penguin function defined in (\ref{GK}). We obtain
\begin{eqnarray}
   \widehat G_V(s) 
   &=& \frac16\,(6+2\alpha_{1,\perp}^V + \alpha_{2,\perp}^V)
    - 4s\,(9+12\alpha_{1,\perp}^V+14\alpha_{2,\perp}^V) \nonumber\\
   &&\mbox{}- 6s^2\,(8\alpha_{1,\perp}^V+35\alpha_{2,\perp}^V)
    + 360s^3\,\alpha_{2,\perp}^V \nonumber\\
   &&\mbox{}+ 12s\sqrt{1-4s} \left[
    1 + (1+4s)\,\alpha_{1,\perp}^V + (1+15s-30s^2)\,\alpha_{2,\perp}^V
    \right] \nonumber\\
   &&\quad\times
    \left( 2\,\mbox{arctanh}\sqrt{1-4s} - i\pi \right) \nonumber\\
   &&\mbox{}- 12s^2 \left[
    1 + (3-4s)\,\alpha_{1,\perp}^V + 2(3-10s+15s^2)\,\alpha_{2,\perp}^V
    \right] \nonumber\\
   &&\quad\times
    \left( 2\,\mbox{arctanh}\sqrt{1-4s} - i\pi \right)^2 + \dots \,.
\end{eqnarray}
For the special cases $s=0$ and $s=1$, this expression reduces to
\begin{eqnarray}
   \widehat G_V(0)
   &=& 1 + \frac{\alpha_{1,\perp}^V}{3} + \frac{\alpha_{2,\perp}^V}{6}
    + \dots \,, \nonumber\\
   \widehat G_V(1)
   &=& -35 + 4\sqrt3\,\pi + \frac{4\pi^2}{3}
    + \left( -\frac{287}{3} + 20\sqrt3\,\pi - \frac{4\pi^2}{3}
    \right) \alpha_{1,\perp}^V \nonumber\\
   &&\mbox{}+ \left( \frac{565}{6} - 56\sqrt3\,\pi
    + \frac{64\pi^2}{3} \right) \alpha_{2,\perp}^V + \dots \,.
\end{eqnarray}

\newpage
\section*{Appendix~C: Summary of experimental results}
\label{sec:exptabs}

In Tables~\ref{tab:data1}--\ref{tab:data4} we compile the available 
experimental data on the CP-averaged branching fractions and CP 
asymmetries in $B\to PP$ and $B\to PV$ decays, distinguishing the two 
classes of decays of the type $\Delta S=1$ and $\Delta D=1$. We also 
present our weighted averages of the data (ignoring correlated errors, 
which are small). Where measurements are inconsistent, the 
combined error is inflated by a factor of $S=\sqrt{\chi^2/(N-1)}$, 
which is shown in parenthesis.

\vfil

\begin{table}[h]
\footnotesize
\centerline{\parbox{14cm}{\caption{\label{tab:data1}
CP-averaged branching ratios (top, in units of $10^{-6}$) and CP 
asymmetries (bottom, in \%) for $\bar B\to PP$ decays with $\Delta S=1$. 
Upper limits are at 90\% confidence level. We show 
$S=\sqrt{\chi^2/(N-1)}$ in cases where $S>1$.}}}
\vspace{0.1cm}
\begin{center}
\begin{tabular}{|l|ccc|c|}
\hline\hline
\multicolumn{1}{|c|}{Mode} & BaBar & Belle & CLEO & Average \\
\hline\hline
\multicolumn{5}{|c|}{Penguin-dominated decays} \\
\hline
$B^-\to\pi^-\bar K^0$
 & $20.0\pm 1.6\pm 1.0$ \cite{Bona}
 & $22.0\pm 1.9\pm 1.1$ \cite{Tomura:2003cb}
 & $18.8_{\,-3.3\,-1.8}^{\,+3.7\,+2.1}$ \cite{Bornheim:2003bv}
 & $20.6\pm 1.3$ \\
$B^-\to\pi^0 K^-$
 & $12.8_{\,-1.1}^{\,+ 1.2}\pm 1.0$ \cite{Aubert:2003qj}
 & $12.8\pm 1.4_{\,-1.0}^{\,+1.4}$ \cite{Tomura:2003cb}
 & $12.9_{\,-2.2\,-1.1}^{\,+2.4\,+1.2}$ \cite{Bornheim:2003bv}
 & $12.8\pm 1.1$ \\
$\bar B^0\to\pi^+ K^-$
 & $17.9\pm 0.9\pm 0.7$ \cite{Aubert:2003qj}
 & $18.5\pm 1.0\pm 0.7$ \cite{Tomura:2003cb}
 & $18.0_{\,-2.1\,-0.9}^{\,+2.3\,+1.2}$ \cite{Bornheim:2003bv}
 & $18.2\pm 0.8$ \\
$\bar B^0\to\pi^0\bar K^0$
 & $10.4\pm 1.5\pm 0.8$ \cite{Aubert:2002jm}
 & $12.6\pm 2.4\pm 1.4$ \cite{Tomura:2003cb}
 & $12.8_{\,-3.3\,-1.4}^{\,+4.0\,+1.7}$ \cite{Bornheim:2003bv}
 & $11.2\pm 1.4$ \\
\hline
$B^-\to\eta K^-$
 & $2.8_{\,-0.7}^{\,+0.8}\pm 0.2$ \cite{Aubert:2003ez}
 & $5.3_{\,-1.5}^{\,+1.8}\pm 0.6$ \cite{Tomura:2003cb}
 & $2.2_{\,-2.2}^{\,+ 2.8}$ ($<6.9$) \cite{Richichi:1999kj}
 & $3.1\pm 0.7$ \\
$\bar B^0\to\eta\bar K^0$
 & $2.6_{\,-0.8}^{\,+0.9}\pm 0.2$ ($<4.6$) \cite{Aubert:2003ez}
 & $<12$ \cite{Tomura:2003cb}
 & $<9.3$ \cite{Richichi:1999kj}
 & $<4.6$ \\
$B^-\to\eta' K^-$
 & $76.9\pm 3.5\pm 4.4$ \cite{Aubert:2003bq}
 & $78\pm 6\pm 9$ \cite{Tomura:2003cb}
 & $80_{\,-\phantom{1}9}^{\,+10}\pm 7$ \cite{Richichi:1999kj}
 & $77.6\pm 4.6$ \\
$\bar B^0\to\eta'\bar K^0$
 & $55.4\pm 5.2\pm 4.0$ \cite{Aubert:2003bq}
 & $68\pm 10_{\,-8}^{\,+ 9}$ \cite{Tomura:2003cb}
 & $89_{\,-16}^{\,+18}\pm 9$ \cite{Richichi:1999kj}
 & $60.6\pm 7.0$ \\[-0.2cm]
 & & & & {\footnotesize ($S=1.3$)} \\
\hline\hline
\end{tabular}
\end{center}
\begin{center}
\begin{tabular}{|l|ccc|c|}
\hline\hline
\multicolumn{1}{|c|}{Mode} & BaBar & Belle & CLEO & Average \\
\hline\hline
\multicolumn{5}{|c|}{Penguin-dominated decays} \\
\hline
$B^-\to\pi^-\bar K^0$
 & $-17\pm 10\pm 2$ \cite{Aubert:2002ng}
 & $7_{\,-8\,-3}^{\,+9\,+1}$ \cite{Tomura:2003cb}
 & $18\pm 24\pm 2$ \cite{Chen:2000hv}
 & $-2\pm 9$ \\[-0.2cm]
 & & & & {\footnotesize ($S=1.4$)} \\[-0.1cm]
$B^-\to\pi^0 K^-$
 & $-9\pm 9\pm 1$ \cite{Aubert:2003qj}
 & $23\pm 11_{\,-4}^{\,+1}$ \cite{Tomura:2003cb}
 & $-29\pm 23\pm 2$ \cite{Chen:2000hv}
 & $1\pm 12$ \\[-0.2cm]
 & & & & {\footnotesize ($S=1.8$)} \\[-0.1cm]
$\bar B^0\to\pi^+ K^-$
 & $-10\pm 5\pm 2$ \cite{Aubert:2002jb}
 & $-7\pm 6\pm 1$ \cite{Tomura:2003cb}
 & $-4\pm 16\pm 2$ \cite{Chen:2000hv}
 & $-9\pm 4$ \\
$\bar B^0\to\pi^0\bar K^0$
 & $3\pm 36\pm 9$ \cite{Aubert:2002jm}
 & ---
 & ---
 & $3\pm 37$ \\
\hline
$B^-\to\eta K^-$
 & $-32_{\,-18}^{\,+22}\pm 1$ \cite{Aubert:2003ez}
 & ---
 & ---
 & $-32\pm 20$ \\
$B^-\to\eta' K^-$
 & $3.7\pm 4.5\pm 1.1$ \cite{Aubert:2003bq}
 & $-1\pm 7\pm 1$ \cite{Aihara}
 & $3\pm 12\pm 2$ \cite{Chen:2000hv}
 & $2\pm 4$ \\
$\bar B^0\to\eta'\bar K^0$
 & $-10\pm 22\pm 3$ \cite{Aubert:2003bq}
 & $26\pm 22\pm 4$ \cite{Abe:2002np}
 & ---
 & $8\pm 18$ \\[-0.2cm]
 & & & & {\footnotesize ($S=1.1$)} \\[-0.1cm]
\qquad $S_{\eta' K_S}$: 
 & $2\pm 34\pm 3$ \cite{Aubert:2003bq} 
 & $71\pm 37_{\,-6}^{\,+5}$ \cite{Abe:2002np} 
 & ---
 & $33\pm 34$ \\[-0.2cm]
 & & & & {\footnotesize ($S=1.4$)} \\
\hline\hline
\end{tabular}
\end{center}
\end{table}

\begin{table}
\footnotesize
\centerline{\parbox{14cm}{\caption{\label{tab:data2}
CP-averaged branching ratios (top, in units of $10^{-6}$) and CP 
asymmetries (bottom, in \%) for $\bar B\to PV$ decays with $\Delta S=1$. 
Upper limits are at 90\% confidence level.}}}
\vspace{0.1cm}
\begin{center}
{\tabcolsep=0.13cm
\begin{tabular}{|l|ccc|c|}
\hline\hline
\multicolumn{1}{|c|}{Mode} & BaBar & Belle & CLEO & Average \\
\hline\hline
\multicolumn{5}{|c|}{Penguin-dominated decays} \\
\hline
$B^-\to\pi^-\bar K^{*0}$
 & $15.5\pm 1.8_{\,-3.2}^{\,+1.5}$ \cite{Aubert:2003dn}
 & $19.4_{\,-3.9\,-2.1\,-6.8}^{\,+4.2\,+2.1\,+3.5}$ \cite{Abe:2002av}
 & $7.6_{\,-3.0}^{\,+3.5}\pm 1.6$ ($<16$) \cite{Jessop:2000bv}
 & $13.0\pm 3.0$ \\[-0.2cm]
 & & & & {\footnotesize ($S=1.4$)} \\[-0.1cm]
$B^-\to\pi^0 K^{*-}$
 & ---
 & ---
 & $<31$ \cite{Jessop:2000bv}
 & $<31$ \\
$\bar B^0\to\pi^+ K^{*-}$
 & ---
 & $14.8_{\,-4.4\,-1.0\,-0.9}^{\,+4.6\,+1.5\,+2.4}$ \cite{Belle_EPS2}
 & $16_{\,-5}^{\,+6}\pm 2$ \cite{Eckhart:2002qr}
 & $15.3\pm 3.8$ \\
$\bar B^0\to\pi^0\bar K^{*0}$
 & ---
 & ---
 & $<3.6$ \cite{Jessop:2000bv}
 & $<3.6$ \\
\hline
$B^-\to\bar K^0\rho^-$
 & ---
 & ---
 & $<48$ \cite{Bona}
 & $<48$ \\
$B^-\to K^-\rho^0$
 & $<6.2$ \cite{Aubert:2003dn}
 & $<12$ \cite{Abe:2002av}
 & $<17$ \cite{Jessop:2000bv}
 & $<6.2$ \\
$\bar B^0\to K^-\rho^+$
 & $7.3_{\,-1.2}^{\,+1.3}\pm 1.3$ \cite{Aubert:2003wr}
 & $15.1_{\,-3.3\,-1.5\,-2.1}^{\,+3.4\,+1.4\,+2.0}$ \cite{Belle_EPS2}
 & $16.0_{\,-6.4}^{\,+7.6}\pm 2.8$ ($<32$) \cite{Jessop:2000bv}
 & $8.9\pm 2.2$ \\[-0.2cm]
 & & & & {\footnotesize ($S=1.4$)} \\[-0.1cm]
$\bar B^0\to\bar K^0\rho^0$
 & ---
 & $<12$ \cite{Huang:2002ev}
 & $<39$ \cite{Bona}
 & $<12$ \\
\hline
$B^-\to\eta K^{*-}$
 & $22.1_{\,-\phantom{1}9.2}^{\,+11.1}\pm 3.3$ \cite{Aihara}
 & $26.5_{\,-7.0}^{\,+7.8}\pm 3.0$ \cite{Aihara}
 & $26.4_{\,-8.2}^{\,+9.6}\pm 3.3$ \cite{Richichi:1999kj}
 & $25.4\pm 5.3$ \\
$\bar B^0\to\eta\bar K^{*0}$
 & $19.8_{\,-5.6}^{\,+6.5}\pm 1.7$ \cite{Aihara}
 & $16.5_{\,-4.2}^{\,+4.6}\pm 1.2$ \cite{Aihara}
 & $13.8_{\,-4.6}^{\,+5.5}\pm 1.6$ \cite{Richichi:1999kj}
 & $16.4\pm 3.0$ \\
$B^-\to\eta' K^{*-}$
 & ---
 & $<90$ \cite{Aihara}
 & $<35$ \cite{Richichi:1999kj}
 & $<35$ \\
$\bar B^0\to\eta'\bar K^{*0}$
 & $<13$ \cite{Aihara}
 & $<20$ \cite{Aihara}
 & $<24$ \cite{Richichi:1999kj} 
 & $<13$ \\
\hline
$B^-\to K^-\omega$
 & $5.0\pm 1.0\pm 0.4$ \cite{Aubert:2003fa}
 & $6.7_{\,-1.2}^{\,+1.3}\pm 0.6$ \cite{Belle_EPS1}
 & $3.2_{\,-1.9}^{\,+2.4}\pm 0.8$ ($<7.9$) \cite{Jessop:2000bv}
 & $5.3\pm 0.8$ \\
$\bar B^0\to\bar K^0\omega$
 & $5.3_{\,-1.2}^{\,+1.4}\pm 0.5$ \cite{Aubert:2003fa}
 & $4.0_{\,-1.6}^{\,+1.9}\pm0.5$ ($<7.6$) \cite{Belle_EPS1}
 & $10.0_{\,-4.2}^{\,+5.4}\pm 1.4$ ($<21$) \cite{Jessop:2000bv}
 & $5.1\pm 1.1$ \\
$B^-\to K^-\phi$
 & $10.0_{\,-0.8}^{\,+0.9}\pm 0.5$ \cite{Aubert:2003tk}
 & $9.4\pm 1.1\pm 0.7$ \cite{Chen:2003jf}
 & $5.5_{\,-1.8}^{\,+2.1}\pm 0.6$ \cite{Briere:2001ue}
 & $9.2\pm 1.0$ \\[-0.2cm]
 & & & & {\footnotesize ($S=1.4$)} \\[-0.1cm]
$\bar B^0\to\bar K^0\phi$
 & $7.6_{\,-1.2}^{\,+1.3}\pm 0.5$ \cite{Aubert:2003tk}
 & $9.0_{\,-1.8}^{\,+2.2}\pm 0.7$ \cite{Chen:2003jf}
 & $5.4_{\,-2.7}^{\,+3.7}\pm 0.7$ ($<12.3$) \cite{Briere:2001ue}
 & $7.7\pm 1.1$ \\
\hline\hline
\end{tabular}}
\end{center}
\begin{center}
\begin{tabular}{|l|ccc|c|}
\hline\hline
\multicolumn{1}{|c|}{Mode} & BaBar & Belle & CLEO & Average \\
\hline\hline
\multicolumn{5}{|c|}{Penguin-dominated decays} \\
\hline
$\bar B^0\to\pi^+ K^{*-}$
 & ---
 & ---
 & $26_{\,-34\,-\phantom{1}8}^{\,+33\,+10}$ \cite{Eisenstein:2003yy}
 & $26\pm 35$ \\
\hline
$\bar B^0\to K^-\rho^+$
 & $28\pm 17\pm 8$ \cite{Aubert:2003wr}
 & $22_{\,-23\,-2}^{\,+22\,+6}$ \cite{Belle_EPS2}
 & ---
 & $26\pm 15$ \\
\hline
%
$B^-\to K^-\omega$
 & $-5\pm 16\pm 1$ \cite{Aubert:2003fa}
 & $6_{\,-18}^{\,+20}\pm 1$ \cite{Belle_EPS1}
 & ---
 & $0\pm 12$ \\
$B^-\to K^-\phi$
 & $3.9\pm 8.6\pm 1.1$ \cite{Aubert:2003tk}
 & $1\pm 12\pm 5$ \cite{Aihara}
 & ---
 & $3\pm 7$ \\
$\bar B^0\to\bar K^0\phi$
 & $80\pm 38\pm 12$ \cite{HamelDeMonchenault:2003pu}
 & $-56\pm 41\pm 16$ \cite{Abe:2002np}
 & ---
 & $19\pm 68$ \\[-0.2cm]
 & & & & {\footnotesize ($S=2.3$)} \\[-0.1cm]
\qquad $S_{\phi K_S}$:
 & $-18\pm 51\pm 7$ \cite{HamelDeMonchenault:2003pu}
 & $-73\pm 64\pm 22$ \cite{Abe:2002np}
 & ---
 & $-38\pm 41$ \\
\hline\hline
\end{tabular}
\end{center}
\end{table}

\begin{table}
\footnotesize
\centerline{\parbox{14cm}{\caption{\label{tab:data3}
CP-averaged branching ratios (top, in units of $10^{-6}$) and CP 
asymmetries (bottom, in \%) for $\bar B\to PP$ decays with $\Delta D=1$. 
Upper limits are at 90\% confidence level.}}}
\vspace{0.1cm}
\begin{center}
\begin{tabular}{|l|ccc|c|}
\hline\hline
\multicolumn{1}{|c|}{Mode} & BaBar & Belle & CLEO & Average \\
\hline\hline
\multicolumn{5}{|c|}{Tree-dominated decays} \\
\hline
$B^-\to\pi^-\pi^0$
 & $5.5_{\,-0.9}^{\,+1.0}\pm 0.6$ \cite{Aubert:2003qj}
 & $5.3\pm 1.3\pm 0.5$ \cite{Tomura:2003cb}
 & $4.6_{\,-1.6\,-0.7}^{\,+1.8\,+0.6}$ \cite{Bornheim:2003bv}
 & $5.3\pm 0.8$ \\
$\bar B^0\to\pi^+\pi^-$
 & $4.7\pm 0.6\pm 0.2$ \cite{Aubert:2002jb} 
 & $4.4\pm 0.6\pm 0.3$ \cite{Tomura:2003cb}
 & $4.5_{\,-1.2\,-0.4}^{\,+1.4\,+0.5}$ \cite{Bornheim:2003bv}
 & $4.6\pm 0.4$ \\
$\bar B^0\to\pi^0\pi^0$
 & $1.6_{\,-0.6\,-0.3}^{\,+0.7\,+0.6}$ \cite{Aubert:2003qj}
 & $1.8_{\,-1.3\,-0.7}^{\,+1.4\,+0.5}$ \cite{Tomura:2003cb}
 & ---
 & $1.6\pm 0.7$ \\[-0.1cm]
 & $<3.6$ \cite{Aubert:2003qj}
 & $<4.4$ \cite{Tomura:2003cb}
 & $<4.4$ \cite{Bornheim:2003bv}
 & $<3.6$ \\
\hline
$B^-\to\pi^-\eta$
 & $4.2_{\,-0.9}^{\,+1.0}\pm 0.3$ \cite{Aubert:2003ez}
 & $5.2_{\,-1.7}^{\,+ 2.0}\pm 0.6$ \cite{Tomura:2003cb}
 & $1.2_{\,-1.2}^{\,+ 2.8}$ ($<5.7$) \cite{Richichi:1999kj}
 & $3.9\pm 0.9$ \\[-0.2cm]
 & & & & {\footnotesize ($S=1.1$)} \\[-0.1cm]
$\bar B^0\to\pi^0\eta$
 & ---
 & ---
 & $<2.9$ \cite{Richichi:1999kj}
 & $<2.9$ \\
$B^-\to\pi^-\eta'$
 & $<12$ \cite{Aubert:2001zf}
 & $<7$ \cite{Aihara}
 & $<12$ \cite{Richichi:1999kj}
 & $<7$ \\
$\bar B^0\to\pi^0\eta'$
 & ---
 & ---
 & $<5.7$ \cite{Richichi:1999kj}
 & $<5.7$ \\
\hline\hline
\multicolumn{5}{|c|}{Penguin-dominated decays} \\
\hline
$B^-\to K^- K^0$
 & $<2.2$ \cite{Bona}
 & $<3.4$ \cite{Tomura:2003cb}
 & $<3.3$ \cite{Bornheim:2003bv}
 & $<2.2$ \\
$\bar B^0\to\bar K^0 K^0$
 & $<1.6$ \cite{Bona}
 & $<3.2$ \cite{Tomura:2003cb}
 & $<3.3$ \cite{Bornheim:2003bv}
 & $<1.6$ \\
\hline\hline
\multicolumn{5}{|c|}{Pure annihilation decays} \\
\hline
$\bar B^0\to K^- K^+$
 & $<0.6$ \cite{Aubert:2002jb}
 & $<0.7$ \cite{Tomura:2003cb}
 & $<0.8$ \cite{Bornheim:2003bv}
 & $<0.6$ \\
\hline\hline
\end{tabular}
\end{center}
\begin{center}
\begin{tabular}{|l|cc|c|}
\hline\hline
\multicolumn{1}{|c|}{Mode} & BaBar & Belle & Average \\
\hline\hline
\multicolumn{4}{|c|}{Tree-dominated decays} \\
\hline
$B^-\to\pi^-\pi^0$
 & $-3_{\,-17}^{\,+18}\pm 2$ \cite{Aubert:2003qj}
 & $-14\pm 24_{\,-4}^{\,+5}$ \cite{Tomura:2003cb}
 & $-7\pm 14$ \\
$\bar B^0\to\pi^+\pi^-$
 & $30\pm 25\pm 4$ \cite{Aubert:2002jb} 
 & $77\pm 27\pm 8$ \cite{Abe:2003ja}
 & $51\pm 23$ \\[-0.2cm]
 & & & {\footnotesize ($S=1.2$)} \\[-0.1cm]
\qquad $S_{\pi^+\pi^-}$: 
 & $2\pm 34\pm 5$ \cite{Aubert:2002jb} 
 & $-123\pm 41_{\,-7}^{\,+8}$ \cite{Abe:2003ja}
 & $-49\pm 61$ \\[-0.2cm]
 & & & {\footnotesize ($S=2.3$)} \\
\hline
$B^-\to\pi^-\eta$
 & $-51_{\,-18}^{\,+20}\pm 1$ \cite{Aubert:2003ez}
 & ---
 & $-51\pm 19$ \\
\hline\hline
\end{tabular}
\end{center}
\end{table}

\begin{table}
\footnotesize
\centerline{\parbox{14cm}{\caption{\label{tab:data4}
CP-averaged branching ratios (top, in units of $10^{-6}$) and CP 
asymmetries (bottom, in \%) for $\bar B\to PV$ decays with $\Delta D=1$. 
Upper limits are at 90\% confidence level.}}}
\vspace{-0.1cm}
\begin{center}
\begin{tabular}{|l|ccc|c|}
\hline\hline
\multicolumn{1}{|c|}{Mode} & BaBar & Belle & CLEO & Average \\
\hline\hline
\multicolumn{5}{|c|}{Tree-dominated decays} \\
\hline
$B^-\to\pi^-\rho^0$
 & $9.3\pm 1.0\pm 0.8$ \cite{Yeche}
 & $8.0_{\,-2.0}^{\,+2.3}\pm 0.7$ \cite{Gordon:2002yt}
 & $10.4_{\,-3.4}^{\,+3.3}\pm 2.1$ \cite{Jessop:2000bv}
 & $9.1\pm 1.1$ \\
$B^-\to\pi^0\rho^-$
 & $11.0\pm 1.9\pm 1.9$ \cite{Yeche}
 & ---
 & $<43$ \cite{Jessop:2000bv}
 & $11.0\pm 2.7$ \\
$\bar B^0\to\pi^+\rho^-$
 & $13.9 \pm 2.7$ \cite{hoecker}
 & ---
 & ---
 & $13.9\pm 2.7$ \\
$\bar B^0\to\pi^-\rho^+$
 & $8.9 \pm 2.5$ \cite{hoecker}
 & ---
 & ---
 & $8.9\pm 2.5$ \\
$\bar B^0\to\pi^\pm\rho^\mp$
 & $22.6\pm 1.8\pm 2.2$ \cite{Aubert:2003wr}
 & $29.1_{\,-4.9}^{\,+5.0}\pm 4.0$ \cite{Abe:2003rj}
 & $27.6_{\,-7.4}^{\,+8.4}\pm 4.2$ \cite{Jessop:2000bv}
 & $24.0\pm 2.5$ \\
$\bar B^0\to\pi^0\rho^0$
 & $<2.5$ \cite{Yeche}
 & $6.0_{\,-2.3}^{\,+2.9}\pm 1.2$ \cite{Abe:2003rj}
 & $<5.5$ \cite{Jessop:2000bv}
 & $<2.5$ \\
\hline
$B^-\to\pi^-\omega$
 & $5.4\pm 1.0\pm 0.5$ \cite{Aubert:2003fa}
 & $5.7_{\,-1.3}^{\,+1.4}\pm 0.6$ \cite{Belle_EPS1}
 & $11.3_{\,-2.9}^{\,+3.3}\pm 1.4$ \cite{Jessop:2000bv}
 & $5.9\pm 1.0$ \\[-0.2cm]
 & & & & {\footnotesize ($S=1.2$)} \\[-0.1cm]
$\bar B^0\to\pi^0\omega$
 & $<3$ \cite{Aubert:2001zf}
 & $<1.9$ \cite{Belle_EPS1}
 & $<5.5$ \cite{Jessop:2000bv}
 & $<1.9$ \\
\hline
$B^-\to\eta\rho^-$
 & $<6.8$ \cite{Aihara}
 & $<6.2$ \cite{Aihara}
 & $<15$ \cite{Richichi:1999kj}
 & $<6.2$ \\
$\bar B^0\to\eta\rho^0$
 & ---
 & $<5.5$ \cite{Aihara}
 & $<10$ \cite{Richichi:1999kj}
 & $<5.5$ \\
$B^-\to\eta'\rho^-$
 & ---
 & ---
 & $<33$ \cite{Richichi:1999kj}
 & $<33$ \\
$\bar B^0\to\eta'\rho^0$
 & ---
 & $<14$ \cite{Aihara}
 & $<12$ \cite{Richichi:1999kj}
 & $<12$ \\
\hline\hline
\multicolumn{5}{|c|}{Penguin-dominated decays} \\
\hline
$B^-\to\pi^-\phi$
 & $<0.41$ \cite{Aubert:2003tk}
 & ---
 & $<5$ \cite{Bergfeld:1998ik}
 & $<0.4$ \\
$\bar B^0\to\pi^0\phi$
 & ---
 & ---
 & $<5$ \cite{Bergfeld:1998ik}
 & $<5$ \\
\hline
$B^-\to K^- K^{*0}$
 & ---
 & ---
 & $<5.3$ \cite{Jessop:2000bv}
 & $<5.3$ \\
\hline\hline
\end{tabular}
\end{center}
\vspace{-0.6cm}
\begin{center}
\begin{tabular}{|l|ccc|c|}
\hline\hline
\multicolumn{1}{|c|}{Mode} & BaBar & Belle & CLEO & Average \\
\hline\hline
\multicolumn{5}{|c|}{Tree-dominated decays} \\
\hline
$B^-\to\pi^-\rho^0$
 & $-17\pm 11\pm 2$ \cite{Yeche}
 & ---
 & ---
 & $-17\pm 11$ \\
$B^-\to\pi^0\rho^-$
 & $23\pm 16\pm 6$ \cite{Yeche}
 & ---
 & --- 
 & $23\pm 17$ \\
\hline
$\bar B^0\to\pi^+\rho^-$
 & $-11_{\,-17}^{\,+16}\pm 4$ \cite{Hamel}
 & ---
 & ---
 & $-11\pm 17$ \\
$\bar B^0\to\pi^-\rho^+$
 & $-62_{\,-28}^{\,+24}\pm 6$ \cite{Hamel}
 & ---
 & ---
 & $-62\pm 27$ \\
$\bar B^0\to\pi^\pm\rho^\mp$: & & & & \\
\qquad $A_{\rm CP}$
 & $-18\pm 8\pm 3$ \cite{Aubert:2003wr}
 & $-38_{\,-21\,-5}^{\,+19\,+4}$ \cite{Abe:2003rj}
 & ---
 & $-21\pm 8$ \\
\qquad $C_{\pi^\pm\rho^\mp}$
 & $36\pm 18\pm 4$ \cite{Aubert:2003wr}
 & ---
 & ---
 & $36\pm 18$ \\
\qquad $S_{\pi^\pm\rho^\mp}$
 & $19\pm 24\pm 3$ \cite{Aubert:2003wr}
 & ---
 & ---
 & $19\pm 24$ \\
\qquad $\Delta C_{\pi^\pm\rho^\mp}$
 & $28_{\,-19}^{\,+18}\pm 4$ \cite{Aubert:2003wr}
 & ---
 & ---
 & $28\pm 19$ \\
\qquad $\Delta S_{\pi^\pm\rho^\mp}$
 & $15\pm 25\pm 3$ \cite{Aubert:2003wr}
 & ---
 & ---
 & $15\pm 25$ \\
\hline
$B^-\to\pi^-\omega$
 & $4\pm 17\pm 1$ \cite{Aubert:2003fa}
 & $48_{\,-20}^{\,+23}\pm 2$ \cite{Belle_EPS1}
 & $-34\pm 25\pm 2$ \cite{Chen:2000hv}
 & $9\pm 21$ \\[-0.2cm]
 & & & & {\footnotesize ($S=1.8$)} \\
\hline\hline
\end{tabular}
\end{center}
\end{table}


\newpage
\begin{thebibliography}{999}

\bibitem{Aubert:2001tu}
B.~Aubert {\it et al.}  [BaBar Collaboration],
Nucl.\ Instrum.\ Meth.\ A {\bf 479} (2002) 1
[hep-ex/0105044].

\bibitem{Mori:2000cg}
S.~Mori {\it et al.}  [Belle Collaboration],
Nucl.\ Instrum.\ Meth.\ A {\bf 479} (2002) 117.

\bibitem{Bauer:1986bm}
M.~Bauer, B.~Stech and M.~Wirbel,
Z.\ Phys.\ C {\bf 34} (1987) 103.

\bibitem{Neubert:1997uc}
M.~Neubert and B.~Stech,
Adv.\ Ser.\ Direct.\ High Energy Phys.\  {\bf 15} (1998) 294
[hep-ph/9705292].

\bibitem{Ali:1997nh}
A.~Ali and C.~Greub,
Phys.\ Rev.\ D {\bf 57} (1998) 2996
[hep-ph/9707251].

\bibitem{Ali:1998eb}
A.~Ali, G.~Kramer and C.~D.~Lu,
Phys.\ Rev.\ D {\bf 58} (1998) 094009
[hep-ph/9804363].

\bibitem{Chen:1999nx}
Y.~H.~Chen, H.~Y.~Cheng, B.~Tseng and K.~C.~Yang,
Phys.\ Rev.\ D {\bf 60} (1999) 094014
[hep-ph/9903453].

\bibitem{BBNS1}
M.~Beneke, G.~Buchalla, M.~Neubert and C.~T.~Sachrajda,
Phys.\ Rev.\ Lett.\ {\bf 83} (1999) 1914
[hep-ph/9905312].

\bibitem{BBNS2}
M.~Beneke, G.~Buchalla, M.~Neubert and C.~T.~Sachrajda,
Nucl.\ Phys.\ B {\bf 591} (2000) 313
[hep-ph/0006124].

\bibitem{BBNS3}
M.~Beneke, G.~Buchalla, M.~Neubert and C.~T.~Sachrajda,
Nucl.\ Phys.\ B {\bf 606} (2001) 245
[hep-ph/0104110].

\bibitem{Muta:2000ti}
T.~Muta, A.~Sugamoto, M.~Z.~Yang and Y.~D.~Yang,
Phys.\ Rev.\ D {\bf 62} (2000) 094020 
[hep-ph/0006022].

\bibitem{Du:2000ff}
D.-s.~Du, D.-s.~Yang and G.-h.~Zhu,
Phys.\ Lett.\ B {\bf 488} (2000) 46
[hep-ph/0005006].

\bibitem{Du:2001hr}
D.-s.~Du, H.-u.~Gong, J.-f.~Sun, D.-s.~Yang and G.-h.~Zhu,
Phys.\ Rev.\ D {\bf 65} (2002) 074001
[hep-ph/0108141].

\bibitem{Yang:2000xn}
M.~Z.~Yang and Y.~D.~Yang,
Phys.\ Rev.\ D {\bf 62} (2000) 114019 
[hep-ph/0007038].

\bibitem{Cheng:2000hv}
H.~Y.~Cheng and K.~C.~Yang,
Phys.\ Rev.\ D {\bf 64} (2001) 074004 
[hep-ph/0012152].

\bibitem{Cheng:2001aa}
H.~Y.~Cheng and K.~C.~Yang,
Phys.\ Lett.\ B {\bf 511} (2001) 40 
[hep-ph/0104090].

\bibitem{diehl}
M.~Diehl and G.~Hiller,
JHEP {\bf 0106} (2001) 067 
[hep-ph/0105194].

\bibitem{Beneke:2002jn}
M.~Beneke and M.~Neubert,
Nucl.\ Phys.\ B {\bf 651} (2003) 225
[hep-ph/0210085].

\bibitem{Du:2002up}
D.-s.~Du, H.-j.~Gong, J.-f.~Sun, D.-s.~Yang and G.-h.~Zhu,
Phys.\ Rev.\ D {\bf 65} (2002) 094025
[Erratum: {\em ibid.\/} D {\bf 66} (2002) 079904]
[hep-ph/0201253].

\bibitem{Du:2002cf}
D.-s.~Du, J.-f.~Sun, D.-s.~Yang and G.-h.~Zhu,
Phys.\ Rev.\ D {\bf 67} (2003) 014023
[hep-ph/0209233].

\bibitem{Sun:2002rn}
J.-f.~Sun, G.-h.~Zhu and D.-s.~Du,
Phys.\ Rev.\ D {\bf 68} (2003) 054003
[hep-ph/0211154].

\bibitem{Aleksan:2003qi}
R.~Aleksan, P.~F.~Giraud, V.~Morenas, O.~Pene and A.~S.~Safir,
Phys.\ Rev.\ D {\bf 67} (2003) 094019
[hep-ph/0301165].

\bibitem{deGroot:2003hp}
N.~de Groot, W.~N.~Cottingham and I.~B.~Whittingham,
preprint hep-ph/0305263.

\bibitem{Chen:2001pr}
C.~H.~Chen, Y.~Y.~Keum and H.-n.~Li,
Phys.\ Rev.\ D {\bf 64} (2001) 112002
[hep-ph/0107165].

\bibitem{Bauer:2001cu}
C.~W.~Bauer, D.~Pirjol and I.~W.~Stewart,
Phys.\ Rev.\ Lett.\  {\bf 87} (2001) 201806
[hep-ph/0107002].

\bibitem{Chay:2003ju}
J.~Chay and C.~Kim,
preprint hep-ph/0301262.

\bibitem{Khodjamirian:2000mi}
A.~Khodjamirian,
Nucl.\ Phys.\ B {\bf 605} (2001) 558
[hep-ph/0012271].

\bibitem{Khodjamirian:2003eq}
A.~Khodjamirian, T.~Mannel and B.~Melic,
Phys.\ Lett.\ B {\bf 571} (2003) 75
[hep-ph/0304179].

\bibitem{Feldmann:1998vh}
T.~Feldmann, P.~Kroll and B.~Stech,
Phys.\ Rev.\ D {\bf 58} (1998) 114006
[hep-ph/9802409].

\bibitem{Grozin:1996pq}
A.~G.~Grozin and M.~Neubert,
Phys.\ Rev.\ D {\bf 55} (1997) 272
[hep-ph/9607366].

\bibitem{Beneke:2000wa}
M.~Beneke and T.~Feldmann,
Nucl.\ Phys.\ B {\bf 592} (2001) 3
[hep-ph/0008255].

\bibitem{Braun:1989iv}
V.~M.~Braun and I.~E.~Filyanov,
Z.\ Phys.\ C {\bf 48} (1990) 239
[Sov.\ J.\ Nucl.\ Phys.\  {\bf 52} (1990) 126].

\bibitem{Ball:1998sk}
P.~Ball, V.~M.~Braun, Y.~Koike and K.~Tanaka,
Nucl.\ Phys.\ B {\bf 529} (1998) 323
[hep-ph/9802299].

\bibitem{Beneke:2002bs}
M.~Beneke,
Nucl.\ Phys.\ Proc.\ Suppl.\  {\bf 111} (2002) 62
[hep-ph/0202056].

\bibitem{Geshkenbein:qn}
B.~V.~Geshkenbein and M.~V.~Terentev,
Yad.\ Fiz.\  {\bf 40} (1984) 758
[Sov.\ J.\ Nucl.\ Phys.\  {\bf 40} (1984) 487].

\bibitem{KP02}
P.~Kroll and K.~Passek-Kumericki,
Phys.\ Rev.\ D {\bf 67} (2003) 054017
[hep-ph/0210045].

\bibitem{Khodjamirian:1997ub}
A.~Khodjamirian, R.~R\"uckl, S.~Weinzierl and O.~I.~Yakovlev,
Phys.\ Lett.\ B {\bf 410} (1997) 275
[hep-ph/9706303].

\bibitem{Ball:1998kk}
P.~Ball and V.~M.~Braun,
Phys.\ Rev.\ D {\bf 58} (1998) 094016
[hep-ph/9805422].

\bibitem{Ball:1998tj}
P.~Ball,
JHEP {\bf 9809} (1998) 005
[hep-ph/9802394].

\bibitem{Bjorken:kk}
J.~D.~Bjorken,
Nucl.\ Phys.\ Proc.\ Suppl.\  {\bf 11} (1989) 325.

\bibitem{Neubert:2001ev}
M.~Neubert,
Int.\ J.\ Mod.\ Phys.\ A {\bf 17} (2002) 2936
[hep-ph/0110301].

\bibitem{Athar:2003yg}
S.~B.~Athar {\it et al.}  [CLEO Collaboration],
preprint hep-ex/0304019.

\bibitem{Luo:2003hn}
Z.-m.~Luo and J.~L.~Rosner,
preprint hep-ph/0305262.

\bibitem{Becher:2001hu}
T.~Becher, M.~Neubert and B.~D.~Pecjak,
Nucl.\ Phys.\ B {\bf 619} (2001) 538
[hep-ph/0102219];\\
%
M.~Neubert and B.~D.~Pecjak,
JHEP {\bf 0202} (2002) 028
[hep-ph/0202128].

\bibitem{Gronau_Rosner}
A.~S.~Dighe, M.~Gronau and J.~L.~Rosner,
Phys.\ Rev.\ D {\bf 54} (1996) 3309
[hep-ph/9604233];\\
%
M.~Gronau and J.~L.~Rosner,
Phys.\ Rev.\ D {\bf 57} (1998) 6843
[hep-ph/9711246].

\bibitem{Fleischer_Mannel}
R.~Fleischer and T.~Mannel,
Phys.\ Rev.\ D {\bf 57} (1998) 2752
[hep-ph/9704423].

\bibitem{Neubert_Rosner}
M.~Neubert and J.~L.~Rosner,
Phys.\ Lett.\ B {\bf 441} (1998) 403
[hep-ph/9808493];
%
Phys.\ Rev.\ Lett.\  {\bf 81} (1998) 5076
[hep-ph/9809311].

\bibitem{Buras_Fleischer}
A.~J.~Buras and R.~Fleischer,
Eur.\ Phys.\ J.\ C {\bf 11} (1999) 93
[hep-ph/9810260].

\bibitem{Neubert}
M.~Neubert,
JHEP {\bf 9902} (1999) 014
[hep-ph/9812396].

\bibitem{Ciuchini:1997hb}
M.~Ciuchini, E.~Franco, G.~Martinelli and L.~Silvestrini,
Nucl.\ Phys.\ B {\bf 501} (1997) 271
[hep-ph/9703353].

\bibitem{Grossman:1996ke}
Y.~Grossman and M.~P.~Worah,
Phys.\ Lett.\ B {\bf 395} (1997) 241
[hep-ph/9612269];\\
%
Y.~Grossman, G.~Isidori and M.~P.~Worah,
Phys.\ Rev.\ D {\bf 58} (1998) 057504
[hep-ph/9708305].

\bibitem{Grossman:2003qp}
Y.~Grossman, Z.~Ligeti, Y.~Nir and H.~Quinn,
Phys.\ Rev.\ D {\bf 68} (2003) 015004
[hep-ph/0303171].

\bibitem{Yoshikawa:2003hb}
T.~Yoshikawa,
Phys.\ Rev.\ D {\bf 68} (2003) 054023
[hep-ph/0306147].

\bibitem{Grossman:1999av}
Y.~Grossman, M.~Neubert and A.~L.~Kagan,
JHEP {\bf 9910} (1999) 029
[hep-ph/9909297].

\bibitem{Lipkin:1998ab}
H.~J.~Lipkin,
Phys.\ Lett.\ B {\bf 445} (1999) 403 
[hep-ph/9810351].

\bibitem{Gronau:1998ep}
M.~Gronau and J.~L.~Rosner,
Phys.\ Rev.\ D {\bf 59} (1999) 113002
[hep-ph/9809384].

\bibitem{Gronau:2003kj}
M.~Gronau and J.~L.~Rosner,
Phys.\ Lett.\ B {\bf 572} (2003) 43
[hep-ph/0307095].

\bibitem{Ciuchini:2002uv}
M.~Ciuchini, E.~Franco, A.~Masiero and L.~Silvestrini,
Phys.\ Rev.\ D {\bf 67} (2003) 075016
[hep-ph/0212397].

\bibitem{Aleksan:1990ts}
R.~Aleksan, I.~Dunietz, B.~Kayser and F.~Le Diberder,
Nucl.\ Phys.\ B {\bf 361} (1991) 141.

\bibitem{Khodjamirian:1998ji}
A.~Khodjamirian and R.~R\"uckl,
Adv.\ Ser.\ Direct.\ High Energy Phys.\  {\bf 15} (1998) 345
[hep-ph/9801443].

\bibitem{hoecker}
A.~H\"ocker, M.~Laget, S.~Laplace and J. v. Wimmersperg-Toeller, 
{\it Using Flavor Symmetry to Constrain $\alpha$ from $B\to\rho 
\pi$}, preprint LAL~03-17.

\bibitem{Chiang:2003rb}
C.~W.~Chiang, M.~Gronau and J.~L.~Rosner,
preprint hep-ph/0306021.

\bibitem{Battaglia:2003in}
M.~Beneke, in: {\em  The CKM Matrix and the Unitarity Triangle\/} 
(edited by M.~Battaglia {\it et al.}), 
preprint hep-ph/0304132, pp.~299.

%
\bibitem{Yeche}
B.~Aubert {\it et al.}  [BaBar Collaboration],
preprint hep-ex/0307087.

\bibitem{Bona}
M. Bona, talk at the Conference on Flavor Physics and CP Violation
(FPCP 2003), Paris, France, June 3--6, 2003, to appear in the 
Proceedings.

\bibitem{Aubert:2003wr}
B.~Aubert  {\it et al.}  [BaBar Collaboration],
preprint hep-ex/0306030.

\bibitem{HamelDeMonchenault:2003pu}
G.~Hamel De Monchenault  [BaBar Collaboration],
preprint hep-ex/0305055.

\bibitem{Hamel}
G.~Hamel de Monchenault, talk at the 38th Rencontres de Moriond: 
Electroweak Interactions and Unified Theories, Les Arcs, France, 
March 15--22, 2003, to appear in the Proceedings.

\bibitem{Aubert:2003bq}
B.~Aubert {\it et al.}  [BaBar Collaboration],
preprint hep-ex/0303046.

\bibitem{Aubert:2003fa}
B.~Aubert {\it et al.}  [BaBar Collaboration],
preprint hep-ex/0303040.

\bibitem{Aubert:2003ez}
B.~Aubert {\it et al.}  [BaBar Collaboration],
preprint hep-ex/0303039.

\bibitem{Aubert:2003tk}
B.~Aubert {\it et al.}  [BaBar Collaboration],
preprint hep-ex/0303029.

\bibitem{Aubert:2003qj}
B.~Aubert {\it et al.}  [BaBar Collaboration],
Phys.\ Rev.\ Lett.\  {\bf 91} (2003) 021801
[hep-ex/0303028].

\bibitem{Aubert:2003dn}
B.~Aubert {\it et al.}  [BaBar Collaboration],
preprint hep-ex/0303022.

\bibitem{Aubert:2002jm}
B.~Aubert {\it et al.}  [BaBar Collaboration],
preprint hep-ex/0207065.

\bibitem{Aubert:2002jb}
B.~Aubert {\it et al.}  [BaBar Collaboration],
Phys.\ Rev.\ Lett.\  {\bf 89} (2002) 281802
[hep-ex/0207055].

\bibitem{Aubert:2002ng}
B.~Aubert {\it et al.}  [BaBar Collaboration],
preprint hep-ex/0206053.


\bibitem{Aubert:2001zf}
B.~Aubert {\it et al.}  [BaBar Collaboration],
Phys.\ Rev.\ Lett.\  {\bf 87} (2001) 221802
[hep-ex/0108017].

%
\bibitem{Belle_EPS2}
K. Abe {\it et al.}  [Belle Collaboration],
conference paper BELLE-CONF-0317 (EPS-ID~535) submitted to the 
International Europhysics Conference on High Energy Physics, Aachen, 
Germany, July 17--23, 2003.

\bibitem{Belle_EPS1}
K. Abe {\it et al.}  [Belle Collaboration],
conference paper BELLE-CONF-0312 (EPS-ID~527) submitted to the 
International Europhysics Conference on High Energy Physics, Aachen, 
Germany, July 17--23, 2003.

\bibitem{Abe:2003rj}
K.~Abe {\it et al.}  [Belle Collaboration],
preprint hep-ex/0307077.

\bibitem{Chen:2003jf}
K.~F.~Chen {\it et al.}  [Belle Collaboration],
preprint hep-ex/0307014.

\bibitem{Aihara}
H. Aihara, talk at the Conference on Flavor Physics and CP Violation
(FPCP 2003), Paris, France, June 3--6, 2003, to appear in the 
Proceedings.

\bibitem{Tomura:2003cb}
T.~Tomura,
preprint hep-ex/0305036.

\bibitem{Abe:2003ja}
K.~Abe {\it et al.}  [Belle Collaboration],
Phys.\ Rev.\ D {\bf 68} (2003) 012001
[hep-ex/0301032].

\bibitem{Abe:2002np}
K.~Abe {\it et al.}  [Belle Collaboration],
Phys.\ Rev.\ D {\bf 67} (2003) 031102
[hep-ex/0212062].

\bibitem{Gordon:2002yt}
A.~Gordon {\it et al.}  [Belle Collaboration],
Phys.\ Lett.\ B {\bf 542} (2002) 183
[hep-ex/0207007].

\bibitem{Huang:2002ev}
H.~C.~Huang  [Belle Collaboration],
preprint hep-ex/0205062.

\bibitem{Abe:2002av}
K.~Abe {\it et al.}  [Belle Collaboration],
Phys.\ Rev.\ D {\bf 65} (2002) 092005
[hep-ex/0201007].

%
\bibitem{Eisenstein:2003yy}
B.~I.~Eisenstein  [CLEO Collaboration],
preprint hep-ex/0304036.

\bibitem{Bornheim:2003bv}
A.~Bornheim {\it et al.}  [CLEO Collaboration],
Phys.\ Rev.\ D {\bf 68} (2003) 052002
[hep-ex/0302026].

\bibitem{Eckhart:2002qr}
E.~Eckhart {\it et al.}  [CLEO Collaboration],
Phys.\ Rev.\ Lett.\  {\bf 89} (2002) 251801
[hep-ex/0206024].

\bibitem{Briere:2001ue}
R.~A.~Briere {\it et al.}  [CLEO Collaboration],
Phys.\ Rev.\ Lett.\  {\bf 86} (2001) 3718
[hep-ex/0101032].

\bibitem{Jessop:2000bv}
C.~P.~Jessop {\it et al.}  [CLEO Collaboration],
Phys.\ Rev.\ Lett.\  {\bf 85} (2000) 2881
[hep-ex/0006008].

\bibitem{Chen:2000hv}
S.~Chen {\it et al.}  [CLEO Collaboration],
Phys.\ Rev.\ Lett.\  {\bf 85} (2000) 525
[hep-ex/0001009].

\bibitem{Richichi:1999kj}
S.~J.~Richichi {\it et al.}  [CLEO Collaboration],
Phys.\ Rev.\ Lett.\  {\bf 85} (2000) 520
[hep-ex/9912059].

\bibitem{Bergfeld:1998ik}
T.~Bergfeld {\it et al.}  [CLEO Collaboration],
Phys.\ Rev.\ Lett.\  {\bf 81} (1998) 272
[hep-ex/9803018].

%
\bibitem{Aubert:2003hf}
B.~Aubert {\it et al.}  [BaBar Collaboration],
preprint hep-ex/0308012.

\bibitem{Abe:2003yy}
K.~Abe {\it et al.}  [Belle Collaboration],
preprint hep-ex/0308040.

\bibitem{belle338}
K. Abe {\it et al.}  [Belle Collaboration],
conference paper BELLE-CONF-0338 (EPS-ID~577) submitted to the
International Europhysics Conference on High Energy Physics, Aachen,
Germany, July 17--23, 2003.

\end{thebibliography}
\end{document}